\documentclass[twoside, onecolumn,11pt]{article}
\usepackage{extsizes}
\usepackage[super,sort&compress,comma]{natbib} 
\usepackage[version=3]{mhchem}
\usepackage[left=1.5cm, right=1.5cm, top=1.785cm, bottom=2.0cm]{geometry}
\usepackage{balance}
\usepackage{times,mathptmx}
\usepackage{sectsty}
\usepackage{graphicx} 
\usepackage{lastpage}
\usepackage[format=plain,justification=justified,singlelinecheck=true,font={stretch=1.125,small,sf},labelfont=bf,labelsep=space]{caption}
\usepackage{float}
\usepackage{fancyhdr}
\usepackage{fnpos}
\usepackage[english]{babel}
\addto{\captionsenglish}{%
  
}
\usepackage{array}
\usepackage{droidsans}
\usepackage{charter}
\usepackage[T1]{fontenc}
\usepackage[usenames,dvipsnames]{xcolor}
\usepackage{setspace}
\usepackage[compact]{titlesec}
\usepackage{hyperref}

\usepackage{epstopdf}

\definecolor{cream}{RGB}{222,217,201}

\usepackage{titlesec}
\titlelabel{\thetitle.\quad}

\usepackage{etoolbox}

\makeatletter
\let\oldcite\cite
\pretocmd{\listoffigures}{\def\cite{\ignorespaces\@gobble}}{}{}
\apptocmd{\listoffigures}{\let\cite\oldcite}{}{}
\pretocmd{\listoftables}{\def\cite{\ignorespaces\@gobble}}{}{}
\apptocmd{\listoftables}{\let\cite\oldcite}{}{}
\makeatother

\begin{document}


\makeFNbottom
\makeatletter
\renewcommand\LARGE{\@setfontsize\LARGE{15pt}{17}}
\renewcommand\Large{\@setfontsize\Large{12pt}{14}}
\renewcommand\large{\@setfontsize\large{10pt}{12}}
\renewcommand\footnotesize{\@setfontsize\footnotesize{7pt}{10}}
\makeatother

\renewcommand{\thefootnote}{\fnsymbol{footnote}}
\renewcommand\footnoterule{\vspace*{1pt}%
\color{cream}\hrule width 3.5in height 0.4pt \color{black}\vspace*{5pt}} 
\setcounter{secnumdepth}{5}

\makeatletter 
\renewcommand\@biblabel[1]{#1}            
\renewcommand\@makefntext[1]%
{\noindent\makebox[0pt][r]{\@thefnmark\,}#1}
\makeatother 
\renewcommand{\figurename}{\small{Figure}}
\sectionfont{\sffamily\Large}
\subsectionfont{\normalsize}
\subsubsectionfont{\bf}
\setstretch{1.125} 
\setlength{\skip\footins}{0.8cm}
\setlength{\footnotesep}{0.25cm}
\setlength{\jot}{10pt}
\titlespacing*{\section}{0pt}{4pt}{4pt}
\titlespacing*{\subsection}{0pt}{15pt}{1pt}

\makeatletter 
\newlength{\figrulesep} 
\setlength{\figrulesep}{0.5\textfloatsep} 

\newcommand{\topfigrule}{\vspace*{-1pt}%
\noindent{\color{cream}\rule[-\figrulesep]{\columnwidth}{1.5pt}} }

\newcommand{\botfigrule}{\vspace*{-2pt}%
\noindent{\color{cream}\rule[\figrulesep]{\columnwidth}{1.5pt}} }

\newcommand{\dblfigrule}{\vspace*{-1pt}%
\noindent{\color{cream}\rule[-\figrulesep]{\columnwidth}{1.5pt}} }

\makeatother

\begin{center}
\noindent\huge\textbf{Advances in 
Honeycomb Layered Oxides:}\\
\noindent\LARGE\centering{Syntheses, Layer Stackings and Material Characterisation Techniques of Pnictogen- and Chalcogen-Based Honeycomb Layered Oxides}

\end{center}

\noindent\large{Godwill Mbiti Kanyolo,\textit{$^{a, c *}$} Titus Masese,\textit{$^{b, c *}$} Abbas Alshehabi,\textit{$^{d}$} and Zhen-Dong Huang \textit{$^{e}$}}\\

\noindent{\textit{$^{a}$Department of Engineering Science, The University of Electro-Communications, 1-5-1 Chofugaoka, Chofu, Tokyo 182-8585, Japan. } Email: gmkanyolo@mail.uec.jp\\
\textit{$^{b}$AIST-Kyoto University Chemical Energy Materials Open Innovation Laboratory (ChEM-OIL), Sakyo-ku, Kyoto 606-8501, Japan.}\\
\textit{$^{c}$Research Institute of Electrochemical Energy, National Institute of Advanced Industrial Science and Technology (AIST), 1-8-31 Midorigaoka, Ikeda, Osaka 563-8577, Japan.} Email: titus.masese@aist.go.jp; gm.kanyolo@aist.go.jp\\
\textit{$^{d}$Department of Industrial Engineering, National Institute of Technology (KOSEN), Ibaraki College, 866 Nakane, Hitachinaka, Ibaraki 312–8508, Japan. }\\
\textit{$^{e}$Key Laboratory for Organic Electronics and Information Displays and Institute of Advanced Materials (IAM), Nanjing University of Posts and Telecommunications (NUPT), Nanjing, 210023, China.}} Email: iamzdhuang@njupt.edu.cn\\

\noindent\normalsize{
\textbf{Advancements in nanotechnology continue to unearth material vistas that presage a new age of revolutionary functionalities replete with unparalleled physical properties and avant-garde chemical capabilities that promise sweeping paradigm shifts in energy, environment, telecommunications and potentially healthcare. At the upper echelons of this realm, the pnictogen and chalcogen class of honeycomb layered oxides have emerged with fascinating crystal chemistry and exotic electromagnetic and topological phenomena that muster multifaceted concepts spanning from materials science to condensed matter physics and potential applications in electrochemistry, quantum mechanics and electronics. In a bid to shed light on the mechanisms governing these biomimetic nanostructures, this review highlights the significant milestones and breakthroughs that have augmented their current fundamental theory, properties, and utilities. Herein, we elucidate the vast promising crystal chemistry space against the backdrop of known synthesis and characterisation techniques employed in the development and optimisation of this class of honeycomb layered oxides.  
Further, we highlight key theoretical models that have reinvigorated the exploration and characterisation of within this class of materials and are 
poised to redefine the frontiers of material research and their applications. We conclude by envisaging future research directions where fascinating physicochemical, topological and electromagnetic properties could be lurking and where valiant efforts ought to be inclined, particularly in the prospective realisation of exotic material compositional space as well as their utility as testing grounds for emergent two-dimensional (2D) topological quantum gravity and conformal field theories.
} \\

\renewcommand*\rmdefault{bch}\normalfont\upshape
\rmfamily
\section*{}
\vspace{-1cm}

\newpage

\tableofcontents

\newpage

\newpage

\newpage

\section{\label{Section: Introduction} Introduction}

Honeybees are lauded as some of the most essential organisms in nature not only for the viscous sugary sap they concoct but also for their role in sustaining ecological balance and biodiversity in nature. However, it is the honeycomb pattern conjured by these miniature societies to store larvae and nectarine exudations that has earned them stature also amongst mathematicians, physicists and philosophers alike on merits of its architectural aesthetics, intricate geometry, structural distinction, and resource optimisation.
\cite{Darwin1859, Zhang2015, Karihaloo2013, raz2013application, lyon2012mathematical}

`{\it He must be a dull man who can examine the exquisite structure of a comb, so beautifully adapted to its end, without enthusiastic admiration}'-- Charles Darwin

\begin{figure*}[!b]
 \centering
 \includegraphics[width=0.95\columnwidth]{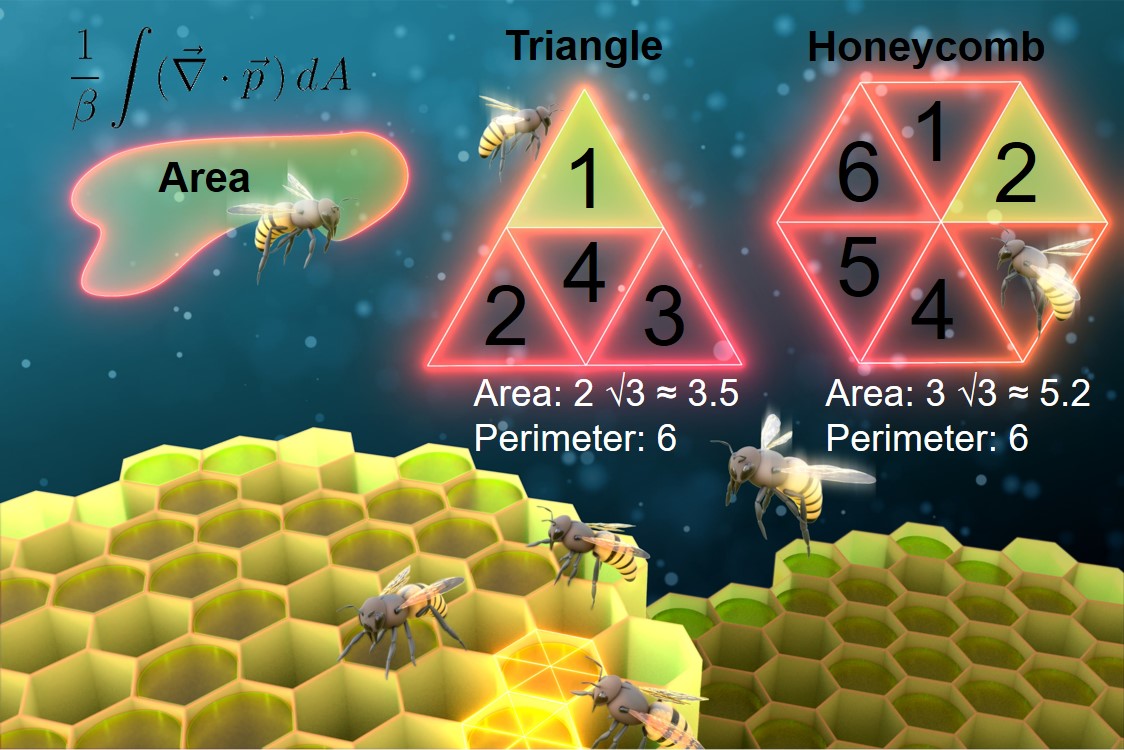}
 \caption{Schematic highlighting the isoperimetric superiority of the hexagon shape (honeycomb). Even though the triangle and hexagon depicted here have the same perimeter, the latter can host more triangles (six) than the triangle (four). Indeed, the honeycomb conjecture states that the honeycomb lattice has an isoperimetric superiority over all other polygons in the maximisation of area within fixed perimetric boundaries.\cite{Hales2001} This conjecture is echoed in evolution by natural selection\cite{raz2013application, lyon2012mathematical} which favoured the bee species/mutations that could store more honey/larvae (maximise storage space) by doing the least amount of work as depicted in the illustration.
 }
 \label{Figure_1}
\end{figure*}

Rooted in theories that date back to antiquity, the medley of honeycomb functionalities has been attributed to the monohedral hexagonal tiling embodying these superstructures. In what was dubbed as the honeycomb conjecture,\cite{Hales2001, raz2013application, lyon2012mathematical} the hexagon has been established to have isoperimetric superiority over other polygons in the maximisation of area within fixed perimetric boundaries. To demonstrate this phenomenon, \textbf{Figure \ref{Figure_1}} illustrates that a hexagon encompasses a larger surface area than a triangle of equal perimeter. In geometric forethought, the hexagonal tiling has been envisioned not only to have significant biomimetic implications but also influence \textit{in situ} processes in highly symmetric and efficient systems in fields such as condensed matter, nanotechnology, and astronomy.\cite{mecklenburg2011spin, georgi2017tuning, allen2010honeycomb, JWT2021} 

\subsection{Honeycomb Layered Oxides}

\begin{figure*}[!b]
\centering
  \includegraphics[width=0.95\columnwidth]{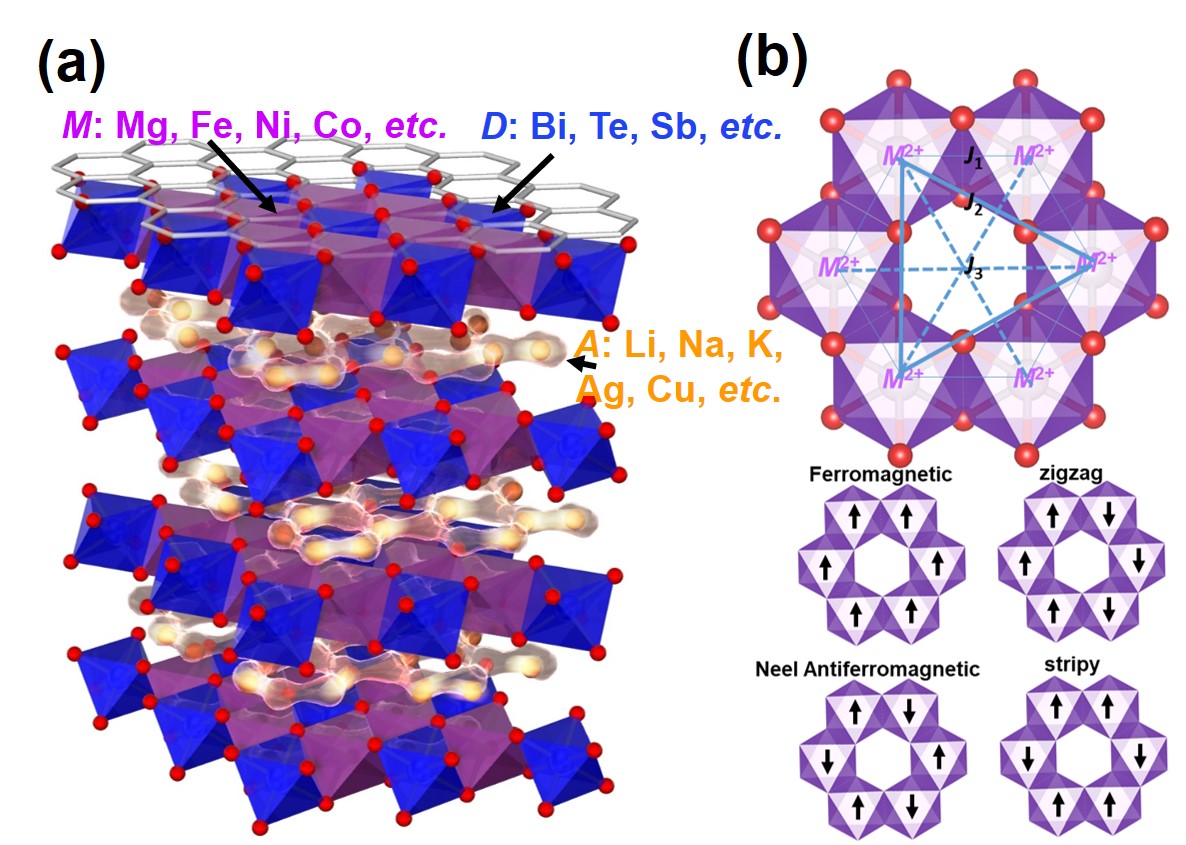}
  \caption{Crystal structural versatility and magnetic functionality of pnictogen- and chalcogen-based honeycomb layered oxides. (a) Elements that constitute pnictogen- and chalcogen-based honeycomb layered oxides. (b) Exemplars of magnetic configurations adopted by pnictogen- and chalcogen-based honeycomb layered oxides. The slabs entail transition metal atoms (divalent ($M^{2+}$) or trivalent) surrounding pnictogen or chalcogen atoms in a honeycomb formation. Such a configuration endows various spin configurations (ferromagnetic, zigzag, stripy, {\it et cetera}) and other exotic magnetic phenomena (such as spin liquid) that have evoked enormous interests particularly in the study of the magnetic interactions occurring in layered materials. Kitaev-type interactions, which principally arise from spins of the adjacent magnetic atoms within the honeycomb lattice, are denoted as $J_1$. Haldane-type interactions, which can also be influenced by spins of magnetic atoms from adjoining layers in the honeycomb configuration, are denoted as $J_2$. Higher-order spin interactions that can originate from distant atoms are denoted as $J_3$. Figure adapted with permission.\cite{Masese2018} Copyright 2018 Springer Nature.} 
  \label{Figure_2}
\end{figure*}

In the realm of nanotechnology, advancements in the discovery and isolation of two-dimensional (2D) materials such as graphene have unveiled new possibilities of exploiting known polyhedral structures to assemble multi-layered structures beyond the mesoscopic regime.\cite{Novoselov2004, Kitagawa2018} At the upper echelons of these frameworks, a novel class of biomimetic heterostructures emulating the honeycomb structural and isoperimetric properties have emerged with fascinating electronic, magnetic, quantum, and chemical properties that promise to bring sweeping paradigm shifts in a cynosure of disciplines.\cite{Kadari2016, Pu2018, Roudebush2015, Kumar2012, Zvereva2013, Derakhshan2007, Viciu2007, Morimoto2007, Derakhshan2008, Koo2008, Zvereva2012, Schmitt2014, Sankar2014, Xiao2019, Zvereva2015a, Itoh2015, Zvereva2015b, Bera2017, Upadhyay2016, Koo2016, Zvereva2016, Kurbakov2017, Karna2017, Zvereva2017, Werner2019, Stavropoulos2019, Korshunov2020, Yao2020, Li2019a, Miura2006, Schmitt2006, Miura2007, Miura2008, Li2010, Kuo2012, Roudebush2013a, Zhang2014, Jeevanesan2014, Lefrancois2016, Wong2016, Werner2017, Scheie2019, Mao2020} These materials embody metal cations (such as Ag, Cu, Ba, K, Na, Li, {\it et cetera}) wedged between arrays of covalently bound honeycomb-ordered transition metal oxides. The intricate combination of honeycomb frameworks in layered nanoarchitectures has been credited for an assortment of piquant electromagnetic behaviour and enigmatic topological properties that have made a significant impact in energy storage.\cite{kanyolo2021honeycomb, tada2022implications, kanyolo2022cationic, masese2023honeycomb, kanyolo2022advances2}

Such materials destined for cathode applications specifically in secondary (rechargeable) battery technologies have the property that facile intercalation/de-intercalation processes of positively charged ions (cations) occur reversibly within $n$-dimensions ($n$D)($n = 2, 3$), corresponding to charge-discharge processes. Whilst the most widely adopted cathode material is \ce{LiCoO2} \cite{mizushima1980lixcoo2}, novel layered materials known as honeycomb layered materials such as $A_2\rm Ni_2TeO_6$ or equivalently as $A_{2/3}\rm Ni_{2/3}Te_{1/3}O_2$ ($A = \rm Li, Na, K,$ \textit{etc}) have recently emerged as contenders destined to augment the otherwise underexplored landscape of functional materials.\cite{kanyolo2021honeycomb}

\subsection{Pnictogen- and Chalcogen-Based Honeycomb Layered Oxides}

\begin{figure*}[!t]
\centering
  \includegraphics[width=0.6\columnwidth]{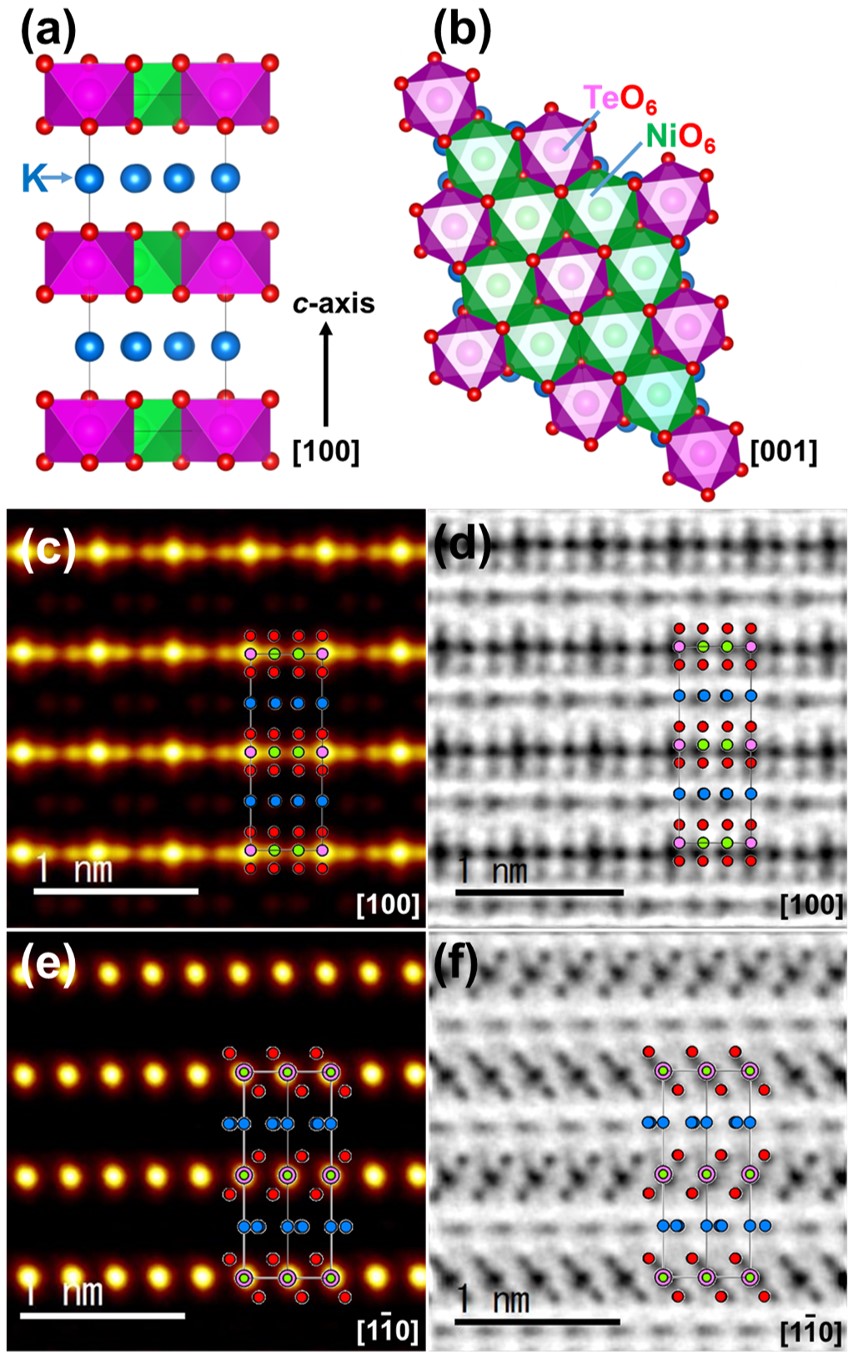}
  \caption{Visualisation of the crystal structural framework of chalcogen-based honeycomb layered $\rm K_2Ni_2TeO_6$. (a) Crystal structure of $\rm K_2Ni_2TeO_6$ along [100] zone axis showing the K atoms sandwiched between honeycomb slab layers in the unit cell (shown in black). O atoms are in red, and the Ni and Te atoms are in green and pink, respectively. K atoms are shown in blue. (b) The honeycomb arrangement of Ni atoms around Te atoms when viewed along the [001] axis. (c) High-resolution electron microscopy image taken along [100] projection showing the ordering sequence of Te and Ni atoms, and (d) High-resolution electron microscopy image of heavy atoms along with lighter atoms such as K and O. (e) High-resolution images of Te and Ni atoms taken when the crystal is rotated at $30^\circ$ ($i.e.$, [1$\overline{1}$0]) and (f) High-resolution imaging of heavy atoms along with lighter atoms along the [1$\overline{1}$0] zone axis. The images taken from the two projections ([100] and [1$\overline{1}$0] zone axes) are used to depict the three-dimensional crystal structure of $\rm K_2Ni_2TeO_6$. Reproduced with permission.\cite{masese2021topological} Copyright 2021 American Chemical Society.}
  \label{Figure_3}
\end{figure*}

\begin{figure*}[!t]
\centering
  \includegraphics[width=0.6\columnwidth]{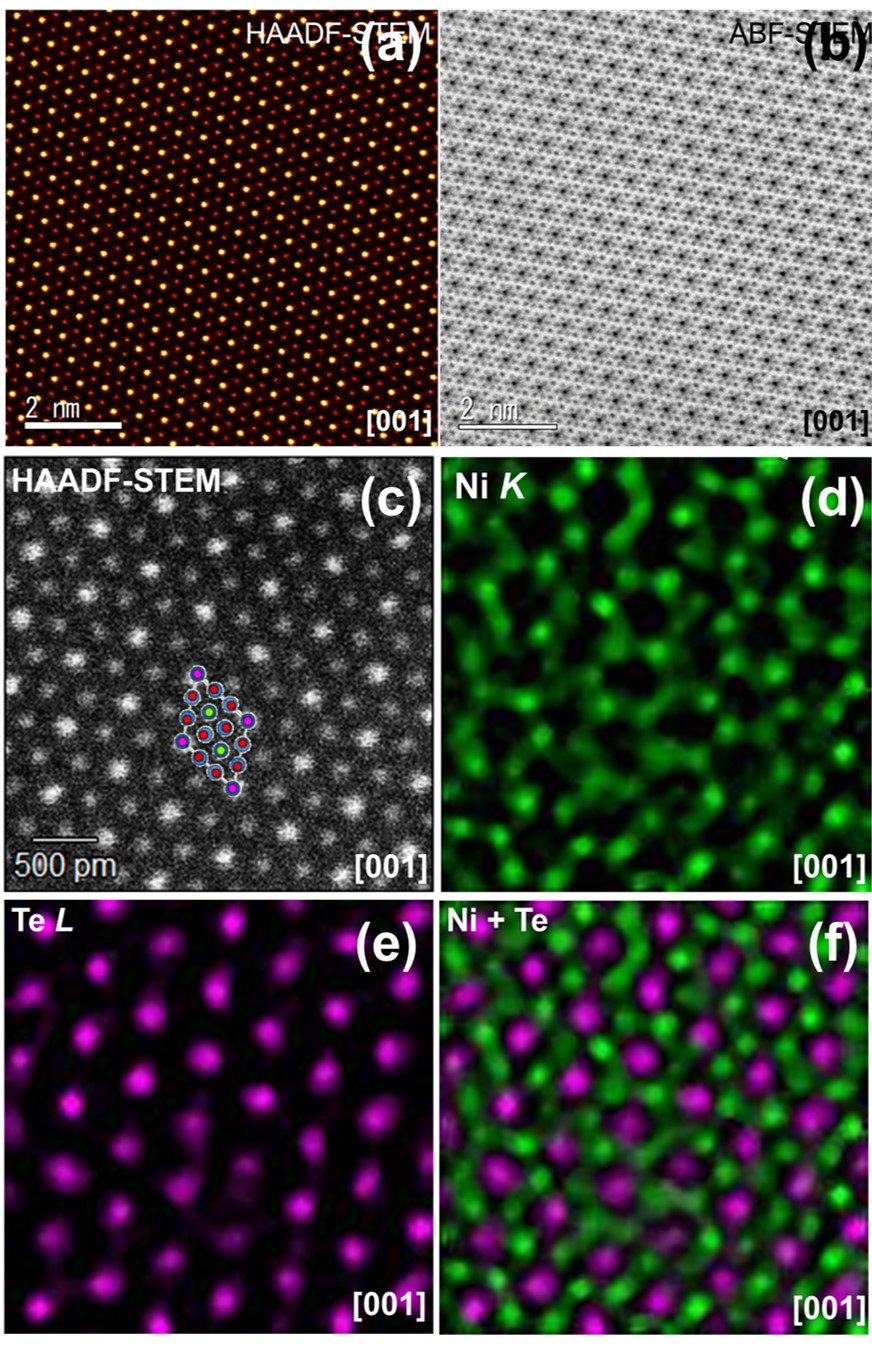}
  \caption{Honeycomb configuration of metal atoms in $\rm K_2Ni_2TeO_6$ along the [001] zone axis ($c$-axis). (a) High-resolution electron microscopy image taken along the [001] projection showing an ordered honeycomb arrangement of Te atoms (shown as bright yellow spots) surrounded by Ni atoms (dark red spots) in an ordered honeycomb arrangement. (b) High-resolution electron microscopy image taken along the [001] projection showing the atomic arrangement of Te and Ni atoms, as well as atoms of lighter atomic mass such as potassium. (c) High-resolution electron microscopy image of a crystallite domain from which elemental mapping of (d) Ni, (e) Te and (f) a combination of Te and Ni was performed. Inset in (c) shows a polyhedral atomic view of the crystal structure of the honeycomb slab, for clarity. Reproduced with permission.\cite{masese2021topological}  
  Copyright 2021 American Chemical Society.}
  \label{Figure_4}
\end{figure*}

The possibility of embedding metal cations within layers of transition metal oxides represents a rich compositional canvas with seemingly endless permutations of avant-garde chemical combinations. Within this compositional space, honeycomb layered oxides (\textbf{Figure \ref{Figure_2}a}) embodied by alkali- or coinage-metal cations such as Cu, Ag, Na, Li, K or Ba, interposed between layers of honeycomb-ordered transition metal (Zn, Cu, Ni, Co, Fe, Mn or Cr) atoms positioned around pnictogen (Bi, Sb or As) or chalcogen (Te) species, have been found to harbour an unparalleled physicochemical properties such as exquisite electrochemical functionalities, fast ionic mobility, exotic phase transitions and piquant magnetic interactions.\cite{kanyolo2021honeycomb} The repetitive 2D layered honeycomb or hexagonal arrangement is characteristic of 3D optimised lattices, which constitute the most densely packed arrangements of atoms possible in nature. In other words, this suggests that the highest possible energy density theoretical limit via packing cations densely within a material is achieved by such honeycomb layered materials. This theoretical limit marginalises efforts by computational and experimental material scientists, steering majority of research to only explore the compositional landscape at the expense of searching for novel lattices with exotic geometries and topologies that may lead to enhanced functionalities. Within this paradigm, geometric and topological features are studied by the energy storage community mainly as far as cataloguing their hindrance to the proper function of the cathode material is concerned. However, recent progress in the field has rendered honeycomb layered oxides (hereafter to be referred to as pnictogen- and chalcogen-based honeycomb layered oxides)} pedagogical arenas for investigating not only intriguing magnetic phenomena (such as those in \textbf{Figure \ref{Figure_2}b}) within its slabs but also serving as testing grounds for 2D conformal field theories and their gravity duals.\cite{kanyolo2021honeycomb, kanyolo2022advances2, kanyolo2022cationic}

These pnictogen-/chalcogen-based honeycomb layered oxides adopt multifarious chemical compositions such as $A_{2}M_{2}D\rm O_{6}$, $A_{3}M_{2}D\rm O_{6}$ or $A_{4}MD\rm O_{6}$, amongst others. Here, $M$ represents the $s$-block metals such as Mg or transition metal species, whereas $A$ depicts coinage- or alkali-metal species suchlike Ag, Cu, K, Na, Li, {\it et cetera}, and $D$ represents a pnictogen or chalcogen elemental species such as Bi, Sb, Te and so forth.
\cite{Viciu2007, Derakhshan2007, Morimoto2007, Derakhshan2008, Koo2008, Zvereva2012, Kumar2012, Zvereva2013, Schmitt2014, Sankar2014, Roudebush2015, Zvereva2015a, Itoh2015, Zvereva2015b, Upadhyay2016, Koo2016, Zvereva2016, Kadari2016, Bera2017, Kurbakov2017, Karna2017, Zvereva2017, Pu2018, Xiao2019, Werner2019, Stavropoulos2019, Korshunov2020, Yao2020, Li2019a, Miura2006, Schmitt2006, Miura2007, Miura2008, Li2010, Kuo2012, Roudebush2013a, Zhang2014, Jeevanesan2014, Lefrancois2016, Wong2016, Werner2017, Scheie2019, Kurbakov2020, motome2020, Nalbandyan2013, Bhardwaj2014, Kumar2013, Gupta2013, Berthelot2012, Laha2013, McCalla2015, Taylor2019, Roudebush2013b, Berthelot2012a, Uma2016, He2017, He2018, Stratan2019, Xu2005, Ramlau2014, Smirnova2005, Schmidt2013, Seibel2013, Seibel2014, Liu2016, Bhange2017, Gyabeng2017, Yan2019, Yadav2019, Smaha2015, Brown2019, Heymann2017, Nalbandyan2013a, Schmidt2014, Sathiya2013, Grundish2019, Yang2017, Yuan2014, Masese2018, yadav2022influence, pal2022insights, yadav2022investigation, song2022influence, masese2023honeycomb, chen2020, Masese2019, Yoshii2019, kanyolo2022advances2, masese2021mixed, tada2022implications} To elucidate the atomistic constitution of the pnictogen-/chalcogen-based honeycomb layered oxides, \textbf{Figure \ref{Figure_3}a} illustrates the crystal structure of honeycomb layered $\rm K_2Ni_2TeO_6$.\cite{Masese2018, masese2021topological} As shown, the K atoms interposed between the transition metal slabs made of $\rm NiO_6$ and $\rm TeO_6$ octahedra are coordinated with oxygen atoms. As shown in \textbf{Figure \ref{Figure_3}b}, the Ni and Te atoms in their respective octahedra are each coordinated to six oxygen atoms to make a honeycomb configuration where six $\rm NiO_6$ octahedra surround each $\rm TeO_6$ octahedron. 

The recent proliferation of layered oxide materials is to some extent exacerbated by advancements in characterisation techniques which allow the visualisation of the respective crystal configurations at an atomic scale. For instance, the layered arrangement of K atoms between the Te and Ni slabs in the $\rm K_2Ni_2TeO_6$ crystal has been validated using high-resolution transmission electron microscopy as illustrated by \textbf{Figures \ref{Figure_3}c}, \textbf{\ref{Figure_3}d}, \textbf{\ref{Figure_3}e} and \textbf{\ref{Figure_3}f}.\cite{masese2021topological} Additionally, elemental mapping performed using high-resolution transmission electron microscopy images can further be used to observe the honeycomb configuration of Ni (transition metal) around Te (chalcogen), as shown in \textbf{Figure \ref{Figure_4}}.

\subsection{Scope of this Review}

\begin{figure*}[!b]
\centering
  \includegraphics[width=0.95\columnwidth]{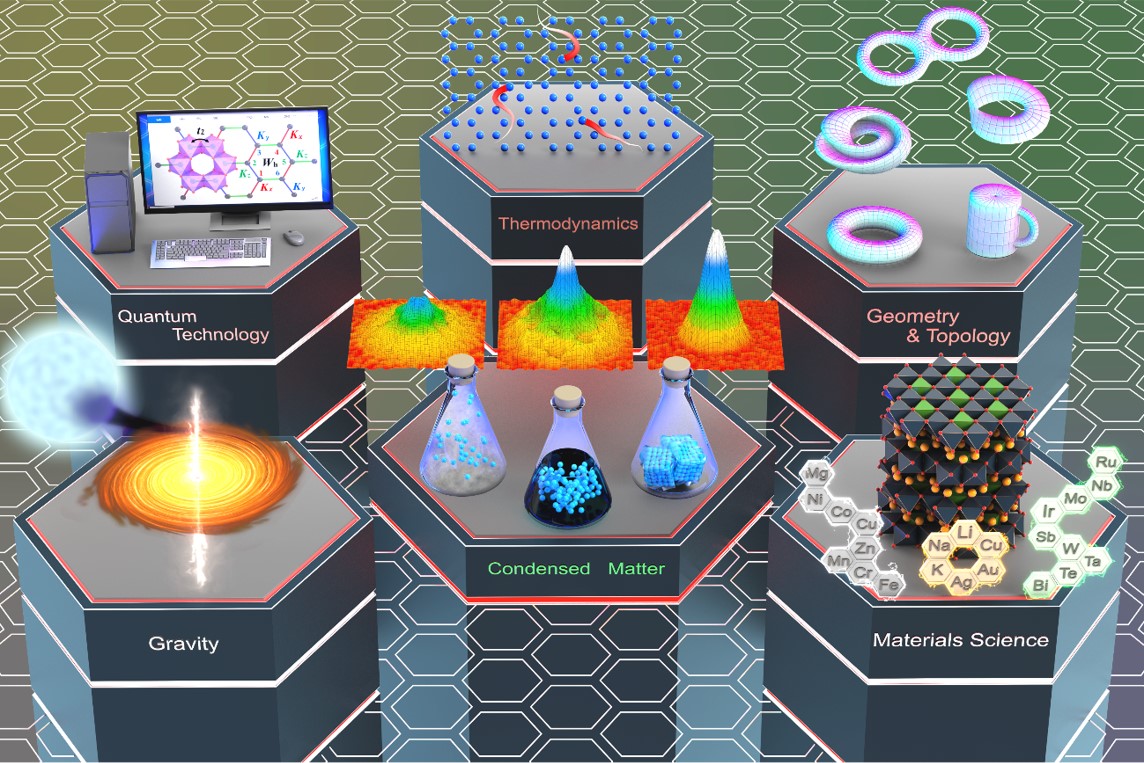}
  \caption{Diverse fields of study of pnictogen- and chalcogen-based honeycomb layered oxides! A comprehensive scope is shown in the Table of Contents as the outline of the review.\cite {kanyolo2022advances2, kanyolo2021partition, kanyolo2022local, kanyolo2021reproducing, kanyolo2022cationic, masese2023honeycomb, Kitaev2006}}
  \label{Figure_5}
\end{figure*}

This review aims to serve as a guideline on the future direction of pnictogen- and chalcogen-based honeycomb layered oxides by highlighting the recent advances made in the development of this class of materials. Since the design of $\rm Li_3Zn_2{\it D}O_6$ ($\rm {\it D} = Bi, Sb$),\cite{Greaves1990} it is our view that the synthesis and experimental characterisation frontier has advanced considerably over the last 32 years, warranting a comprehensive review. Here in, we comprehensively discuss the significant milestones made in preparation and characterisation techniques, and potential avenues of application for this class of materials in the context of their fundamental chemistry. In this vein, we critically evaluate the significant milestones made over the last three decades with particular focus on authoritative literature pertaining to the synthesis and characterisation techniques that have been instrumental in unveiling the physicochemical, electromagnetic, and topological properties of this class of honeycomb layered oxides. Further, we critique the challenges and knowledge gaps ingrained in this ever-expanding literature space and highlight new scopes for the future research development of honeycomb layered oxides. \textbf{Figure \ref{Figure_5}} shows a schematic illustrating the diverse fields of study and related functionalities affiliated with pnictogen- and chalcogen-based honeycomb layered oxides. Further commentary on the explicit connection of honeycomb layered oxides to these fields of study has been availed in \textbf{Section} \ref{Section: Summary}.  

To keep the readers abreast on the strides made in the development of pnictogen-and chalcogen-based honeycomb layered oxides, a summary of material compositions is provided in \textbf{Section \ref{Section: Milestone}}. We further adumbrate the synthesis techniques employed in efforts to expand the vast crystal chemistry space entailing the pnictogen and chalcogen class of honeycomb layered oxides in \textbf{Section \ref{Section: Synthesis}}. The multitude of layer arrangement sequences inherent in this class of honeycomb layered oxides are detailed in \textbf{Section \ref{Section: Stacking}}.
Further, \textbf{Section \ref{Section: Characterisation}} outlines characterisation techniques utilised in delineating various functionalities in this class of materials. We conclude this review in \textbf{Section \ref{Section: Summary}} by envisaging future research directions where fascinating physicochemical, topological and electromagnetic properties could be lurking, particularly in the prospective realisation of exotic material compositional space as well as their utility as testing grounds for ideas in topological field/2D emergent quantum gravity theories.

Despite our attempts to exhaustively delineate the advancements made in the present class of honeycomb layered oxides, we acknowledge the vastness and diversity of concepts behind the underlying properties and behaviours exhibited by these materials. As such, whilst we also mention in passing novel concepts that remain uncovered in 
research literature or those that arc beyond the target of concision of this review, this nonetheless, no way diminishes the quality, utility, and broad scope of this review, which we expect to be a worthwhile publication to recent graduates and frontier experts alike not only in the chemistry community but also in other fields that scope beyond thereof. Further, due to the universality of the concepts presented herein, the properties and functionalities represented are expected to arc beyond the scope of pnictogen- and chalcogen-based honeycomb layered compounds spilling over into the entire field of materials science and condensed matter. Therefore, the facets covered by our work will bestead a wide community of experimentalists and theoreticians delving in condensed matter (solid-state physics, materials science, solid state chemistry and solid-state ionics) and electromagnetism (photonics, electromagnetic dynamics) as well as in fundamental fields of physics (topology, quasi-particle physics, modelling and simulation techniques).

\newpage

\section{\label{Section: Milestone} 
Development of pnictogen- and chalcogen-based honeycomb layered oxides}

To elucidate the advancements made in the development of the present class of honeycomb layered oxides, it is important to first demarcate their vast chemical compositional space. This pnictogen and chalcogen family of honeycomb layered oxides generally comprises stoichiometric chemical compositions of $\rm {\it A}^{+}_{4}{\it M}^{3+}{\it D}^{5+}O^{2-}_{6}$ (or equivalently as $\rm {\it A}^{+}_{4/3}{\it M}^{3+}_{1/3}{\it D}^{5+}_{1/3}O^{2-}_{2}$), $\rm {\it A}^{+}_{4}{\it M}^{2+}$
$\rm {\it D}^{6+}O^{2-}_{6}$ ($\rm {\it A}^{+}_{8}{\it M}^{2+}_{2}{\it D}^{6+}_{2}O^{2-}_{12}$ or $\rm {\it A}^{+}_{4/3}{\it M}^{2+}_{1/3}{\it D}^{6+}_{1/3}O^{2-}_{2}$), $\rm {\it A}^{+}_{3}{\it M}^{2+}_{2}{\it D}^{5+}O^{2-}_{6}$ ($\rm {\it A}^{+}{\it M}^{2+}_{2/3}{\it D}^{5+}_{1/3}O^{2-}_{2}$), $A^{+}_{2}M^{2+}_{2}D^{6+}\rm O^{2-}_{6}$ (or equivalently as $\rm {\it A}^{+}_{2/3}{\it M}^{2+}_{2/3}{\it D}^{6+}_{1/3}O^{2-}_{2}$), $\rm {\it A}^{+}_{3}{\it A'}^{+}{\it M}^{3+}{\it D}^{5+}O^{2-}_{6}$ ($\rm {\it A}^{+}{\it A'}^{+}_{1/3}{\it M}^{3+}_{1/3}{\it D}^{5+}_{1/3}O^{2-}_{2}$), $\rm {\it A}^{+}_{2}{\it A'}^{+}{\it M}^{3+}{\it D}^{6+}O^{2-}_{6}$ ($\rm {\it A}^{+}{\it A'}^{+}_{1/3}{\it M}^{3+}_{1/3}{\it D}^{6+}_{1/3}O^{2-}_{2}$), $\rm {\it A}^{+}{\it A'}^{+}{\it M}^{2+}{\it D}^{6+}O^{2-}_{6}$ ($\rm {\it A}^{+}_{1/3}{\it A'}^{+}_{1/3}{\it M}^{2+}_{1/3}{\it D}^{6+}_{1/3}O^{2-}_{2}$) or $\rm {\it A}^{+}_{4.5}{\it M}^{3+}_{0.5}{\it D}^{6+}O^{2-}_{6}$\\
($\rm {\it A}^{+}_{3/2}{\it M}^{3+}_{1/6}{\it D}^{6+}_{1/3}O^{2-}_{2}$),\cite{Kumar2012, Zvereva2013, Kurbakov2020, motome2020, Nalbandyan2013, Zvereva2017, Bhardwaj2014, Yao2020, Kumar2013, Morimoto2007, Gupta2013, Berthelot2012, Laha2013, McCalla2015, Taylor2019, Roudebush2013b, Berthelot2012a, Uma2016, He2017, He2018, Stratan2019, Xu2005, Ramlau2014, Bera2017, Smirnova2005, Schmidt2013, Seibel2013, Seibel2014, Liu2016, Bhange2017, Gyabeng2017, Yan2019, Yadav2019, Smaha2015, Brown2019,Szillat1995, Greaves1990, Skakle1997, Nagarajan2002, Gupta2015, zvereva2016d, Mather2000, Mather1995, nguyen2020, navaratnarajah2019, zheng2018, koch2015, tamaru2013, gazizova2018, mogare2004, lu2019, bastow1994, wang2013, robertson2003, boulineau2009, zuo2018, strobel1988, ma2014, shinova2005, okada1999, asakura1999, omalley2008, hermann2019, luo2013, Todorova2011b, kimber2014, yu2019, jang2020, james1988} provided charge electro-neutrality is maintained. In this medley of chemical compositions, $\rm {\it A}$ and $A'$ represent alkali metal species such as Li, Na, K, {\it et cetera} or coinage-metal species such as Cu, Ag, {\it et cetera} (with $A \neq A'$); $M$ can be divalent or trivalent transition metal species such as Cr, Mn, Fe, Co, Ni, Cu, Zn, s-block elements such as Mg or combinations thereof; and $\rm {\it D}$ denotes predominantly hexavalent chalcogen atoms (Te) or pentavalent pnictogen atoms such as Sb, Bi, As, amongst others. Recent augmentation of this class of honeycomb layered oxides has also unearthed new compositional configurations such as $A^{2+}M^{2+}_{2}D^{6+}\rm O^{2-}_{6}$ (or equivalently as $\rm {\it A}^{2+}_{1/3}{\it M}^{2+}_{2/3}{\it D}^{6+}_{1/3}O^{2-}_{2}$) which has been exemplified by $\rm BaNi_2TeO_6$\cite{karati2021band}; alkali-excess honeycomb layered oxides (such as $\rm Na_3Ni_{1.5}TeO_6$,\cite{Grundish2019} $\rm Na_{2.4}Zn_2TeO_6$ and $\rm Na_{2{\it x}}Zn_2TeO_6$\cite{itaya2021effect, itaya2021sintering}); and multicomponent layered oxides such as $\rm Na_3Cu_{2/3}Ni_{2/3}Co_{2/3} SbO_6$,\cite{karati2021band} $\rm Na_3Cu_{1/2}Ni_{1/2}Co_{1/2}$\\$\rm Fe_{1/2}SbO_6$\cite{karati2021band} and $\rm Na_3Cu_{2/5}Ni_{2/5}Co_{2/5}Fe_{2/5}Mn_{2/5}SbO_6$.\cite{karati2021band}

Although pnictogen- and chalcogen-based honeycomb layered oxides date back to 1990 in a report which focused on the layered crystal structure and Li-ionic conductivity measurements of $\rm Li_3Zn_2SbO_6$ and $\rm Li_3Zn_2BiO_6$,\cite{Greaves1990} explorations into this family of materials point back to $\rm Li_3Mg_2SbO_6$\cite{castellanos1982crystal} which was reported in 1982. Although 
$\rm Li_3Mg_2SbO_6$ does not crystallise in a honeycomb formation, its discovery in 1982 is viewed as instrumental breakthrough in the development of layered antimonates and bismuthates chemical compositions in 1990.\cite{Greaves1990} Following this discovery, the exploration into this family of materials mainly involved the development of new varieties of honeycomb layered lithium antimonates and bismuthates achieved through the incorporation of the various transition metal species available. However, characterisation of the emergent materials revealed that Li compositions have a proclivity towards the formation of disordered structures. Continued pursuit of ordered layered materials in this compositional space led to the development of Cu- and Ag-based antimonates in 2002\cite{Nagarajan2002} and Na-based antimonates and tellurates shortly thereafter (2005)\cite{Smirnova2005}, both of which catapulted the further development of ordered layered materials with intriguing magnetic properties. It is worth mentioning that the development of Cu- and Ag-based honeycomb layered oxides, which are related to the Delafossite family of compounds, was in-part triggered by a report on a Delafossite-type antimonate adopting the composition of $\rm Cu_3Mg_{1.4}SbO_6$.\cite{Szillat1995}  
Nonetheless, the utility of these coinage metal species in honeycomb layered oxides has mainly been focused on compositions such as $\rm {\it A}_3Cu_2SbO_6$ ($A = \rm Li, Na, Cu$) and  $\rm Na_2Cu_2TeO_6$,\cite{Mather1995, Skakle1997, Nagarajan2002, Smirnova2005, ma2007synthesis, Morimoto2007, Miura2007, Nalbandyan2013} owing to their intriguing spin gap magnetic behaviour engendered by the pseudo-octahedral coordination geometry of divalent Cu. In particular, $\rm Na_2Cu_2TeO_6$ and $\rm Na_3Cu_2SbO_6$ have been found to possess a spin gap that endows them with a spin-singlet magnetic ground state separated from magnetic excited states by an energy gap.\cite{Morimoto2007, Miura2007} This property allows the magnetic suseptibility to vanish below a critical temperature, thereby heralding their possible application in quantum mechanical disciplines. As such, the peculiar magnetic behaviour of honeycomb layered antimonates, bismuthates and tellurates has been a key impetus in the advancement of this class of materials.

\begin{figure*}[!t]
\centering
  \includegraphics[width=0.95\columnwidth]{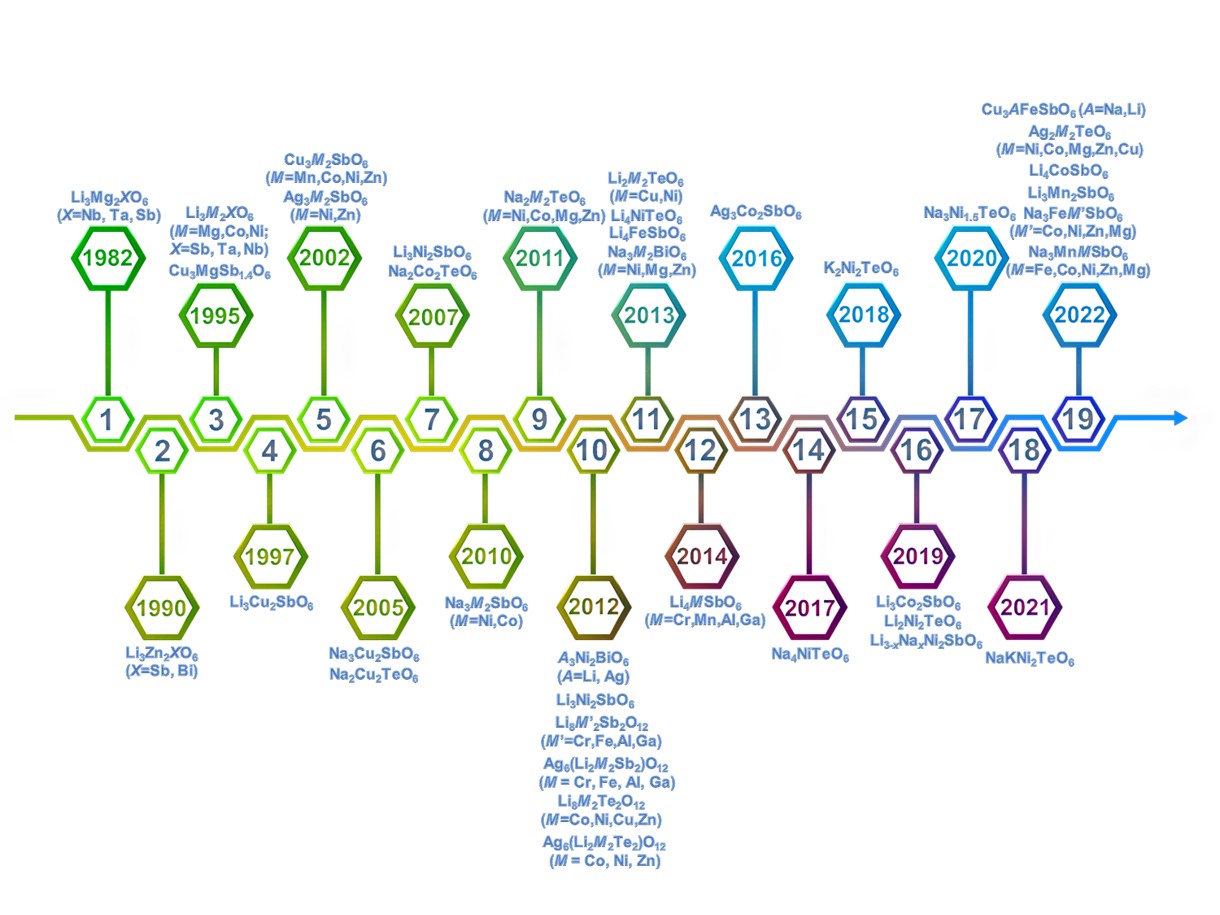}
  \caption{Material compositions that were instrumental in the development of pnictogen- and chalcogen-based honeycomb layered oxides.\cite{castellanos1982crystal, Greaves1990, Nagarajan2002, Smirnova2005, Szillat1995, Mather1995, Skakle1997, ma2007synthesis, Morimoto2007, Miura2007, Nalbandyan2013, Evstigneeva2011, Sathiya2013, Grundish2019, McCalla2015, McCalla2015a, Yuan2014, Ma2015, Gupta2013, He2017, He2018, Yang2017, Masese2018, grundish2020structural, Vallee2019, masese2021mixed, berthelot2021stacking, song2022influence, karati2021band, yadav2022influence, sethi2022cuprous, Kumar2012, Uma2016, Bhardwaj2014, Kumar2013, Yadav2019, Gupta2015, mandujano2021, yadav2022investigation, masese2023honeycomb, chen2020, Masese2019, Yoshii2019, kanyolo2022advances2, kanyolo2021honeycomb, Bhange2017, Dubey2020}}
  \label{Figure_6}
\end{figure*}

Besides these intriguing possibilities of quantum applications, the exploration of this class of honeycomb layered oxides was accelerated by the reports on fast Na-ion conductivity in $\rm Na_2{\it M}_2TeO_6$ ($\rm {\it M} = Ni, Co, Mg$ and $\rm Zn$) in 2011.\cite{Evstigneeva2011} Subsequent traction in this class of honeycomb layered oxides was exacerbated by their crystal structural versatility (as was noted in the following years) and their high-voltage electrochemistry cognate to those observed in Li-excess compositions such as $\rm Li_4NiTeO_6$,\cite{Sathiya2013} $\rm Li_2Ni_2TeO_6$\cite{Grundish2019} and $\rm Li_4FeSbO_6$\cite{McCalla2015, McCalla2015a} (exhibiting oxygen redox electrochemistry) when used as cathode materials for rechargeable Li-ion batteries. Considering the ongoing shift towards sustainable energy materials, reports (in 2013, 2014, 2016 and 2017) on the high-voltage electrochemistry of $\rm Na_3Ni_2SbO_6$,\cite{Yuan2014, Ma2015} $\rm Na_3Ni_2BiO_6$,\cite{Liu2016, Bhange2017} $\rm Na_2Ni_2TeO_6$,\cite{Gupta2013} and subsequently in Na-excess compositions such as $\rm Na_4{\it M}TeO_6$ ($\rm {\it M} = Ni, Co$ and $\rm Cu$)\cite{He2017, He2018} (which do not assume the honeycomb layered framework) were considered pivotal not only to next-generation energy materials but also to the expansion of this material space. Most importantly, the report on the high-voltage $\rm Na_4NiTeO_6$\cite{Yang2017} cathode material for rechargeable Na-ion batteries portended the development of $\rm K_2Ni_2TeO_6$ (reported in 2018)\cite{Masese2018} and related compositions which augur well as a high-voltage layered oxide cathode materials for the nascent K-ion batteries.

The discovery of Na-excess $\rm Na_3Ni_{1.5}TeO_6$\cite{grundish2020structural} has also been seen as a key milestone amongst these materials due to its role in unearthing the excellent electrochemical performance of honeycomb layered tellurates for high-capacity energy storage systems as well as their high compositional flexibility. In particular, the compositional flexibility evinced the possibility to replace some of the alkali metal atoms with other species to create exotic compositions such as $\rm Na_{1.6}Sr_{0.2}Ni_2TeO_6$\cite{Gupta2013} and $\rm Li_{1.5}Na_{1.5}Ni_2SbO_6$\cite{Vallee2019}. Further advancement on this development route gave rise to mixed-alkali metal compounds such as $\rm NaKNi_2TeO_6$ (2021)\cite{masese2021mixed, berthelot2021stacking} which augur the emergence of a new class of hybrid rechargeable batteries that can exploit the advantages of multiple mobile cation chemistries (namely, dual-ion batteries). Although the pnictogen- and chalcogen-based honeycomb layered oxides have seen tremendous advancements over the years, their near infinite compositions and applications promise enormous prospects in the future. In fact, a recent inquest on the honeycomb layered tellurate, $\rm BaNi_2TeO_6$ (2022)\cite{song2022influence} unmasked new prospects of developing a new class of honeycomb layered oxides with divalent cation atoms sandwiched between the transition metal slabs. This new class which has been dubbed as `high-entropy' honeycomb layered oxides not only promises to augment the compositional space of these materials but also the possible design of antimonates with multi-transition metal constituents such as $\rm Na_3Cu_{2/5}Ni_{2/5}Co_{2/5}Fe_{2/5}Mn_{2/5}SbO_6$ which was also reported in 2021.\cite{karati2021band} \textbf{Figure \ref{Figure_6}} shows a timeline of the milestones achieved in the development of pnictogen- and chalcogen-based honeycomb layered oxides to date. Based on chronology, more compositions should be expected in the coming decades.

\newpage

\section{\label{Section: Synthesis} Synthesis: pnictogen- and chalcogen-based honeycomb layered oxides}

As outlined in the previous section, the functionalities of pnictogen- and chalcogen-based honeycomb layered oxides are far-reaching and bear significant impact on a wide range of disciplines and realms. In general, the choice of synthesis technique is considered a crucial step in the design and optimisation of the chemical, structural, electrochemical, and magnetic properties of these materials. Although the simplicity and scalability of solid-state calcination methods makes them optimal for synthesising most layered oxide materials, the vast compositional space provided by the pnictogen- and chalcogen-based honeycomb layered oxides implies varied thermal dynamic stability and physicochemical properties that may not be suitable for high-temperature synthesis routes. To clearly delineate the synthetic processes available and their plausible integration into commercialised production processes of the present family of honeycomb materials, this section highlights the commonly employed pnictogen- and chalcogen-based honeycomb layered oxides preparation methods which include solid-state reaction, topochemical ion-exchange and crystal growth. Further, we classify the synthetic process according to their compositional configurations to highlight the various functionalities and application routes availed by the pnictogen and chalcogen class of honeycomb layered oxides.

\subsection{High-temperature solid-state reactions}
The solid reagents utilised in the synthesis of most polycrystalline materials generally tend to be thermodynamically stable and relatively inert at room temperatures. Therefore, high temperatures are usually required to activate the interdiffusion of atoms between the individual precursor compounds and to remove impurity traces in the resultant compounds. 

The fabrication of most pnictogen- and chalcogen-based honeycomb layered oxides entails high-temperature solid-state reactions (commonly referred to as `shake and bake' or `heat and beat' processing) where the powdered starting reagents are intimately mixed using a pestle and mortar or ball milling to obtain the desired phase. Subsequently, the mixtures are pressed into pellets to increase their contact surface area and subjected to high temperatures under controlled environments to develop well-defined crystal structures. In these synthesis processes, parameters such as the calcination temperatures, heat-treatment duration and reaction procedures are typically determined by physicochemical properties such as particle sizes, atomic mobility, volatility, and reactivity of the starting materials. For instance, some materials may require multiple grinding and/or heating steps to facilitate complete solid-state diffusion between the initial precursors. In such cases, the initial heat treatment triggers the diffusion process between the starting reagents. However, when the initial grinding cannot fully guarantee atomic-level mixing, subsequent grinding can be applied to break up the interfaces between the product and residual starting materials at the core of the particles. This creates new interfaces for further reaction during successive annealing steps where the previously formed products act as a foundation that steers the reaction process towards the target end products.

The heat-treatment temperatures are typically determined by the thermodynamical properties of the starting materials/precursors. Some starting reagents such as NiO (for compositions such as  $\rm {\it A}_2Ni_2TeO_6$,\cite{Evstigneeva2011, Grundish2019, Masese2018, Kumar2013}  $\rm {\it A}_3Ni_2SbO_6$,\cite{Yuan2014, Ma2015, Masese2018} and $\rm {\it A}_3Ni_2BiO_6$\cite{Liu2016, Bhange2017, Masese2018} ($A = \rm Li, Na, K$)) are highly inert and require elevated temperatures for reactions to occur, whereas reagents such as $\rm Li_2CO_3$, $\rm Na_2CO_3$ or $\rm K_2CO_3$ are volatile at higher temperatures and could be lost during a reaction. To circumvent this, an excess amount of volatile carbonate starting reagents are usually mixed at the onset of the synthesis, to compensate for possible material loss during heat treatment.

It is worth noting that even though more volatile reactants such as $\rm Sb_2O_3$, $\rm Bi_2O_3$ and $\rm TeO_2$ are utilised in some synthesis processes, their evaporation at the high synthesis temperatures is generally not observed and thus does not factor when determining the temperature parameters. Nonetheless, the reactivity of such materials with the reaction vessel holding the materials in the furnace is an important factor when selecting the appropriate crucibles. Alumina crucibles are commonly used as reaction vessels, but they can be swapped with platinum or gold trays if they are incompatible with the reactants. In the same vein, the firing atmosphere can also influence the formulation of products due to the propensity of some precursors to react with gaseous components in the air. Most compounds in this class of materials may be adequately prepared in air or oxygen environments, but some materials containing readily oxidisable transition metals such as Fe and Mn may require firing under controlled atmospheres such as under various flowing gases (CO, $\rm H_2$, Ar, $\rm N_2$, and so forth) or in vacuum. For example, some reports have shown $\rm N_2$ to be suitable in the preparation of $\rm Na_2Co_2TeO_6$ and $\rm Na_3Co_2SbO_6$; whilst others have reported the use of Ar in the syntheses of $\rm Li_3Mn_2SbO_6$,\cite{yadav2022influence} $\rm Li_3Co_2SbO_6$, $\rm Na_3Fe_2SbO_6$\cite{Yadav2019} and $\rm Na_3Mn_2SbO_6$,\cite{Yadav2019} including their solid-solution derivatives ($i.e.$, $\rm Li_3Mn{\it M}SbO_6$ ($\rm {\it M} = Zn, Ni, Co, Mg $),\cite{pal2022insights} $\rm Li_3Co{\it M}SbO_6$ ($\rm {\it M} = Zn, Ni, Mg $),\cite{pal2022insights} $\rm Na_3Fe{\it M}SbO_6$ ($\rm {\it M} = Zn, Ni, Mg $)\cite{yadav2022investigation} and $\rm Na_3Mn{\it M}SbO_6$ ($\rm {\it M} = Fe, Co, Zn, Ni, Mg $)\cite{yadav2022investigation}).

In the following sub-sections, we detail a variety of compositional formations of pnictogen- and chalcogen-based honeycomb layered oxides that employ high-temperature syntheses techniques to outline the functionalities that motivate their exploration and accordingly, their syntheses.

\subsubsection{Lithium-based materials}
\begin{figure*}[!t]
\centering
  \includegraphics[width=0.95\columnwidth]{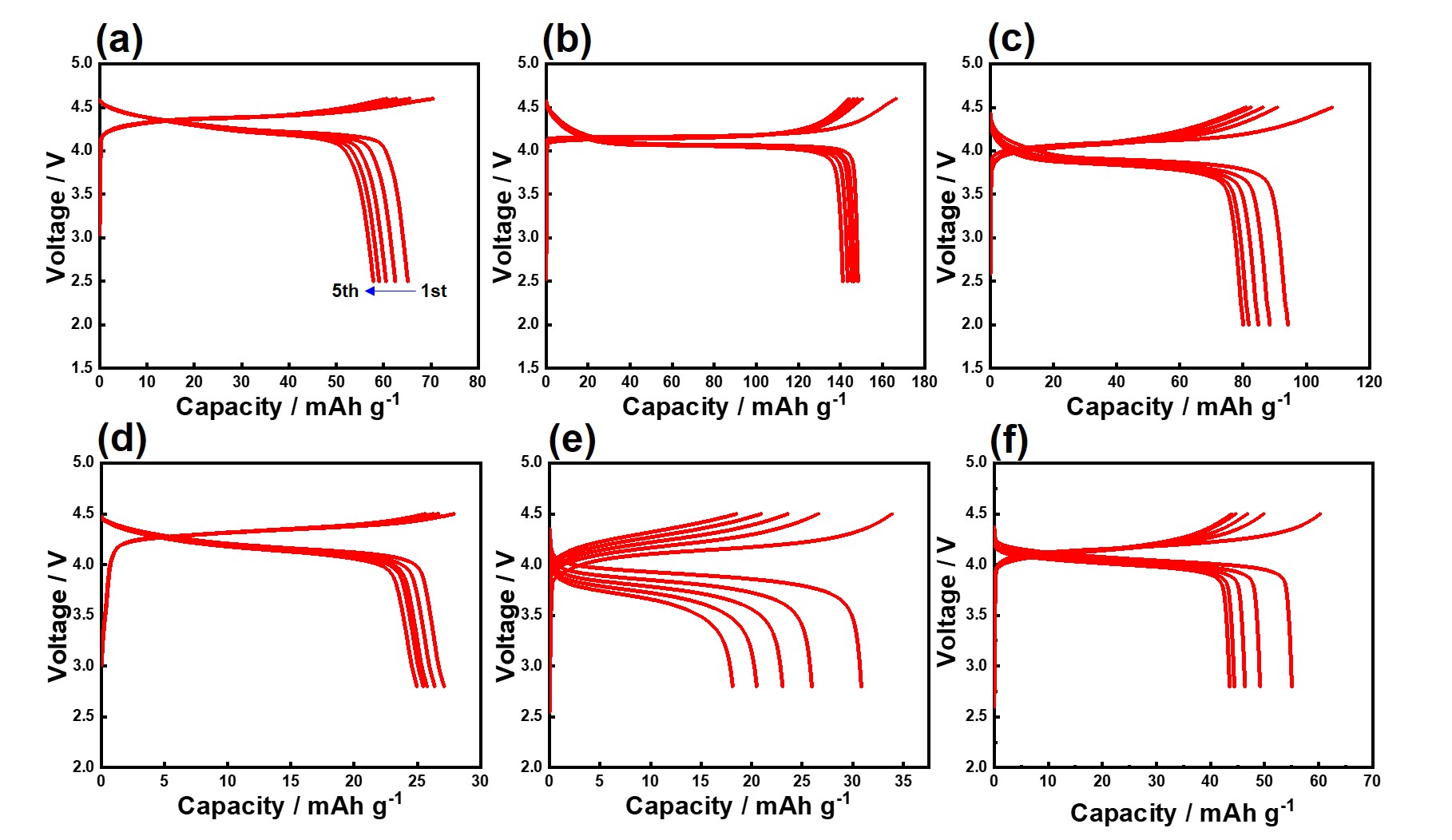}
  \caption{Plots of voltage against capacity for selected pnictogen-and chalcogen-based honeycomb layered oxides containing Li atoms sandwiched between the slabs. Voltage-capacity plots of (a) $\rm Li_4CoTeO_6$, (b) $\rm Li_4NiTeO_6$, (c) $\rm Li_3Ni_2SbO_6$, (d) $\rm Li_2Co_2TeO_6$ (prepared via topochemical ion-exchange route), (e) $\rm Li_2Ni_2TeO_6$ (prepared via topochemical ion-exchange route) and (f) $\rm Li_2Ni_2TeO_6$ (prepared via high-temperature solid-state reaction of $\rm LiNiO_2$ and $\rm TeO_2$). A current density equivalent to (dis)charging for 20 hours was used (technically denoted as C/20 rate). The $x$-axis (capacity scale) has been adjusted accordingly to show the voltage regime operable using the materials.}
  \label{Figure_7}
\end{figure*}

Since their inception, pnictogen- and chalcogen-based honeycomb layered oxides embodying lithium cations have vastly been investigated for their diverse magnetic states that present possible utilities in the emergent fields of condensed matter physics and quantum mechanics. However, it is the discovery of the high-voltage properties established in compositions such as $\rm Li_4NiTeO_6$,\cite{Bao2014, Zvereva2015a, Sathiya2013} $\rm Li_2Ni_2TeO_6$,\cite{Grundish2019} $\rm Li_3Ni_2SbO_6$,\cite{ma2007synthesis, twu2015designing} $\rm Li_4FeSbO_6$,\cite{McCalla2015, McCalla2015a} {\it et cetera}. that has sparked renewed traction in their exploration as high-performance energy materials. In fact, the Li-containing materials have become the conventional cathode materials for lithium-ion battery technologies owing to their ability to facilitate intercalation/deintercalation redox processes using cationic species such as transition metals. \textbf{Figure \ref{Figure_7}} shows the voltage-capacity profiles of some exemplar pnictogen- and chalcogen-based honeycomb layered oxides such as $\rm Li_4CoTeO_6$, $\rm Li_4NiTeO_6$, $\rm Li_2Ni_2TeO_6$, $\rm Li_2Co_2TeO_6$ and $\rm Li_3Ni_2SbO_6$. Most of these materials exhibit voltage plateaux beyond 4 V, placing them as suitable 4V-class positive electrode (cathode) materials for rechargeable lithium-ion batteries.

Considering the unique position occupied by these materials in energy storage applications, it is important to briefly highlight the precursors and protocols employed for their syntheses. \textbf{Table \ref{Table_1}} shows the preparative conditions of Li-based compounds. Most materials in this class utilise $\rm Li_2CO_3$ as a starting reagent for their preparation, although $\rm LiNiO_2$ and $\rm LiCoO_2$ can also be used in the preparation of honeycomb layered tellurates such as $\rm Li_2Ni_2TeO_6$ and $\rm Li_2Co_2TeO_6$, respectively, based on the following reaction: 

\begin{align}
    2\,{\rm Li{\it M}O_2} ({\it M}= \rm Co, Ni) + \,{\rm TeO_2} \rightarrow {\rm Li_2{\it M}_2TeO_6}.
\end{align}

\begin{table*}
\caption{High-temperature solid-state synthetic conditions of selected pnictogen-and chalcogen-based honeycomb layered oxides containing Li atoms sandwiched between the slabs. \cite{castellanos1982crystal, Greaves1990, ma2007synthesis, Kumar2012, Uma2016, Bhardwaj2014, Kumar2013, pal2022insights, mandujano2021, yadav2022influence, Nalbandyan2013, Berthelot2012, berthelot2012studies, Koo2016, Zvereva2012, Sathiya2013, Skakle1997, Zvereva2013, Gupta2015} Preliminary investigations are underscored by asterisks (*).}\label{Table_1}
\begin{center}
\scalebox{0.7}{
\begin{tabular}{lcll} 
\hline
\textbf{Compound} & \textbf{Synthesis technique} & \textbf{Precursors} & \textbf{Firing condition }\\ 
 & & & \textbf{(Temperature, atmosphere)}\\ 
\hline\hline
$\rm Li_8{\it M}_2Te_2O_{12}$ ($\rm {\it M} = Ni, Co, Cu$ and $\rm Zn$)\cite{Kumar2012} & Solid-state reaction & $\rm Li_2CO_3$, $\rm {\it M}O$ ($\rm {\it M} = Ni, Co, Cu$ and $\rm Zn$) or $\rm Co_3O_4$, $\rm TeO_2$ & 650-900$^\circ$C; 12h\\
$\rm Li_8{\it M}_2Sb_2O_{12}$ ($\rm {\it M} = Cr, Fe, Al$ and $\rm Ga$)\cite{Kumar2012} & Solid-state reaction & $\rm Li_2CO_3$, $\rm {\it M}_2O_3$ ($\rm {\it M} = Cr, Fe, Al$ or $\rm Ga$) or $\rm Sb_2O_3$ & 650-850$^\circ$C; 24h\\
$\rm Li_4ZnTeO_6$\cite{Nalbandyan2013} & Solid-state reaction & $\rm Li_2CO_3$, $\rm ZnO$, $\rm TeO_2$ & 600-850$^\circ$C; 12h\\
$\rm Li_4NiTeO_6$\cite{Sathiya2013} & Solid-state reaction & $\rm Li_2CO_3$, $\rm NiC_4H_6O_4 \cdot 4H_2O$, $\rm TeO_2$ & 1000-1050$^\circ$C; 6-15h\\
$\rm Li_3Cu_2SbO_6$\cite{Nalbandyan2013, Skakle1997, Koo2016} & Solid-state reaction & $\rm Li_2CO_3$, $\rm CuO$, $\rm Sb_2O_5$ & 1000-1025$^\circ$C; 12h in $\rm O_2$\\
$\rm Li_3Zn_2SbO_6$\cite{Greaves1990} & Solid-state reaction & $\rm Li_2CO_3$, $\rm ZnO$, $\rm \alpha-Bi_2O_3$ & 600$^\circ$C; 12h in $\rm O_2$\\
$\rm Li_3Ni_2SbO_6$\cite{Zvereva2012} & Solid-state reaction & $\rm Li_2CO_3$, $\rm Ni(OH)_2$, $\rm H_3O_4Sb$ & 700-1150$^\circ$C; 1-2h\\
$\rm Li_3Ni_2SbO_6$ & Solid-state reaction & $\rm Li_3SbO_4$, $\rm NiCO_3$ & 1000$^\circ$C; 5 days\\
$\rm Li_3Ni_2SbO_6$ ($\rm LiNi_{2/3}Sb_{1/3}O_2$) & Solid-state reaction & $\rm Li_2CO_3$, $\rm NiO$, $\rm Sb_2O_3$ & 1100$^\circ$C; 24h in air\\
$\rm Li_3Zn_2SbO_6$ ($\rm LiZn_{2/3}Sb_{1/3}O_2$) & Solid-state reaction & $\rm Li_2CO_3$, $\rm ZnO$, $\rm Sb_2O_3$ & 950$^\circ$C; 12h in $\rm O_2$\\
$\rm Li_3Zn_2BiO_6$ & Solid-state reaction & $\rm Li_2CO_3$, $\rm ZnO$, $\rm Bi_2O_3$ & 550-900$^\circ$C; 12-48h in $\rm O_2$\\
$\rm Li_3Ni{\it M}BiO_6$ ($\rm {\it M} = Mg, Cu$ and $\rm Zn$) & Solid-state reaction & $\rm Li_2CO_3$, $\rm {\it M}O$ ($\rm {\it M} = Mg, Cu$ and $\rm Zn$), $\rm Bi_2O_3$ & 550-900$^\circ$C; 12-48h in $\rm O_2$\\
$\rm Li_3Co_2SbO_6$ & Solid-state reaction & $\rm Li_3SbO_4$, $\rm CoO$ & 1100$^\circ$C; 90min in $\rm Ar$\\
$\rm Li_4FeSbO_6$\cite{Zvereva2013} & Solid-state reaction & $\rm Li_2CO_3$, $\rm Fe_2O_3$), $\rm Sb_2O_5 \cdot 2.74H_2O$ & 747-1047$^\circ$C; 30min-3h\\
$\rm Li_4{\it M}SbO_6$ ($\rm {\it M} = Cr, Fe, Al$ and $\rm Ga$)\cite{Bhardwaj2014} & Solid-state reaction & $\rm Li_2CO_3$, $\rm {\it M}_2O_3$ ($\rm {\it M} = Cr, Fe, Al$ and $\rm Ga$), $\rm Sb_2O_3$ & 650-850$^\circ$C; 12-48h\\
$\rm Li_3Mn_2SbO_6$\cite{yadav2022influence} & Solid-state reaction & $\rm Li_3SbO_4$, $\rm MnO$ & 770$^\circ$C; 12-48h in Ar\\
$\rm Li_3Mg_2SbO_6$\cite{castellanos1982crystal} & Solid-state reaction & $\rm Li_2CO_3$, $\rm MgO$, $\rm Sb_2O_3$ & 1225-1350$^\circ$C; 5h in air\\
$\rm Li_4CoSbO_6$\cite{yadav2022influence} & Solid-state reaction & $\rm Li_2CO_3$, $\rm Co_3O_4$, $\rm Sb_2O_3$ & 1000$^\circ$C; 12h in air\\
$\rm Li_{4.5}{\it M}_{0.5}TeO_6$ ($\rm {\it M} = Cr, Mn, Al$ and $\rm Ga$)\cite{Uma2016} & Solid-state reaction & $\rm Li_2CO_3$, $\rm {\it M}_2O_3$ ($\rm {\it M} = Cr, Al$ and $\rm Ga$) or $\rm MnO$, $\rm TeO_2$ & 950-1100$^\circ$C; 12-48h in air\\
$\rm Li_{4.5}Fe_{0.5}TeO_6$\cite{Gupta2015} & Solid-state reaction & $\rm Li_2CO_3$, $\rm Fe_2O_3$, $\rm TeO_2$ & 650-950$^\circ$C; 12-24h\\
$\rm Li_2Ni_2TeO_6$ (disordered)\cite{Kumar2013} & Solid-state reaction & $\rm Li_2CO_3$, $\rm NiO$, $\rm TeO_2$ & 600-900$^\circ$C; 24-48h\\
$\rm Li_3Mn{\it M}SbO_6$ ($\rm {\it M} = Co, Ni, Mg$ and $\rm Zn$) & Solid-state reaction & $\rm Li_3SbO_4$, $\rm MnO$, $\rm {\it M}O$ ($\rm {\it M} = Co, Ni, Mg$ and $\rm Zn$) & 700$^\circ$C; 12h in $\rm Ar$\\
$\rm Li_3Co{\it M}SbO_6$ ($\rm {\it M} = Ni, Mg$ and $\rm Zn$) & Solid-state reaction & $\rm Li_3SbO_4$, $\rm CoO$, $\rm {\it M}O$ ($\rm {\it M} = Ni, Mg$ and $\rm Zn$) & 700$^\circ$C; 12h in $\rm Ar$\\
$\rm Li_4CrTeO_6$ & Solid-state reaction & $\rm Li_2CO_3$, $\rm Cr_2O_3$, $\rm Te_2O_5$ & 900$^\circ$C; 12h\\
$\rm Li_4CrSbO_6$\cite{Bhardwaj2014} & Solid-state reaction & $\rm Li_2CO_3$, $\rm Cr_2O_3$, $\rm Sb_2O_5$ & 900$^\circ$C; 12h\\
$\rm Li_4{\it M}TeO_6$ ($\rm {\it M} = Ni, Co, Cu, Zn$ and $\rm Mg$)* & Solid-state reaction & $\rm Li_2CO_3$, $\rm {\it M}O$ ($\rm {\it M} = Ni, Co, Cu, Zn$ or $\rm Mg$), $\rm TeO_2$ & 800$^\circ$C; 24h in air\\
$\rm Li_2{\it M}_2TeO_6$ ($\rm {\it M} = Ni$ and $\rm Co$)* & Solid-state reaction & $\rm Li{\it M}O_2$ ($\rm {\it M} = Ni$ and $\rm Co$), $\rm TeO_2$ & 800$^\circ$C; 24h in air\\
$\rm Li_3Ni_2Bi_{0.5}Sb_{0.5}O_6$* & Solid-state reaction & $\rm Li_2CO_3$, $\rm NiO$, $\rm Bi_2O_3$, $\rm Sb_2O_3$ & 900$^\circ$C; 48h in air\\
$\rm Li_4Ni_{0.5}{\it M}_{0.5}TeO_6$ ($\rm {\it M} = Co, Cu, Zn$ and $\rm Mg$)* & Solid-state reaction & $\rm Li_2CO_3$, $\rm NiO$, $\rm {\it M}O$ ($\rm {\it M} = Co, Cu, Zn, Mg$ or $\rm Mg$), $\rm TeO_2$ & 800$^\circ$C; 24h in air\\
\hline
\end{tabular}}
\end{center}
\end{table*}

\subsubsection{Sodium and potassium-based materials}
\begin{figure*}[!t]
\centering
  \includegraphics[width=0.8\columnwidth]{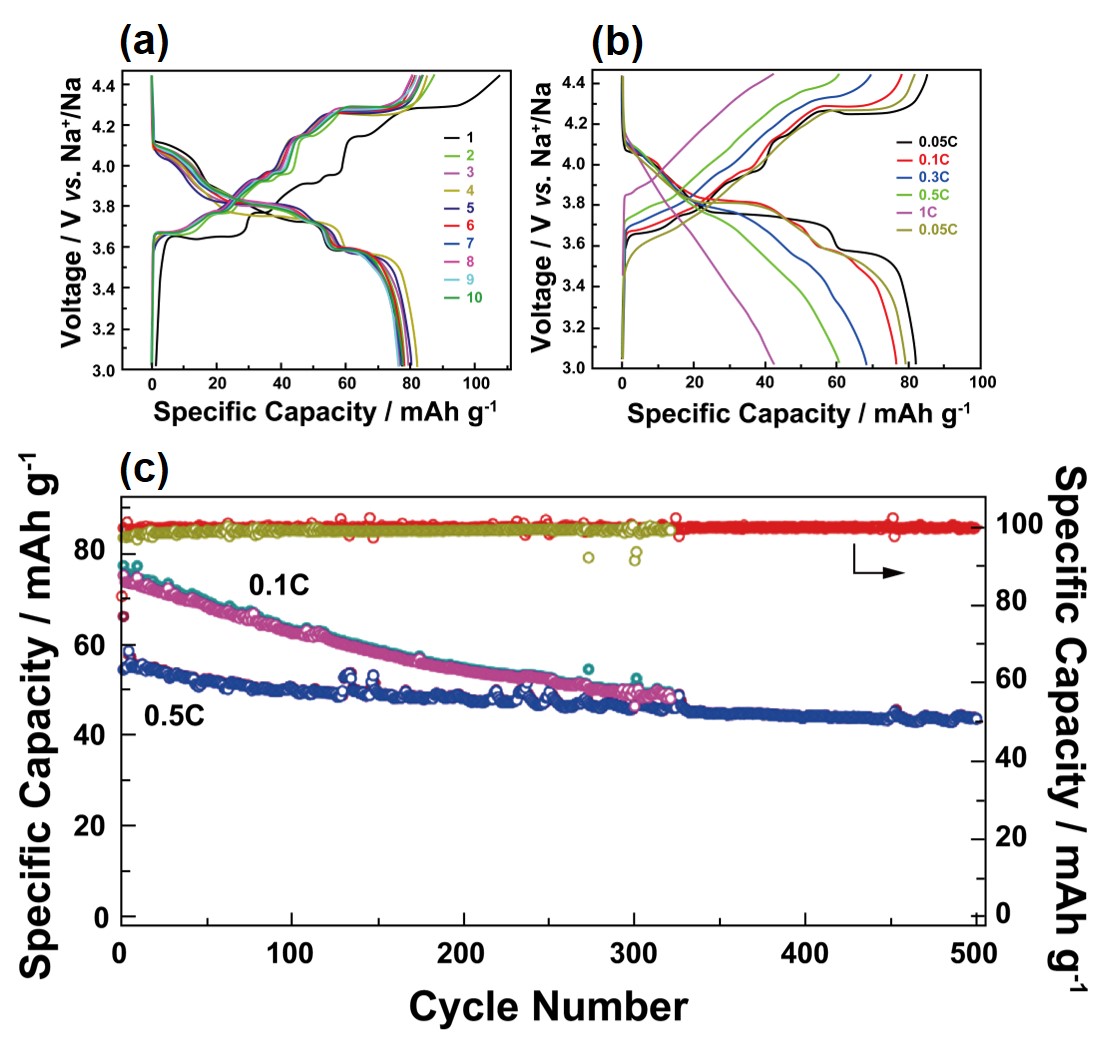}
  \caption{(a) Voltage-capacity plots of $\rm Na_2Ni_2TeO_6$ at a C/20 rate. (b) Rate performance of $\rm Na_2Ni_2TeO_6$ and (c) cycle performance of $\rm Na_2Ni_2TeO_6$. Reproduced with permission.\cite{pati2022unraveling} Copyright 2022 Royal Society of Chemistry.}
  \label{Figure_8}
\end{figure*}

\begin{figure*}[!t]
\centering
  \includegraphics[width=0.8\columnwidth]{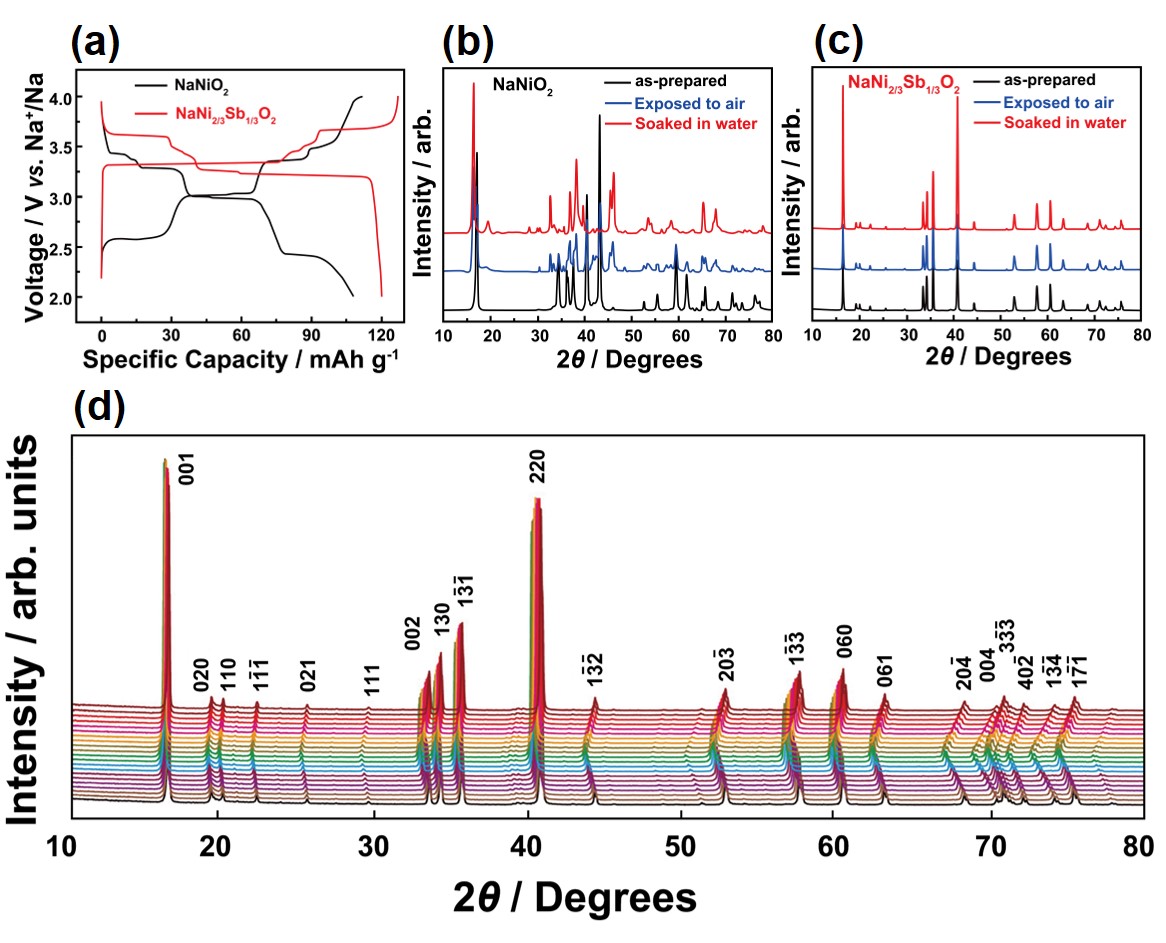}
  \caption{(a) Voltage plots of $\rm Na_3Ni_2SbO_6$ ($\rm NaNi_{2/3}Sb_{1/3}O_2$) and $\rm NaNiO_2$. (b) X-ray diffraction (XRD) patterns of as-synthesised (pristine) and aged $\rm NaNiO_2$. (c) XRD patterns of as-synthesised and aged $\rm NaNi_{2/3}Sb_{1/3}O_2$, highlighting its moisture stability. (d) {\it In situ} XRD patterns of $\rm NaNi_{2/3}Sb_{1/3}O_2$ material during thermal treatment process. Figures reproduced with permission.\cite{Wang2019b}  
  Copyright 2019 Wiley-VCH.}
  \label{Figure_9}
\end{figure*}

Sodium- and potassium-based honeycomb layered oxides are considered more suited to high temperature solid-state synthesis techniques due to their tendency to form ordered structures unlike lithium-based materials which have a propensity towards disordered structures. Despite this, the sodium- and potassium-based systems have displayed numerous similarities with their lithium-based counterparts with respect to the exceptional magnetic properties and high-voltage electrochemistry. As shown by the voltage-capacity profiles in \textbf{Figure \ref{Figure_8}}, the voltage-capacity profiles of selected sodium-based compounds in this class of honeycomb layered oxides ($\rm Na_2Ni_2TeO_6$,\cite{Gupta2015, chen2020} $\rm Na_3Ni_2BiO_6$\cite{Liu2016, Bhange2017} and $\rm Na_3Ni_2SbO_6$\cite{Yuan2014, Ma2015}) have been found to exhibit higher voltages, better cyclability and superior rate performance than conventional layered oxides such as $\rm NaNiO_2$ (\textbf{Figure \ref{Figure_9}a}). The superior performance is accredited to the partial substitution of 1/3 of the Ni in $\rm NaNiO_2$ with pnictogens, such as Sb, to form $\rm NaNi_{2/3}Sb_{1/3}O_2$ ($\rm Na_3Ni_2SbO_6$) honeycomb layered oxide.\cite{Wang2019b} In addition, these materials exhibit enhanced air (\textbf{Figure \ref{Figure_9}b} and \textbf{Figure \ref{Figure_9}c}) and thermal (\textbf{Figure \ref{Figure_9}d}) stability which has been ascribed to the honeycomb-type ordered superlattice (`$\rm Ni_6$-rings') formed by six Ni atoms surrounding the Sb atoms. Besides the enhanced stability established by the ordered `$\rm Ni_6$-rings', the honeycomb frameworks have also been envisioned to accommodate a super-exchange interaction with symmetric atomic configurations and degenerate electronic orbitals that also raise the redox potential (and subsequently high operating voltage) and curtail the phase-transition processes during subsequent charging and discharging. The superior stability of the honeycomb layered oxides in moist air is further substantiated through X-ray diffraction (XRD) as demonstrated by the XRD patterns obtained from $\rm NaNiO_2$ (\textbf{Figure \ref{Figure_9}b}) and $\rm Na_3Ni_2SbO_6$ (\textbf{Figure \ref{Figure_9}c}). Additionally, the thermal stability of $\rm Na_3Ni_2SbO_6$ is further ascertained by the high-temperature XRD patterns displayed in \textbf{Figure \ref{Figure_9}d}.

Apropos of synthesis, $\rm Na_2CO_3$ is typically employed as the starting reagent (Na source) for the preparation of most pnictogen- and chalcogen-based honeycomb layered oxides embodying Na atoms. \textbf{ Table \ref{Table_2}} shows the preparative conditions of Na-based compounds along with K-based compounds.

\begin{table*}
\caption{High-temperature solid-state synthetic conditions of selected pnictogen-and chalcogen-based honeycomb layered oxides containing Na or K atoms sandwiched between the slabs.\cite{Evstigneeva2011, Grundish2019, Masese2018, berthelot2012studies, Berthelot2012a, Schmidt2013, Schmidt2014, Yuan2014, Ma2015, Liu2016, Bhange2017, yadav2022investigation, Yang2017, Nalbandyan2013a, Seibel2013, Viciu2007, grundish2020structural, Evstigneeva2011, Smirnova2005, Miura2006, Miura2008, yadav2023, Yadav2019}}\label{Table_2}
\begin{center}
\scalebox{0.7}{
\begin{tabular}{lcll} 
\hline
\textbf{Compound} & \textbf{Synthesis technique} & \textbf{Precursors} & \textbf{Firing condition }\\ 

 & & & \textbf{(Temperature, atmosphere)}\\ 
\hline\hline
$\rm Na_2{\it M}_2TeO_6$ ($\rm {\it M} = Cu, Co$ and $\rm Ni$)\cite{Grundish2019, Evstigneeva2011} & Solid-state reaction & $\rm Na_2CO_3$, $\rm {\it M}O$ ($\rm {\it M} = Cu, Ni$) or $\rm Co_3O_4$, $\rm TeO_2$ & 600-900$^\circ$C; 24-48h\\
$\rm Na_2{\it M}_2TeO_6$ ($\rm {\it M} = Co, Ni$)\cite{Miura2006} & Solid-state reaction & $\rm Na_2CO_3$, $\rm {\it M}O$ ($\rm {\it M} = Co, Ni$), $\rm TeO_2$ & 860$^\circ$C; 12h in air\\
$\rm Na_2{\it M}_2TeO_6$ ($\rm {\it M} = Zn, Mg$)\cite{Evstigneeva2011} & Solid-state reaction & $\rm Na_2CO_3$, $\rm NaNO_3$, $\rm {\it M}CO_3$ ($\rm {\it M} = Zn, Mg$), $\rm TeO_2$ & 800-820$^\circ$C; 2-4h\\
$\rm Na_2Ni_2TeO_6$\cite{Evstigneeva2011} & Solid-state reaction & $\rm Na_2CO_3$, $\rm NaNO_3$, hydrous nickel oxide, $\rm TeO_2$ & 800-820$^\circ$C; 2-4h\\
$\rm Na_2Co_2TeO_6$\cite{Evstigneeva2011} & Solid-state reaction & $\rm Na_2CO_3$, $\rm NaNO_3$, hydrous cobalt oxide, $\rm TeO_2$ & 800-820$^\circ$C; 2-4h\\
$\rm Na_2Cu_2TeO_6$\cite{Miura2006} & Solid-state reaction & $\rm Na_2CO_3$, $\rm CuO$, $\rm TeO_2$ & 600-900$^\circ$C; 24-72h\\
$\rm Na_2Zn_2TeO_6$\cite{Berthelot2012a} & Solid-state reaction & $\rm Na_2CO_3$, $\rm ZnO$, $\rm TeO_2$ & 900$^\circ$C; 24h in air\\
$\rm Na_2Co_2TeO_6$\cite{Viciu2007} & Solid-state reaction & $\rm Na_2CO_3$, $\rm Co_3O_4$, $\rm TeO_2$ & 800$^\circ$C; 8 days in $\rm N_2$\\
$\rm Na_4NiTeO_6$\cite{Yang2017} & Solid-state reaction & $\rm Na_2CO_3$, $\rm NiO$, $\rm TeO_2$ & 900$^\circ$C; 10h\\
$\rm Na_3Ni_{1.5}TeO_6$\cite{grundish2020structural} & Solid-state reaction & $\rm Na_2CO_3$, $\rm NiO$, $\rm TeO_2$ & 900-1000$^\circ$C; 12-24h\\
$\rm Na_3{\it M}_2SbO_6$ ($\rm {\it M} = Zn, Cu, Ni$ and $\rm Mg$)\cite{Schmidt2013} & Solid-state reaction & $\rm Na_2CO_3$, $\rm {\it M}O$ ($\rm {\it M} = Cu, Ni, Zn$ and $\rm Mg$) or $\rm Co_3O_4$, $\rm Sb_2O_3$ & 900$^\circ$C; 12h in air\\
$\rm Na_3Ni_2SbO_6$\cite{Yuan2014} & Solid-state reaction & $\rm Na_2CO_3$, $\rm NiO$, $\rm Sb_2O_3$ & 900$^\circ$C; 12h\\
$\rm Na_3Ni_2SbO_6$\cite{Ma2015} & Solid-state reaction & $\rm Na_2CO_3$, $\rm Ni(OH)_2$, $\rm Sb_2O_3$ & 750-1000$^\circ$C; 5h\\
$\rm Na_3Ni_2SbO_6$ polytype\cite{Ma2015} & Solid-state reaction & $\rm Na_2CO_3$, $\rm Ni(OH)_2$, $\rm Sb_2O_3$ & 1200$^\circ$C; 4h\\
$\rm Na_3Co_2SbO_6$\cite{Viciu2007} & Solid-state reaction & $\rm Na_2CO_3$, $\rm Co_3O_4$, $\rm Sb_2O_4$ & 800$^\circ$C; 8 days in $\rm N_2$\\
$\rm Na_3Cu_2SbO_6$\cite{Miura2006, Miura2008} & Solid-state reaction & $\rm Na_2CO_3$, $\rm CuO$, $\rm Sb_2O_3$ or $\rm Sb_2O_5$ & 850-1150$^\circ$C; 2h in air\\
$\rm Na_3Cu_2SbO_6$\cite{Smirnova2005} & Solid-state reaction & $\rm Na_2CO_3$, $\rm Cu(OH)_2CO_3$, $\rm Sb_2O_5 \cdot {\it x}H_2O$ & 850-1050$^\circ$C; 2-3h in air\\
$\rm Na_3Mn_2SbO_6$ (disordered)\cite{yadav2023} & Solid-state reaction & $\rm Na_2CO_3$, $\rm MnO$, $\rm NaSbO_3$ & 900$^\circ$C; 24h in Ar\\
$\rm Na_3Mn_2SbO_6$\cite{Yadav2019} & Solid-state reaction & $\rm Na_2CO_3$, $\rm MnO$, $\rm NaSbO_3$ & 820$^\circ$C; 16h in Ar\\
$\rm Na_3Fe_2SbO_6$\cite{Yadav2019} & Solid-state reaction & $\rm Na_2CO_3$, $\rm FeO$, $\rm NaSbO_3$ & 760$^\circ$C; 24h in Ar\\
$\rm Na_3MnFeSbO_6$\cite{Yadav2019} & Solid-state reaction & $\rm Na_2CO_3$, $\rm MnO$, $\rm FeO$, $\rm NaSbO_3$ & 770$^\circ$C; 12h in Ar\\
$\rm Na_3MnCoSbO_6$\cite{Yadav2019} & Solid-state reaction & $\rm Na_2CO_3$, $\rm MnO$, $\rm Co_3O_4$, $\rm NaSbO_3$ & 900$^\circ$C; 12h in Ar\\
$\rm Na_4CuTeO_6$ & Solid-state reaction & $\rm Na_2CO_3$, $\rm CuCO_3 \cdot Cu(OH)_2 \cdot H_2O$, $\rm TeO_2$, $\rm Na_2MoO_4$ & 1000$^\circ$C; 24h\\
$\rm Na_3Ni_2BiO_6$\cite{Liu2016, Bhange2017} & Solid-state reaction & $\rm Na_2CO_3$, $\rm NiO$, $\rm NaBiO_3$ & 900$^\circ$C; 12h in $\rm O_2$\\
$\rm Na_3Ni_2BiO_6$ (disordered)\cite{Liu2016} & Solid-state reaction & $\rm Na_2CO_3$, $\rm NiO$, $\rm Bi_2O_3$ & 700-750$^\circ$C; 2-20h in $\rm O_2$\\
$\rm Na_3{\it M}_2BiO_6$ ($\rm {\it M} = Ni, Zn, Mg$)\cite{Seibel2013} & Solid-state reaction & $\rm Na_2CO_3$, $\rm {\it M}O$ ($\rm {\it M} = Ni, Zn$) or $\rm Mg(OH)_2$, $\rm NaBiO_3$ & 700-750$^\circ$C; 12-40h in $\rm O_2$\\
$\rm Na_3Mn{\it M}SbO_6$ ($\rm {\it M} = Ni, Zn$ and $\rm Mg$)\cite{yadav2022investigation} & Solid-state reaction & $\rm Na_2CO_3$, $\rm {\it M}O$ ($\rm {\it M} = Ni, Mg, Zn$), $\rm MnO$, $\rm NaSbO_3$ & 820-900$^\circ$C; 12-16h in $\rm Ar$\\
$\rm Na_3Fe{\it M}SbO_6$ ($\rm {\it M} = Co, Ni, Zn$ and $\rm Mg$)\cite{yadav2022investigation} & Solid-state reaction & $\rm Na_2CO_3$, $\rm {\it M}O$ ($\rm {\it M} = Co, Ni, Mg, Zn$), $\rm FeO$, $\rm NaSbO_3$ & 770-900$^\circ$C; 12-16h in $\rm Ar$\\
$\rm Na_2LiFeTeO_6$\cite{Nalbandyan2013a} & Solid-state reaction & $\rm Fe_2TeO_6$, $\rm Li_2CO_3$, $\rm Na_2CO_3$ & 630-700$^\circ$C; 1-13h in air\\
$\rm Na_2LiFeTeO_6$\cite{Nalbandyan2013a} & Solid-state reaction & $\rm Na_2CO_3$, $\rm LiFeO_2$, $\rm NaNO_3$, $\rm TeO_2$ & 630-700$^\circ$C; 1-13h\\
$\rm Na_4FeSbO_6$\cite{Schmidt2014} & Solid-state reaction & $\rm Na_2CO_3$, $\rm Fe_2O_3$, $\rm Sb_2O_3$ & 1000$^\circ$C; 48h\\
$\rm Na_3LiFeSbO_6$\cite{Schmidt2014} & Solid-state reaction & $\rm Na_2CO_3$, $\rm Li_2CO_3$, $\rm Fe_2O_3$, $\rm Sb_2O_3$ & 1000$^\circ$C; 48h\\
$\rm Na_3Ni_2Bi_{0.5}Sb_{0.5}O_6$* & Solid-state reaction & $\rm Na_2CO_3$, $\rm NiO$, $\rm Bi_2O_3$, $\rm Sb_2O_3$ & 900$^\circ$C; 48h in air\\
$\rm K_2{\it M}_2TeO_6$ ($\rm {\it M} = Ni, Mg$ and $\rm Co$)\cite{Masese2018} & Solid-state reaction & $\rm K_2CO_3$, $\rm {\it M}O$ ($\rm {\it M} = Ni, Mg$) or $\rm Co_3O_4$, $\rm TeO_2$ & 700-900$^\circ$C; 24-48h\\
$\rm K_3Ni_2Bi_{0.5}Sb_{0.5}O_6$* & Solid-state reaction & $\rm K_2CO_3$, $\rm NiO$, $\rm Bi_2O_3$, $\rm Sb_2O_3$ & 900$^\circ$C; 48h in air\\
$\rm K_3{\it M}_2BiO_6$ ($\rm {\it M} = Ni, Cu, Zn, Mg$ and $\rm Co$)* & Solid-state reaction & $\rm K_2CO_3$, $\rm {\it M}O$ ($\rm {\it M} = Co, Ni, Cu, Mg, Zn$), $\rm Bi_2O_3$ & 800-900$^\circ$C; 15h in air\\
$\rm K_3Ni{\it M}BiO_6$ ($\rm {\it M} = Co, Cu, Zn, Mg$ and $\rm Co$)* & Solid-state reaction & $\rm K_2CO_3$, $\rm NiO$, $\rm {\it M}O$ ($\rm {\it M} = Co, Cu, Mg, Zn$), $\rm Bi_2O_3$ & 800-900$^\circ$C; 15h in air\\
\hline
\end{tabular}}
\end{center}
\end{table*}

\subsubsection{Copper-based materials}

Layered oxide compositions entailing coinage metal atoms (such as Cu) interspersed between arrays of transition metal atoms have gained traction in a manifold of fields ranging from fundamental studies to applied utilities. In particular, these materials have evoked momentous pedagogical interest in solid-state chemistry on account of the exquisite properties emerging from the peculiar coordination bonds formed between the copper atoms and the oxygen atoms in the honeycomb slabs. For instance, the remarkable photoelectrochemical water splitting capabilities observed in compositions such as $\rm CuFeO_2$ and $\rm CuRhO_2$ have driven their utility in photocathode applications, whilst compositions such as $\rm CuAlO_2$, $\rm CuGaO_2$, and other members of the broad Delafossite family show superb conductivities that have made them prolific as $p$-type transparent conducting oxides in thin film assembly.\cite{kawazoe1997p, snure2007cu, marquardt2006crystal} Related phases such as $\rm CuCrO_2$ have also been found to host exotic ferroelectric and thermoelectric properties stemming from their transition metal octahedra which are aligned in a triangular geometry: a typical structural motif for studying magnetic frustration. 

The fabrication of honeycomb layered antimonates such as $\rm Cu_3Ni_2SbO_6$ and $\rm Cu_3Co_2SbO_6$ typically entails calcination performed at high temperatures (1000$^{\circ}$ C) in air.\cite{Nagarajan2002} Most of these compositions have been reported as disordered Delafossites with randomly distributed Sb and Co (or Ni) atoms within the metal slabs. The metal slabs were also found to crystallise in a variety of metal slab arrangements (stacking variants) that form a mixture of layered phases. However, it is worth mentioning that other synthesis attempts have found that calcination performed at higher temperatures just below the starting reagent melting points resulted in stacking variants that are nearly fully ordered, as has been affirmed by high-resolution electron microscopy and synchrotron powder X-ray diffraction. \textbf{ Table \ref{Table_3}} shows the preparative conditions of Cu-based compounds.

\begin{table*}
\caption{High-temperature solid-state synthetic conditions of selected pnictogen-and chalcogen-based honeycomb layered oxides containing Cu atoms sandwiched between the slabs.\cite{Nagarajan2002, roudebush2015rhombohedral, Roudebush2015, Ramlau2014, Roudebush2013b, climent2012spin}}\label{Table_3}
\begin{center}
\scalebox{0.9}{
\begin{tabular}{lcll} 
\hline
\textbf{Compound} & \textbf{Synthesis technique} & \textbf{Precursors} & \textbf{Firing condition }\\ 

 & & & \textbf{(Temperature, atmosphere)}\\ 
\hline\hline
$\rm Cu_3{\it M}_2SbO_6$ ($\rm {\it M} = Co$ and $\rm Ni$)\cite{Nagarajan2002} & Solid-state reaction & $\rm Cu_2O$, $\rm {\it M}O$ ($\rm {\it M} = Co, Ni$), $\rm Sb_2O_3$ & 1100$^\circ$C; 24h\\
$\rm Cu_3Mn_2SbO_6$\cite{Nagarajan2002} & Solid-state reaction & $\rm Cu_2O$, $\rm Mn_2O_3$, $\rm Sb_2O_3$ & 900$^\circ$C; 24h in vacuum\\
$\rm Cu_3{\it M}_2SbO_6$ ($\rm {\it M} = Zn$ and $\rm Mg$)\cite{Nagarajan2002} & Solid-state reaction & $\rm Cu_2O$, $\rm {\it M}O$ ($\rm {\it M} = Zn, Mg$), $\rm Sb_2O_3$ & 1100-1150$^\circ$C; 24h\\
$\rm Cu_3Zn_2SbO_6$\cite{Roudebush2013b} & Solid-state reaction & $\rm Cu_2O$, $\rm ZnO$, $\rm Sb_2O_5$ & 1150$^\circ$C\\
$\rm Cu_3Ni_2SbO_6$\cite{roudebush2015rhombohedral, Roudebush2013b} & Solid-state reaction & $\rm Cu_2O$, $\rm NiO$, $\rm Sb_2O_5$ & 800-1360$^\circ$C; 24-72h in air\\
$\rm Cu_3Co_2SbO_6$\cite{roudebush2015rhombohedral, Roudebush2013b} & Solid-state reaction & $\rm Cu_2O$, $\rm Co_3O_4$, $\rm Sb_2O_5$ & 800-1260$^\circ$C; 24-72h in air\\
$\rm Cu_5SbO_6$\cite{climent2012spin} & Solid-state reaction & $\rm CuO$, $\rm Sb_2O_3$ & 950-1130$^\circ$C; 40h in air\\
\hline
\end{tabular}}
\end{center}
\end{table*}

\subsubsection{Mixed alkali-metal-based materials}

\begin{figure*}[!t]
\centering
  \includegraphics[width=0.8\columnwidth]{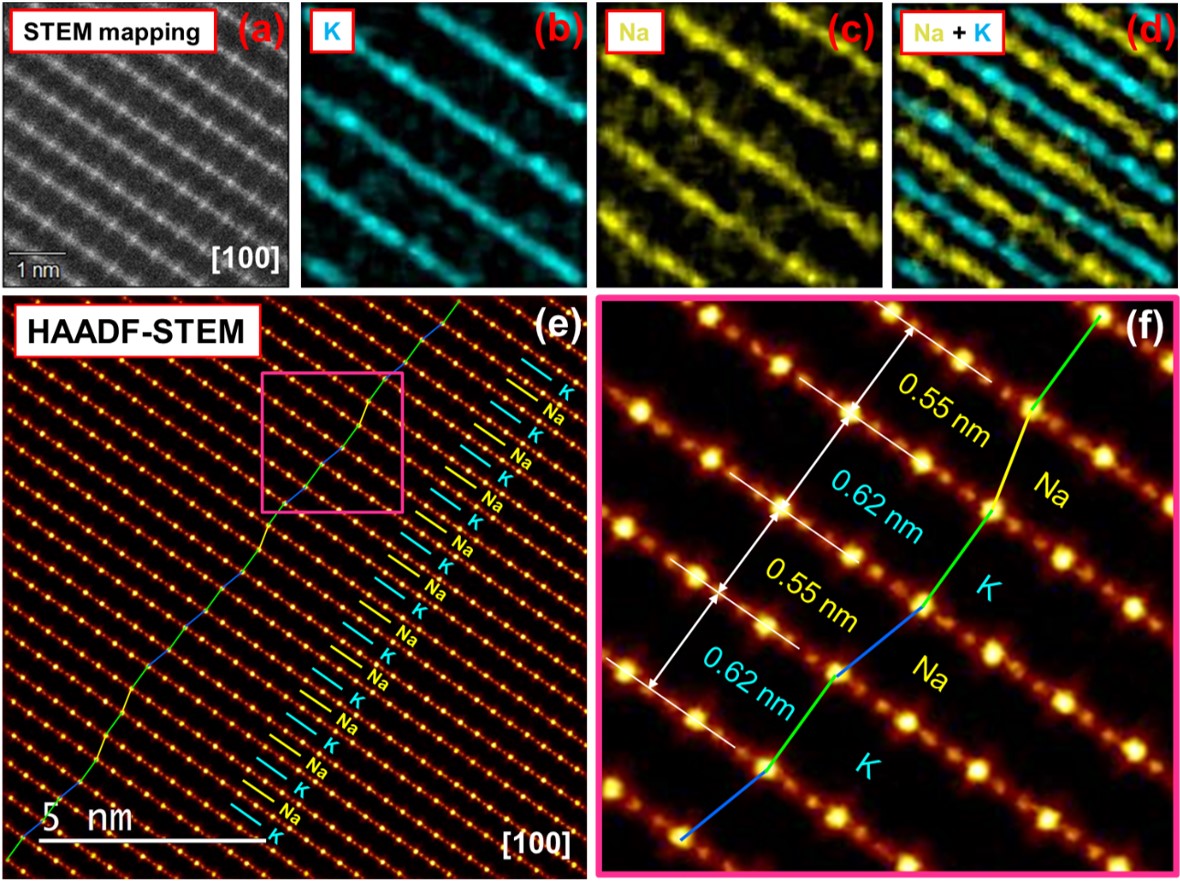}
  \caption{Arrangement of K and Na atoms in mixed alkali honeycomb layered $\rm NaKNi_2TeO_6$. (a) HAADF-STEM image of $\rm NaKNi_2TeO_6$ taken along the [100] zone axis, revealing bright planes corresponding to the layers consisting of Ni and Te. (b,c,d) STEM-EDX mapping of the area shown in (a), where the colours explicitly visualise the distribution of K and Na atoms. The EDX maps show that the layers occupied by K alternate with those of Na. (e) HAADF-STEM image illustrating the unique stacking sequence in $\rm NaKNi_2TeO_6$. In the layers where Na atoms occupy the interlayer space, shifts of the Ni / Te slabs are observed. The blue and yellow lines show shifts in different directions. Note the aperiodicity in the stacking sequence. However, for layers where K atoms reside, Ni / Te slabs are not shifted with respect to each other (marked by a green line). Theoretical computations have revealed a multitude of configurations that can be adopted in the arrangement of Na and K atoms within the slab.\cite {berthelot2021stacking} (f) Enlarged view of the domain highlighted in (e), showing that the interlayer distance is dependent on the alkali atom species (K or Na) sandwiched between the Ni / Te layers. Figures reproduced with permission.\cite{masese2021mixed} 
  Copyright 2021 Springer Nature.}
  \label{Figure_10}
\end{figure*}

\begin{figure*}[!t]
\centering
  \includegraphics[width=0.8\columnwidth]{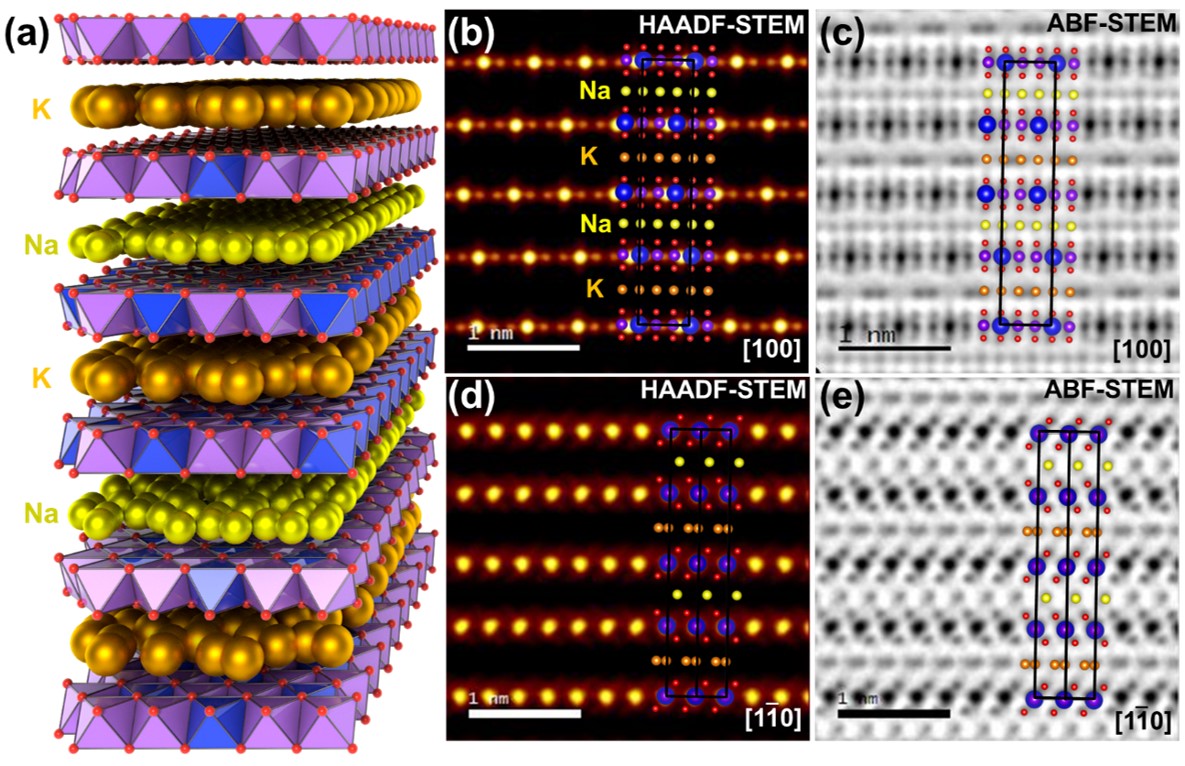}
  \caption{High-resolution STEM imaging of the aperiodic atomic structure of $\rm NaKNi_2TeO_6$ taken along [100] and [1$\overline{1}$0] zone axes. O atoms are in red, and the Ni and Te atoms are in purple and blue, respectively. K and Na atoms are shown in brown and yellow, respectively. (a) Atomistic model of the average aperiodic structure of $\rm NaKNi_2TeO_6$ acquired based on STEM analyses along the [100] zone axis. (b) Superimposition of the model on a HAADF-STEM image taken along the [100] zone axis, showing an excellent overlap between the positions of the Te and Ni atoms in the model that is in line with the intensity distribution of the atom spots observed in the image. (c) Superimposition of the model on an annular dark-field (ADF) image taken along the [100] zone axis, confirming the atomic positions of O, K and Na. (d) Superimposition of the model on a HAADF-STEM image taken along the [1$\overline{1}$0] zone axis, confirming the atomic positions of Te and Ni. (e) Superimposition of the model on an ABF-STEM image taken along the [1$\overline{1}$0] zone axis, also showing an excellent overlap between the positions of the O, K and Na atoms in the model that is in line with the intensity distribution of the atom spots observed in the image. Figures reproduced with permission.\cite{masese2021mixed} Copyright 2021 Springer Nature.}
  \label{Figure_11}
\end{figure*}

To a marked extent, the scientific rigour surrounding the pnictogen and chalcogen class of honeycomb layered oxides is driven by their structural versatility which offers ample possibilities for developing neoteric crystal structures such as honeycomb layered compositions with multiple alkali metal cations. For instance, high-resolution transmission electron microscopy (\textbf{Figure \ref{Figure_10}} and \textbf{Figure \ref{Figure_11}}) performed on a mixed alkali honeycomb layered tellurate ($\rm NaKNi_2TeO_6$) demonstrated that the commixture of Na and K atoms formed unique cation configurations within the layers, yielding versatile stacking structures.\cite{masese2021mixed, berthelot2021stacking} This structural versatility has further facilitated the design of various crystal structures entailing the $\rm NaK{\it M}_2TeO_6$ ($\rm {\it M} = Co, Zn, Mg$ and $\rm Cu$) mixed-alkali compositions,\cite{masese2021mixed} opening avenues for the discovery of new functionalities amongst these materials. Within this framework, $\rm NaKNi_2TeO_6$ has found great utility not only as a promising high-voltage cathode material for dual-ion battery application, but also as a variant material to the typical $A_2\rm Ni_2TeO_6$ ($A = \rm Na, K$) for testing higher order effects such as shear and dislocations in the emergent 2D quantum geometries.\cite{kanyolo2022cationic, masese2023honeycomb, Kanyolo2020, kanyolo2022advances2, kanyolo2021honeycomb, masesemaths, kanyolo2021reproducing, kanyolo2022local, kanyolo2021partition} \textbf{ Table \ref{Table_4}} shows the preparative conditions of mixed alkali-metal-based materials. For example, $\rm NaKNi_2TeO_6$ can be prepared via solid-state reaction of $\rm Na_2Ni_2TeO_6$ and $\rm K_2Ni_2TeO_6$ as follows:

\begin{align}
    \,{\rm Na_2 Ni_2TeO_6} + \,{\rm K_2 Ni_2TeO_6} \rightarrow 2{\rm NaKNi_2TeO_6}.
\end{align}

Note that compositions such as $\rm Na_2LiFeTeO_6$\cite{Nalbandyan2013a} and $\rm Na_3LiFeSbO_6$\cite{Schmidt2014} possess alkali metal atoms (in this case Li) in the transition metal slab but not sandwiched between the slab together with Na; therefore, such compositions are not considered `mixed-alkali' in this context. Nevertheless, their syntheses conditions along with related compositions (such as $\rm Cu_3NaFeSbO_6$,\cite{sethi2022cuprous} $\rm Ag_3NaFeSbO_6$,\cite{politaev2009subsolidus} $\rm Ag_3Li{\it M}SbO_6$ ($M  = \rm  Fe, Mn, Cr$)\cite{Bhardwaj2014} and $\rm Ag_3Li{\it M}TeO_6$ ($M  = \rm  Ni, Co$)\cite{Kumar2012} have been availed in \textbf{ Table \ref{Table_7}} and \textbf{ Table \ref{Table_8}}, for completeness.

\begin{table*}
\caption{High-temperature solid-state synthetic conditions of selected pnictogen-and chalcogen-based honeycomb layered oxides comprising mixed alkali atoms sandwiched between the slabs.\cite{masese2021mixed, berthelot2021stacking, Vallee2019} Preliminary investigations are underscored by asterisks (*).}\label{Table_4}
\begin{center}
\scalebox{0.7}{
\begin{tabular}{lcll} 
\hline
\textbf{Compound} & \textbf{Synthesis technique} & \textbf{Precursors} & \textbf{Firing condition }\\ 

 & & & \textbf{(Temperature, atmosphere)}\\ 
\hline\hline
$\rm NaKNi_2TeO_6$\cite{masese2021mixed} & Solid-state reaction & $\rm Na_2CO_3$, $\rm K_2CO_3$, $\rm NiO$, $\rm TeO_2$ & 800-840$^\circ$C; 24h in air\\
$\rm NaKNi_2TeO_6$\cite{masese2021mixed, berthelot2021stacking} & Solid-state reaction & $\rm Na_2Ni_2TeO_6$, $\rm K_2Ni_2TeO_6$ & 200-840$^\circ$C; 9h-4 days in air\\
$\rm NaK{\it M}_2TeO_6$ ($\rm {\it M} = Co, Mg, Zn$ and $\rm Cu$)\cite{masese2021mixed} & Solid-state reaction & $\rm Na_2CO_3$, $\rm K_2CO_3$, $\rm {\it M}O$ ($\rm {\it M} = Co, Mg, Zn, Cu$), $\rm TeO_2$ & 800-840$^\circ$C; 24h in air\\
$\rm Li_{3-{\it x}}Na_{\it x}Ni_2SbO_6$\cite{Vallee2019} & Solid-state reaction & $\rm Na_2CO_3$, $\rm Li_2CO_3$, $\rm NiO$, $\rm Sb_2O_3$ & 900$^\circ$C; 5h in air\\
$\rm Li_{1.5}Na_{1.5}Ni_2SbO_6$\cite{Vallee2019} & Solid-state reaction & $\rm Li_2CO_3$, $\rm Na_2CO_3$, $\rm NiO$, $\rm Sb_2O_3$ & 800-840$^\circ$C; 24h in air\\
$\rm Li_{1.5}Na_{1.5}Ni_2SbO_6$* & Solid-state reaction & $\rm Li_3Ni_2SbO_6$, $\rm Na_3Ni_2SbO_6$ & 800-900$^\circ$C; 24h in air\\
$\rm NaLiNi_2TeO_6$* & Solid-state reaction & $\rm Na_2CO_3$, $\rm Li_2CO_3$, $\rm NiO$, $\rm TeO_2$ & 800-840$^\circ$C; 24h in air\\
$\rm NaLiNi_2TeO_6$* & Solid-state reaction & $\rm Na_2Ni_2TeO_6$, $\rm Li_2Ni_2TeO_6$ & 800-840$^\circ$C; 24h in air\\
$\rm KLiNi_2TeO_6$* & Solid-state reaction & $\rm K_2CO_3$, $\rm Li_2CO_3$, $\rm NiO$, $\rm TeO_2$ & 800-840$^\circ$C; 24h in air\\
$\rm KLiNi_2TeO_6$* & Solid-state reaction & $\rm K_2Ni_2TeO_6$, $\rm Li_2Ni_2TeO_6$ & 800-840$^\circ$C; 24h in air\\
$\rm K_{2/3}Na_{2/3}Li_{2/3}Ni_2TeO_6$* & Solid-state reaction & $\rm K_2CO_3$, $\rm Na_2CO_3$, $\rm Li_2CO_3$, $\rm NiO$, $\rm TeO_2$ & 800-840$^\circ$C; 24h in air\\
$\rm K_{2/3}Na_{2/3}Li_{2/3}Ni_2TeO_6$* & Solid-state reaction & $\rm K_2Ni_2TeO_6$, $\rm Na_2Ni_2TeO_6$, $\rm Li_2Ni_2TeO_6$ & 800-840$^\circ$C; 24h in air\\
\hline
\end{tabular}}
\end{center}
\end{table*}

\subsubsection{Alkali-metal-excess materials}
\begin{figure*}[!t]
\centering
  \includegraphics[width=0.8\columnwidth]{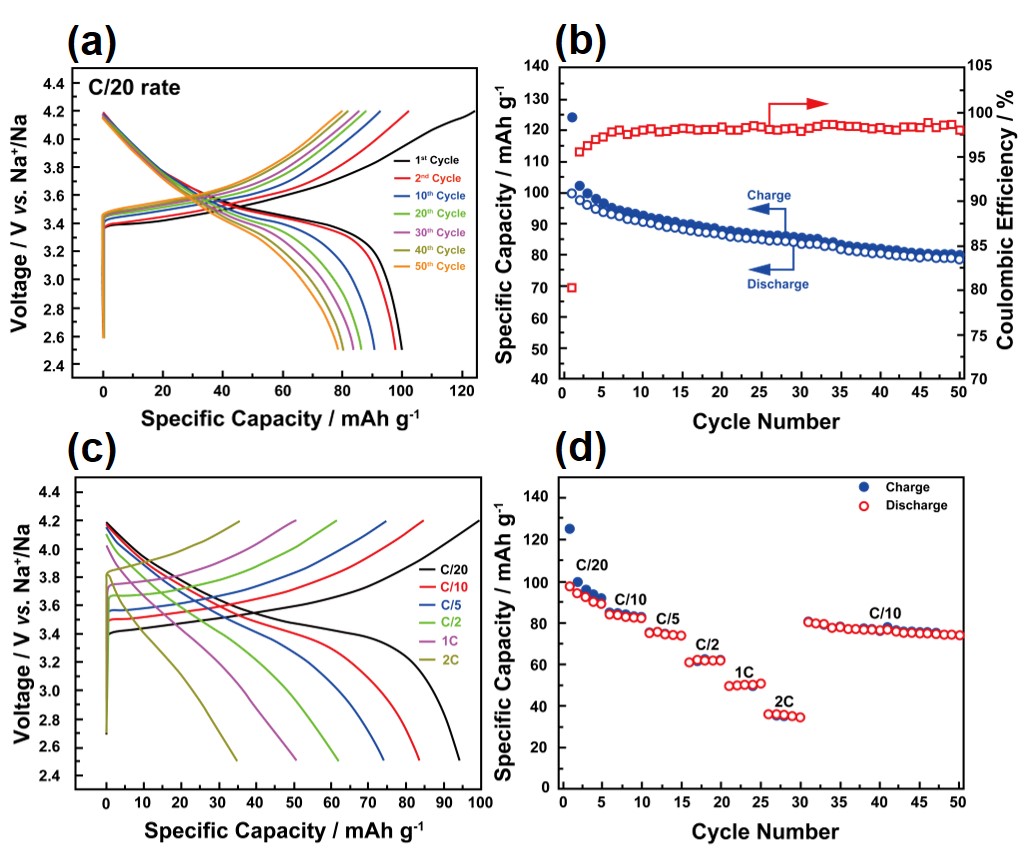}
  \caption{Electrochemical performance of alkali-excess $\rm Na_3Ni_{1.5}TeO_6$. (a) Voltage-capacity plots. (b) Cyclability plots. (c,d) Rate performance plots. Figures reproduced with permission.\cite{grundish2020structural}  
  Copyright 2019 American Chemical Society.}
  \label{Figure_12}
\end{figure*}

Besides their ability to host multiple cation species, the versatile crystal frameworks engendered by the pnictogen and chalcogen class of honeycomb layered oxides have also been found to have the flexibility to accommodate surplus amounts of alkali metal atoms to form exotic compositions such as in $\rm Na_{2.4}Ni_2TeO_6$.\cite {samarakoon2021static} Inquests on this class of materials have reported interesting magnetic dynamics that render them an ideal platform for fine-tuning the dimensionality of magnetic lattices and evaluating bond-dependent anisotropic interactions.\cite {samarakoon2021static}
Moreover, Na-rich compositions such as $\rm Na_{2{\it x}}Zn_2TeO_6$\cite{itaya2021effect, itaya2021sintering} have been found to propagate high ionic conductivities that make them promising electrode and electrolyte materials for rechargeable Na-ion batteries. As clearly shown in the electrochemical performance plots of $\rm Na_3Ni_{1.5}TeO_6$\cite{Grundish2019} in \textbf{Figure \ref{Figure_12}}, this class of materials engenders voltage profiles comparable to those of commercialised lithium-based layered oxides such as $\rm LiCoO_2$. Additionally, $\rm Na_3Ni_{1.5}TeO_6$ displays fast ${\rm Na^{+}}$ kinetics and good capacity retention even though the constituent particles are in the micrometric range which tends to inhibit fast ${\rm Na^{+}}$ diffusion. The structural nuances observed in such Na-rich materials and their electrochemical implications can be envisioned as guidelines for designing new layered oxide cathode materials. In the preparation of materials in this class, the following solid-state reaction used to prepare $\rm Na_3Ni_{1.5}TeO_6$ can be applied: 

\begin{align}
    1.5\,{\rm Na_2CO_3} + 1.5\,{\rm NiO} + {\rm TeO_2} \rightarrow {\rm Na_3Ni_{1.5}TeO_6}.
\end{align}

\subsubsection{Alkali-earth metal-based materials}
Efforts to delineate the influence of divalent atoms on the crystal lattice and magnetic coupling of inter- and intra-honeycomb layers and their emergent magnetic properties have stimulated research interests in the design and development of pnictogen- and chalcogen-based honeycomb layered oxides that can accommodate alkaline-earth metal atoms suchlike Ca, Mg, Ba, {\it et cetera}. between the transition metal slabs. This quest for alkaline-earth metal-based honeycomb layered oxides (such as those consisting of Mg and Ca cations) is further propounded by the scarcity of high-voltage materials for rechargeable multivalent batteries. This class of honeycomb layered oxides present excellent prospects as high-voltage multivalent cathode materials due to the high volumetric energy densities proffered by their high material densities as has been ascertained using measurements such as pycnometry. Against this backdrop, the successful design of $\rm BaNi_2TeO_6$ can be viewed as a significant landmark, showing the feasibility to design related honeycomb layered oxides which can incorporate Mg or Ca within the metal slabs. Single-crystal samples of $\rm BaNi_2TeO_6$\cite{song2022influence} can be obtained via calcination of reagents at high temperatures using molten flux, a topic that will be tackled in a later section. $\rm BaNi_2TeO_6$ can be prepared via the following solid-state reaction:

\begin{align}
    \,{\rm BaCO_3} + 2\,{\rm NiO} + \,{\rm TeO_2} \rightarrow {\rm BaNi_2TeO_6}.
\end{align}

\subsubsection{Multi-transition metal-based (high-entropy oxides) materials}
\begin{figure*}[!t]
\centering
  \includegraphics[width=0.8\columnwidth]{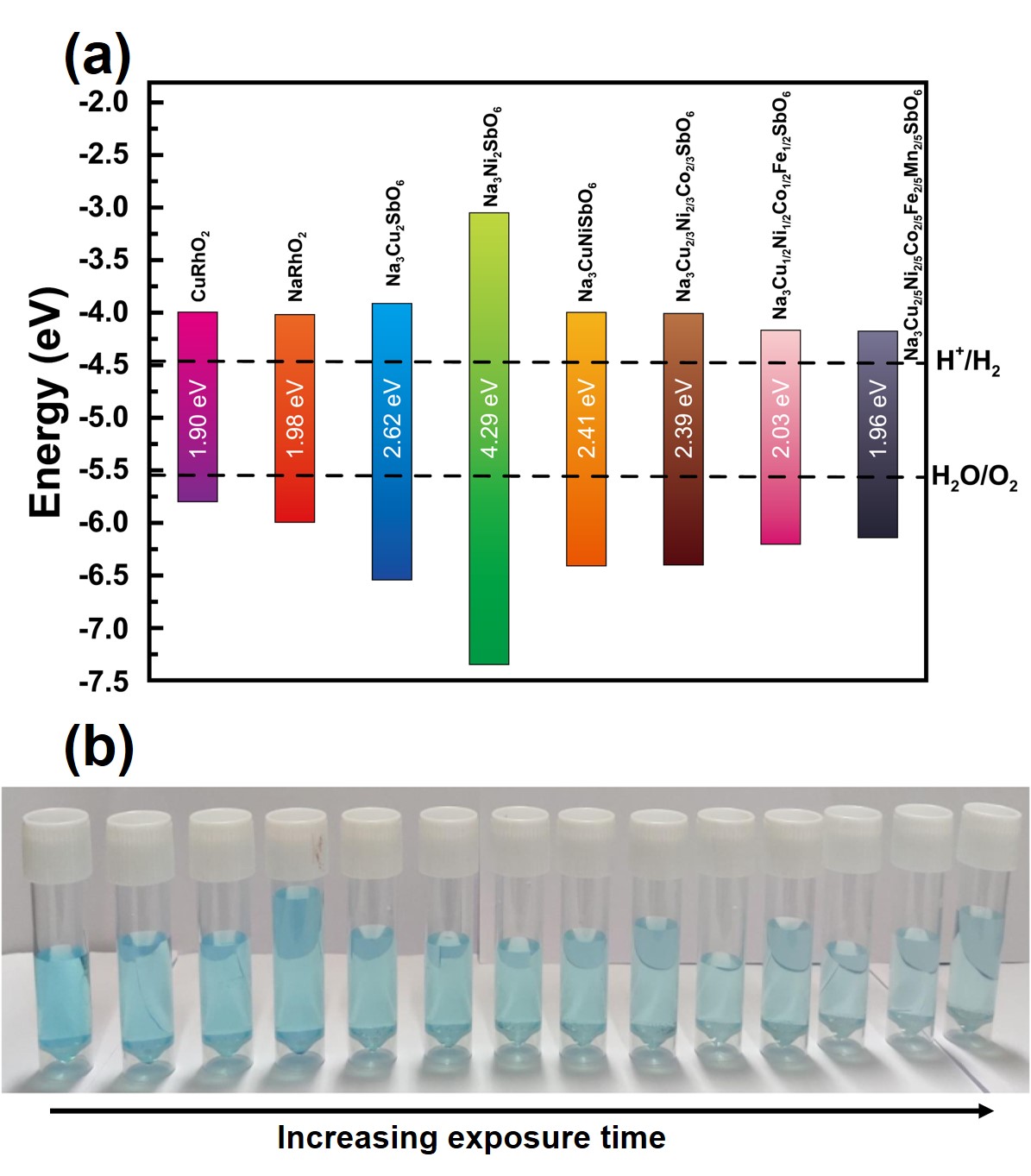}
  \caption{(a) Band gaps calculated for representative layered oxides along with multicomponent honeycomb layered oxides. (b) Degradation of methylene blue solutions upon photocatalysis using $\rm Na_3Cu_{2/5}Ni_{2/5}Co_{2/5}Fe_{2/5}Mn_{2/5}SbO_6$. Figures reproduced with permission.\cite{karati2021band} 
  Copyright 2021 Elsevier.}
  \label{Figure_13}
\end{figure*}

Recent years have witnessed the emergence of a new oxide development direction that entails the combination of multi-principal atomic elements in equimolar amounts to attain compounds dubbed as `high-entropy oxides'. Beyond the scope of oxides, the term `high-entropy materials' has become an umbrella term for describing a plethora of diverse high-entropy compounds in materials science. Materials in this class have gained significant traction especially in energy-related applications such as electrocatalysis, sensors and energy storage. In an interesting discovery, the number of transition metal atoms has been found to have a correlation with the optical band-gap (\textbf{Figure \ref{Figure_13}a}), whereby the band-gap decreases with increasing
number of elements leading to intriguing optical properties. As illustrated by the photocatalytic measurements of $\rm Na_3Cu_{2/5}Ni_{2/5}Co_{2/5}Fe_{2/5}Mn_{2/5}SbO_6$\cite{karati2021band} in \textbf{Figure \ref{Figure_13}b}, high-entropy honeycomb layered antimonates, such as $\rm Na_3Cu_{2/3}Ni_{2/3}Co_{2/3}SbO_6$,\cite{karati2021band} $\rm Na_3Cu_{1/2}Ni_{1/2}Co_{1/2}Fe_{1/2}SbO_6$\cite{karati2021band} and $\rm Na_3Cu_{2/5}Ni_{2/5}Co_{2/5}Fe_{2/5}Mn_{2/5}SbO_6$,\cite{karati2021band} show good photocatalytic properties, uniquely positioning them as excellent candidates for band-gap tunability via the `high-entropy' concept to engineer visible light absorbing photocatalysts. Their utility in photocatalysis is bound to gain increased prominence in applications such as water purification, due to the abundance of sunlight in most parts of the world. \textbf{Table \ref{Table_5}} shows the preparative conditions of $\rm Na_3Cu_{2/3}Ni_{2/3}Co_{2/3}SbO_6$, $\rm Na_3Cu_{1/2}Ni_{1/2}Co_{1/2}Fe_{1/2}SbO_6$ and $\rm Na_3Cu_{2/5}Ni_{2/5}Co_{2/5}Fe_{2/5}Mn_{2/5}SbO_6$.

\begin{table*}
\caption{High-temperature solid-state synthetic conditions of multi-transition metal-based (high entropy) honeycomb layered oxides comprising alkali atoms sandwiched between the slabs.\cite{karati2021band}}\label{Table_5}
\begin{center}
\scalebox{0.75}{
\begin{tabular}{lcll} 
\hline
\textbf{Compound} & \textbf{Synthesis technique} & \textbf{Precursors} & \textbf{Firing condition }\\ 

 & & & \textbf{(Temperature, atmosphere)}\\ 
\hline\hline
$\rm Na_3Cu_{2/3}Ni_{2/3}Co_{2/3}SbO_6$\cite{karati2021band} & Solid-state reaction & $\rm Na_2CO_3$, $\rm CuO$, $\rm NiO$, $\rm CoO$, $\rm Sb_2O_3$ & 1100$^\circ$C; 2h in air\\
$\rm Na_3Cu_{1/2}Ni_{1/2}Co_{1/2}Fe_{1/2}SbO_6$\cite{karati2021band} & Solid-state reaction & $\rm Na_2CO_3$, $\rm CuO$, $\rm NiO$, $\rm CoO$, $\rm Fe_2O_3$, $\rm Sb_2O_3$ & 1100$^\circ$C; 2h in air\\
$\rm Na_3Cu_{2/5}Ni_{2/5}Co_{2/5}Fe_{2/5}Mn_{2/5}SbO_6$\cite{karati2021band} & Solid-state reaction & $\rm Na_2CO_3$, $\rm CuO$, $\rm NiO$, $\rm CoO$, $\rm Fe_2O_3$, $\rm MnO$, $\rm Sb_2O_3$ & 1100$^\circ$C; 2h in air\\
\hline
\end{tabular}}
\end{center}
\end{table*}

\subsection{Low-temperature preparative methods}
Considering the vast compositional space occupied by pnictogen- and chalcogen-based honeycomb layered oxides, the success of exploratory syntheses can be envisioned as a vital determinant in the discovery of new structural variants. In material design, factors such as the thermodynamic and morphological properties of the starting reagents are seen to play prescriptive roles when determining the synthesis route of choice. For instance, the conventional high-temperature (> 800$^{\circ}$ C) preparation methods are not only unsuitable for materials with low thermal stability but also present other challenges such as unpredictable reaction courses that make it difficult to target specific products. Therefore, efforts to circumvent the thermodynamic impositions have shifted focus towards other alternate low-temperature synthesis reactions (sometimes referred to as `soft chemistry') in attempts to design and develop honeycomb layered oxide materials with specific structures and properties. In fact, recent years have witnessed a proliferation in the utilisation of low-temperature reactions that have opened access to a series of new compounds (some of which are metastable) unattainable by conventional high-temperature synthetic routes. As such, this subsection details low-temperature synthetic routes employed to design pnictogen- and chalcogen-based honeycomb layered oxides.

\subsubsection{Topochemical ion-exchange}
Topochemical or topotactic ion-exchange synthesis methods refer to techniques that enable the manipulation of the structural formations of materials after their crystalline lattices have been established. To facilitate these methodologies, starting compounds (commonly referred to as hosts) that provide the lattice frameworks of the final product are selected and combined with salts (cation source) to allow exchange of atoms. In one of the protocols under this methodology, the host compounds are tacitly manipulated by reacting them with molten salts which act as a cation source at temperatures corresponding to the melting point of the salts. For instance, $\rm Ag_3Co_2SbO_6$ can be prepared via topochemical ion-exchange of $\rm Na_3Co_2SbO_6$ using $\rm AgNO_3$ as the molten salt at 300$^{\circ}$ C.\cite{zvereva2016d} $\rm Cu_3{\it M}_2SbO_6$ ({\it M} = Ni, Co)\cite {Roudebush2015}can be prepared via topochemical ion-exchange of $\rm Na_3{\it M}_2SbO_6$ using $\rm CuCl$ as the molten salt under argon atmosphere at 400$^{\circ}$ C based on the following reaction:

\begin{align}
    \,{\rm Na_3{\it M}_2SbO_6} ({\it M}= \rm Co, Ni) + 3\,{\rm CuCl} \rightarrow {\rm Cu_3{\it M}_2SbO_6} + 3\,{\rm NaCl}.
\end{align}

A similar approach can be used to prepare honeycomb layered tellurates and bismuthates suchlike $\rm Cu_3{\it M}_2BiO_6$ ({\it M} = Ni, Mg, Zn), $\rm Ag_3{\it M}_2BiO_6$ ({\it M} = Ni, Mg, Zn), $\rm Cu_2{\it M}_2TeO_6$ ({\it M} = Ni, Cu, Co, Mg, Zn) and $\rm Ag_2{\it M}_2TeO_6$ ({\it M} = Ni, Cu, Co, Mg, Zn)\cite {masese2023honeycomb}:

\begin{subequations}
\begin{align}
    \,{\rm Na_3{\it M}_2BiO_6} ({\it M}= \rm Ni, Mg, Zn) + 3\,{\rm CuCl} \rightarrow {\rm Cu_3{\it M}_2BiO_6} + 3\,{\rm NaCl},\\
    \,{\rm Na_2{\it M}_2TeO_6} ({\it M}= \rm Co, Cu, Ni, Mg, Zn) + 2\,{\rm CuCl} \rightarrow {\rm Cu_2{\it M}_2TeO_6} + 2\,{\rm NaCl},\\
    \,{\rm Na_3{\it M}_2BiO_6} ({\it M}= \rm Ni, Mg, Zn) + 3\,{\rm AgNO_3} \rightarrow {\rm Ag_3{\it M}_2BiO_6} + 3\,{\rm NaNO_3},\\
    \,{\rm Na_2{\it M}_2TeO_6} ({\it M}= \rm Co, Cu, Ni, Mg, Zn) + 2\,{\rm AgNO_3} \rightarrow {\rm Ag_2{\it M}_2TeO_6} + 2\,{\rm NaNO_3}.
\end{align}
\end{subequations}

In the design of topotactic ion-exchange reactions, the selection of a host structure that can allow its constituent ions to be substituted is a key requisite for successful synthesis. For the cation source, a nitrate salt (or salt with low melting temperature) of the desired metal is usually portioned in excess (in most cases), mixed with the host material through grinding to break down the particles into smaller sizes that facilitate better exchange efficiency. Subsequently, the mixture is heated to a temperature equivalent to or slightly above the melting point of the metal salt to allow the metal in the molten salt to diffuse into the host lattice and push out the original metal (mobile cation) in the structure. The ejected metal ions form part of the molten mixture outside of the solid structure to be washed away using a suitable solvent ($e.g.$, deionised water, ammonium hydroxide, ethanol, $etc.$) leaving behind a new solid structure. It is worth noting that as the interlayer cations are exchanged, the host structure retains its integrity but may undergo a simple transition to accommodate the new ions.

The topochemical reaction routes proffer crucial low-temperature (<500$^{\circ}$ C) approaches to rationally design new honeycomb layered oxides, providing access to materials that either tend to be disordered when prepared at high-temperatures or materials that cannot be prepared by standard high-temperature methods (such as most of the coinage-metal-based materials). In the following subsections, we highlight the topochemical ion-exchange techniques for selected pnictogen- and chalcogen-based honeycomb layered oxides. 

\paragraph{Alkali-metal-based materials}
Although lithium systems such as $\rm Li_2Ni_2TeO_6$ can be prepared through the conventional high-temperature solid-state reactions,\cite{Kumar2013} their resultant products tend to crystallise in disordered structures. The formation of disordered frameworks is ascribed to the comparable Shannon-Prewitt\cite{shannon1970} ionic radii of Li atoms ($0.76$ \AA\, in $\rm LiO_6$ octahedra) and Ni atoms ($0.69$ \AA\, in $\rm NiO_6$ octahedra) which leads to swapping of their crystallographic atomic positions (also referred to as `cationic mixing') during the high-temperature solid-state reactions, thus creating a disordered crystal framework. The disordered $\rm Li_2Ni_2TeO_6$ has been found to exhibit poor electrochemical performance which has been posited to result from obstructed Li diffusion resulting from the occupation of Ni atoms in the Li cation diffusion planes.\cite{Grundish2019} 

To avert such structural deformities, the topochemical ion-exchange (metathesis reaction) method is considered an expedient strategy for synthesising ordered lithium-based compositions such as ordered $\rm Li_2Ni_2TeO_6$.\cite{Grundish2019}  Here, $\rm Na_2Ni_2TeO_6$ is employed as the host material and subjected to a low-temperature heat-treatment alongside a molten Li salt such as $\rm LiNO_3$. 
Despite this, the attained $\rm Li_2Ni_2TeO_6$ still tends to be metastable,\cite{Grundish2019}  requiring additional annealing process to obtain a stable variant. \textbf{ Table \ref{Table_6}} shows the preparative techniques of lithium-based materials along with their precursors and molten salt used. Polytypes of sodium-based materials can be prepared using potassium-based materials, as outlined by the design conditions provided therein.

\begin{table*}
\caption{Low-temperature topochemical ion-exchange synthetic conditions of selected pnictogen-and chalcogen-based honeycomb layered oxides containing alkali metal atoms sandwiched between the slabs.\cite{Grundish2019, Koo2016, Zvereva2012, ma2007synthesis, Yadav2019} Preliminary investigations are underscored by asterisks (*).}\label{Table_6}
\begin{center}
\scalebox{0.67}{
\begin{tabular}{lcll} 
\hline
\textbf{Compound} & \textbf{Synthesis technique} & \textbf{Precursors} & \textbf{Firing condition }\\ 

 & & & \textbf{(Temperature, atmosphere)}\\ 
\hline\hline
$\rm Li_2Ni_2TeO_6$ polytypes\cite{Grundish2019} & Topochemical ion-exchange & $\rm LiNO_3$, $\rm Na_2Ni_2TeO_6$ & 250-300$^\circ$C; 4-12h\\
$\rm Li_2{\it M}_2TeO_6$ ($\rm {\it M} = Co, Ni, Cu, Zn$ and $\rm Mg$) polytypes* & Topochemical ion-exchange & $\rm LiNO_3$, $\rm Na_2{\it M}_2TeO_6$ ($\rm {\it M} = Co, Ni, Cu, Zn$ and $\rm Mg$) & 250-300$^\circ$C; 4-12h\\
$\rm Li_3Cu_2SbO_6$ polytypes\cite{Koo2016} & Topochemical ion-exchange & $\rm LiNO_3$, $\rm Na_3Cu_2SbO_6$ & 290$^\circ$C; 4h\\
$\rm Li_3Ni_2SbO_6$\cite{ma2007synthesis} polytypes & Topochemical ion-exchange & $\rm LiNO_3$ + $\rm LiNO_3$ eutectic melt, $\rm Na_3Ni_2SbO_6$ & 280$^\circ$C; 8h in air\\
$\rm Li_3Ni_2SbO_6$\cite{Zvereva2012} & Topochemical ion-exchange & $\rm LiNO_3$, $\rm Na_3Ni_2SbO_6$ & 300$^\circ$C; 12h\\
$\rm Li_3{\it M}_2SbO_6$ ($\rm {\it M} = Fe$ and $\rm Mn$)\cite{Yadav2019} & Topochemical ion-exchange & $\rm LiNO_3$, $\rm Na_3{\it M}_2SbO_6$ ($\rm {\it M} = Fe$ and $\rm Mn$) & 300$^\circ$C; Ar\\
$\rm Li_2Ni_2TeO_6$ polytypes* & Topochemical ion-exchange & $\rm LiNO_3$, $\rm K_2Ni_2TeO_6$ & 250-300$^\circ$C; 4-12h\\
$\rm Li_3Ni_2SbO_6$ polytypes* & Topochemical ion-exchange & $\rm LiNO_3$, $\rm Na_3Ni_2SbO_6$ & 300$^\circ$C; 2h\\
$\rm Na_2Ni_2TeO_6$ polytypes* & Topochemical ion-exchange & $\rm NaNO_3$, $\rm K_2Ni_2TeO_6$ & 250-300$^\circ$C; 4-12h\\

\hline
\end{tabular}}
\end{center}
\end{table*}

\begin{figure*}[!b]
\centering
  \includegraphics[width=0.8\columnwidth]{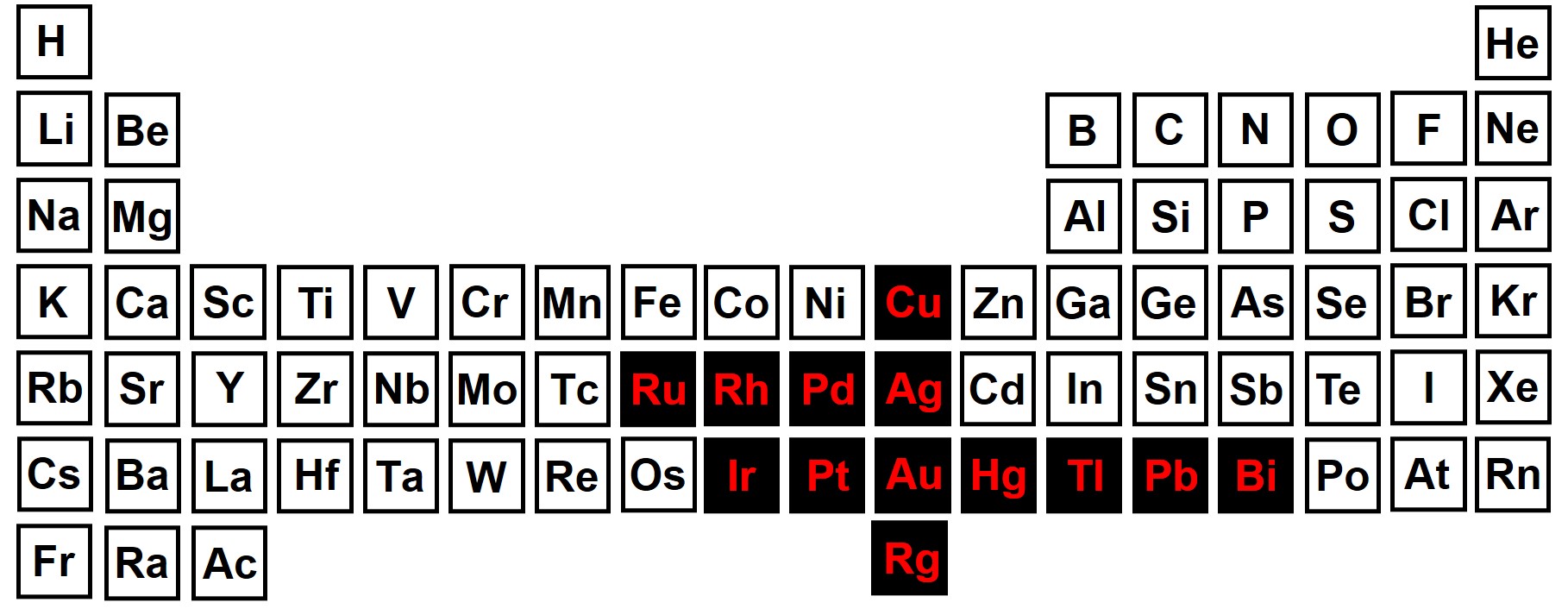}
  \caption{Elements that tend to show metallophilicity. Elements that tend to show metallophilicity are shown in red with a black background. Elements, such as Rg, are projected to also show metallophilicity and are also highlighted in red with a black background.}
  \label{Figure_14}
\end{figure*}

\paragraph{Coinage-metal-based materials}
Honeycomb layered antimonates, bismuthates and tellurates embedded with coinage metal atoms (suchlike Cu and Ag) have been found to present piquant crystal versatility with projected applications in optics, magnetism, and catalysis amongst others. As has been documented in several early observations, the remarkable crystal properties of these materials are ascribed to the unique coordination chemistry between coinage metal atoms and ligands (such as oxygen) where the coinage metal atoms (Ag, Cu, Au) form close metal-metal contacts referred to as `numismophilicity' (or `metallophilicity' in general).\cite{vicente1993synthesis} Besides their ability to induce enigmatic changes in the redox, spectroscopic, and charge transport properties of this class of materials, the metal-metal interactions have also been noted to bear significant functionalities in the `bottom-up' fabrication techniques for complex self-assembled architectures.\cite{schmidbaur2008briefing, schmidbaur2000aurophilicity, schmidbaur2015argentophilic} As such, the discovery and design of coinage-metal-based compounds with metallophilic bonding is of paramount importance in the augmentation of pnictogen- and chalcogen-based honeycomb oxide functionalities. In particular, gold chemistry has attracted increasing attention stemming from the intriguing variations in the electronic characteristics of metallophilic gold contacts (aurophilicity) that have been noted to engender exotic physical, photophysical and photochemical properties.\cite{schmidbaur2008briefing, schmidbaur2000aurophilicity, schmidbaur2015argentophilic} \textbf{Figure \ref{Figure_14}} shows a list of elements that exhibit metallophilicity.

Given the auspicious prospects presented by coinage metal systems in the pnictogen and chalcogen class of honeycomb layered oxides, it is prudent to explore the synthesis protocols applied during their development. Although Cu-based materials such as $\rm Cu_3{\it M}_2SbO_6$ ($\rm {\it M} = Co, Ni $) are attainable via high-temperature solid-state routes,\cite{Nagarajan2002} such techniques are unsuitable for compositions such as $\rm Cu_3{\it M}FeSbO_6$ ($\rm {\it M} = Na, Li $) due to their low thermodynamic stability. Therefore, for the fabrication of Cu-based Delafossite oxides entailing the compositions such as $\rm Cu_3NaFeSbO_6$ and $\rm Cu_3LiFeSbO_6$, topotactic ion-exchange reactions are performed using $\rm Na_4FeSbO_6$ and $\rm Na_3LiFeSbO_6$ as the respective host precursors in combination with CuCl halide salt as the cation source (for both cases) at around 400$^{\circ}$ C.\cite{sethi2022cuprous} The reaction typically takes place in an argon atmosphere to avert the oxidation of CuCl. Even though Cu-based Delafossite oxides can be synthesised via calcination, low-temperature ion-exchange reactions have been instrumental in the discovery of new polytypes and even new phases or properties which have been unattainable using the conventional high-temperature methods. For instance, $\rm Cu_3{\it M}_2SbO_6$ ($\rm {\it M} = Co, Ni $) honeycomb layered oxides prepared using the low-temperature ion-exchange reaction were reported to crystallise with different stacking variations (polytypes) that were previously unattainable at high temperatures.\cite{Roudebush2015} Details relating to the polytypes will be mentioned in a latter section. \textbf{ Table \ref{Table_7}} shows a summary of the preparative conditions of Cu-based materials.

Analogous to their Cu-based counterparts, several Ag-based Delafossite oxides are routinely prepared via the conventional solid-state reactions at high temperatures (usually beyond 1000$^{\circ}$ C). However, for compounds utilising $\rm Ag_2O$ as the Ag source precursor, high temperatures are unsuitable due to the decomposition of $\rm Ag_2O$ to metallic silver and oxygen in air at above 300$^{\circ}$ C. In this light, low-temperature ion-exchange routes have become expedient strategies for attaining such compositions with reasonable efficiency. The topotactic ion-exchange syntheses of honeycomb layered oxides such as $\rm Ag_3Li{\it M}SbO_6$ ($\rm {\it M} = Al, Cr, Fe, Ga $),\cite{Bhardwaj2014} $\rm Ag_3Li{\it M}TeO_6$ ($\rm {\it M} = Zn, Co, Ni $)\cite{Kumar2012} and $\rm Ag_3Ni_2BiO_6$\cite{Berthelot2012} are typically conducted using a Li-based honeycomb layered oxide precursor in combination with $\rm AgNO_3$ molten salt. Others such as $\rm Ag_3NaFeSbO_6$\cite{politaev2009subsolidus} and $\rm Ag_3{\it M}_2SbO_6$ ($\rm {\it M} = Zn, Ni, Co $)\cite{zvereva2016d} have been prepared using Na-based materials as precursors. \textbf{ Table \ref{Table_8}} shows the preparative conditions of Ag-based materials.

\begin{table*}
\caption{Low-temperature topochemical ion-exchange synthetic conditions of selected pnictogen-and chalcogen-based honeycomb layered oxides containing Cu atoms sandwiched between the slabs.\cite{sethi2022cuprous, Roudebush2015} In this table, ``TIE'' stands for \textit{topochemical ion-exchange}. Preliminary investigations are underscored by asterisks (*).}\label{Table_7}
\begin{center}
\scalebox{0.8}{
\begin{tabular}{lcll} 
\hline
\textbf{Compound} & \textbf{Synthesis} & \textbf{Precursors} & \textbf{Firing condition }\\ 

 & technique & & \textbf{(Temperature, atmosphere)}\\ 
\hline\hline
$\rm Cu_3LiFeSbO_6$\cite{sethi2022cuprous} & TIE & $\rm CuCl$, $\rm Na_3LiFeSbO_6$ & 400$^\circ$C; 1h in argon\\
$\rm Cu_3NaFeSbO_6$\cite{sethi2022cuprous} & TIE & $\rm CuCl$, $\rm Na_4FeSbO_6$ & 400$^\circ$C; 1h in argon\\
$\rm Cu_3{\it M}_2SbO_6$ ($\rm {\it M} = Co, Ni $)\cite{Roudebush2015} & TIE & $\rm CuCl$, $\rm Na_3{\it M}_2SbO_6$ ($\rm {\it M} = Co, Ni $) & 400$^\circ$C; 24h in argon\\
$\rm Cu_2{\it M}_2TeO_6$ ($\rm {\it M} = Mg, Zn, Co, Ni, Cu $)* & TIE & $\rm CuCl$, $\rm Na_2{\it M}_2TeO_6$ ($\rm {\it M} = Mg, Zn, Co, Ni, Cu $) & 430$^\circ$C; 24h in argon\\
$\rm Cu_2Ni_2TeO_6$* & TIE & $\rm CuCl$, $\rm K_2Ni_2TeO_6$ & 430$^\circ$C; 24h in argon\\
$\rm Cu_2{\it M}_2TeO_6$ ($\rm {\it M} = Mg, Zn, Ni, Co, Cu $)* & TIE & $\rm CuCl$, $\rm Na_2{\it M}_2TeO_6$ ($\rm {\it M} = Mg, Zn, Ni, Co, Cu $) &430$^\circ$C; 24h in argon\\
$\rm Cu_3{\it M}_2BiO_6$ ($\rm {\it M} = Mg, Zn, Ni $)* & TIE & $\rm CuCl$, $\rm Na_3{\it M}_2BiO_6$ ($\rm {\it M} = Mg, Zn, Ni $) & 430$^\circ$C; 24h in argon\\

\hline
\end{tabular}}
\end{center}
\end{table*}

\begin{table*}
\caption{Low-temperature topochemical ion-exchange synthetic conditions of selected pnictogen-and chalcogen-based honeycomb layered oxides containing Ag atoms sandwiched between the slabs. \cite{Kumar2012, Uma2016, Bhardwaj2014, Kumar2013, kanyolo2022advances2, masese2023honeycomb, Berthelot2012, zvereva2016d, politaev2010, politaev2009subsolidus} In this table, ``TIE'' stands for \textit{topochemical ion-exchange}. Preliminary investigations are underscored by asterisks (*).}\label{Table_8}
\begin{center}
\scalebox{0.75}{
\begin{tabular}{lcll} 
\hline
\textbf{Compound} & \textbf{Synthesis} 
& \textbf{Precursors} & \textbf{Firing condition }\\ 

 & \textbf{technique} & & \textbf{(Temperature, atmosphere)}\\ 
\hline\hline
$\rm Ag_3(Li_{1.5}{\it M}_{0.5}Te)O_6$ ($\rm {\it M} = Cr, Mn, Fe, Al$ and $\rm Ga$)\cite{Uma2016} & TIE & $\rm AgNO_3$, $\rm Li_{4.5}{\it M}_{0.5}TeO_6$ & 250-300$^\circ$C; 3h\\
$\rm Ag_3Li{\it M}SbO_6$ ($\rm {\it M} = Cr, Fe, Al$ and $\rm Ga$)\cite{Bhardwaj2014} & TIE & $\rm AgNO_3$, $\rm Li_4{\it M}SbO_6$ ($\rm {\it M} = Cr, Fe, Al$ and $\rm Ga$) & 250$^\circ$C; 1h\\
$\rm Ag_3Ni_2BiO_6$\cite{Berthelot2012} & TIE & $\rm AgNO_3$, $\rm Li_3Ni_2BiO_6$ & 230-250$^\circ$C; overnight\\
$\rm Ag_3Co_2SbO_6$\cite{zvereva2016d, politaev2010} & TIE & $\rm AgNO_3$, $\rm Na_3Co_2SbO_6$ & 260$^\circ$C\\
$\rm Ag_6(Li_2{\it M}_2Sb_2)O_{12}$ ($\rm {\it M} = Al, Fe, Cr$ and $\rm Ga$)\cite{Kumar2012} & TIE & $\rm AgNO_3$, $\rm Li_8{\it M}_2Sb_2O_{12}$ ($\rm {\it M} = Al, Fe, Cr$ and $\rm Ga$) &230-250$^\circ$C; 1h\\
$\rm Ag_6(Li_2{\it M}_2Te_2)O_{12}$ ($\rm {\it M} = Co, Ni$ and $\rm Zn$)\cite{Kumar2012} & TIE & $\rm AgNO_3$, $\rm Li_8{\it M}_2Te_2O_{12}$ ($\rm {\it M} = Co, Ni$ and $\rm Zn$) &230-250$^\circ$C; 1h\\
$\rm Ag_3{\it M}_2SbO_6$ ($\rm {\it M} = Ni$ and $\rm Zn$) & TIE & $\rm AgNO_3$, $\rm Li_3{\it M}_2SbO_6$ ($\rm {\it M} = Ni$ and $\rm Zn$) & 250$^\circ$C; 5 days in air\\
$\rm Ag_3NaFeSbO_6$\cite{politaev2009subsolidus} polytype & TIE & $\rm AgNO_3$, $\rm Na_4FeSbO_6$ & 240$^\circ$C\\
$\rm Ag_2{\it M}_2TeO_6$ ($\rm {\it M} = Ni, Mg, Co, Cu$ and $\rm Zn$)\cite{masese2023honeycomb} & TIE & $\rm AgNO_3$, $\rm Na_2{\it M}_2TeO_6$ ($\rm {\it M} = Ni, Mg, Co, Cu$ and $\rm Zn$) & 250$^\circ$C; 5 days in air\\
$\rm Ag_2{\it M}_2TeO_6$ ($\rm {\it M} = Ni, Mg, Co, Cu$ and $\rm Zn$)\cite{masese2023honeycomb} & TIE & $\rm AgNO_3$, $\rm NaK{\it M}_2TeO_6$ ($\rm {\it M} = Ni, Mg, Co, Cu$ and $\rm Zn$) & 250$^\circ$C; 5 days in air\\
$\rm Ag_3Ni_2BiO_6$ polytype* & TIE & $\rm AgNO_3$, $\rm Na_3Ni_2BiO_6$ & 250$^\circ$C; 4 days in air\\
$\rm Ag_3{\it M}_2BiO_6$ ($\rm {\it M} = Mg, Co$ and $\rm Zn$)* & TIE & $\rm AgNO_3$, $\rm Na_3{\it M}_2BiO_6$ ($\rm {\it M} = Mg, Co$ and $\rm Zn$) & 250$^\circ$C; 5 days in air\\
$\rm Ag_3LiMgTeO_6$ ($\rm Ag_6Li_2Mg_2Te_2O_{12}$)* & TIE & $\rm AgNO_3$, $\rm Li_4MgTeO_6$* & 250$^\circ$C; 4 days in air\\
$\rm Ag_3{\it M}_2SbO_6$ ($\rm {\it M} = Ni, Co, Mg, Cu$ and $\rm Zn$)* & TIE & $\rm AgNO_3$, $\rm Na_3{\it M}_2SbO_6$ ($\rm {\it M} = Ni, Co, Mg, Cu$ and $\rm Zn$) & 250$^\circ$C; 5 days in air\\
\hline
\end{tabular}}
\end{center}
\end{table*}

\subsubsection{Electrochemical ion-exchange}

\begin{figure*}[!t]
\centering
  \includegraphics[width=0.8\columnwidth]{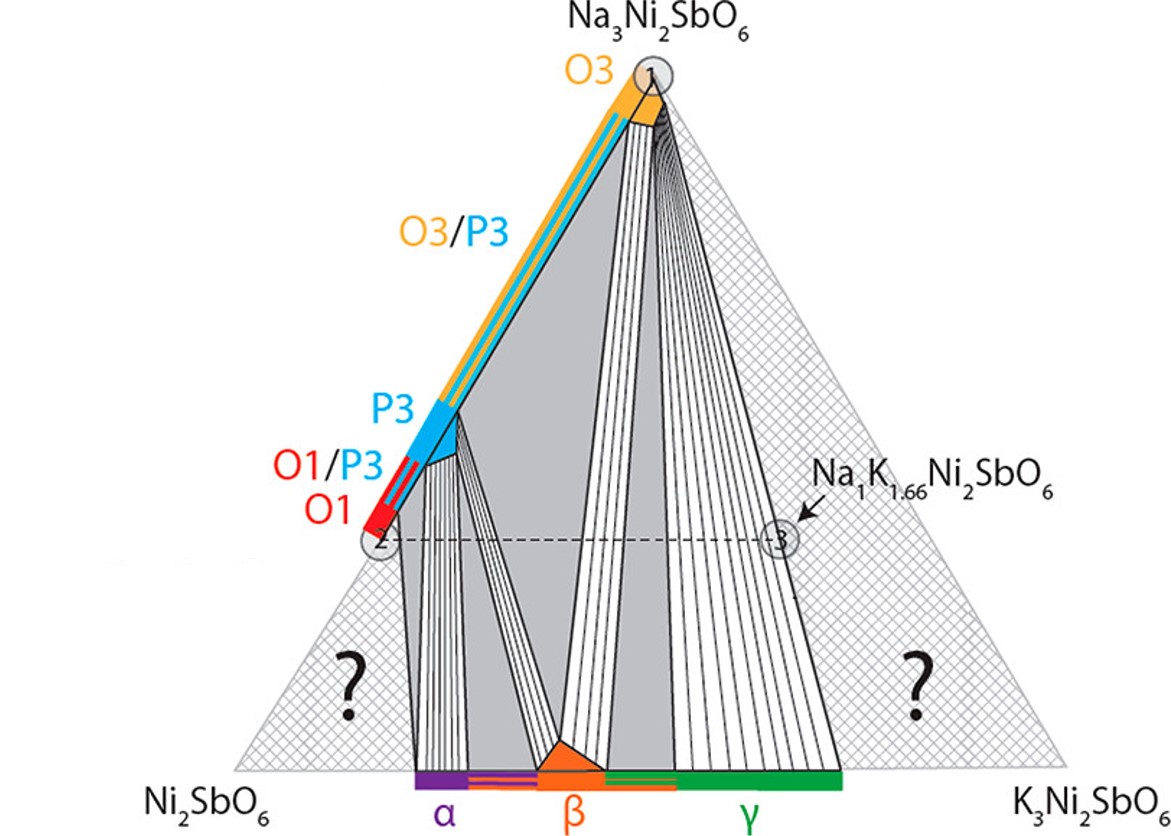}
  \caption{Ternary phase diagram for the desodiation and subsequent potassiation of $\rm Na_3Ni_2SbO_6$. The stacking evolution of the various phases attained upon Na-ion extraction from $\rm Na_3Ni_2SbO_6$ (point 1 $\rightarrow$ point 2 are shown in Hagenmuller-Delmas notation. Details relating to this notation are provided in \textbf{Section \ref{Section: Stacking}}. K-ion insertion into $\rm NaNi_2SbO_6$ to attain $\rm NaK_{1.66}Ni_2SbO_6$ is shown in dashed lines (point 2 $\rightarrow$ point 3. Figures reproduced with permission.\cite{Kim2020} Copyright 2021 American Chemical Society.}
  \label{Figure_15}
\end{figure*}

As mentioned earlier, the low thermodynamic stability exhibited by metastable products, coupled with poor stoichiometric precision renders high-temperature synthesis methods unsuitable for the fabrication of some nanomaterials. Electrochemical intercalation presents an auspicious synthesis strategy especially for the exploration and design of pnictogen- and chalcogen-based honeycomb materials with anomalous stoichiometric configurations. Intercalation or deintercalation is the process of introducing or removing a stringently predetermined amount of guest species (\textit{i.e.}, cations sandwiched between the metal slab (interlayer cations)) into/from a solid host lattice employed as an electrode in an electrochemical cell. In this method, voltage is applied on an electrochemical cell comprising the host material placed alongside the cation source material in a suitable electrolyte to mobilise the cations into the host lattice structure without imposing major structural changes. On completion of the electrochemical process, the cell is disassembled and further characterised to confirm the final stoichiometry of the material. This technique has been instrumental in the fabrication of mixed-alkali compositions such as $\rm Na_{0.5}K_{1.37}Ni_2SbO_6$ and $\rm NaK_{1.66}Ni_2SbO_6$ achieved through the electrochemical ${\rm K^{+}}$ / ${\rm Na^{+}}$ ion-exchange with $\rm Na_3Ni_2SbO_6$ as the starting material.\cite{Kim2020} \textbf{Figure \ref{Figure_15}} shows the ternary phase diagram for the desodiation and subsequent potassiation of $\rm Na_3Ni_2SbO_6$.

\subsection{Single crystal growth: molten-flux method}

Single crystals are three-dimensional arrays of atoms (usually uninterrupted) with repeating geometry present in a single piece of a material. In experimental studies, these atomic arrays are indispensable particularly when determining the intrinsic physical properties of materials because they enable the elimination of subtle influences caused by impurity phases and grain boundaries. Besides, they additionally allow the scrutiny of emergent anisotropic physical properties along various crystallographic axes, facilitating the characterisation of honeycomb layered oxide materials through various spectroscopic and transport techniques. As such, the ability to grow high-quality single crystals of honeycomb layered oxides is critical to the advancement of both their fundamental chemistry and physics despite the considerable time and effort required for these processes. 

Amongst the numerous crystal growth techniques accessible to material scientists, chemists and physicists, the metal-flux method has become commonplace in the growth of single crystals for a variety of pnictogen- and chalcogen-based honeycomb layered oxides such as $\rm Na_3Cu_2SbO_6$, $\rm Na_2Cu_2TeO_6$, $\rm Na_3Co_2SbO_6$, $\rm Na_2Ni_2TeO_6$ and more recently $\rm BaNi_2TeO_6$.\cite{song2022influence, murtaza2021magnetic, Sankar2014, reya2011crystal, Kumar2012, Seibel2013, Yan2019, Nagarajan2002, Bhardwaj2014, Miura2008, Yao2020, Xiao2019, xiao2021magnetic} The choice of the technique to be applied to grow single crystals is contingent on a confluence of factors suchlike the particle size, volatility, thermodynamic stability and extent of doping induced on the crystal. Therefore, to elaborate how these factors influence synthesis parameters, the following subsection highlights the utilisation of the metal flux technique for the growth of single crystals of the pnictogen and chalcogen class of honeycomb layered oxides.

\subsubsection{Metal flux method}
\begin{figure*}[!b]
\centering
  \includegraphics[width=0.8\columnwidth]{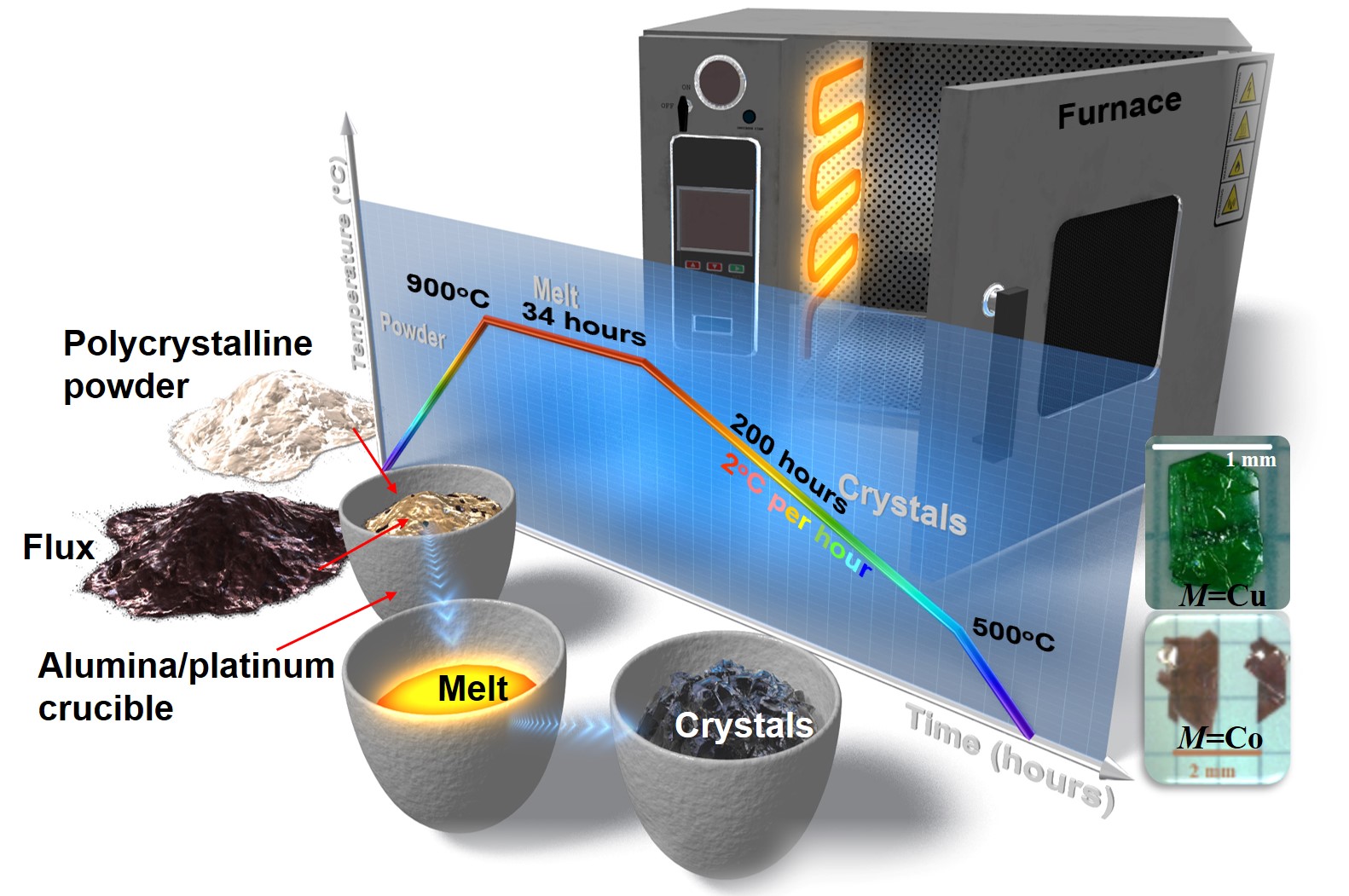}
  \caption{Schematic representation of the flux method of growing single crystals of $\rm Na_2{\it M}_2TeO_6$ ($\rm {\it M} = Co, Ni, Cu $) along with the temperature profile for crystal growth. Figure adapted with permission.\cite{murtaza2021magnetic} Copyright 2021 Elsevier.}
  \label{Figure_16}
\end{figure*}

This synthesis method presents a simple and versatile approach for growing single crystals of layered oxides. Here, a low-melting metal flux (solvent) is employed to dissolve the constituent elements at relatively low temperatures. For enhanced atomic diffusion, the flux is subjected to relatively high temperatures (slightly above saturation temperature) to allow the molten flux to dissolve the reactant elements and form a homogeneous solution. Controlled cooling is initiated to allow the formation of single crystals of the desired material, once the solution reaches its saturation limit. The process continues in the liquid medium until the temperatures reach the melting point of the flux. The excess flux can be removed through centrifugation at high temperature or simple decanting. \textbf{Figure \ref{Figure_16}} shows a typical setup used for the flux growth technique.

The primary advantage for utilising this synthesis method (as opposed to others such as Czochralski methods, Bridgman-Stockbarger growth and optical floating zone techniques\cite{koohpayeh2008optical, bridgman2013certain, stockbarger1936production, tomaszewski2002jan, uecker2014historical, mueller2007czochralski})
is that it does not require specialised equipment other than a temperature-controllable furnace with homogeneous temperature distribution and a suitable crucible.\cite{kanatzidis2005metal, canfield1992growth}
Additionally, the abundance of fluxes that can be used for crystal growth makes this technique easy to replicate at a laboratory scale. 

For successful single crystal growth, binary phase diagrams can be referenced to give vital information relating to the temperature ranges for crystal growth and the composition of the starting material. A typical cooling rate of 1–10$^{\circ}$ C/h is normally applied although it can change depending on the slope of the concentration–temperature phase and the growth kinetics. The chemical composition of the flux may also be a determining factor in the design of synthesis protocols. For instance, if the flux provides at least one of the constituents required to form the desired compound, a technique known as the self-flux method is employed. This method has been utilised in the growth of single crystals for $\rm Na_2Ni_2TeO_6$,\cite{Sankar2014, murtaza2021magnetic} $\rm Na_2Cu_2TeO_6$,\cite{Sankar2014, murtaza2021magnetic} $\rm Na_2Co_2TeO_6$,\cite{Yao2020, Xiao2019} $\rm Na_3Co_2SbO_6$,\cite{Yan2019} and $\rm Na_3Cu_2SbO_6$\cite{Miura2008} compositions. Otherwise, an external flux can be chosen depending on the individual solubilities of the constituent elements. An example of such materials fabricated using this technique is $\rm BaNi_2TeO_6$ whose crystals were grown in a $\rm PbCl_2$ flux.\cite{song2022influence}

\begin{figure*}[!t]
\centering
  \includegraphics[width=0.75\columnwidth]{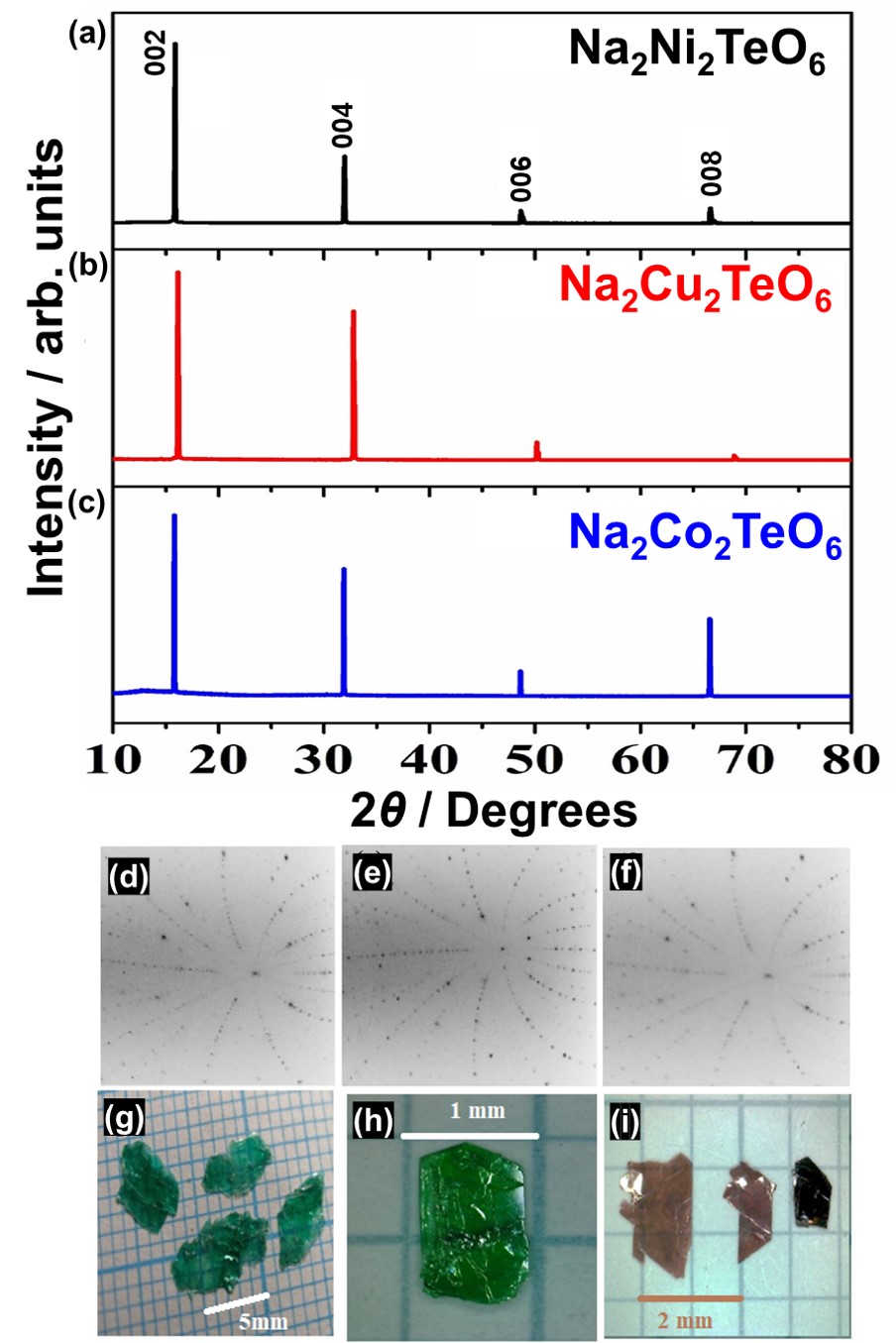}
  \caption{Single crystal growth of  $\rm Na_2{\it M}_2TeO_6$ ($\rm {\it M} = Ni, Cu $ and $\rm Co$) using flux method. (a, b, c) XRD patterns of single crystals of $\rm Na_2{\it M}_2TeO_6$ ($\rm {\it M} = Co, Cu $ and $\rm Ni$). (d, e, f) the Laue patterns and (g, h, i) their corresponding crystal images. Figures adapted with permission.\cite{murtaza2021magnetic} Copyright 2021 Elsevier.}
  \label{Figure_17}
\end{figure*}

\textbf{Figures \ref{Figure_17}a}, \textbf{\ref{Figure_17}b} and \textbf{\ref{Figure_17}c} show the XRD patterns of single crystals of $\rm Na_2{\it M}_2TeO_6$ ($\rm {\it M} = Ni, Co $ and $\rm Cu$).\cite{murtaza2021magnetic} The conspicuous Bragg peaks of 00$l$ reflections displayed in the patterns indicate the formation of materials with good crystallinity. The Laue patterns depicted in \textbf{Figures \ref{Figure_17}d}, \textbf{\ref{Figure_17}e} and \textbf{\ref{Figure_17}f} further validate the crystallinity of the as-grown samples of $\rm Na_2Ni_2TeO_6$, $\rm Na_2Cu_2TeO_6$ and $\rm Na_2Co_2TeO_6$, respectively. The well-developed facets of optimally grown crystals using metal-flux technique can readily be recognised from their shapes (See \textbf{Figures \ref{Figure_17}g}, \textbf{\ref{Figure_17}h} and \textbf{\ref{Figure_17}i}). Crystals can take the forms of polygons, ribbons and wires, depending on the bonding anisotropies in a compound. It is worth noting that single crystals of pnictogen- and chalcogen-based honeycomb layered oxides have a propensity towards polygon-like shapes.

A vast majority of single crystals of topological materials are grown via flux growth (as the method of choice), due to its relative simplicity compared to other crystal growth techniques. Despite this expediency, there are some limitations that constrain its use for some materials. For example, flux growth is only suitable for materials that exist in equilibrium with the liquid. Another demerit of the process is that the metal fluxes often enter crystals as inclusions, thereby influencing the crystal properties of materials. Additionally, in most cases, the crystals sizes are usually not large enough to facilitate bulk transport measurements, such as thermal conductivity, superconductivity and Nernst effect measurements. One strategy for circumventing this issue is to employ other crystal growth techniques such as chemical vapour transport. \textbf{ Table \ref{Table_9}} shows the preparative conditions of single crystals of representative pnictogen- and chalcogen-based honeycomb layered oxides.

\begin{table*}
\caption{Single crystal growth synthetic conditions of selected pnictogen-and chalcogen-based honeycomb layered oxides. \cite{song2022influence, murtaza2021magnetic, Sankar2014, reya2011crystal, Kumar2012, Seibel2013, Yan2019, Nagarajan2002, Bhardwaj2014, Miura2008, Yao2020, Xiao2019, xiao2021magnetic}}\label{Table_9}
\begin{center}
\scalebox{0.73}{
\begin{tabular}{llll} 
\hline
\textbf{Compound} & \textbf{Synthesis technique} & \textbf{Precursors} & \textbf{Firing condition }\\ 

 & & & \textbf{(Temperature, atmosphere)}\\ 
\hline\hline
$\rm Na_2{\it M}_2TeO_6$ ($\rm {\it M} = Ni, Cu, Co$)\cite{murtaza2021magnetic} & Self-flux method & Polycrystalline $\rm Na_2{\it M}_2TeO_6$ ($\rm {\it M} = Ni, Cu$), $\rm Na_2O$, $\rm TeO_2$ & 900$^\circ$C$\rightarrow$500$^\circ$C\\
$\rm Na_2{\it M}_2TeO_6$ ($\rm {\it M} = Ni, Cu$)\cite{Sankar2014} & Self-flux method & Polycrystalline $\rm Na_2{\it M}_2TeO_6$ ($\rm {\it M} = Ni, Cu$), $\rm Na_2O$, $\rm TeO_2$ & 800$^\circ$C$\rightarrow$500$^\circ$C\\
$\rm Na_2Co_2TeO_6$\cite{murtaza2021magnetic,Xiao2019} & Self-flux method & Polycrystalline $\rm Na_2Co_2TeO_6$, $\rm Na_2O$, $\rm TeO_2$ & 900$^\circ$C$\rightarrow$500$^\circ$C in air\\
$\rm Na_3Co_2SbO_6$\cite{Yan2019} & Self-flux method & Polycrystalline $\rm Na_3Co_2SbO_6$, $\rm Na_2CO_3$ & 1100$^\circ$C$\rightarrow$800$^\circ$C\\
$\rm Na_3Cu_2SbO_6$\cite{Miura2008} & Self-flux method & Polycrystalline $\rm Na_3Cu_2SbO_6$, $\rm Na_2CO_3$ & 800-1260$^\circ$C; 12-48 in air\\
$\rm Cu_5SbO_6$\cite{reya2011crystal} & Chemical vapour transport & $\rm CuO$, $\rm Sb_2O_3$, $\rm PdCl_2$ (transport agent) & 800-900$^\circ$C in vacuum\\
$\rm Cu_3Mn_2SbO_6$\cite{Nagarajan2002} & Solid-state reaction & $\rm Cu_2O$, $\rm Sb_2O_3$, $\rm Mn_2O_3$ & 800$^\circ$C$\rightarrow$500$^\circ$C in vacuum\\
$\rm Cu_3Mg_2SbO_6$\cite{Nagarajan2002} & Solid-state reaction & $\rm Cu_2O$, $\rm Sb_2O_3$, $\rm MgO$ & 1100$^\circ$C; 24h in vacuum\\
$\rm Li_8Cu_2Te_2O_{12}$ ($\rm Li_4CuTeO_6$)\cite{Kumar2012} & Solid-state reaction & Polycrystalline $\rm Li_2Cu_2TeO_6$ & 800-1260$^\circ$C; 12-48 in air\\
$\rm Li_8Co_2Te_2O_{12}$ ($\rm Li_4CoTeO_6$)\cite{Kumar2012} & Self-flux method & Polycrystalline $\rm Li_8Co_2Te_2O_{12}$ & 1125$^\circ$C$\rightarrow$1000$^\circ$C$\rightarrow$800$^\circ$C in air\\
$\rm Li_4CrSbO_6$\cite{Bhardwaj2014} & Self-flux method & Polycrystalline $\rm Li_4CrSbO_6$ & 1100$^\circ$C$\rightarrow$950$^\circ$C$\rightarrow$850$^\circ$C\\
$\rm Na_3Ni_2BiO_6$\cite{Seibel2013} & Self-flux method & Polycrystalline $\rm Na_3Ni_2BiO_6$, $\rm Na_2CO_3$:$\rm Bi_2CO_3$ (1:1) flux & 850$^\circ$C$\rightarrow$750$^\circ$C in oxygen\\
$\rm BaNi_2TeO_6$\cite{song2022influence} & Self-flux method & $\rm BaCO_3$, $\rm NiO$, $\rm TeO_2$, $\rm PbCl_2$ flux & 1050$^\circ$C; 5h in air\\
\hline
\end{tabular}}
\end{center}
\end{table*}

\newpage

\section{\label{Section: Stacking} Honeycomb layer stacking}
At their structural front, pnictogen- and chalcogen-based honeycomb layered oxides generally crystallise in layered frameworks comprising alkali, alkaline-earth or coinage metal atoms wedged between parallel slabs consisting of transition metals or s-block metal atoms interlinked with pnictogen or chalcogen atoms via oxygen atoms in a honeycomb profile. The oxygen atoms within the slabs further coordinate with the alkali, alkaline-earth or coinage metal atoms forming interlayer bonds whose strength is substantially weaker than the covalent in-plane bonds within the slabs. The nature and magnitude of the interlayer coordination is typically prescribed by the Shannon-Prewitt\cite{shannon1970} radii of the alkaline-earth, alkali or coinage metal atoms which determine the interlayer distance of resultant heterostructures and their emergent configurations and stacking variations. 

\begin{figure*}[!b]
\centering
  \includegraphics[width=0.8\columnwidth]{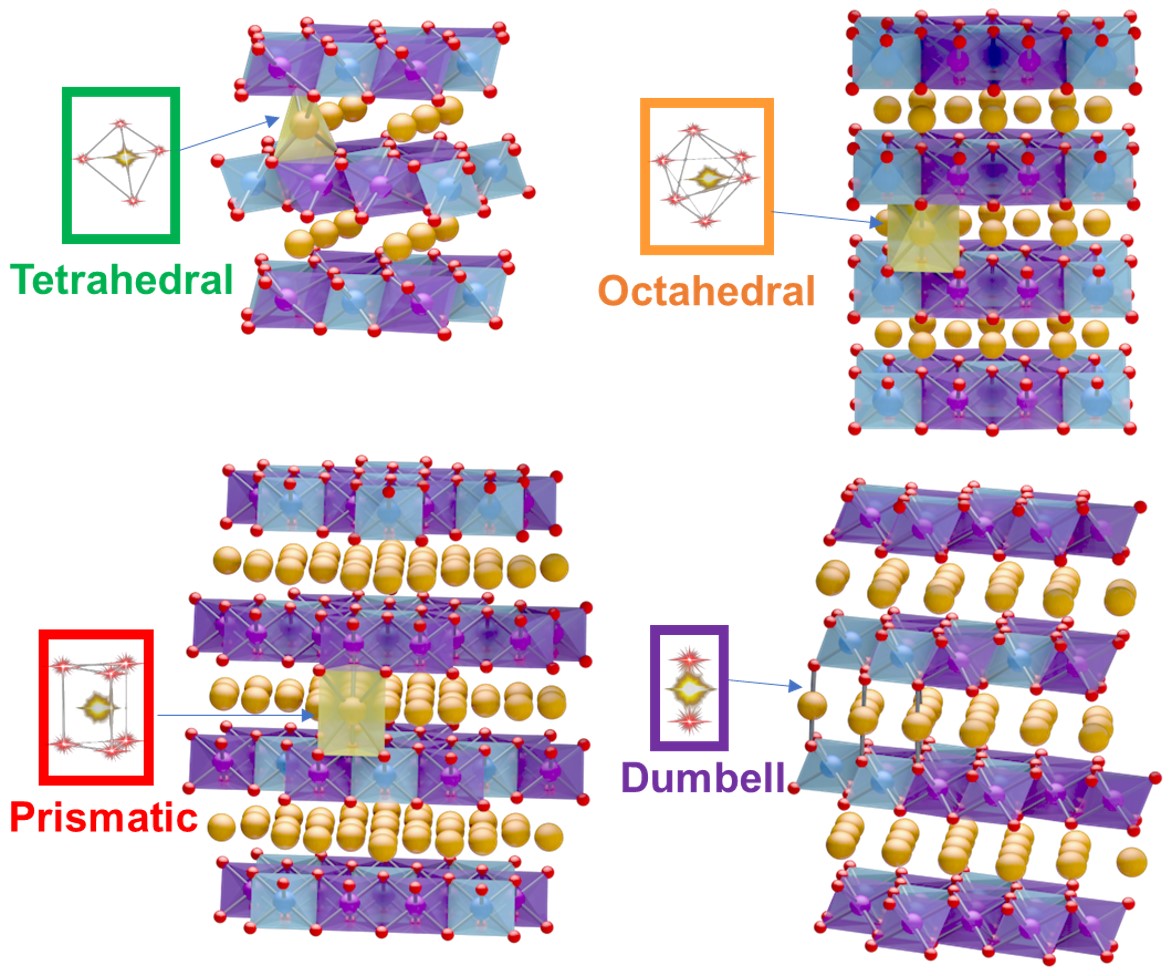}
  \caption{Stacking arrangement of coinage or alkali metal atoms in representative pnictogen- and chalcogen-based honeycomb layered oxides. Coinage, alkali or alkaline-earth metal atoms are shown in brown. O atoms are in red, whereas pnictogen or chalcogen atoms are in pale blue. Transition metal 3$d$ atoms are in purple. Figures adapted with permission.\cite{kanyolo2021honeycomb} Copyright 2021 Royal Society of Chemistry.}
  \label{Figure_18}
\end{figure*}

Depending on the atomic properties of the metal species (that is, K, Na, Li, Ag, Cu, Ba and so forth) embedded within the transition metal arrays, several variations in the nature of stacking formations of the pnictogen and chalcogen class of honeycomb layered oxides as well as their pertinent synthesis procedures are generally observed. For instance, the smaller-sized Li atoms in honeycomb layered oxide compositions suchlike $\rm Li_2Ni_2TeO_6$\cite{Grundish2019} and $\rm Li_3Ni_2BiO_6$\cite{Berthelot2012} have been found to coordinate with four oxygen atoms or six oxygen atoms from the adjacent slabs to form tetrahedral ($\rm LiO_4$) or octahedral ($\rm LiO_6$) coordination geometry respectively. Conversely, atoms with larger radii such as Na atoms, {\it vide infra}, have shown a strong propensity for coordination with six oxygen atoms ($\rm NaO_6$) to adopt prismatic or octahedral coordination geometries, as has been found in $\rm Na_2Ni_2TeO_6$\cite{Evstigneeva2011} and $\rm Na_3Ni_2BiO_6$\cite{Seibel2013, Liu2016}, respectively. \textbf{Figure \ref{Figure_18}} shows the coordination geometry of alkali, alkaline-earth and coinage metal atoms sandwiched between the slabs in pnictogen- and chalcogen-based honeycomb layered oxides along with their stacking sequences. 

Honeycomb layered oxides are indexed according to the arrangement of honeycomb stackings within their structures. In a notation known as Hagenmuller-Delmas notation,\cite{Delmas1976, delmas1981} a letter is typically assigned to represent the coordination geometry formed by the bonds between the alkaline-earth, alkali or coinage metal atoms and the surrounding oxygen atoms (generally, D for dumbbell (linear), O for octahedral, P for prismatic or T for tetrahedral). Further, a digit or numeral is used to indicate the number of repetitive honeycomb slabs (layers) per unit cell (usually, 1, 2 or 3). For example, $\rm K_2Ni_2TeO_6$ structure (illustrated in \textbf{Figure \ref{Figure_3}}) is classified as a P2-type framework,\cite{Masese2018, kanyolo2021honeycomb, masese2021topological} a notation that arises from the prismatic (P) coordination between the K atoms and oxygen atoms in the interlayer region and the two (2) repetitive honeycomb layers in each unit cell. A compendium (in Hagenmuller-Delmas' notation) outlining the various coordination geometries formed by the alkaline-earth, alkali and coinage metal atoms sandwiched between the transition metal slabs in pnictogen- and chalcogen-based honeycomb layered oxides and their stacking sequences is provided in \textbf{ Table \ref{Table_10}}. In the following subsection, we discuss the various coordination of alkali, alkaline-earth or coinage metal atoms found in pnictogen- and chalcogen-based honeycomb layered oxides ($i.e.$, honeycomb layered bismuthates, antimonates and tellurates).

\begin{table*}
\caption{Stacking sequences (in Hagenmuller-Delmas' notation) adopted by selected pnictogen- and chalcogen-based honeycomb layered oxides.\cite {Zvereva2013, Kurbakov2020, motome2020, Nalbandyan2013, Zvereva2017, Bhardwaj2014, Yao2020, Kumar2013, Morimoto2007, Gupta2013, Berthelot2012, Laha2013, McCalla2015, Kumar2012, Taylor2019, Roudebush2013b, Berthelot2012a, He2017, He2018, Stratan2019, Xu2005, Ramlau2014, Bera2017, Smirnova2005, Schmidt2013, Seibel2013, Seibel2014, Liu2016, Bhange2017, Gyabeng2017, Yan2019, Yadav2019, Smaha2015, Brown2019, Greaves1990, Nagarajan2002, Gupta2015, zvereva2016d, Mather2000, Mather1995, Schmidt2014, Uma2016, reya2011crystal, song2022influence, Masese2018, masese2021unveiling, masese2021topological, Grundish2019, Kimber2010} Table adapted with permission.\cite {kanyolo2021honeycomb} Copyright 2021 Royal Society of Chemistry.}
\label{Table_10}

\begin{center}
\scalebox{0.8}{
\begin{tabular}{ccl} 
\hline
\textbf{Coordination of alkali atoms} & \textbf{Hagenmuller-Delmas' notation} & \textbf{Honeycomb layered oxide compositions} \\
\textbf{with oxygen} & \textbf{(slab stacking sequence)} &  \\
\hline\hline
& & \\
Tetrahedral & T2 & $\rm Li_2Ni_2TeO_6$\cite{Grundish2019}\\
& &\\
Octahedral & O1 & $\rm NaNi_2BiO_{6-\delta}$\cite{Seibel2014}\\
 & O3 & $\rm Li_2Ni_2TeO_6$\cite{Grundish2019}, $\rm Li_3Ni_2BiO_6$\cite{Berthelot2012}, $\rm Na_3Ni_2BiO_6$\cite{Seibel2013, Liu2016, Bhange2017}, $\rm Na_2Cu_2TeO_6$\cite{Xu2005},\\
 & & ${\rm Na}_3M_2\rm SbO_6$ ($M = \rm Ni, Cu, Cr, Co, Mg, Zn$)\cite{Stratan2019, Smirnova2005, Schmidt2013, Yan2019, Yadav2019}, $\rm Li_2Cu_2TeO_6$,\\
 & & ${\rm Li}_3M_2\rm SbO_6$ ($M = \rm Ni, Cu, Co, Zn$)\cite{Stratan2019, Brown2019, Greaves1990},\\
 & & $\rm Na_3LiFeSbO_6$\cite{Schmidt2014}, $\rm Li_4{\it M}TeO_6$ ($M = \rm Co, Ni, Cu, Zn$), \\ 
 & & $\rm Li_4{\it M}SbO_6$ ($M = \rm Cr, Fe, Al, Ga, Mn$)\cite{Zvereva2013, Bhardwaj2014}\\
 & & \\
Prismatic & P2 & $\rm K_2Ni_2TeO_6$\cite{Masese2018, masese2021topological}, $\rm BaNi_2TeO_6$\cite{song2022influence}, \\
 & & $\rm Na_2{\it M}_2TeO_6$ ($M = \rm Mg, Zn, Co, Ni$)\cite{Kurbakov2020, Bera2017, masese2021unveiling}\\
 & P3 & $\rm K_2Ni_2TeO_6$\cite{masese2021topological}\\ 
 & & \\
Dumbbell & D & $\rm Ag_3NaFeSbO_6$, ${\rm Ag_3Li}M\rm TeO_6$ ($M = \rm Co, Ni$),\\
(linear) & & ${\rm Ag_3Li}M\rm SbO_6$ ($M = \rm Cr, Mn, Fe$)\cite{Bhardwaj2014}, ${\rm Ag_3Li}M_2\rm O_6$ ($M = \rm Ir, Ru, Rh$)\cite{Kimber2010, bette2019a},\\
 & & $A_3M_2D\rm O_6$ ($M = {\rm Ni, Co, Mg, Zn, Mn}$; $A = \rm Cu, Ag$; $D = \rm Bi, Sb$)\cite{Berthelot2012, Ramlau2014, Nagarajan2002, zvereva2016d}\\
 & & $\rm Cu_3{\it M}FeSbO_6$ ($\rm {\it M} = Li, Na$), $\rm Cu_5SbO_6$\cite{reya2011crystal}\\
 & & \\
\hline
\end{tabular}}
\end{center}
\end{table*}

\subsection{Tetrahedral coordination}
Cations comprising alkali metal atoms with small-sized atomic radii such as Li atoms in $\rm Li_2Ni_2TeO_6$ have been noted to form T2-type structures,\cite{Grundish2019} wherein the Li atoms form tetrahedral coordinations with oxygen atoms with 2 repetitive honeycomb layers (consisting of $\rm TeO_6$ and $\rm NiO_6$ octahedra) for each unit cell.  
Nonetheless, $\rm Li_2Ni_2TeO_6$ synthesised through topochemical ion-exchange route employing a P2-type $\rm Na_2Ni_2TeO_6$ host structure yields a metastable T2-type crystal structure which transforms into an O3-type $\rm Li_2Ni_2TeO_6$ (polytype) when subjected to high temperatures.\cite{Grundish2019} This indicates that Li atoms are more inclined to form octahedral coordinations over tetrahedral coordinations. This change in coordination for Li can be ascribed to the proximity of the atomic numbers of Li and O which allows their 2$p$ and 2$s$ orbitals to take part in the coordination chemistry. Since the radial distribution of the orbitals tend to hybridise in close proximity, the orthogonality of $p_{z}$, $p_{y}$ and $p_{x}$ orbitals is reflected, and thus the octahedral coordination of $\rm LiO_6$ becomes predominant.\cite{tada2022implications}

\begin{figure*}[!b]
\centering
  \includegraphics[width=0.8\columnwidth]{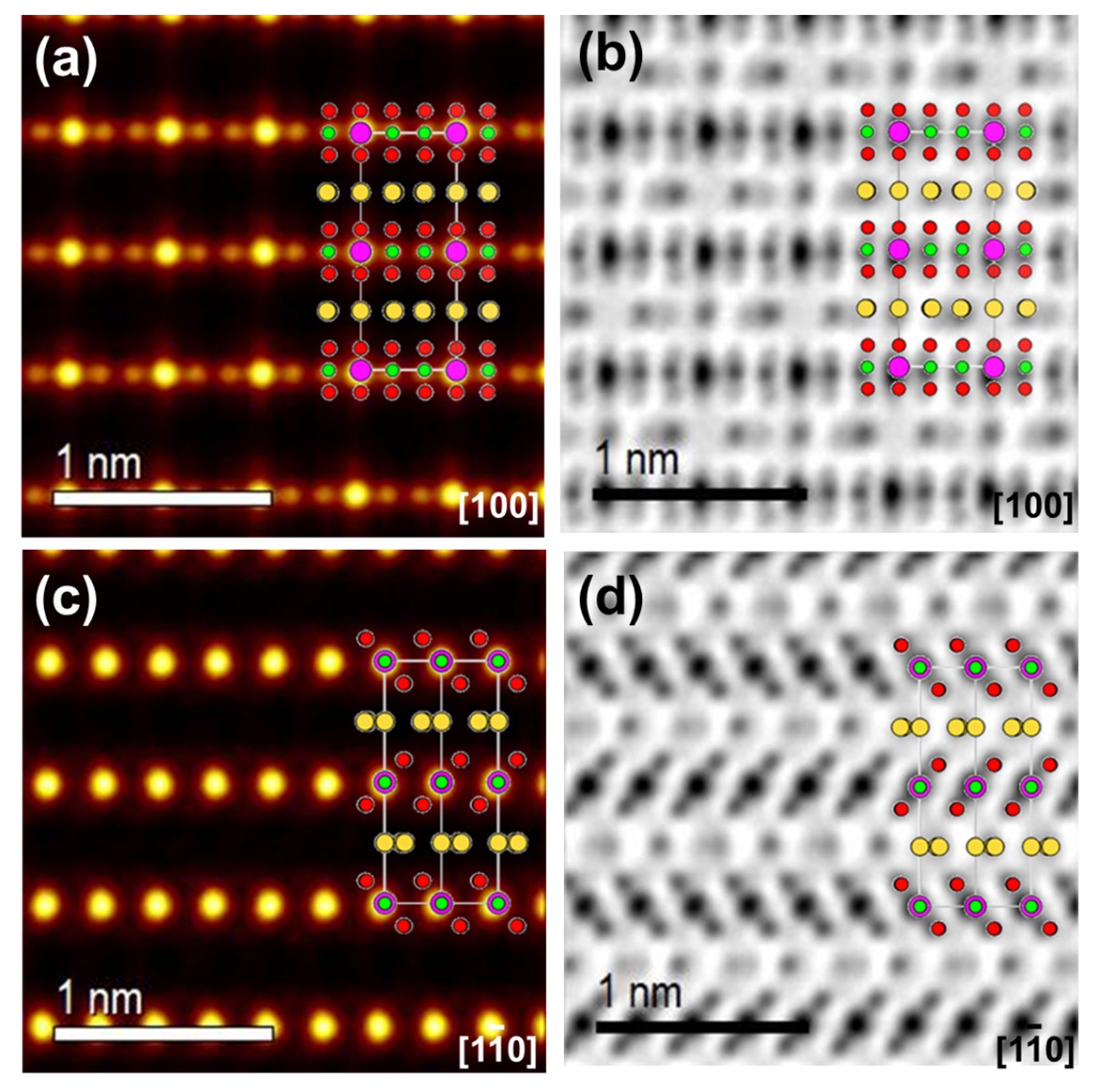}
  \caption{High-resolution TEM imaging of the stacking sequences of Na and transition metal atoms in P2-type $\rm Na_2Ni_2TeO_6$ along multiple zone axes. O atoms are in red, and the Ni and Te atoms are in green and pink, respectively. Na atoms are shown in yellow. (a) High-resolution electron microscopy image taken along [100] zone axis showing the ordering sequence of Te and Ni atoms, and (b) High-resolution electron microscopy image of heavy atoms along with lighter atoms such as Na and O. (c) High-resolution images of Te and Ni atoms taken when the crystal is rotated by 30$^{\circ}$ ($i.e.$, 
  [1$\overline{1}$0]) and (d) High-resolution imaging of heavy atoms along with lighter atoms along the 
  [1$\overline{1}$0] zone axis. The images taken from the two projections ([100] and 
  [1$\overline{1}$0] zone axes) are used to depict the three-dimensional crystal structure of P2-type $\rm Na_2Ni_2TeO_6$. Figures reproduced with permission.\cite{masese2021unveiling}
  Copyright 2021 Elsevier.}
  \label{Figure_19}
\end{figure*}

\subsection{Octahedral coordination}
Amongst the pnictogen- and chalcogen-based honeycomb layered oxides reported so far, a number of compounds entailing Li and Na alkali metal species have been found to form octahedral coordinations. Na-based compounds form a majority of the materials encompassing this geometry as outlined by the list of materials in \textbf{ Table \ref{Table_10}}. It is worth noting that electrochemical cation extraction (deintercalation) can also produce O-type frameworks. For instance, electrochemical delithiation of T2-type $\rm Li_2Ni_2TeO_6$ yields O2- and O6-type frameworks.\cite{Grundish2019}

\begin{figure*}[!b]
\centering
  \includegraphics[width=0.8\columnwidth]{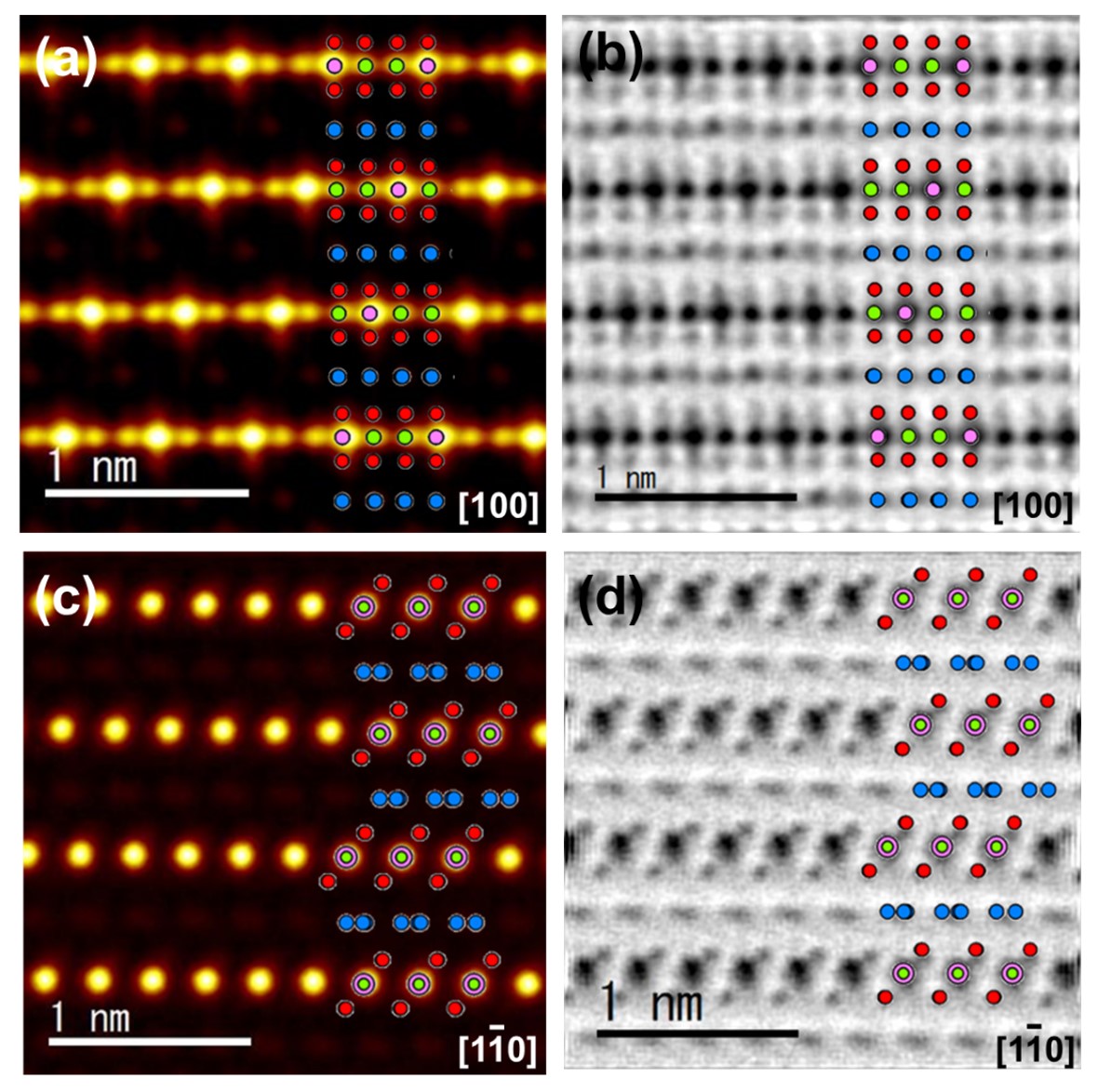}
  \caption{High-resolution TEM imaging of the stacking sequences of Na and transition metal atoms in P3-type $\rm K_2Ni_2TeO_6$ along multiple zone axes. O atoms are in red, and the Ni and Te atoms are in green and pink, respectively. K atoms are shown in blue. (a) High-resolution electron microscopy image taken along [100] zone axis showing the ordering sequence of Te and Ni atoms, and (b) High-resolution electron microscopy image of heavy atoms along with lighter atoms such as K and O. (c) High-resolution images of Te and Ni atoms taken when the crystal is rotated at 30$^{\circ}$ ($i.e.$, [1$\overline{1}$0]) and (d) High-resolution imaging of heavy atoms along with lighter atoms along the [1$\overline{1}$0] zone axis. The images taken from the two projections ([100] and [1$\overline{1}$0] zone axes) are used to depict the three-dimensional crystal structure of P3-type $\rm K_2Ni_2TeO_6$. Figures reproduced with permission.\cite{masese2021topological} Copyright 2021 American Chemical Society.}
  \label{Figure_20}
\end{figure*}

\subsection{Prismatic coordination}
The formation of prismatic coordinations (P-type), requires the materials to embody interlayer distances wider than octahedrally coordinated (O-type) layered oxides. As such, alkali atoms with larger ionic radii are considered a requisite for fabricating honeycomb layered oxides with this structure. It is worth noting that cation species utilised in these structures exclusive to alkali metal atoms (such as K and Na); some alkaline-earth metal atoms such as Ba in $\rm BaNi_2TeO_6$\cite{song2022influence} have been reported to engender similar configurations. Further, in an interesting finding, the electrochemical desodiation of O3-$\rm Na_3Ni_2SbO_6$ was also found to engender unique P3-type frameworks that were previously thought to be exclusive for atoms with atomic radii equal to or larger than K atoms. To explicitly visualise the slab stacking sequences in P-type honeycomb layered oxides, synchrotron XRD has been used in tandem with transmission electron microscopy (TEM) as shown in \textbf{Figures \ref{Figure_19}} and \textbf{\ref{Figure_20}}. Here, the high-resolution TEM images of P2-type $\rm Na_2Ni_2TeO_6$\cite{masese2021unveiling} and P3-type $\rm K_2Ni_2TeO_6$\cite{masese2021topological} highlight their P-type slab sequences within the layered frameworks.

\begin{figure*}[!b]
\centering
  \includegraphics[width=0.8\columnwidth]{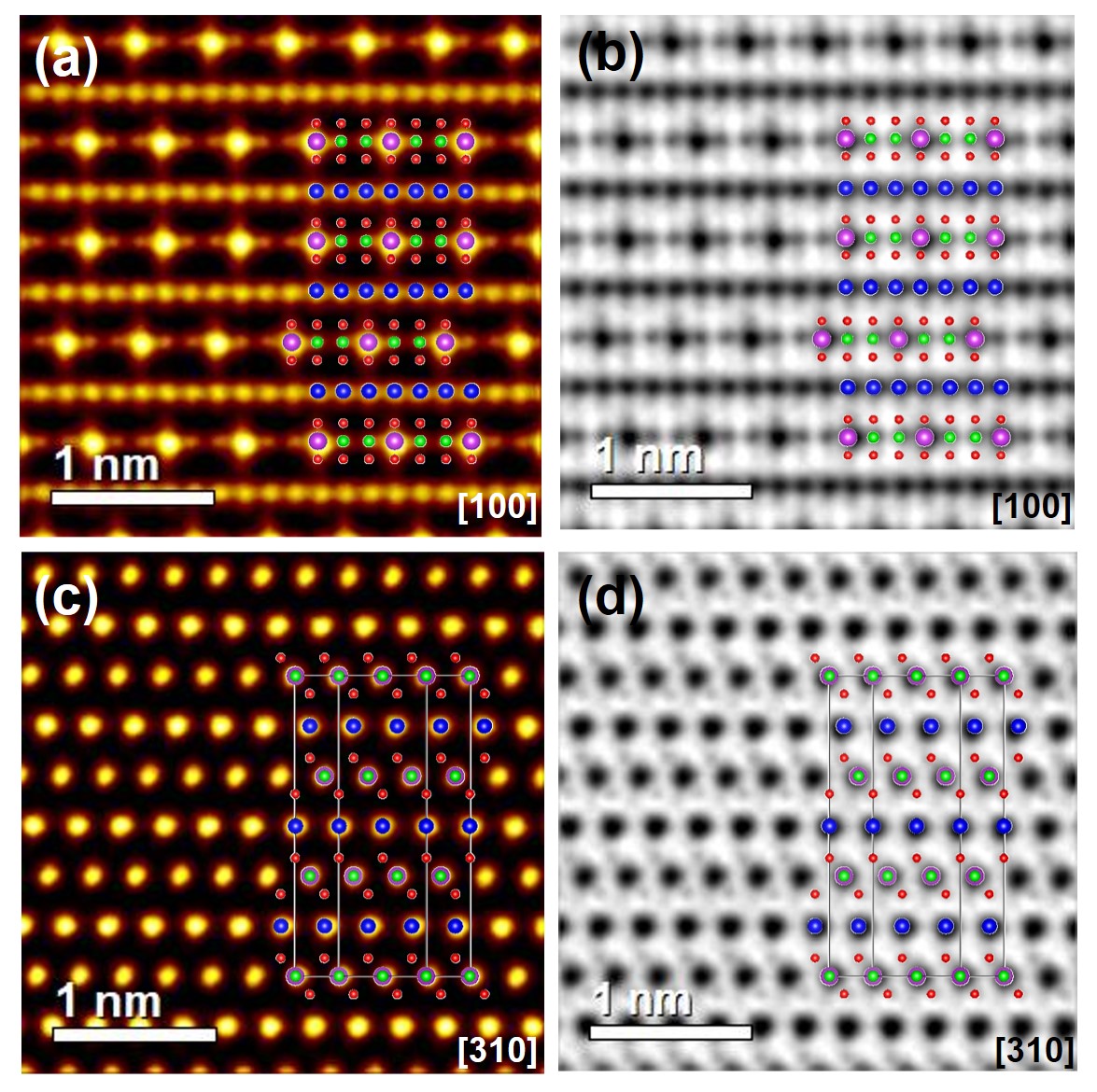}
  \caption{High-resolution TEM imaging of the stacking sequences of Na and transition metal atoms in D3-type $\rm Ag_3Ni_2BiO_6$ along multiple zone axes. O atoms are in red, and the Ni and Te atoms are in green and pink, respectively. Ag atoms are shown in blue. (a) High-resolution electron microscopy image taken along [100] zone axis showing the ordering sequence of Bi, Ni and Ag atoms, and (b) High-resolution electron microscopy image of heavy atoms along with lighter atoms such as O. (c) High-resolution images of Bi, Ni and Ag atoms taken along the [310] zone axis and (d) High-resolution imaging of heavy atoms along with lighter atoms along the [310] zone axis. The images taken from the two projections ([100] and [310] zone axes) are used to depict the three-dimensional crystal structure of D3-type $\rm Ag_3Ni_2BiO_6$.}
  \label{Figure_21}
\end{figure*}

\subsection{Linear (Dumbbell) coordination}
Coinage metal atoms such as Cu and Ag embedded between the sheets of transition metal slabs in the honeycomb layered frameworks have been found to form linear (dumbbell-like) coordinations with oxygen atoms. This proclivity towards linear coordinations between coinage-metal atoms with oxygen atoms is ascribed to the hybridised $d_{z^2}$ orbitals of the group 11 ($A^{+}$, where $A = \rm Cu, Ag, Au$ {\it et cetera}) monovalent cations.\cite{tada2022implications} Even though these group 11 elements are known $s$-orbital systems with fully closed $d$-valence electrons, they are also classified as transition metals and are thus expected to undergo ionisation leading to the formation of compounds manifesting $sd$ hybridisation.\cite{pettifor1978theory, lacroix1981density, manh1987electronic, gallagher1983positive, maiti1980improved, horn1979adsorbate}
It is also worth noting that the dumbbell coordinations in the Ag-based honeycomb layered oxides have been found to crystallise with even larger interlayer distances than their Na analogues, in consonance with a theoretical prediction which stipulated that the Ag atoms in the tellurate subject of the study had a propensity for a dumbbell-like linear coordination instead of a prismatic coordination.\cite{tada2022implications}
As depicted by the high-resolution TEM image of $\rm Ag_3Ni_2BiO_6$ in \textbf{Figure \ref{Figure_21}}, the crystal forms a linear arrangement which can be viewed along multiple zone axes.

\begin{figure*}[!t]
\centering
  \includegraphics[width=0.8\columnwidth]{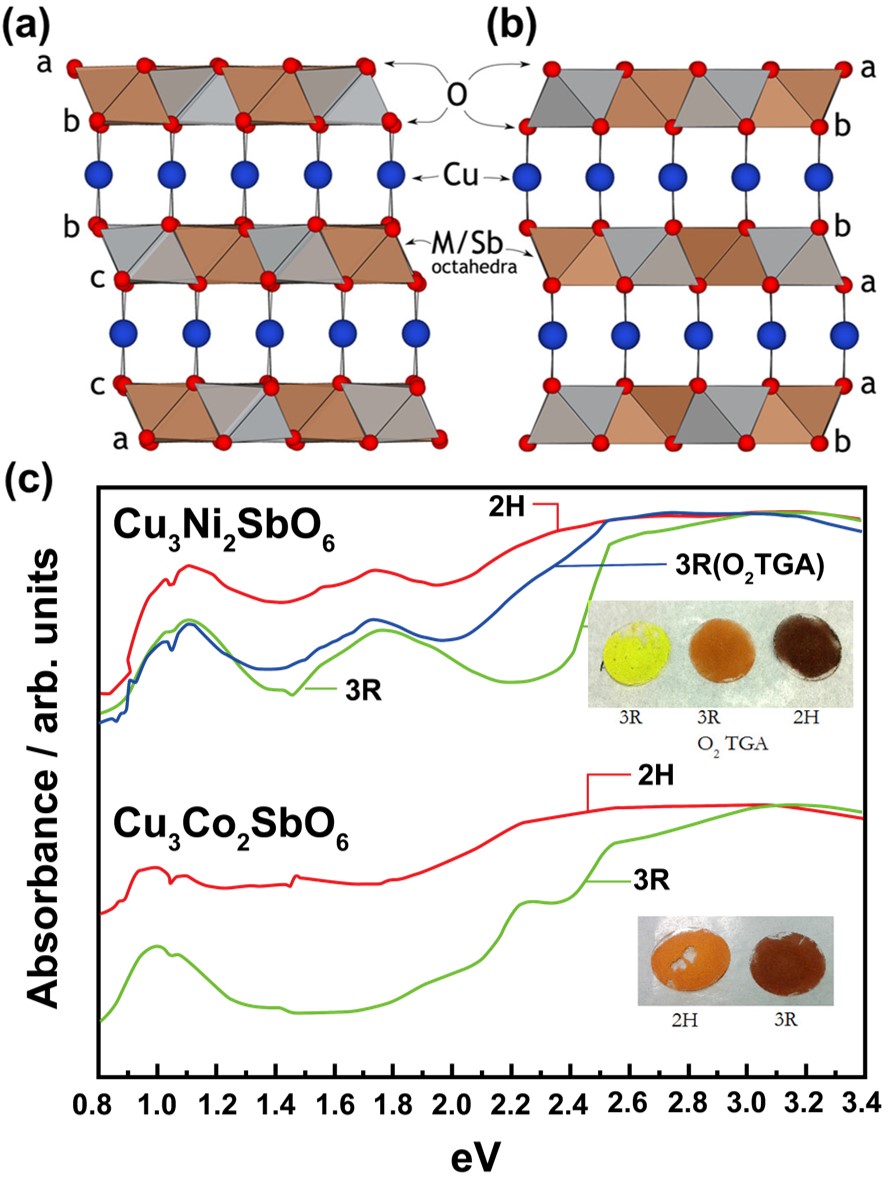}
  \caption{Copper-based Delafossites $\rm Cu_3{\it M}_2SbO_6$ ($\rm {\it M} = Co, Ni $). Depiction of the stacking sequence of (a) 3R polytype and (b) 2H polytype. (c) Diffuse reflectance spectroscopic measurement of 3R- and 2H-polytypes of $\rm Cu_3Co_2SbO_6$ and $\rm Cu_3Ni_2SbO_6$ measured between 0.8 and 5.4 eV. The spectra have been normalised to their maximum values. A colour bar marks the part of the spectrum in the visible regime. Photographs of the samples used for the measurements are shown for visual reference. Figures reproduced with permission.\cite{roudebush2015rhombohedral} Copyright 2015 American Chemical Society.}
  \label{Figure_22}
\end{figure*}

As mentioned in a previous section, copper-based layered oxides belong to the broader class of Delafossite compounds. In the Delafossite family, compounds have been noted to crystallise in two polytypes: (i) a three-layer rhombohedral polytype (denoted as 3R) which has all octahedra axes oriented in the same direction (shown in \textbf{Figure \ref{Figure_22}a}) and close-packed cubic stacking; and (ii) a two-layer hexagonal polytype (denoted as 2H), wherein all the octahedra axes are oriented at 180$^{\circ}$ with respect to neighbouring layers that are stacked in a close-packed hexagonal fashion (\textbf{Figure \ref{Figure_22}b}).

Inquests into the Delafossite polytypes of pnictogen- and chalcogen-based honeycomb layered oxides have unearthed intriguing optical and topological characteristics that bear significance in a wide variety of contemporary applications. For example, 3R polytypes such as that of $\rm Cu_3{\it M}_2SbO_6$ ($\rm {\it M} = Co, Ni $) are noted to display distinct optical absorbance spectra compared to their 2H counterparts.\cite{Roudebush2015, roudebush2015rhombohedral} Additionally, thermal gravimetric analyses have further revealed 3R→2H polytypic transformations (of second order) and a first-order structural transition that entailed the rearrangement of the honeycomb slabs. The optical spectrum of the 3R polytype of $\rm Cu_3Ni_2SbO_6$ was noted to resemble that of its 2H polytype variant (\textbf{Figure \ref{Figure_22}c}) after heating in flowing $\rm O_2$, implying that the difference between the samples was due to oxygen intercalation. This intercalation of oxygen is a prominent feature of the rare-earth Delafossites that has been observed in several other investigations. Similar colour changes have also been observed when 3R polytypes were heated in Ar gas, even though the changes were not as dramatic as when oxygen is employed. Altogether, these results suggest the further investigation of this broad class of honeycomb layered antimonates as gas sensors since they have a high colouration efficiency upon exposure to oxygen.

In accordance with the Hagenmuller-Delmas notation, the polytypes mentioned above can be denoted as D2 and D3, where the number denotes the number of layers in the unit cell and ``D'' denotes the dumbbell configuration of ${\rm Cu^{+}}$. $\rm Cu_3Co_2SbO_6$\cite{Roudebush2015, roudebush2015rhombohedral} and $\rm Cu_3Ni_2SbO_6$\cite{Roudebush2015, roudebush2015rhombohedral} were noted to display the D2-type frameworks whereas D3-type frameworks have been observed in Delafossites such as $\rm Ag_3Co_2SbO_6$,\cite{zvereva2016d} $\rm Ag_3Ni_2BiO_6$\cite{Berthelot2012} and so forth.

\subsection{Bilayered coordinations}

Although not experimentally prevalent within pnictogen- and chalcogen-based honeycomb layered oxides, bilayered arrangements of cations coordinated with oxygen atoms have been found in Ag-based materials.\cite{masese2023honeycomb} These materials exhibit the universality of comprising a bifurcated bipartite honeycomb lattice into bilayers comprising its hexagonal sub-lattices. Generally in group 11-based materials ($A = \rm Cu, Ag, Au$ \textit{et cetera}), it has been theorised that $sd_{ z^2}$ hybridisation leads to the degeneracy of their $(n + 1)s^1$ and $nd_{z^2}^2$ orbitals ($n = 3, 4, 5$) which in turn affects their valence states leading to three types of cations, $A^{1-}, A^{2+}$ and $A^{1+}$. 
The first two valence states ($A^{1-}, A^{2+}$) correspond to the two-fold degenerate $nd_{z^2}^1, (n + 1)s^2$ state. In other words, to participate in bonding, this state can either gain an electron in the $d_{z^2}$ orbital (valence band) or lose two electrons from the $s^2$ orbital (conduction band). This degeneracy corresponds to the left or right Ag chiral states distinguished by a degree of freedom known as pseudo-spin, 
analogous to the pseudo-spin of carbon atoms in graphene.\cite{mecklenburg2011spin, georgi2017tuning, allen2010honeycomb} Like the electrons in graphene, diffusing cations are treated as mass-less Dirac fields at the Dirac points (honeycomb lattice).\cite{mecklenburg2011spin, georgi2017tuning, allen2010honeycomb} In this case, the chirality of the cation corresponds to the helicity state (the electron spin) of the $d_{z^2}^1$ electron. The last valence state ($A^{1+}$) 
corresponds to the original $nd_{z^2}^2, (n + 1)s^1$ state before $sd$ hybridisation. 

\begin{figure*}[!b]
\centering
  \includegraphics[width=0.75\columnwidth]{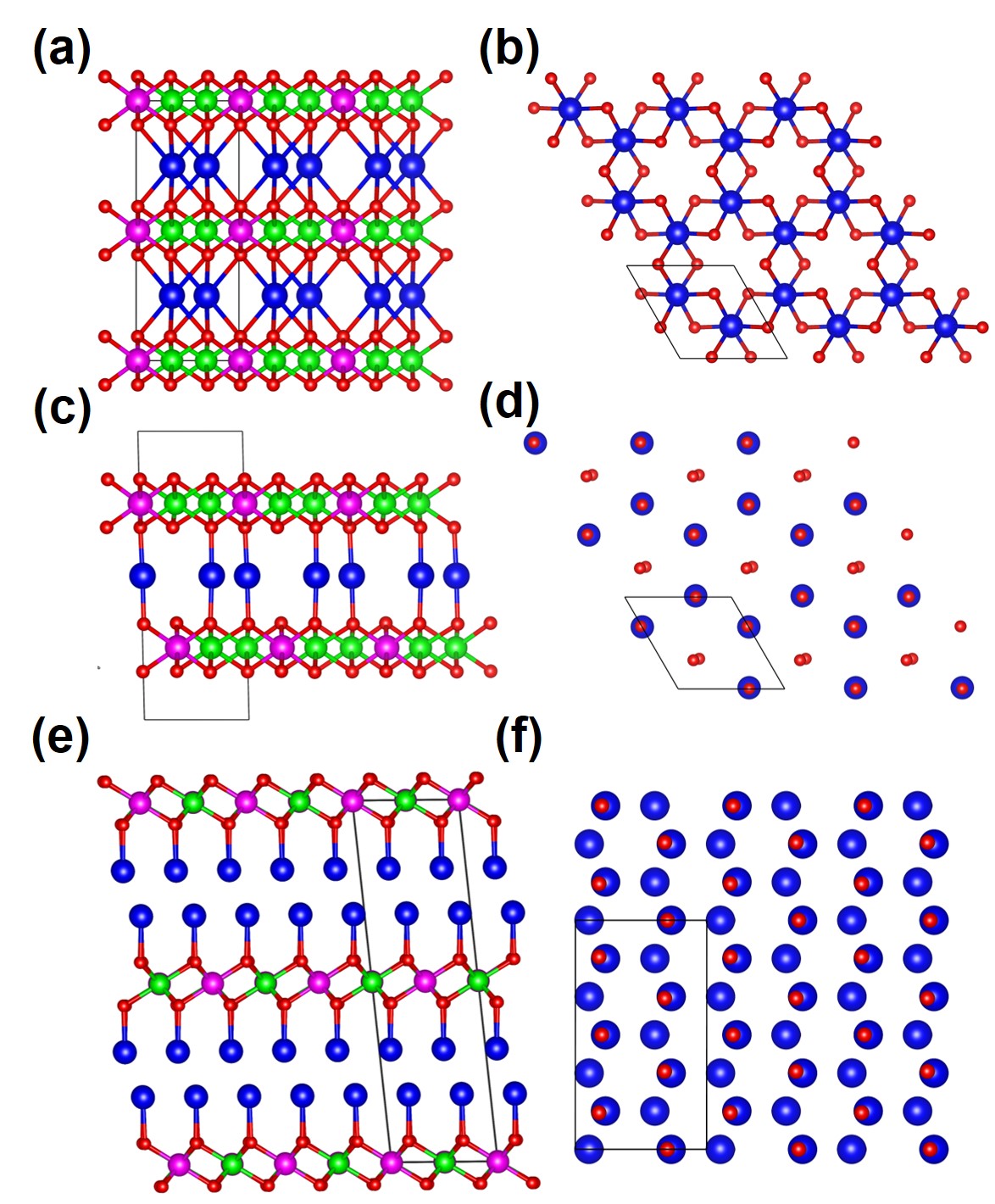}
  \caption{Coordination geometry of silver atoms in honeycomb layered tellurates.\cite{tada2022implications, masese2023honeycomb} Ag atoms are in blue, O atoms are in red, and the Ni and Te atoms are in green and pink, respectively. (a) Prismatic coordination of Ag with O atoms in simulated $\rm Ag_2Ni_2TeO_6$, as viewed along the $c$-axis. (b) Honeycomb arrangement of Ag atoms prismatically coordinated with O atoms in $\rm Ag_2Ni_2TeO_6$, as viewed on the $ab$-plane. All O atoms are bonded with only two Ag atoms. 
  (c) Linear coordination of Ag with O atoms in simulated $\rm Ag_2Ni_2TeO_6$ polytype, as viewed along the $c$-axis. (d) Honeycomb arrangement of Ag atoms linearly coordinated with O atoms in $\rm Ag_2Ni_2TeO_6$ polytype, as viewed on the $ab$-plane. O atoms are bonded either with one Ag atom or none.
  (e) The expected crystal structure of $\rm Ag_6Ni_2TeO_6$ after monolayer-bilayer phase transition and saturation, already 
  designed in preliminary experiments.\cite{masese2023honeycomb} (f) Honeycomb arrangement of Ag atoms linearly coordinated with O atoms in $\rm Ag_6Ni_2TeO_6$, as viewed on the $ab$-plane.}
  \label{Figure_48}
\end{figure*}

Nonetheless, since $A^{1-}, A^{1+}$ and $A^{2+}$ are related to each other by $sd$ hybridisation,  
the interactions of the cations within the lattice can be modelled by SU($2$)$\times$U($1$) interactions, analogous to electro-weak theory\cite{Zee2010}, which leads to a monolayer-bilayer phase transition by generating a mass term, 
\begin{subequations}
\begin{align}
   H_{\rm Ising} \propto -T_{\rm c}^2K(R)\left \langle S_1 + S_2 \right\rangle - J_{\rm RKKY}(R)\,\left\langle S_1S_2 \right\rangle \propto (T_{\rm c} - T(R))\Bar{\psi}(R)\psi(R), 
\end{align}
where $H_{\rm Ising}$ is the 1D Ising Hamiltonian (energy density), $K(R) = 1/R^2$ is a pseudo-magnetic field (Gaussian curvature on the two-sphere of radius, $R$), $\psi$ is the Dirac spinor, $\psi^{\rm T} = (A^{1-},\,\, A^{2+})$ is the transpose showing the right-handed and left-handed components, $\overline{\psi} = \psi^{*\rm T}\gamma^0$ is the adjoint spinor ($\gamma^0$ is the time component Dirac matrix), $T_{\rm c} \propto m$ is a critical temperature proportional to the acquired mass of the cations, $T(R)$ is a temperature field playing the role of the Higgs, with a finite temperature gradient ($\Vec{\nabla}T(R) \neq 0$) $S_1$ and $S_2$ are the pseudo-spin operators, and $J_{\rm RKKY}$ is the Ruderman–Kittel–Kasuya–Yosida (RKKY) exchange interaction\cite{ruderman1954indirect, kasuya1956prog, yosida1957magnetic}, \begin{align}
    J_{\rm RKKY}(R) \sim \frac{T_{\rm c}k_{\rm F}^3}{(2\pi)^3}\frac{(2k_{\rm F}R\cos(2k_{\rm F}R) - \sin(2k_{\rm F}R))}{(2k_{\rm F}R)^4},
\end{align}
\end{subequations}
Consequently, the mass term corresponds to a metallophilic (numismophilic) interaction which leads to a monolayer-bilayer phase transition in cationic honeycomb lattices for finite pseudo-spin magnetisation, $S_1 + S_1 = 1$. 
This phase transition is analogous to Peierls instability\cite{pytte1974peierls, peierls1955quantum}, whose potential energy, $U = K + 2J\cos(2kr)$ has two terms consistent with H\"{u}ckel method\cite{bendazzoli2004huckel}:, \textit{i.e.} a mechanical term,

\begin{subequations}
\begin{align}
    K = \int dr_1 \phi_1(r_1)V(r_1)\phi_1(r_1) + \int dr_2\phi_2(r_2)V(r_2)\phi_2(r_2) \equiv \alpha_1 + \alpha_2,
\end{align}
due to Coulomb interaction $V(r)$ responsible for dimerisation and the exchange interaction term,
\begin{align}
    J = \int dr_1dr_2\phi_1^*(r_1)\phi_2(r_2)V(r_1 - r_2)\phi_1(r_2)\phi(r_1) \sim \alpha_1\alpha_2,
\end{align}
\end{subequations}
where $k$ ($0 \leq k \leq k_{\rm F}$) is the one-dimensional wave vector, $\phi_j$ ($j = 1,2$) are the spatial electron wave functions (at positions $r_j$) of adjacent $\pi$ bond $p_z$ electrons of \textit{e.g.} carbon atoms in polyacetylene, and $R$ defines the relative positions of carbon atoms on the lattice.\cite{bendazzoli2004huckel, garcia1992dimerization} 

Thus, a theorem analogous to Peierls theorem\cite{peierls1955quantum, peierls1979surprises} in one-dimension and Jahn-Teller theorem\cite{jahn1937stability} in three dimensions can be stated for two-dimensional honeycomb lattices of $A = \rm Cu, Ag$ or $\rm Au$ in layered materials: \textit{given a 2D bipartite honeycomb lattice with degenerate pseudo-spin states, the system can always lower its energy by geometrically (topologically) lifting this degeneracy by distortion}. In the case of Ag bilayers, the degeneracy is between $\rm Ag^{2+}$ and $\rm Ag^{-}$ valence states. In the simple case of the bilayered $\rm Ag_2NiO_2$, which requires the existence of the subvalent state $\rm Ag^{1/2+}$ to be electronically neutral is replaced by $\rm Ag_2NiO_2 = Ag^{2+}Ag^{1-}Ni^{3+}O_2^{2-}$ instead, which already implies bifurcation of the honeycomb lattice. Hybrids with a stable honeycomb monolayer and hexagonal bilayer arranged along the [001] plane in an alternating fashion can also be explained,  
$\rm Ag_3Ni_2O_4 = \frac{1}{2}(Ag_2^{1+}Ni_2^{3+}O_4^{2-})(Ag_2^{2+}Ag_2^{1-}Ni_2^{3+}O_4^{2-})$ with (an effective) subvalent state, $\rm Ag^{2/3+}$. Thus, summarising the possible fractional subvalent states of Ag in these materials is a matter of considering the various ratios of coinage metal atoms in the possible lattices. In this case, the lattice with $1:1$ left-right chiral ($A^{2+}, A^{1+}$) is bilayered (bifurcated honeycomb) with sub-valency $1/2+$, whilst the lattice with $1:1$ left chiral ($A^{1+}, A^{1+}$) is hexagonal with valency $1+$. These hexagonal lattices tend to form linear coordinations with oxygen atoms. Note that, $sd_{z^2}$ hybridisation tends to occur efficiently whenever the $d_z^2$ orbital is isolated from the rest of the $d$ orbitals by crystal field splitting. 
Thus, the bifurcation mechanism is favoured in layered crystal structures whose Ag atoms at the monolayer-bilayer critical point ($T = T_{\rm c}$, with $T_{\rm c}$ the critical temperature/effective cationic mass on the honeycomb lattice) of the lattice exhibit prismatic or linear coordinations to O atoms, since these systems would have an isolated $nd_{z^2}$ orbital according to crystal field splitting theory.\cite{burns1993mineralogical, ballhausen1963introduction, jager1970crystal, de19902} 

Bilayers with prismatic or linear coordination are expected to be under-saturated with Ag cations at the critical point (point where the separation distance of the sub-lattices vanishes, $T = T_{\rm c}$), leaving either all O atoms bonded with only two Ag atoms in the prismatic case, or bonded with only one Ag atom with some bonded to none in the linear case, as shown in \textbf{Figures \ref{Figure_48}a} and \textbf{\ref{Figure_48}b}.  
Saturating the prismatic and linear types with $\rm Ag^{1+}$ ought to create a bifurcated Ag lattice with linear coordination, whereby each Ag atom is bonded to a single O atom as shown in \textbf{Figure \ref{Figure_48}c}.\cite{tada2022implications} 
In this case, the lattice behaves like a hybrid with the ratio $1:1:1$ ($\rm Ag^{2+}, Ag^{1-}, Ag^{1+}$) corresponding to valency $2/3+$. Indeed, such a saturated pnictogen- and chalcogen-based honeycomb layered oxide, $\rm Ag_6Ni_2^{2+}Te^{4+}O_6^{2-}$ with Ag sub-valency $2/3+$ has been reported.\cite{kanyolo2022advances2, masese2023honeycomb} From this, the under-saturated material is expected to be given by $\rm Ag_4Ni_2^{3+}Te^{4+}O_6^{2-}$ or $\rm Ag_4Ni_2^{2+}Te^{6+}O_6^{2-}$ (prismatic) with Ag sub-valency $1/2+$, whereby the saturation is achieved either by the $\rm Ni^{3+} \rightarrow Ni^{2+}$ or $\rm Te^{6+} \rightarrow Te^{4+}$ reaction, 
\begin{align}
    {\rm Ag^{2+}_2Ag^{1-}_2Ni}_2^{x+}{\rm Te}^{y+}{\rm O_6^{2-} + 2Ag^{1+}} \underset{\rm de-sat.}{\stackrel{\rm sat.}{\rightleftharpoons}} {\rm Ag^{2+}_2Ag^{1-}_2Ag_2^{1+}Ni}_2^{x'+}{\rm Te}^{y'+}{\rm O_6^{2-}},
\end{align}
where $\Vec{r} = (3, 4) \rightarrow \Vec{r'} = (2,4)$ or $\Vec{r} = (2,6) \rightarrow \Vec{r'} = (2,4)$ respectively with $\Vec{r} = (x, y)$ and $\Vec{r'} = (x', y')$. This mechanism to explain the linear coordination in saturated bilayers is essential since structurally the cation lattice of $\rm Ag_6Ni_2^{2+}Te^{4+}O_6^{2-}$ with a linear coordination of Ag to O and a sub-valency of $+2/3$ cannot be directly obtained by a bifurcation mechanism without saturation. In fact, at the critical point, $\rm Ag_6Ni_2^{2+}Te^{4+}O_6^{2-}$ has to revert back to $\rm Ag_4Ni_2TeO_6$ by expelling the excess $\rm 2Ag^{1+}$ cations, before the bifurcation can be undone. Meanwhile, the reported Ag structure of $\rm Ag_6O_2$ instead of $\rm Ag_4O_2$ with a subvalency of $2/3+$ (instead of $1/2+$) is bilayered, whereby the Ag atoms are octahedrally coordinated with O atoms.\cite{beesk1981x} The octahedral structure can be understood to be as a result of Ag saturation, $\rm Ag_4O_2 + 2Ag^{1-} \rightarrow Ag_6O_2$, which transforms the coordination from the simple prismatic structure similar to $\rm Ag_2F = Ag^{2+}Ag^{1-}F$ to the observed octahedral structure.\cite{beesk1981x}

Finally, more complicated structures may have different ratios and combinations leading to sub-valency states, $1/3+$ ($\rm 2Ag^{1+}, Ag^{1-}$) or $4/5+$ ($\rm Ag^{2+}, Ag^{1-}, 3Ag^{1+}$), provided $sd_{z^2}$ hybridisation is guaranteed.\cite{pettifor1978theory, lacroix1981density, manh1987electronic, gallagher1983positive, maiti1980improved, horn1979adsorbate} 
In principle, subvalent Ag cations have been reported in Ag-rich oxide compositions such as $\rm Ag_5SiO_4$, $\rm Ag_5GeO_4$, $\rm Ag_5Pb_2O_6$, $\rm Ag_{13}OsO_6$, $\rm Ag_3O$, $\rm Ag_{16}B_4O_{10}$, alongside the halides such as $\rm Ag_2F$, and the theoretically predicted $\rm Ag_6Cl_4$.\cite{derzsi2021ag, kovalevskiy2020uncommon, ahlert2003ag13oso6, jansen1992ag5geo4, jansen1990ag5pb2o6, argay1966redetermination, beesk1981x, bystrom1950crystal}

\subsection{Cation potential map}

\begin{figure*}[!t]
\centering
  \includegraphics[width=0.8\columnwidth]{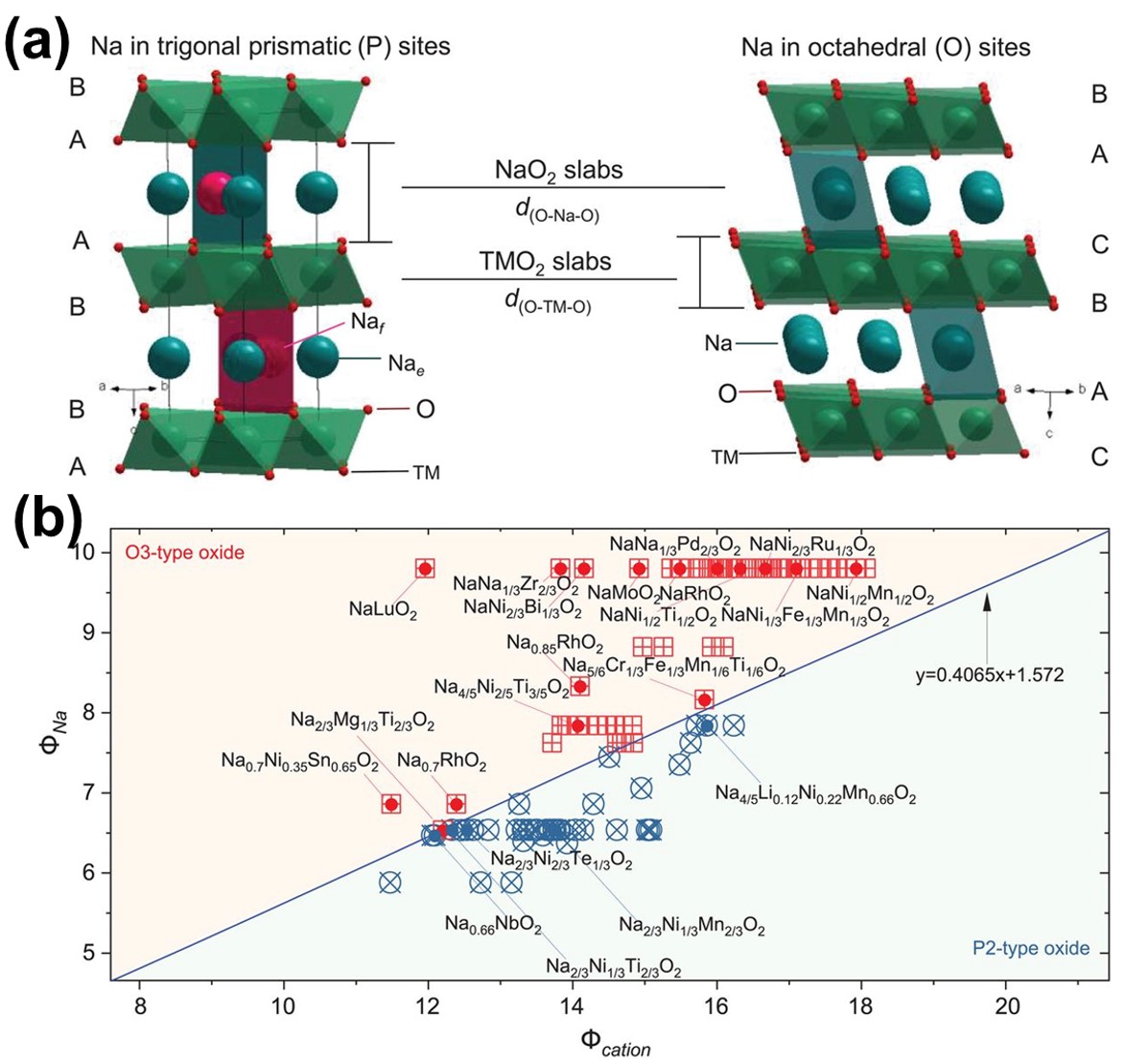}
  \caption{Cationic potential map of Na-based layered oxides. (a) Crystal structural framework of representative O3-type (rhombohedral) and P2-type (hexagonal) layered oxides.The stacking sequences of oxygen atoms are shown respectively as ABCABCABC and AABBAABB. The interlayer distance of transition metal layer and Na metal layer is denoted as $\rm {\it d}_{(O-TM-O)}$ and $\rm {\it d}_{(O-Na-O)}$, respectively. In the P2-type structure, Na atoms reside in trigonal prismatic sites that either share faces with transition metal octahedra ($\rm Na_{\it f}$) or share edges ($\rm Na_{\it e}$). (b) Cationic potential of representative O3- and P2-type Na-ion layered oxides, considering the transition metal (TM) composition, Na content and TM oxidation state. The cationic potential ($\rm \Phi_{cation}$) is attained by normalising the ionic potential of the transition metal ($\rm \Phi_{TM}$) and the ionic potential of the alkali-metal ($\rm \Phi_{A}$) relative to the ionic potential of oxygen ($\rm \Phi_{O}$). O3-type $\rm Na_3Ni_2BiO_6$ ($\rm NaNi_{2/3}Bi_{1/3}O_2$) and P2-type $\rm Na_2Ni_2TeO_6$ ($\rm Na_{2/3}Ni_{2/3}Te_{1/3}O_2$) are shown in the plots. Figures reproduced with permission.\cite{zhao2020rational} Copyright 2020 American Association for the Advancement of Science.}
  \label{Figure_23}
\end{figure*}

As previously mentioned, P-type layered frameworks present wider interslab distances than O-type frameworks even though new P-type frameworks can be attained through the electrochemical cation deintercalation of certain O-type frameworks. Electrosynthesis presents a feasible route for synthesising new P-type frameworks, although the reaction may produce metastable crystal structures. Given that the performance of layered oxides is heavily influenced by their crystal structures, synthesis control over their final crystal structures and compositions becomes an important aspect for consideration during the design and development of suchlike materials. This issue is further propounded by the economic aspects surrounding synthesis techniques such as conventional solid-state techniques which can pose significant inhibitions to exploratory studies. As such, predicting the stacking mode of the target layered oxides in advance can be envisioned to not only hasten the discovery of this class of materials but also cut down the costs involved in exploratory synthesis. Although theoretical computations have been found to sufficiently predict the stability of simple oxides, their accuracy in modelling oxides with complex transition metal combinations is still inadequate.

An empirical formula (based on `cationic potential' ($\rm \Phi_{cation}$) has been broached to give an accurate, and a very effective criterion to predict the stacking mode of layered oxides.\cite{zhao2020rational} The cationic potential ($\rm \Phi_{cation}$) is attained by normalising the ionic potential of the transition metal ($\rm \Phi_{TM}$) and the ionic potential of the alkali-metal ($\rm \Phi_{A}$) relative to the ionic potential of oxygen ($\rm \Phi_{O}$), which reflects the interaction between the alkali atoms and the transition metal atoms through their coordinations with oxygen atoms. The cationic potential method provides more accurate predictions, in comparison with the Rouxel diagram methodology,\cite {rouxel1976diagramme, delmas1980structural} for both the materials with complex transition metals as well as those with the same transition metal in different oxidation states. By plotting a phase map (cationic potential phase map) of layered oxides based on their cationic potential values ($\rm \Phi_{cation}$) against the ionic potential of alkali atoms ($\rm \Phi_{A}$), a boundary line that distinguishes P-type and O-type structures can be observed. \textbf{Figure \ref{Figure_23}}  shows the plot for Na-based layered oxides, indicating the cationic potential values of O3-type $\rm Na_3Ni_2BiO_6$ ($\rm NaNi_{2/3}Bi_{1/3}O_2$) and P2-type $\rm Na_2Ni_2TeO_6$ ($\rm Na_{2/3}Ni_{2/3}Te_{1/3}O_2$). 

Based on the phase map, a higher cationic potential indicates higher electron density in the transition metal layers and stronger covalency between the transition metal and oxygen, leading to P-type stacking, and for those with similar transition metal constituents, the average ion potential of alkali ions can increase with an increasing content of alkali atoms, which is conducive to O-type stacking. In that vein, the stacking mode of a layered oxide in this range of compositions can be predicted by examining the region of its potential point in the phase map once its composition has been determined. Therefore, new pnictogen- and chalcogen-based honeycomb layered oxides can rationally be designed under the guidance of the cationic potential phase map. Thus, opening up new uncharted chemical composition territories for this class of honeycomb layered oxides.

\newpage

\section{\label{Section: Characterisation} Material characterisation techniques}
Pnictogen- and chalcogen-based honeycomb layered oxides constitute a burgeoning class of nanostructures with vast compositional spectra and diverse structural configurations. Despite the prospects envisioned in this class of materials, their augmentation is still encumbered by knowledge gaps that obscure the current understanding of their atomistic properties and behaviours. In this context, characterisation techniques are considered an efficient approach for accelerating the pace of discovery and optimisation of these materials through the elucidation of the atomic-level origins of material functionality and performance. In this section, we introduce preeminent analytical techniques based on diffraction and spectroscopy that have been applied to unravel various structural and electronic aspects of pnictogen- and chalcogen-based honeycomb layered oxides. Although the fundamental principles of these techniques are well-established, we shall briefly introduce, where appropriate, the fundamental principles to provide sufficient context for their use with the present honeycomb layered oxides.  
Finally, relevant literature will be appropriately cited for a detailed scope of the fundamental physics of the various techniques detailed herein.

\subsection{Diffraction techniques}
Crystal structural analysis is considered a fundamental predicate to major breakthroughs involving crystal materials. In general, for the elucidation of material structural properties, crystallographic techniques based on either electron beam, neutron or X-ray radiation are employed to determine the spatially averaged periodic structures of crystalline materials via the analyses of their diffraction patterns. In principle, the patterns, which correspond to the crystal structure factors of the material, provide the amplitude and the phase of diffraction to determine properties such as atomic species and positions in a unit cell. In the following subsections, we detail the applications of X-rays and neutron scattering, two of the most established crystallographic methods, and their roles in unravelling some noteworthy structural dynamics and functionalities of honeycomb layered oxides.

\subsubsection{X-ray diffraction (XRD)}
X-ray diffraction (XRD) has become the {\it de facto} approach for ascertaining the atomic structural framework of honeycomb layered oxides. The widespread utilisation of this method is principally based on the efficacy of X-rays in creating pair correlation functions between a given element and its neighbours in its entirety. Despite this advantage, X-ray scattering intensity is known to increase with heavier atoms, rendering light atoms such as Li extremely challenging or even impossible to study with XRD. 
Nonetheless, it is worth mentioning that variants of this diffraction methodology ({\it in situ} XRD) have still been instrumental in providing new insights into honeycomb layered bismuthates and antimonates (such as $\rm Li_4FeSbO_6$,\cite{McCalla2015a} $\rm Na_3Ni_2BiO_6$,\cite{Bhange2017, Wang2017} $\rm Na_3Ni_2SbO_6$,\cite{Wang2019b,Wang2018} $\rm Na_3Ni_{1.5}TeO_6$,\cite{grundish2020structural} and so forth) through the real-time observation of their electrochemical mechanistic behaviour. 

Notably, XRD has become instrumental in analyses pertaining to energy materials. In these disciplines, a few variations that depend on the XRD measurement conditions have emerged. Improvements in XRD technology have made it possible to investigate compounds inside a battery during electrochemical operations through protocols referred to as ``{\it in situ}'' and ``{\it in operando}'' XRD studies. The distinction between the two protocols is based on whether the measurements of the samples are done under a specific condition, such as voltage, state-of-charge, or temperature ({\it in situ}) or whether data of the sample were collected synchronically with the sample transformations in a real and dynamic reaction processes ({\it in operando}). Conversely, XRD analyses can also be performed after dismantling the battery upon the termination of the targeted functionality. These protocols performed on the extracted compounds are referred to as ``{\it ex situ}'' or ``{\it postmortem }'' XRD studies. Typically, high-intensity XRD data can be obtained from {\it ex situ} measurements, although {\it ex situ} measurements that involve washing the electrodes after dismantling the battery, in some instances, may alter the structural status of the electrodes and thus compromise the reliability of the attained results.

\begin{figure*}[!t]
\centering
  \includegraphics[width=0.8\columnwidth]{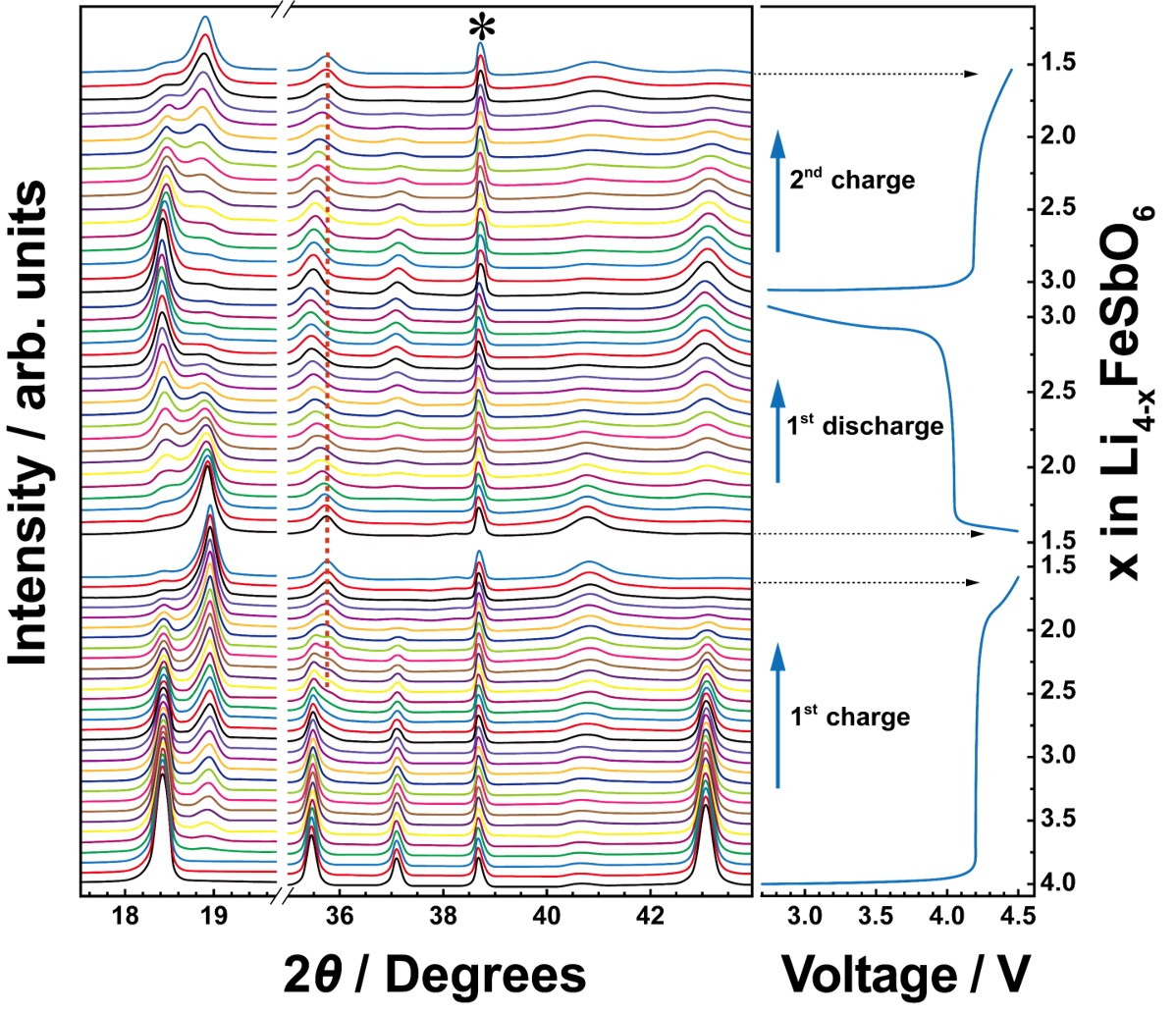}
  \caption{{\it In situ} XRD patterns of $\rm Li_4FeSbO_6$ during charging and discharging. The current density used was commensurate to (dis)charging for 40 hours. Technically denoted as C/40. Bragg reflections emanating from the battery component are marked in asterisks (*). The {\it in situ} XRD data affirms two-phase plateau during (dis)charge. Figures reproduced with permission.\cite{McCalla2015a} Copyright 2015 American Chemical Society.}
  \label{Figure_24}
\end{figure*}

To shed light on the {\it modus operandi} of the XRD analyses, \textbf{Figure \ref{Figure_24}} shows {\it in situ} XRD patterns obtained from $\rm Li_4FeSbO_6$ to analyse the electrochemical behaviour during successive ${\rm Li^{+}}$ extraction and insertion.\cite{McCalla2015a} During ${\rm Li^{+}}$ extraction (charging), some Bragg peaks (for instance, those located at around 18.4$^{\circ}$) were observed to slightly shift in the initial charging stages before gradually changing in intensity and eventually disappearing in lieu of new ones, which become sharper as the charging step progresses. This suggests the occurrence of a two-phase (biphasic) intercalation process in accord with the plateau formed by the voltage-(dis)charge plots. During the subsequent discharge process, the reverse peak amplitude variation can also be observed. This indicates the full reversibility of the (dis)charge process as was substantiated by the remarkable similarity between the charged/discharged electrodes and the pristine electrode (bottom XRD).

\begin{figure*}[!t]
\centering
  \includegraphics[width=0.8\columnwidth]{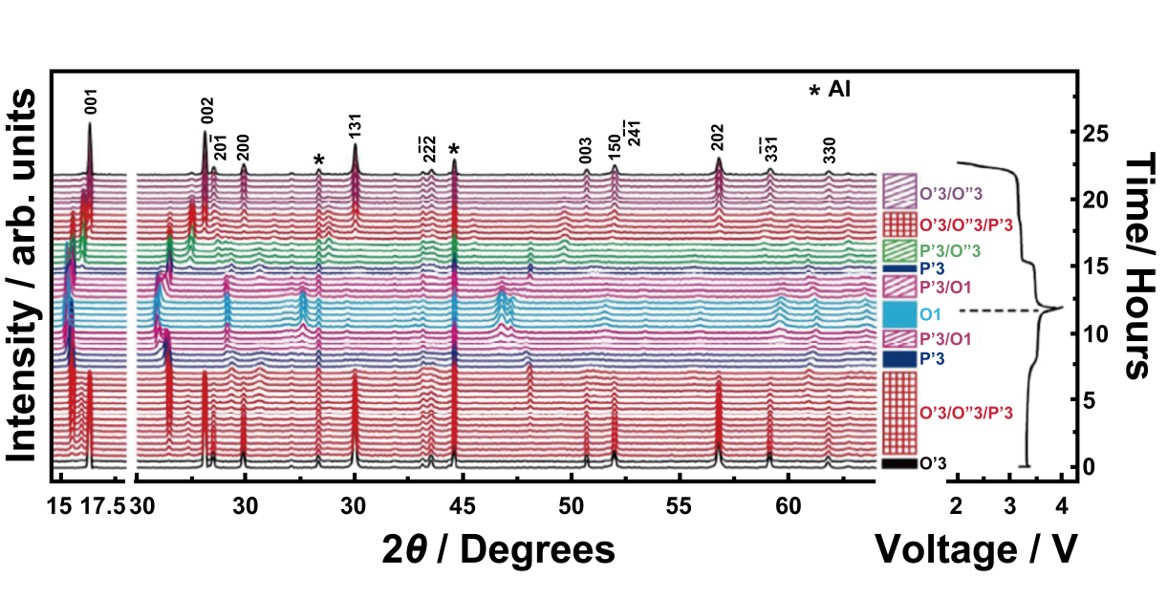}
  \caption{{\it In situ} XRD patterns of $\rm Na_3Ni_2BiO_6$ collected during the initial charging and discharging process. The current density used was C/10 and at a voltage range of 2 -4 V. Asterisks denote peaks emanating from Al used as a component in the electrode assembly. Bragg reflections arising from the Al foil used for casting the electrode are highlighted in asterisks (*). Figures reproduced with permission.\cite{Wang2017} Copyright 2017 American Chemical Society.}
  \label{Figure_25}
\end{figure*}

The applicability of {\it in situ} XRD in depicting phase transformations during electrochemical operations is further demonstrated by \textbf{Figure \ref{Figure_25}}, which illustrates the crystal characterisation of a honeycomb layered bismuthate, $\rm Na_3Ni_2BiO_6$. Here, the electrode material undergoes two reversible biphasic transition mechanisms (namely, initial O'3-type phase → P'3-type intermediate phase → O1-type final phase in the Hagenmuller-Delmas' notation) during ${\rm Na^{+}}$ extraction. Note that a prime symbol (') is added to the letter in its notation, when the crystal lattice exhibits in-plane distortion with respect to the hexagonal lattice. {\it In situ} XRD has been used to unveil a multitude of phase transitions innate to honeycomb layered oxides,\cite{McCalla2015a, Bhange2017, Wang2017, Wang2019b,Wang2018} enriching our understanding of their alkali-ion extraction and insertion mechanisms.

Although XRD is a popular technique, its low sensitivity to light elements such as Li presents a few constraints that make the complementary use of neutron diffraction or electron microscopy mandatory when analysing the crystal structures of certain materials. Moreover, the presence of numerous stacking defects in layered oxide materials makes it difficult to obtain reliable structural characterisation from XRD patterns using the Rietveld technique, although programs such as DIFFaX or FAULTS are now able to provide support when quantifying the stacking faults presented by the XRD data.\cite{casas2016faults, treacy1991general, casas2007microstructural, casas2006faults}

\subsubsection{Neutron diffraction (ND)}
Compared to XRD, neutron scattering is far superior due to its greater sensitivity towards lighter elements ($e.g.$, Li, Na, H and O). Moreover, its ability to offer sufficient elemental contrast between the neighbouring elements makes it ideal for distinguishing elements suchlike Mn, Ni, Co, and Fe) in place of XRD, which does not provide elemental contrast with such atoms (Fe and Mn) due to their similar X-ray scattering length density. Despite these capabilities, neutron diffraction is ineffective for analyses pertaining to mixed alkali honeycomb layered $\rm NaKNi_2TeO_6$\cite{masese2021mixed} because neutron scattering cannot easily distinguish the atomic positions of Na and K owing to their similar neutron scattering lengths. As such, neutron diffraction is typically employed in combination with XRD for honeycomb layered oxide analysis to fill the gaps left by X-ray deficiencies. For instance, X-ray scattering can be utilised precisely to determine the atomic positions of Na and K in compositions such as $\rm NaKNi_2TeO_6$, whereas neutron scattering can aid in distinguishing compositions that entail lighter elements such as Li or the neighbouring atoms such as Fe and Mn. Indeed, the synergism between neutron scattering and X-ray scattering is poised to play a pivotal role in the effective characterisation of honeycomb layered oxide materials.

\begin{figure*}[!t]
\centering
  \includegraphics[width=0.8\columnwidth]{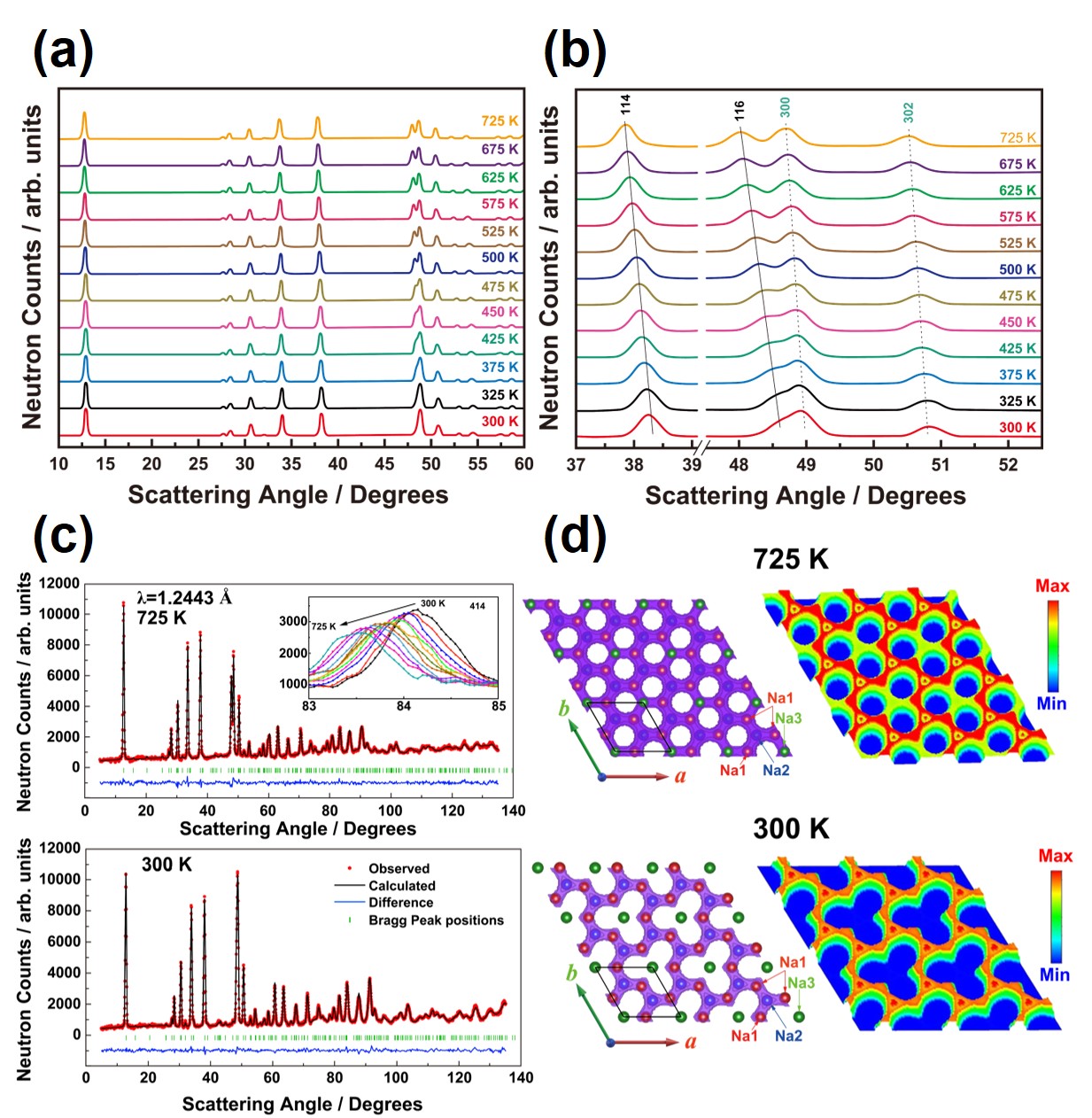}
  \caption{Neutron powder diffraction patterns for $\rm Na_2Ni_2TeO_6$. (a) Evolution of the of the neutron diffraction patterns for $\rm Na_2Ni_2TeO_6$ with temperature, taken over a selected scattering angular range of 10–60$^\circ$. (b) Temperature dependences of selected Bragg peaks at (302), (300), (116), and (114). The lines are guides to the eye. (c) Rietveld refinement plots of the neutron diffraction patterns for $\rm Na_2Ni_2TeO_6$ taken at 725 K and 300 K. The calculated and observed patterns are shown by solid black lines and filled circles, respectively. The difference between calculated and observed patterns is shown by the thin line at the bottom of each panel. The vertical bars are the allowed Bragg peak positions. The inset shows the temperature dependence of a representative Bragg peak (414), revealing the effect of the thermal displacement parameter on the width and intensity of the Bragg peak. Broadening of the peaks concomitant with the decline in peak intensities are evident, with increasing temperature. (d) Bond valence energy landscape (BVEL) map (${\rm \Delta}$E = 0.6 eV) of $\rm Na_2Ni_2TeO_6$ at 725 and 300 K. Right: A bird's-eye view of the Na-ion conduction pathways (BVEL map). The colour bar represents the depth profile of the conduction pathways. The enhanced contribution of the Na3 sites to the conduction pathways is noticeable with increasing temperature. Figures reproduced with permission.\cite{Bera2020} Copyright 2020 American Chemical Society.}
  \label{Figure_26}
\end{figure*}

The neutron diffraction (ND) technique fundamentally relies on the elastic scattering of neutrons to determine crystallographic as well as magnetic structures of pnictogen- and chalcogen-based honeycomb layered oxides. As such, this method has become a potent tool for investigating the piquant magnetic orders and spin-wave excitations exhibited by honeycomb layered tellurates (such as $\rm Na_2Co_2TeO_6$ and $\rm Na_2Ni_2TeO_6$),\cite{Viciu2007, Bera2017} antimonates (such as $\rm {\it A}_3Cu_2SbO_6$ ($A = \rm Li, Na $))\cite{Miura2008} and bismuthates (such as $\rm Na_3Ni_2BiO_6$)\cite{Seibel2013}. Despite the alluring mechanisms presented by this technique, the fundamental physics of neutron scattering falls outside the scope of the present review. Therefore, we recommend interested readers to gather this information from well-established literature. 

Besides its functionality in crystallographic and magnetic studies, ND has also found utility in ion dynamics examinations employing the Fourier difference maps analysis or advanced maximum entropy method.\cite{Karna2017, Bera2020} These methods provide an analytical platform for investigating concepts such as the residual scattering caused by mobile ions, which cannot be obtained through the applied structural (static) models. For instance, residual scattering can describe the time-averaged trace of a mobile ion inside the crystallographic lattice, making it possible to deduce the diffusion paths being favoured by the ions at an atomic scale. The prominent bond valence sum (BVS) method, which captures the differences in ionic size and formal charge of the refined structures, can also be used to trace the ionic diffusion pathways. In fact, neutron diffraction has been applied alongside the BVS approach to successfully visualise the microscopic ${\rm Na^+}$ conduction pathways in $\rm Na_2Ni_2TeO_6$ as a function of temperature.\cite{Karna2017, Bera2020} \textbf{Figures \ref{Figure_26}a} and \textbf{\ref{Figure_26}b} show the evolution of the ND patterns of $\rm Na_2Ni_2TeO_6$ when subjected to high temperatures.\cite{Bera2020} The ND reflection patterns are noted to shift to lower scattering angles with the increase in temperature, which is to be expected in consideration of the lattice thermal expansion. However, a comparison of the ND patterns attained at 300 K, and 725 K (shown in \textbf{Figure \ref{Figure_26}c}) reveals significant differences in the 2D ${\rm Na^+}$ conduction pathways formed by the material at the different temperatures. As shown in \textbf{Figure \ref{Figure_26}d}, the Na ions occupy different crystallographic sites in $\rm Na_2Ni_2TeO_6$ (denoted as Na1, Na2 and Na3) when the material is at 300 K. At temperatures below 500 K, the ionic conduction appears to be dominated by the Na ions residing at the Na2 and Na1 sites. However, all the Na ions located in all the three Na sites are seen to contribute to the conduction process via assuming a honeycomb pathway when the $\rm Na_2Ni_2TeO_6$ is subjected to a temperature above 500 K.

\subsubsection{Electron diffraction (ED)}
Compared to X-rays and neutrons, electrons interact more strongly with a material, allowing the probing of nanoscopic structures in resolutions unmatched by neutron diffraction or X-ray. In principle, the scattering factor of electron diffraction is very roughly proportional to the atomic number $Z$ ($i.e.$, ${\rm {\it Z}^{1/3}}$), whereas, for X-ray scattering factor, the proportionality is to an order of about 1. This implies that the scattering of electrons is less sensitive to the atomic number of an element giving electron diffraction better efficacy than X-ray diffraction in detecting light elements, such as oxygen in layered oxides and, accordingly, superstructures based on oxygen ordering.

Strong interactions between sample atoms and high-energy electrons result in large scattering cross-sections, which imply that very small sample amounts are required to attain diffraction patterns with sufficient signal-to-noise ratios. Electron diffraction is usually used to complement neutron and X-ray diffraction data during characterisation studies pertaining to intricate crystal structures. Although a collection of XRD patterns is straightforward, their data can often be difficult to interpret owing to overlapping or broadening of reflections with similar phase problems or diffraction angles, particularly when the structure has a low symmetry or large unit cell. Therefore, to determine a complex structure, the XRD intensity data are typically combined with structure factor phase information obtained through crystallographic techniques, $e.g.$, precession electron diffraction (PED),\cite{midgley2015precession} rotation electron diffraction (RED) or high-resolution TEM images.

\begin{figure*}[!t]
\centering
  \includegraphics[width=0.8\columnwidth]{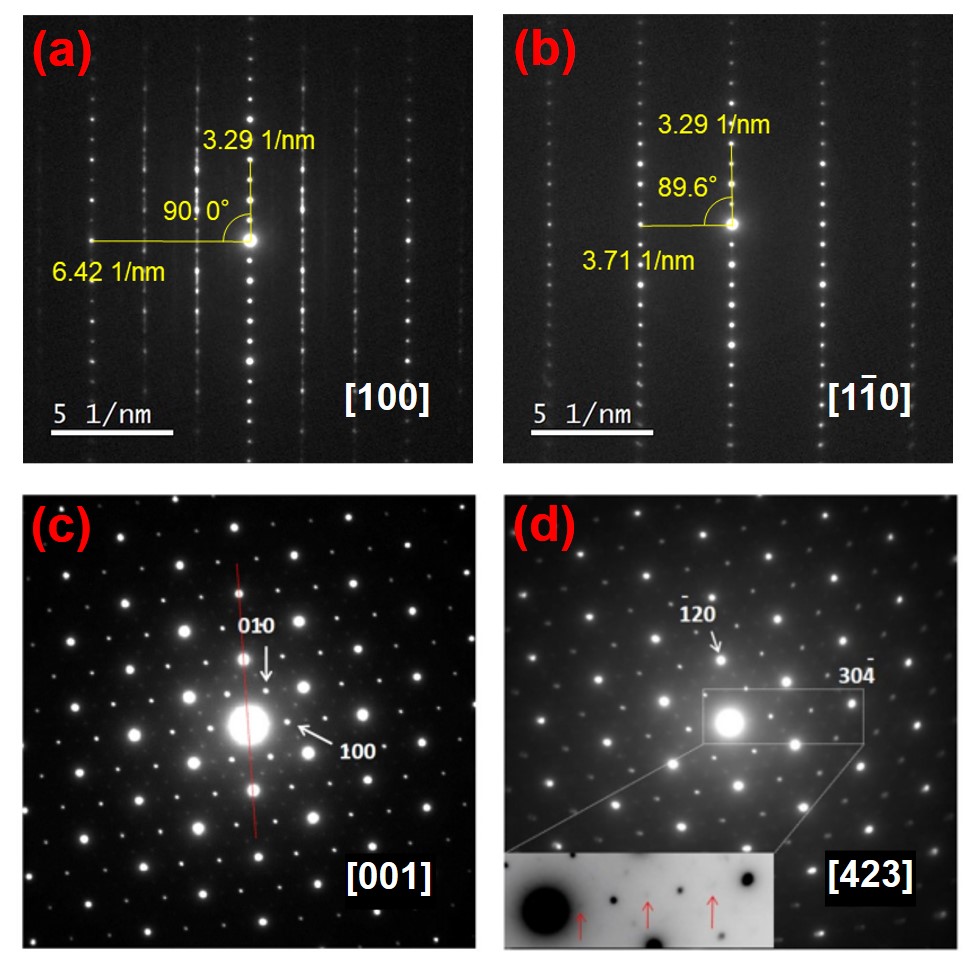}
  \caption{Electron diffraction patterns for $\rm NaKNi_2TeO_6$. (a) [100] zone axis. (b) [1$\overline{1}$0] zone axis. (c) [001] zone axis. (d) [423] zone axis. Figures (a, b) and (c, d) reproduced with permission.\cite{masese2021mixed, berthelot2021stacking}. Copyright 2021 Springer Nature and 2021 American Chemical Society, respectively.}
  \label{Figure_27}
\end{figure*}

ED has been found to be effective in ascertaining the possible ordering of layered slabs and alkali atoms, for instance, in honeycomb layered oxides such as $\rm NaKNi_2TeO_6$,\cite{masese2021mixed, berthelot2021stacking}$\rm Na_3Ni_2SbO_6$,\cite{Xiao2020, Wang2019b} $\rm Na_3Ni_2BiO_6$,\cite{Liu2016, Seibel2013} $\rm Li_4FeSbO_6$,\cite{McCalla2015} $\rm K_2Ni_2TeO_6$\cite{Masese2018} and $\rm Na_2Ni_2TeO_6$\cite{masese2021unveiling}. \textbf{Figure \ref{Figure_27}} shows the ED patterns collected for $\rm NaKNi_2TeO_6$ at various zone axes.\cite{masese2021mixed, berthelot2021stacking} Streaks indicative of the presence of defects along with additional diffraction spots not indexable to the crystal structure adopted can be observed.

\subsubsection{Electron pair distribution functions}
Routine crystallographic techniques, such as powder X-ray and single crystal diffraction, have been found to rely on structural periodicity solely and thus avail scant insight into structural properties such as defects that fall beyond the scope of such structures. Although these techniques are adequate for crystalline structures whose unit cells are replicated throughout the structure, they are unsuitable for structures that lack periodicity (aperiodic structures) or are disordered because the information they provide is limited to local atomic structures that may not necessarily apply to the rest of the material. 

Given the limitations mentioned above, diffraction-based techniques may appear to be ineffective for characterising amorphous and disordered materials. As a solution, crystallographers rely on pair distribution function (PDF), a potent analytical technique for extracting important structural information from neutron, X-ray, or electron diffraction patterns. PDF gives insight into the local structure by showing the prevalence of pairwise distances between atoms, thus providing a picture of the bond-length distributions. When plotted, the graphs exhibit peaks as a function of distance in real space, amongst other information based on the size and shape of the peaks. The fundamental physics of PDF falls beyond the scope of the present review; therefore, readers are advised to explore authoritative reports for more information.\cite{perdew1992pair, billinge2019rise, terban2021structural}

\begin{figure*}[!t]
\centering
  \includegraphics[width=0.8\columnwidth]{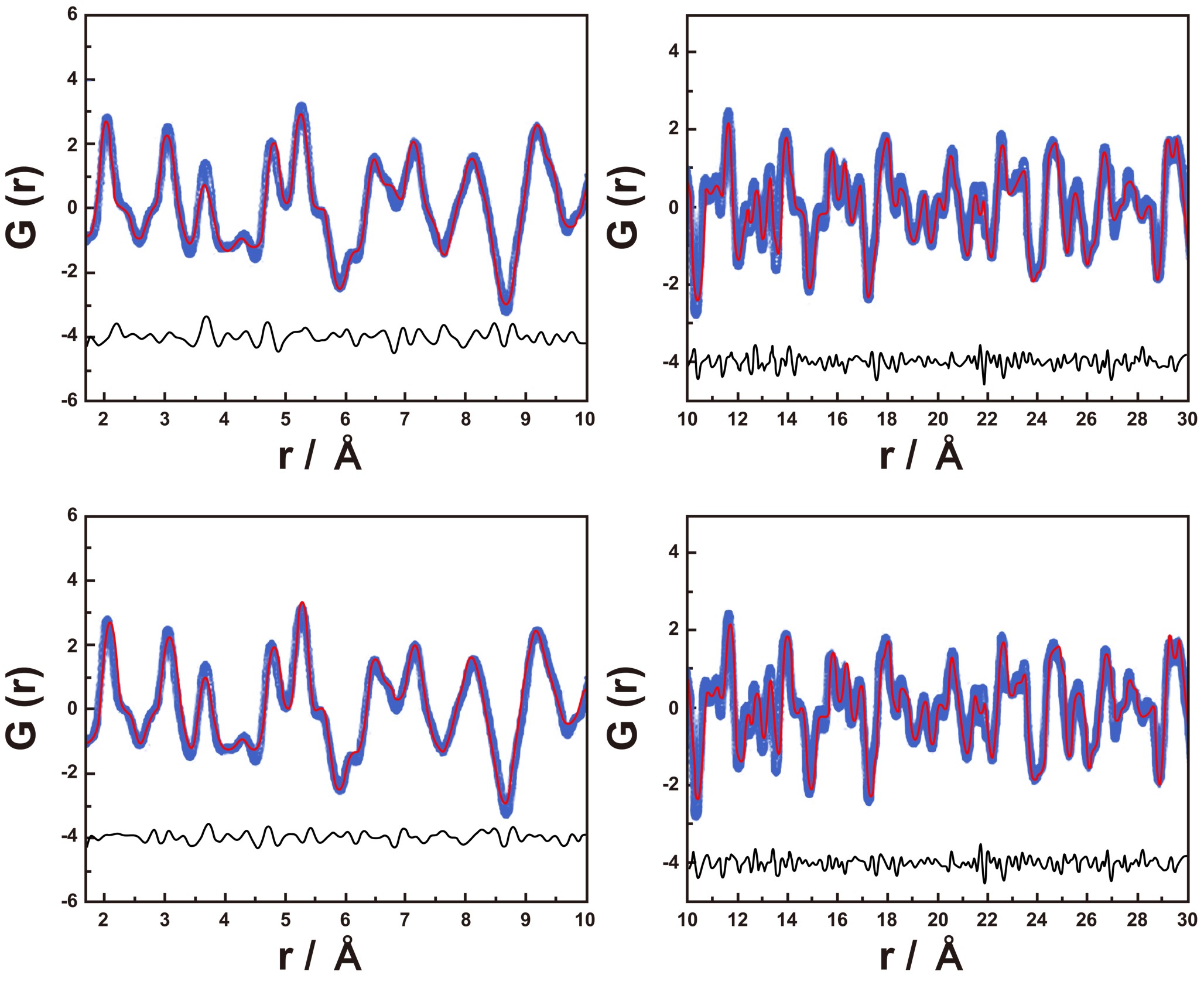}
  \caption{Neutron pair distribution function (PDF) data for a disordered $\rm Na_3Ni_2SbO_6$ ($\rm NaNi_{2/3}Sb_{1/3}O_2$). Fits of neutron PDF data for a disordered $\rm NaNi_{2/3}Sb_{1/3}O_2$ sample over longer (right) and shorter (left) length scales using either the honeycomb-ordered $P$3${\rm _1}$12 supercell (bottom) or the disordered $R$-3$m$ subcell (top). Although there is negligible effect at longer distances, the use of the honeycomb supercell substantially improves the fit at distances below 1 nm (10 \AA\,). Figures reproduced with permission.\cite{Ma2015} Copyright 2015 American Chemical Society.}
  \label{Figure_28}
\end{figure*}

It is worth mentioning that the analysis of the structural relationship through electron pair distribution functions is uniquely suited for nanoscale disordered materials. \textbf{Figure \ref{Figure_28}} shows PDF analyses applied to accurately attain average crystal structural models of the disordered structures of $\rm Na_3Ni_2SbO_6$.\cite{Ma2015}

\subsection{Microscopic imaging techniques}
Electron microscopy facilitates the extraction of multi-dimensional spatiotemporally correlated structural information of myriad materials down to atomic resolution – crucial information for establishing complex structure-property relationships. In studies aimed at delineating the atomistic mechanisms of honeycomb layered oxides, both conventional transmission electron microscopy (TEM) and scanning transmission electron microscopy (STEM) have been gaining widespread utility with spectroscopic and concomitant analytical techniques, suchlike energy-dispersive X-ray spectroscopy (EDS) and electron energy loss spectroscopy (EELS), due to their spatial resolution. These techniques have been seen to open new opportunities not only for identifying the local chemical compositions of materials but also for elucidating their electronic and magnetic fluctuations of such materials. In this subsection, we will discuss the electron microscopy techniques and the insights obtained within the context of pnictogen- and chalcogen-based honeycomb layered oxide materials.

\subsubsection{Scanning electron microscopy (SEM)}
Scanning electron microscopy (SEM) has been instrumental in providing valuable insight into surface morphology for a wide spectrum of pnictogen- and chalcogen-based honeycomb layered oxide materials. Amongst the structural features detectable by electron microscopy, their crystal size and morphology are by far the easiest to determine. These analyses typically entail direct observation from the SEM images. Honeycomb layered oxides tend to show lamellar-like morphologies with varying particle sizes depending on the composition.

Scanning electron microscopy can also be used alongside energy-dispersive X-ray analysis (SEM-EDX) to acquire information on the chemical composition and the amount of the constituent elements in pnictogen- and chalcogen-based honeycomb layered oxides. However, EDX cannot detect materials with lighter elements such as Li and also suffers from a high background depth that results in attenuated elemental sensitivity compared to other techniques such as X-ray fluorescence spectroscopy. Altogether, SEM is an apposite technique to show particle morphologies of as-prepared pnictogen- and chalcogen-based honeycomb layered oxide materials that may assume various lamellar-like microstructures.

\subsubsection{Transmission electron microscopy (TEM)}
Transmission electron microscopy (TEM) is a long-established characterisation technique not only for identifying the morphology but also directly imaging the crystal structures of honeycomb layered oxides. TEM characterisation presents a great advantage in its ability to directly image structures of materials in real space with exceedingly high resolution. However, it should still be noted that scanning probe methods suchlike scanning tunnelling microscopy are also based in real space, but they tend to be restrictively surface-sensitive. On the other hand, TEM measurements average the through-thickness structure of material samples, allowing ``bulk'' characterisation and the study of buried defects and interphases to be conducted effectively. In honeycomb layered oxides, the different phases/stackings can be distinguished intuitively by measuring the lattice distance of the exposed lattice plane through high-resolution transmission electron microscopy (HRTEM) imaging in combination with their corresponding selected-area electron diffraction. For a complete explanation of the fundamental principles of the technique, readers are advised to refer to authoritative literature at their discretion.\cite{cowley1969image, wall1974scanning}

In contemporary investigations, the atomic structures for most crystals can be evaluated through imaging techniques based on TEM and scanning TEM (typically abbreviated as STEM). It is worth noting that these techniques have now reached sub-\AA ngstr\"{o}m resolutions after the introduction of powerful aberration correctors and field-emission electron sources. In the aberration-corrected STEM technique, electrons are accelerated and focused into a sub-\AA ngstr\"{o}m-sized probe that scans across the surface of a thin material sample. The incident electrons interact with local electrostatic fields in the sample as they propagate through the material and produce a scattered exit beam of electrons that forms a diffraction pattern on the detector plane. The detectors collect the electrons to form a segment of the diffraction pattern whose integrated intensity corresponds with the electrostatic field of the material. 

\begin{figure*}[!t]
\centering
  \includegraphics[width=0.8\columnwidth]{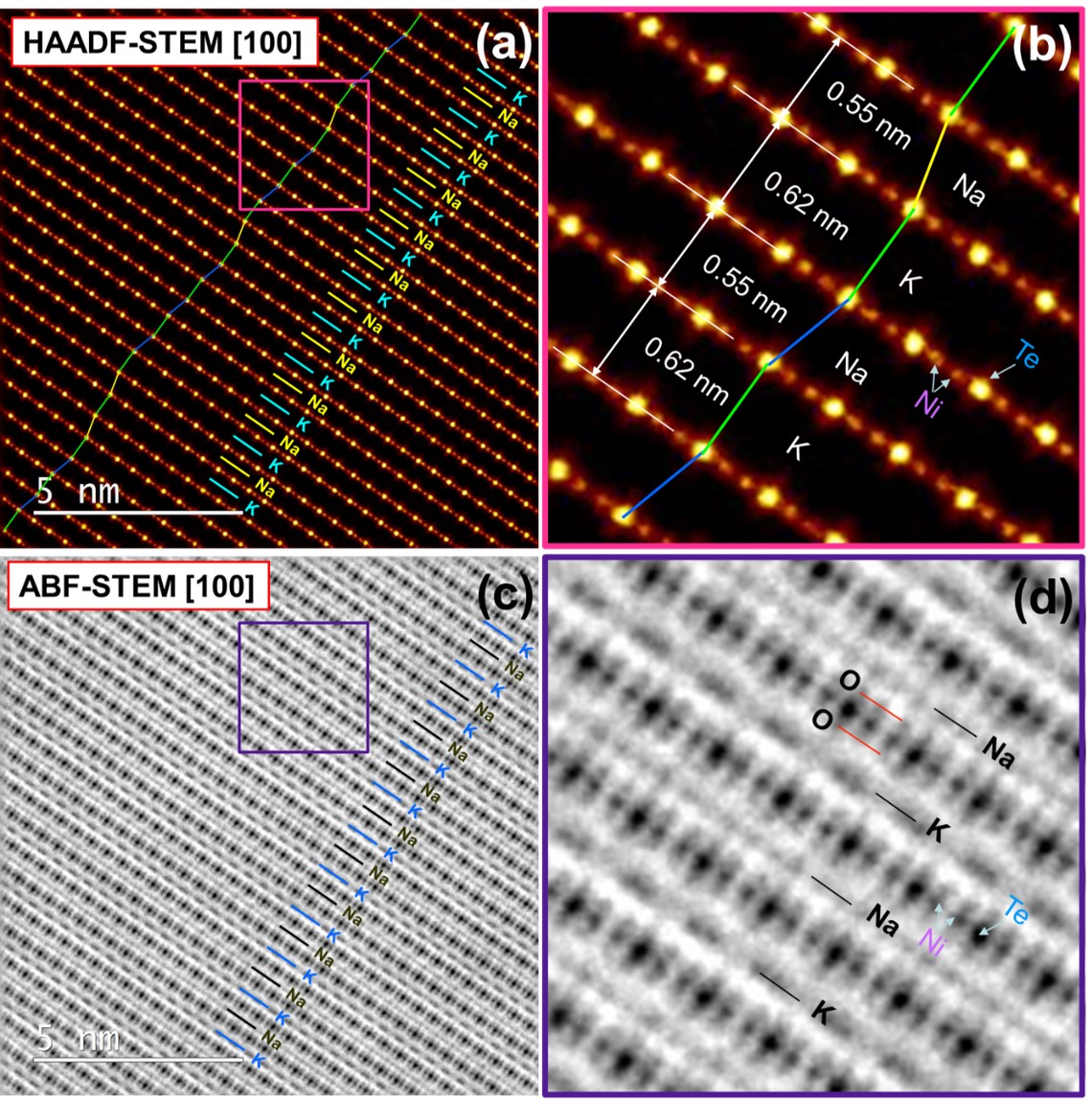}
  \caption{Visualisation of the arrangement of atoms in mixed alkali honeycomb layered $\rm NaKNi_2TeO_6$ along the [100] zone axis using aberration-corrected scanning transmission electron microscopy (STEM). (a) HAADF-STEM image illustrating the unique stacking sequence in $\rm NaKNi_2TeO_6$. In the layers where Na atoms occupy the interlayer space, shifts of the Ni / Te slabs are observed. The blue and yellow lines show shifts in different directions. Note the aperiodicity in the stacking sequence. However, for layers where K atoms reside, Ni / Te slabs are not shifted with respect to each other (marked by a green line). (b) Magnified view of the domain highlighted in (a), showing that the interlayer distance is dependent on the alkali atom species (K or Na) sandwiched between the Ni / Te layers. (c) Annular bright-field (ABF)-STEM image, where also lighter elements (O, K and Na) can readily be visualised. (d) An enlarged view of the domain highlighted in (c), showing the arrangement of lighter elements such as O, Na and K, along with heavy atoms such as Te and Ni. Figures reproduced with permission.\cite{masese2021mixed} Copyright 2021 Springer Nature.}
  \label{Figure_29}
\end{figure*}

The widely-used high-angle annular dark-field (HAADF) STEM technique uses an annular detector to collect the electrons scattered to high angles. A key benefit to this technique is that it is not affected by the wave character of the electrons and thus does not suffer from interferences that complicate the interpretation of images. Therefore, HAADF-STEM images are easily interpretable, whereby the atomic columns in a crystal appear as bright dots with a dark background. The brightness of the dot is usually scaled to the average atomic number $Z$ in the atomic column (typically ${\rm {\it Z}^{1.6-2.0}}$).\cite{pennycook2006scanning, pennycook1988chemically, pennycook2006materials} Utilising this technique, transition metal atoms such as Ni, Co, $etc.$, and heavy elements such as Bi, Te, and Sb, amongst others, can be distinguished since the scattered intensity is directly proportional to their atomic number. \textbf{Figure \ref{Figure_29}a} shows a HAADF-STEM micrograph of $\rm NaKNi_2TeO_6$,\cite{masese2021mixed} wherein the atomic arrangement of heavier elements (Ni ($Z$ = 28) and Te ($Z$ = 52)) is succinctly visualised by spots of differing intensities. The bright yellow spots represent Te atoms columns, and the red dots correspond to the Ni atom columns. HAADF-STEM imaging also enables the visualisation of the intricate aperiodic stacking sequence structure of $\rm NaKNi_2TeO_6$,\cite{masese2021mixed} as highlighted in \textbf{Figure \ref{Figure_29}b}.

Despite the visualisation expedience afforded by HAADF-STEM imaging, the structural configuration of pnictogen- and chalcogen-based honeycomb layered oxides present a key hurdle to its utility. As mentioned in an earlier section, this class of materials comprise elements with significant differences in their atomic numbers. The close proximity between heavy metal atoms (Te, Bi, Sb, $etc.$) and light atoms like (Li, O, $etc.$) makes it difficult to pinpoint the positions occupied by the lighter elements and quantify them accordingly because their scattering strength can be too low to be detected when compared to that of the adjacent heavy elements. As such, these challenges can be posited to be higher in materials with closer proximity between light elements and their heavier counterparts.

Conversely, the contrast of annular bright-field (ABF) imaging has a weak ${\rm {\it Z}^{1/3}}$ dependency, making it a more suitable approach for imaging light elements.\cite{pennycook2006scanning, pennycook1988chemically, pennycook2006materials} In ABF-STEM, the collection angles are drastically diminished to obtain adequate signal from light elements such as lithium and oxygen, as illustrated by the ABF-STEM image of $\rm NaKNi_2TeO_6$ (\textbf{Figure \ref{Figure_29}c}). Here, the positions of K ($Z$ = 19), Na ($Z$ = 11) and O ($Z$ = 8) can be distinguished unequivocally (\textbf{Figure \ref{Figure_29}d}) even alongside heavy atoms such as Te and Ni. This capability has proved to be highly instrumental in garnering sufficient structural insights into the aperiodicity in this material. 

High-resolution STEM has become an indispensable tool in the characterisation of honeycomb layered oxide crystal structures at the nanoscale. Other techniques such as electron energy-loss spectroscopy (EELS) can also be combined with STEM to provide a holistic mapping of atomic species, as has been done for $\rm Na_3Ni_2SbO_6$.\cite{Xiao2020} Pair distribution functions attained from electron diffraction (ePDF) and EELS are generally complementary techniques since they are contingent on distinct aspects of electron scattering. EELS uses inelastically scattered electrons to measure physical and chemical properties (such as oxidation states and electronic structures, whereas ePDF and diffraction utilise elastically scattered electrons to obtain spatial information about a material sample. Thus, these approaches can be combined to garner a more holistic understanding of various intricate structures of pnictogen- and chalcogen-based honeycomb layered oxide materials.

\subsection{Spectroscopic characterisation techniques}
For wholesome and accurate elucidation of crystal structural information, microscopic imaging is generally applied alongside more refined characterisation techniques. Spectroscopic characterisation is highly sensitive to the electronic structure, atomic arrangement, chemical state, and elemental composition of materials and can therefore cover the analytical voids left by microscopic techniques. Spectroscopic techniques utilised in the characterisation of pnictogen- and chalcogen-based honeycomb layered oxides encompass {\it inter alia}: inductively coupled plasma atomic emission spectroscopy (ICP-AES), X-ray photoelectron spectroscopy (XPS) and electron energy loss spectroscopy (EELS), muon spin rotation spectroscopy, differential electrochemical mass spectrometry (DEMS), electrochemical impedance spectroscopy (EIS), magnetic susceptibility measurements, X-ray absorption spectroscopy (XAS), electron paramagnetic resonance (EPR) spectroscopy, nuclear magnetic resonance (NMR) spectroscopy, M\"{o}ssbauer spectroscopy, photoluminescence spectroscopy and ultraviolet-visible (UV-Vis) diffuse reflectance spectroscopy.

Although the spectroscopic techniques mentioned above each offer distinctive characterisation merits for utilisation, ICP-AES, EELS, EIS, XPS and magnetic susceptibility measurements have become a commonplace in a vast majority of studies on honeycomb layered oxides. In the following subsections, we delve into the following techniques: XPS, XAS, NMR, EPR, M\"{o}ssbauer spectroscopy, photoluminescence spectroscopy and UV-Vis diffuse reflectance spectroscopy. We highlight the various insights/functionalities attained via the use of these spectroscopic techniques in the context of the pnictogen-and chalcogen-based honeycomb layered oxides. We shall refer the reader to authoritative literature for in-depth coverage of the other techniques.

\subsubsection{X-ray photoelectron spectroscopy (XPS)}
\begin{figure*}[!b]
\centering
  \includegraphics[width=0.8\columnwidth]{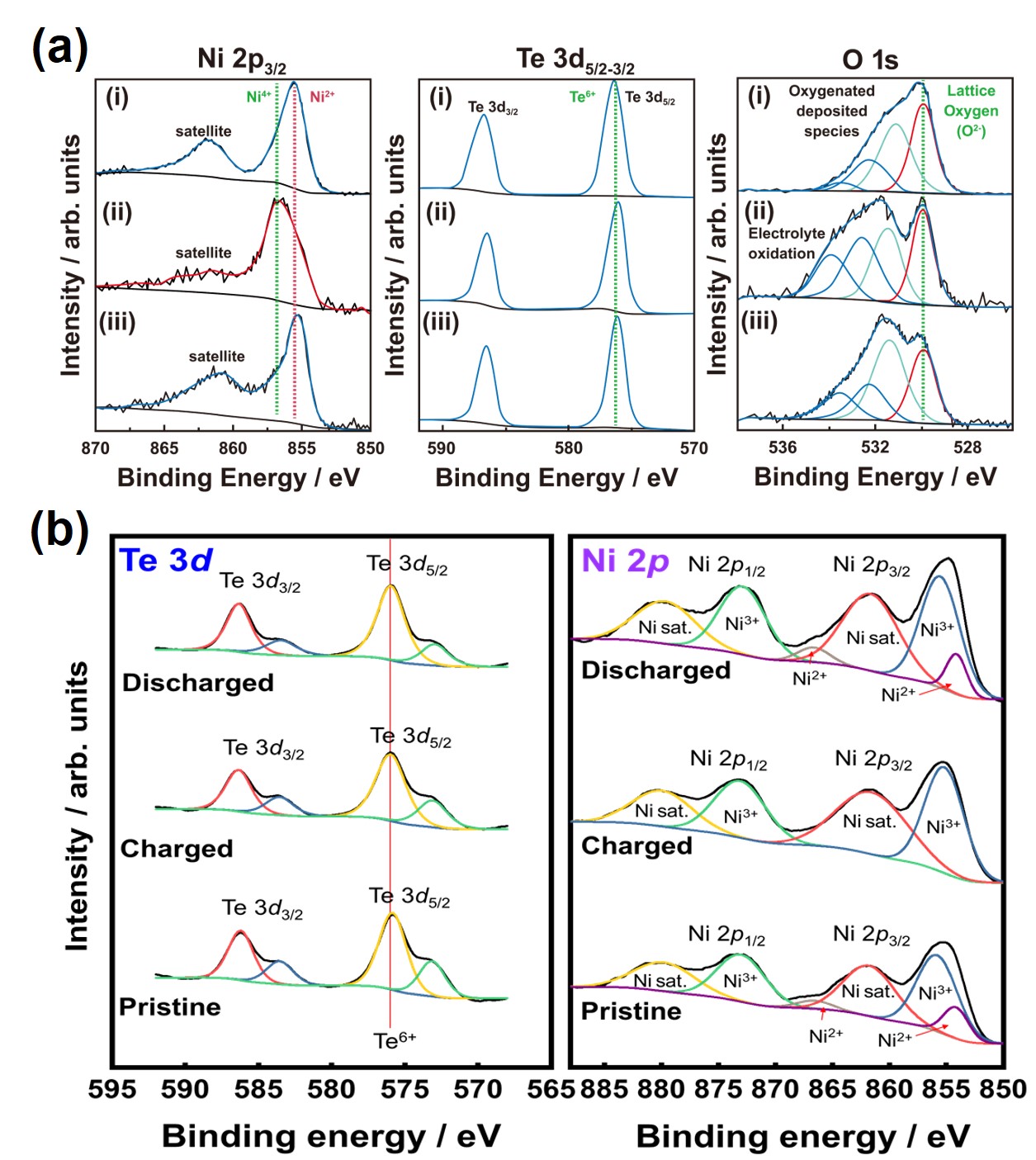}
  \caption{X-ray photoelectron spectroscopy (XPS) spectra of honeycomb layered tellurates. (a) XPS spectra of $\rm Li_4NiTeO_6$ collected at the binding energies of Ni-2$p$, Te-3$d$ and O-1$s$ core for (i) the $\rm Li_4NiTeO_6$ pristine electrode, (ii) charged to 4.6 V, (iii) further discharged to 2.0 V. (b) Te-3$d$ and Ni-2$p$ XPS spectra taken for the pristine, charged and discharged electrodes of $\rm NaKNi_2TeO_6$ cycled in a NaK battery setup. The satellite peaks are marked as `Ni sat.'. Figure (a) and Figure (b) reproduced with permission.\cite{Sathiya2013, masese2021mixed} Copyright 2013 Royal Society of Chemistry and 2021 Springer Nature respectively.}
  \label{Figure_30}
\end{figure*}

In principle, X-ray photoelectron spectroscopy (XPS) is based on a process whereby a solid sample is bombarded with photons of a particular energy to excite some of its constituent elements, eventually ejecting them from the sample. XPS is initiated by irradiating a sample with monoenergetic soft X-rays (most commonly being Al $K \,\rm \alpha$ or Mg $K \,\rm \alpha$). Subsequently, the kinetic energy of electrons emitted from the sample surface is analysed to yield information on the electronic states of atoms in the surface region. During characterisation, the small variations in binding energies of the satellite peaks, Auger lines, photoelectron lines, and multiple splitting can be utilised to assess the chemical states of the resident elements in the material. XPS is prominent in its utility, versatility, and popularity compared with many other spectroscopic techniques to be mentioned hereafter. However, the fundamental physics principles governing XPS mechanisms fall beyond the scope of this review but can be found in existing literature.\cite{fadley2010x}

By virtue of its surface sensitivity, XPS has been widely employed in various studies aimed at establishing the elemental composition and oxidation states of elements at the surface of pnictogen- and chalcogen-based honeycomb layered oxides. For instance, this technique has been applied to elucidate the electrochemical reaction occurring during charging and discharging processes of $\rm Li_4NiTeO_6$ electrode materials in Li battery configurations (see \textbf{Figure \ref{Figure_8}a} for the voltage capacity plots of the material).\cite{Sathiya2013} Although XPS is fundamentally a surface-sensitive technique, the changes occurring in the valency of Ni (${\rm Ni^{2+}}$→${\rm Ni^{4+}}$) atoms during charging were captured in the Ni 2${\rm {\it p}_{3/2}}$ core spectra (\textbf{Figure \ref{Figure_30}a}). During analysis, the inspection of the Te 3$d$ core spectra did not reveal any substantial changes, indicating that Te does not participate in the redox process.
The O 1$s$ core spectra were also investigated to ascertain the presence of peroxo-like species that have been observed in other layered oxides with oxygen anion redox chemistry. However, only peaks ascribable to oxygenated species arising from oxidation of the electrolyte were observed, with a notable absence of peroxo-like species. Despite the efficacy demonstrated by XPS, caution still needs to be exercised during characterisation due to the broadness of the spectra collected from the charged samples. This can be solved by adopting bulk sensitive spectroscopic techniques such as XAS or resonant inelastic X-ray scattering (RIXS) for detailed analyses of any possible reaction taking place in the oxygen lattice. The difficulty in identifying the presence of peroxo-like species was also noted in the analyses of $\rm Li_4FeSbO_6$,\cite{McCalla2015a} owing to the overlap between Sb and O transitions, necessitating the combined use of magnetic susceptibility measurements along with M\"{o}ssbauer spectroscopy to indirectly observe the presence of peroxo species. XPS analyses have also been done on mixed alkali $\rm NaKNi_2TeO_6$,\cite{masese2021mixed} revealing the changes in Ni valence states (${\rm Ni^{2+}}$→${\rm Ni^{3+}}$) during the charging process whilst ascertaining the dormant role of Te previously observed in $\rm Li_4NiTeO_6$\cite{Sathiya2013} (\textbf{Figure \ref{Figure_30}b}) and $\rm K_2Ni_2TeO_6$\cite{Masese2018}.

\subsubsection{X-ray absorption spectroscopy (XAS)}
Current developments in synchrotron radiation techniques have undraped X-ray absorption spectroscopy (XAS) as an efficacious structural characterisation tool for garnering in-depth insight about pnictogen- and chalcogen-based honeycomb layered oxides. This technique utilises X-ray radiation to provide information on the structural, magnetic and electronic properties of a material. This information is usually proffered using an X-ray absorption spectrum of the sample, which entails a plot of the linear absorption coefficient against the incident photon energies. The absorption remarkably increases at an element-specific energy (denoted hereafter as ${\rm {\it E}_0}$), leading to an absorption edge. At the high energy side of the absorption edge, the curve can display a species-specific oscillation, $i.e. $, the so-called ``X-ray absorption fine structure (XAFS)''. XAFS consists of: (i) XANES (X-ray absorption near edge structure), where the energy of the incident X-ray beam is ${\rm {\it E}}$ = ${\rm {\it E}_0}$ ± 10 eV; (ii) NEXAFS (near-edge X-ray absorption fine structure), in the region between 10 and 50 eV above the edge; and (iii) EXAFS (extended X-ray absorption fine structure), in the region ranging from 50 eV up to 1000 eV above the edge. 

Although the terms XAS and XAFS are often used interchangeably, some variations in their connotations do exist where XAS is used as a general term, whereas XAFS denotes specific modulations of the absorption coefficient by the chemical environment. These differences in some mechanisms are employed by both techniques. For instance, XAS, which is carried out at synchrotron facilities, can be used to determine the coordination environment, site symmetry and the electronic configuration of the absorbing atom. Moreover, the absolute position of the edge further provides information about the oxidation state of the absorbing atom. On the other hand, XAFS represents varied modulations: $i.e.$, NEXAFS, XANES, and EXAFS, with different characterisation capabilities. NEXAFS is sensitive to bond angles and therefore offers data on the intramolecular bond lengths and orientation. XANES provides information pertaining the site symmetry, electronic configuration, vacant orbits, and oxidation states of the absorbing atom, whereas EXAFS is capable of attaining information about coordination number, bond distances and atomic number of the atoms surrounding the element, whose absorption edge is under examination. It should be pointed out that the information obtained from EXAFS spectra typically necessitates rigorous modelling using best-fit models derived from well-developed scattering theories to determine the atomic species of the neighbours, nearest neighbour bond distances, local coordination number, and the possible existence of local disorders. 

Worthy of note is that the chemical species to be studied must generally dominate the absorptance at the absorption edge energy for successful XAS measurements. As such, inorganic elements in target materials can easily be studied using this methodology. In any case, trace elements can also be studied if their absorption edges are well separated from the absorption edges of other elements. Hard X-rays are essentially used to probe the bulk properties of the materials in transmission mode. Nonetheless, for material films on thick substrates or thicker samples, collection modes other than transmission (for example, total electron, fluorescence yield modes $etc. $) are often used, and calibrations are generally applied to cater for self-absorption effects. In such cases, XAS measurements using soft X-rays (having short absorption lengths) are often made in yield modes. We refer readers to the following bibliography that detail the fundamental principles and applications of XAS.\cite{de2001high, bressler2004ultrafast}

\begin{figure*}[!t]
\centering
  \includegraphics[width=0.8\columnwidth]{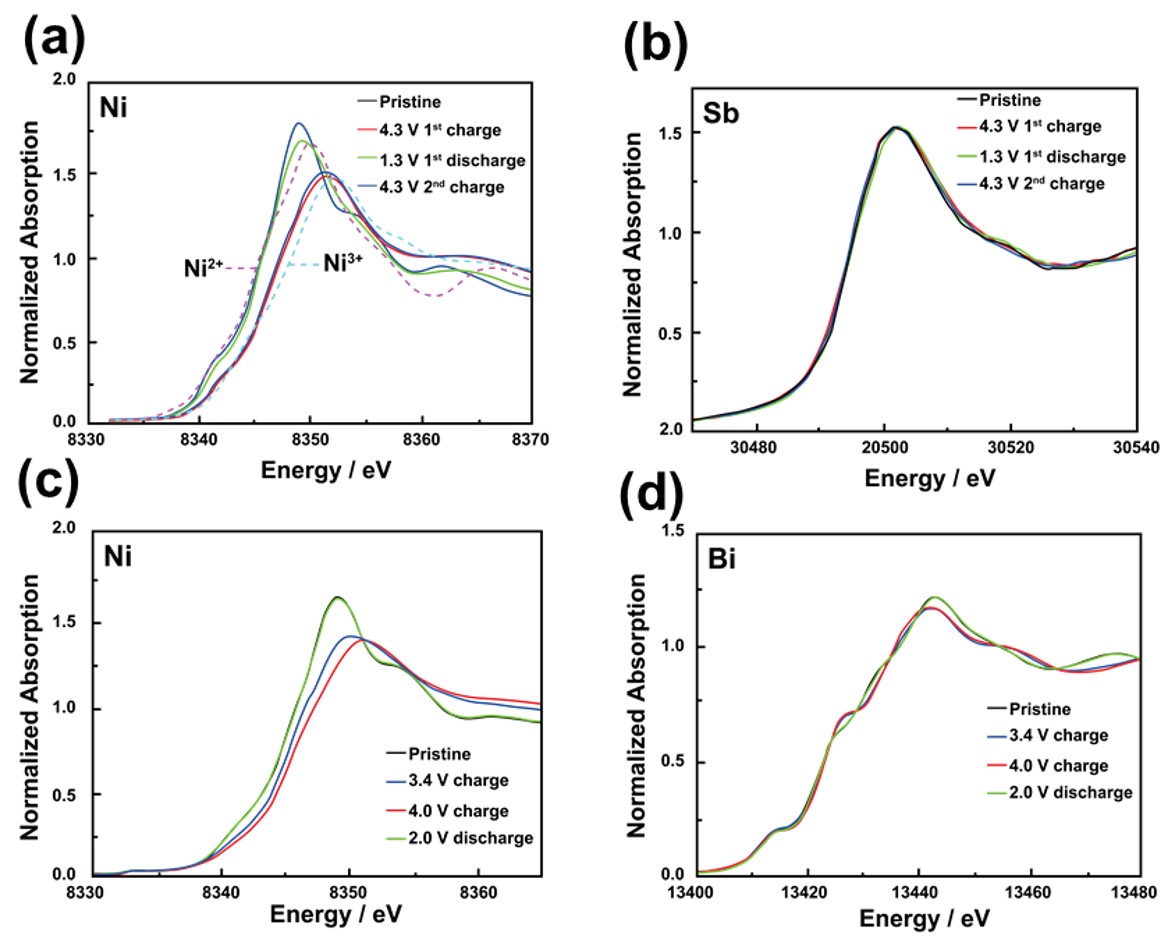}
  \caption{Hard X-ray absorption spectroscopy (XAS) spectra of honeycomb layered $\rm Na_3Ni_2SbO_6$ and $\rm Na_3Ni_2BiO_6$. (a,b) Ni $K$-edge and Sb $K$-edge X-ray absorption near edge structure (XANES) spectra attained upon (dis)charging of $\rm Na_3Ni_2SbO_6$: pristine and after the first charge, first discharge, and second charge. (c,d) Ni $K$-edge and  Bi ${\rm {\it L}_{3}}$-edge {\it ex situ} XANES spectra attained upon (dis)charging $\rm Na_3Ni_2BiO_6$ samples; pristine, halfway charged (3.4 V charge), fully charged (4.0 V charge), and fully discharged (2.0 V discharge). Figures (a, b) and Figures (c, d) reproduced with permission.\cite{Kim2020, Bhange2017} Copyright from 2020 American Chemical Society and 2017 Royal Society of Chemistry, respectively.}
  \label{Figure_31}
\end{figure*}

Given the vast capabilities presented by this technique, hard X-ray XAS measurements have been used to investigate the electrochemical reactions occurring in honeycomb layered tellurates ($e.g.$, $\rm K_2Ni_2TeO_6$)\cite{Masese2018} and bismuthates (such as $\rm Na_3Ni_2BiO_6$ and $\rm Na_3Ni_{1.5}Cu_{0.5}BiO_6$)\cite{Bhange2017, Wang2017}. Additionally, hard XAS has been performed on honeycomb layered antimonates such as $\rm Na_3Ni_2SbO_6$ to reveal the redox reaction (charge compensation mechanism) occurring during charge and discharge processes.\cite{Wang2019b, Kim2020} As visualised by the Ni $K$-edge XANES spectra in \textbf{Figure \ref{Figure_31}a}, the reversible ${\rm Ni^{3+}}$/${\rm Ni^{2+}}$ is noted to be the active redox couple, whilst ${\rm Bi^{5+}}$ remains inactive (\textbf{Figure \ref{Figure_31}b}) during cycling of $\rm Na_3Ni_2SbO_6$.\cite{Kim2020} Upon charging from 2.0 to 4.0 V, the shifts in the XANES spectra (\textbf{Figure \ref{Figure_31}a}) at Ni $K$-edge resemble the edge position of the reference Ni oxides: $\rm LiNi^{3+}O_2$ and ${\rm Ni^{2+}}$O, thus evincing the sequential oxidation of ${\rm Ni^{2+}}$ → ${\rm Ni^{3+}}$. In addition, the weak intensity of the pre-edge peak at Ni $K$-edge centred at around 8335 eV was maintained, indicating that the nickel ions remain in octahedral sites ($\rm NiO_6$) upon Na extraction and insertion of $\rm Na_{3-{\it x}}Ni_2SbO_6$. Conversely, no apparent changes are observed for the Sb $K$-edges (\textbf{Figure \ref{Figure_31}b}), revealing that ${\rm Sb^{5+}}$ do not participate in the electrochemical reaction.

Correspondingly, the Bi ${\rm {\it L}_{3}}$-edge and Ni $K$-edge XAS performed on $\rm Na_3Ni_2BiO_6$\cite{Bhange2017} further show Ni to be actively involved as the redox centre during the (dis)charge process. The Ni $K$-edge XANES spectra show a clear edge shift towards higher energies when the electrode was charged to 4.0 V (\textbf{Figure \ref{Figure_31}c}), indicating an increase in the average Ni oxidation state that approaches 3+. A comparison between the edge positions of the reference Ni oxides $\rm LiNi^{3+}O_2$ and ${\rm Ni^{2+}}$O and those of known oxidation states confirmed that the average oxidation states of Ni in the fully charged samples and pristine correspond to ${\rm Ni^{3+}}$ and ${\rm Ni^{2+}}$, respectively. The edge position completely reverts to its original position when the electrode is discharged to 2.0 V, indicating the reversibility of the ${\rm Ni^{3+/2+}}$ redox reactions during charge and discharge processes. Note also that the very weak intensity of the pre-edge peak centred at around 8335 eV remains unchanged during the entire process, revealing that the Ni ions in $\rm Na_{3-{\it x}}Ni_2BiO_6$ remain in their octahedral sites ($\rm NiO_6$) during charging and discharging processes. Conversely, the Bi ${\rm {\it L}_{3}}$-edge XANES spectra (\textbf{Figure \ref{Figure_31}d}) do not show any conspicuous edge shifts during charging and discharging processes. Comparison with the XANES spectrum of a reference $\rm NaBi^{5+}O_3$ compound further affirms the oxidation state of ${\rm Bi^{5+}}$ in the $\rm Na_{3-{\it x}}Ni_2BiO_6$ electrode remains invariant during charging and discharging.\cite{Bhange2017} 

\begin{figure*}[!t]
\centering
  \includegraphics[width=0.6\columnwidth]{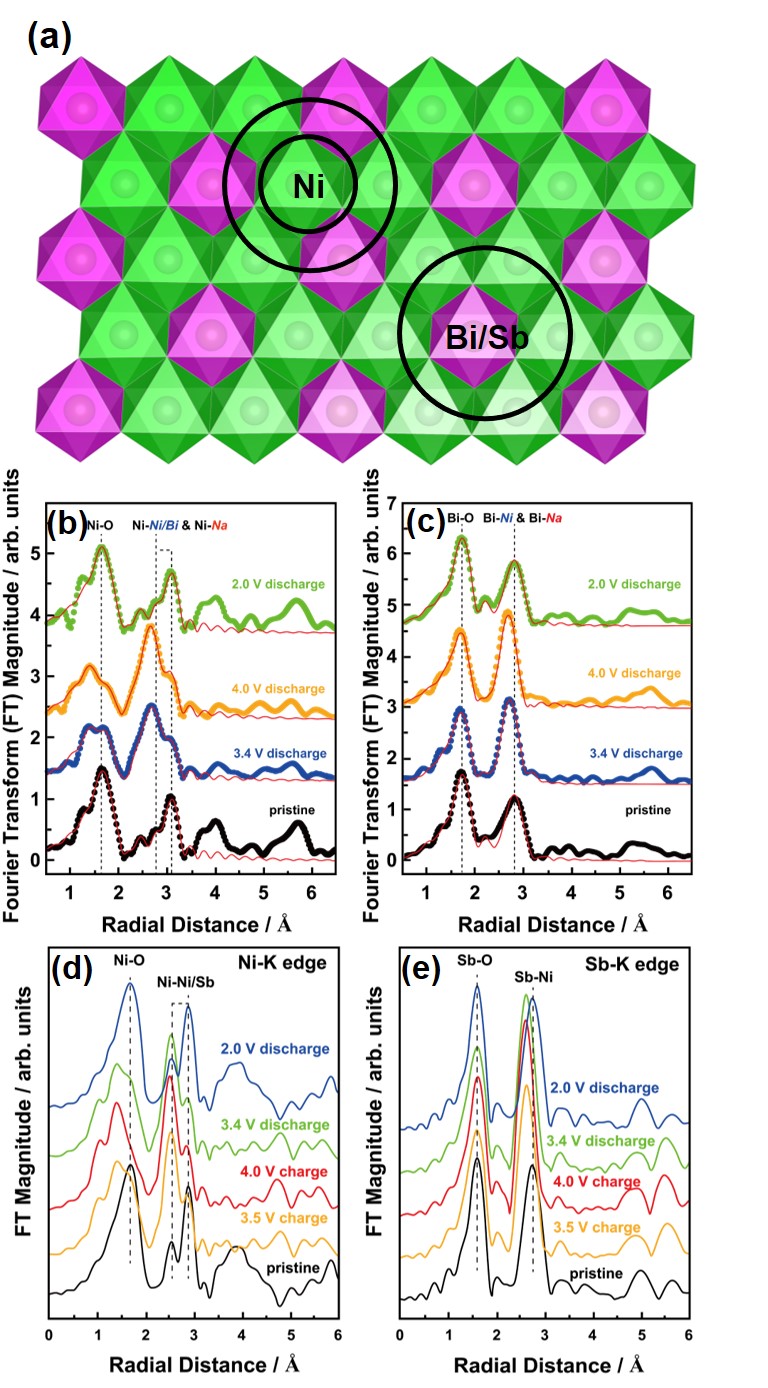}
  \caption{Extended X-ray absorption fine structure (EXAFS) spectra of honeycomb layered $\rm Na_3Ni_2SbO_6$ and $\rm Na_3Ni_2BiO_6$. (a) Honeycomb slab of $\rm Na_3Ni_2BiO_6$/$\rm Na_3Ni_2SbO_6$ highlighting the local arrangement of transition metal Ni atoms and pnictogens (Bi, Sb) in their first neighbouring shells. (b) Ni $K$-edge, and (c) Bi ${\rm {\it L}_{3}}$-edge EXAFS spectra of $\rm Na_3Ni_2BiO_6$ samples; pristine, half-way charged to 3.4 V, fully charged to 4.0 V, and fully discharged to 2.0 V. (d,e) Corresponding {\it ex situ} EXAFS spectra taken for $\rm Na_3Ni_2SbO_6$ electrodes collected at (d) the Ni $K$-edge and (e) Sb $K$-edge. Figures (b, c) and Figures (d, e) reproduced with permission.\cite{Bhange2017, Wang2019b} Copyright 2017 Royal Society of Chemistry and 2019 Wiley-VCH respectively.}
  \label{Figure_32}
\end{figure*}

The local structural changes around Bi or Sb and Ni atoms in the $\rm Na_{3-{\it x}}Ni_2BiO_6$ and $\rm Na_{3-{\it x}}Ni_2SbO_6$ electrodes during charging and discharging have also been investigated using EXAFS (\textbf{Figure \ref{Figure_32}}) combined with curve fitting analysis (\textbf{Figures \ref{Figure_32}b} and \textbf{\ref{Figure_32}c}) in some cases.\cite{Bhange2017, Wang2019b} EXAFS has also been used to provide quantitative structural information on changes in degree of disorder and bond lengths in honeycomb layered oxides. \textbf{Figures \ref{Figure_32}b} and \textbf{\ref{Figure_32}c} show {\it ex situ} Ni $K$-edge and Bi ${\rm {\it L}_{3}}$-edge EXAFS spectra of $\rm Na_{3-x}Ni_2BiO_6$ electrode during charging and discharging.\cite{Bhange2017} The first peak centered around $1.7$ \AA\, seen in the Fourier-transform (FT) magnitudes of the Ni and Bi EXAFS spectra (\textbf{Figures \ref{Figure_32}b} and \textbf{\ref{Figure_32}c}) corresponds to the first $M$–O ($M$ = Ni or Bi) bonding in the distorted $\rm NiO_6$ or $\rm BiO_6$ octahedra. The second FT peak at around $2.8$ \AA\, is associated with both the second nearest $M$–$M$ bonds within the layer slabs as well as the $M$–Na bonds. 

The EXAFS spectra were well-fitted using a model that assumes that: (i) the Bi atoms are coordinated by six Ni atoms only; (ii) the Ni atoms are surrounded by three Ni, and three Bi atoms, and (iii) the layers have a perfect honeycomb ordering, as shown in \textbf{Figure \ref{Figure_32}a}. During charging and discharging, dramatic changes are observed at the local structure around the Ni atom, whereas the local structure around Bi atoms displays no significant changes. This observation is in consonance with the XANES results (\textbf{Figures \ref{Figure_32}c} and \textbf{\ref{Figure_32}d}), which demonstrated the occurrence of an active ${\rm Ni^{2+/3+}}$ redox reaction with non-existent participation of the ${\rm Bi^{5+}}$ ions during the charge compensation processes. On an interesting note, the changes in the second FT peak in the Ni $K$-edge reveal very distinct features compared to those observed in the typical layered oxide materials such as $\rm Na_{2/3}Ni_{1/3}Mn_{2/3}O_2$. This behaviour is ascribed to the strain-induced disorder in Ni–Ni/Bi bonding caused by the size mismatch between the large ${\rm Bi^{5+}}$ and ${\rm Ni^{2+}}$ ions.\cite{Bhange2017} The Bi and Ni EXAFS results during charging and discharging show highly reversible local structural changes around Bi and Ni atoms in the $\rm Na_3Ni_2BiO_6$ electrode, which validates the stable cycling performance of the honeycomb-layered $\rm Na_3Ni_2BiO_6$ when utilised as a cathode material. 

\textbf{Figures \ref{Figure_32}d} and \textbf{\ref{Figure_32}e} show the EXAFS spectra at the Ni and Sb $K$-edges of $\rm Na_3Ni_2SbO_6$.\cite{Wang2019b}  In the first $M$–O coordination shell, the Ni $K$-edge delivers lowered Fourier transform (FT) magnitude and shortened interatomic distances concomitant with the oxidation of ${\rm Ni^{2+}}$ to ${\rm Ni^{3+}}$, whilst almost no change is observed at the Sb $K$-edge upon further charging. In the $M$–$M$ second coordination shells, there are three Sb and three Ni atoms surrounding one Ni atom, whilst only six Ni atoms are coordinated to every Sb atom, as shown in \textbf{Figure \ref{Figure_32}a}. There are two kinds of Ni–Sb and Ni–Ni environments and only one Ni–Sb peak in the second coordination shells, which further affirms a perfect honeycomb ordering within the transition metal slabs.

\begin{figure*}[!t]
\centering
  \includegraphics[width=0.8\columnwidth]{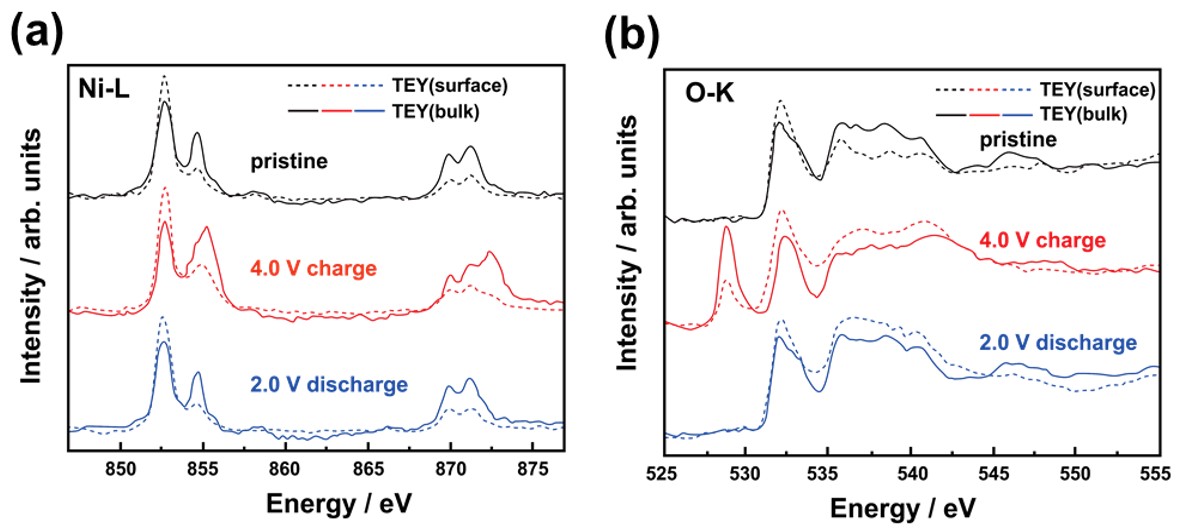}
  \caption{Soft X-ray absorption spectroscopy (XAS) spectra of honeycomb layered $\rm Na_3Ni_2SbO_6$. (a) Ni $L$-edge and (b) O $K$-edge {\it ex situ} soft XAS spectra of $\rm Na_3Ni_2SbO_6$ electrodes collected at different (dis)charge states in total fluorescence yield (TFY (solid line)) and total electron yield (TEY (shown in dotted line)). Figures reproduced with permission.\cite{Wang2019b} Copyright 2019 Wiley-VCH.}
  \label{Figure_33}
\end{figure*}

On another front, soft X-ray absorption spectra collected in total electron yield (TEY) and total fluorescence yield (TFY) modes provide enhanced contrast between the surface and bulk regions, respectively.\cite{abbate1992probing, amemiya2020development} In effect, this technique has been used to reveal the various electronic changes occurring in honeycomb layered oxides such as $\rm Na_3Ni_2SbO_6$ ($\rm NaNi_{2/3}Sb_{1/3}O_2$).\cite{Wang2019b} \textbf{Figures \ref{Figure_33}a} and \textbf{\ref{Figure_33}b} present the TFY (bulk-sensitive) and TEY (surface-sensitive) signals of $\rm Na_{1-{\it x}}Ni_{2/3}Sb_{1/3}O_2$ at O $K$-edge and Ni $L$-edge upon Na (de)intercalation are displayed as dotted and solid lines, respectively. The Ni $L$-edge TFY spectrum obtained from $\rm NaNi_{2/3}Sb_{1/3}O_2$ (\textbf{Figure \ref{Figure_33}a}) displays two peaks: a satellite peak at 854.8 eV and a main peak at 852.5 eV – typical patterns for high-spin ${\rm Ni^{2+}}$. The intensity of the satellite peak reversibly increases (decreases) upon Na extraction and insertion, indicating that the oxidation state of nickel ions change reversibly in the electronic state of the [$\rm NiO_6$]${\rm ^{n-}}$ cluster in $\rm Na_{1-{\it x}}Ni_{2/3}Sb_{1/3}O_2$. In a stark difference with the TFY (bulk-sensitive) spectra, the TEY (surface-sensitive) signals reveal the presence of a stable ${\rm Ni^{2+}}$ layer on the surface of $\rm Na_{1-{\it x}}Ni_{2/3}Sb_{1/3}O_2$. The Ni-$L$ edge XAS results further reveal that the Ni redox couple involves losing (oxidising) or adding (reducing) electrons in the spin-up ${\rm {\it e}_{g}}$ orbitals.\cite{thole1988branching} In the O $K$-edge XAS (\textbf{Figure \ref{Figure_33}b}), the pre-edge feature observed around 532 eV is attributed to the transition from O 1$s$ to 2$p$. Further, the intensity of this feature is associated with the density of unoccupied Sb 4$d$–O 2$p$ and Ni 3$d$–O 2$p$ hybridised states.\cite{de1989oxygen} The pre-edge intensity at 529 eV exhibits an increase upon Na extraction from the lattice, which is presumed to originate from the stronger hybridisation between the oxygen 2$p$ and the higher-valent transition metal nickel ions. The pre-edge peak shape and intensity are restored during discharging, indicating reversible Ni oxidation/reduction process. It is worth noting that similar spectral features were observed in the O $K$-edge XAS spectra taken during the charging and discharging of $\rm K_2Ni_2TeO_6$.\cite{Masese2018}

The ability to visualise changes in the electronic state of constituent elements during electrochemical reaction of honeycomb layered oxide materials have made hard and soft X-ray absorption spectroscopic techniques invaluable in the elucidation of ion mechanisms that govern the operations in this class of materials in rechargeable battery systems. By revealing the dormant role of pnictogen ions such as ${\rm Bi^{5+}}$ and ${\rm Sb^{5+}}$ as well as chalcogen ions such as ${\rm Te^{6+}}$,\cite{Bhange2017, Kim2020, Masese2018} soft X-rays have established that the transition metal atoms or anions (oxygen) are exclusively accountable for the charge compensation processes. Since honeycomb layered antimonates such as $\rm Li_4FeSbO_6$ demonstrate oxygen redox reaction,\cite{McCalla2015a} it will be worthwhile to augment O $K$-edge XAS along with recent techniques such as resonant inelastic X-ray scattering (RIXS)\cite{chen2014recent, bergmann2009x, rovezzi2014hard} to further elucidate their underlying oxygen redox mechanisms.

\subsubsection{Nuclear magnetic resonance (NMR) spectroscopy}
Nuclear magnetic resonance (NMR) has become a preeminent spectroscopic technique for distinguishing local and isotope-specific structural information of materials. During characterisation, information is obtained through the observation of nuclear spin interactions when in the presence of a magnetic field. NMR is typically seen as a local probing technique because it presents some inherent spectra complications that only allow the interpretation of short-range interactions. Nonetheless, it can also provide quantitative dynamical information, for instance, the cation diffusional constants of energy materials such as those used in batteries. Here, we provide a number of reference sources for readers interested in detailed explanation of the NMR methodology.\cite{gunther2013nmr, gunther1994nmr}

\begin{figure*}[!t]
\centering
  \includegraphics[width=0.8\columnwidth]{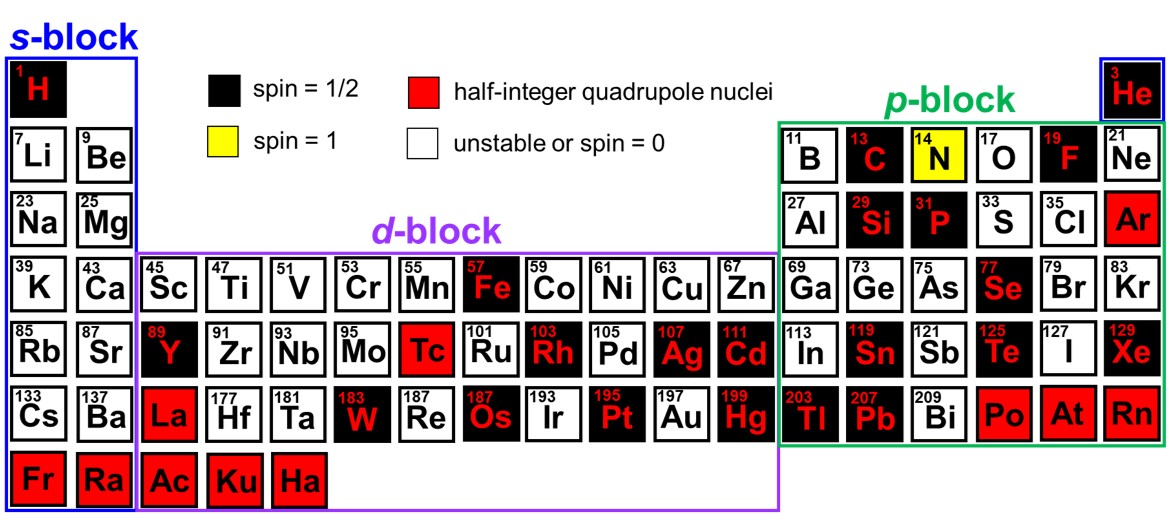}
  \caption{Periodic table of nuclear magnetic resonance (NMR) isotopes. Nuclei with even charge, and even mass number, for instance, ${\rm ^{12}}$C and ${\rm ^{16}}$O are non-spinning and do not exhibit NMR response. Nuclei with even mass number but odd charge have integral spins. Nuclei with odd mass number have half integral spins. The latter two exhibit NMR signals. The nuclei with half integral spins ($I$=1/2) give the best resolved spectra as they behave as charged spinning spherical bodies. Nuclei with spins greater than 1/2 behave like charged rotating spheroids and often show broadening of their NMR signals.}
  \label{Figure_34}
\end{figure*}

There exists a wide variety of NMR-active nuclides (shown in \textbf{Figure \ref{Figure_34}}) with intriguing properties; however, the subsection shall only focus on nuclei pertinent to the present class of pnictogen- and chalcogen-based honeycomb layered oxides: $i.e.$, ${\rm ^{6/7}}$Li, ${\rm ^{23}}$Na, ${\rm ^{25}}$Mg, ${\rm ^{39/40/41}}$K, ${\rm ^{53}}$Cr, ${\rm ^{59}}$Co, ${\rm ^{61}}$Ni, ${\rm ^{67}}$Zn, ${\rm ^{121/123}}$Sb, ${\rm ^{135/137}}$Ba, ${\rm ^{209}}$Bi, ${\rm ^{125}}$Te, amongst others. Amongst the NMR-active isotopes (spin $I \neq $0) mentioned above, some (such as ${\rm ^{209}}$ Bi, ${\rm ^{39}}$K, ${\rm ^{135}}$Ba, ${\rm ^{137}}$Ba, ${\rm ^{75}}$As, ${\rm ^{67}}$Zn, ${\rm ^{121/123}}$Sb) are much less amenable and receptive to NMR and are thus under-reported in literature. This variety of isotopes (most of which are quadrupolar $i.e.$, have a quantum spin number ($I$) greater than 1/2 ($I$> 1/2)) are usually referred to as {\it exotic nuclei}. The challenges posed in studying such nuclei stem from intrinsic isotopic properties which include low natural abundance, low gyromagnetic ratio, high quantum spin number ($I$) and large quadrupole moments; or some combination thereof. As such, spectra from these nuclides are typically plagued with prohibitively broad line widths and/or low receptivities that make it difficult to interpret the NMR data collected. For clarity, `receptivity' refers to the ease with which a signal of a given isotope may be acquired relative to another one (it scales with the natural abundance of the isotope, the gyromagnetic ratio and the quantum spin number).  For instance, the receptivity of ${\rm ^{43}}$Ca with respect to ${\rm ^{13}}$C (at natural abundance of 1.11\%) is only 0.05,\cite{leroy2018recent} rendering ${\rm ^{43}}$Ca a somewhat difficult nucleus to study.

Nonetheless, most constituent elements of the pnictogen and chalcogen class of honeycomb layered oxides fall under the NMR-active nuclei, making this class of materials an ideal platform for cutting-edge NMR studies. It is worth noting that a vast majority of the honeycomb layered oxides in the present class are promising battery materials and thus provide high compatibility between solid-state NMR and battery technologies. In particular, the ${\rm ^{6/7}}$Li and ${\rm ^{23}}$Na nuclei are frequently probed due to their direct participation as charge carriers in lithium-ion and sodium-ion battery technologies, respectively.

\begin{figure*}[!b]
\centering
  \includegraphics[width=0.8\columnwidth]{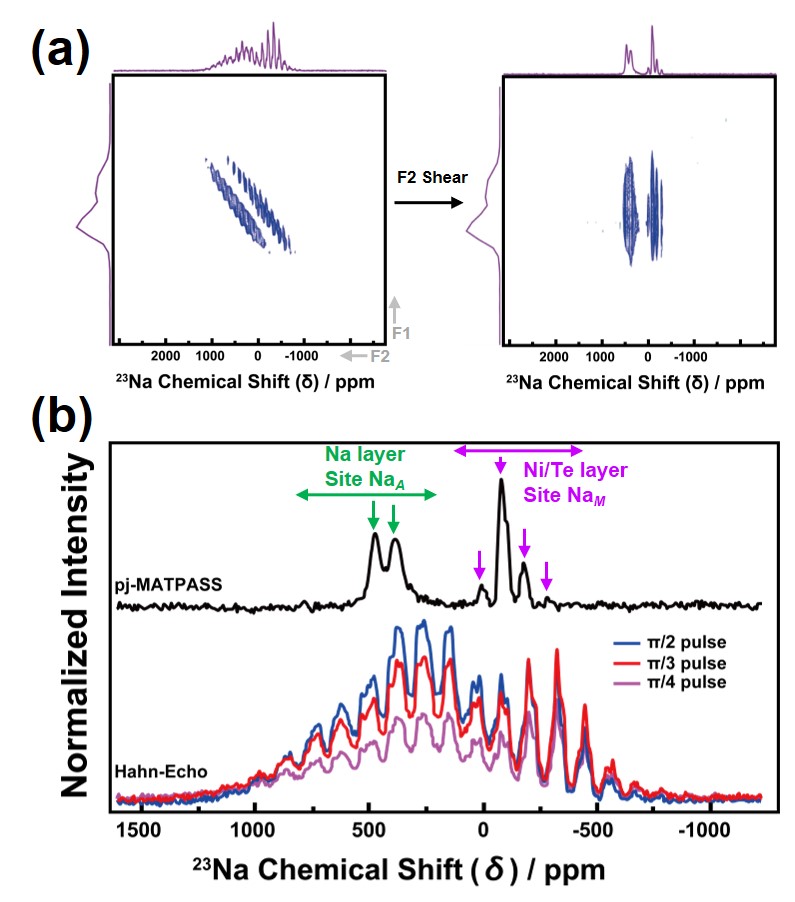}
  \caption{(a) 2D pj-MATPASS ${\rm ^{23}}$Na NMR spectrum of $\rm Na_3Ni_{1.5}TeO_6$ attained at 13 kHz and 400 MHz MAS. Spectra before processing (left) and after processing (right), entailing a shear in the F2 dimension are shown with F1 and F2 projections (in purple). (pj-MATPASS: projection-(Magic Angle Turning – Phase Adjusted Spinning Sideband) Dimension F2 out of 3D spectroscopy is called F2 shear, which is a cross section from F2 surface.) (b) ${\rm ^{23}}$Na Hahn-Echo NMR spectra of $\rm Na_3Ni_{1.5}TeO_6$ acquired at various pulse lengths (${\rm \pi}$/2, ${\rm \pi}$/3 and ${\rm \pi}$/4) relative to an aqueous NaCl solution as reference at 13 kHz MAS. The isotropic pj-MATPASS spectrum is also included to distinguish the position of the isotropic peaks from the spinning sidebands. Looking at the ${\rm \pi}$/4 pulses of the Hahn-Echo NMR spectra, the relative intensity of the peaks arising from Na residing within the Na layer interposed between the transition metal slabs ($\rm Na_{\it A} $ sites) and Na residing within the transition metal slabs ($\rm Na_{\it M} $ sites) are different, suggesting differences in the quadrupole coupling constants. Figures reproduced with permission.\cite{grundish2020structural} Copyright 2020 American Chemical Society.}
  \label{Figure_35}
\end{figure*}

Considering the sensitivity of solid-state NMR with respect to the microstructure, both quantitative and qualitative information pertaining the short-range, and electronic structure of the material sample can be obtained by analysing the line intensity, line shape and chemical shift of the NMR signal; to provide a paramount basis for further theoretical analyses aimed at optimising existing materials and/or the design of novel conformations. Generally, the ability to analyse both stationary and magic-angle spinning (MAS) powders is considered useful during the characterisation of chemical shifts and quadrupolar coupling tensors. With the recent progress in advanced pulse methods (such as the projection of magic-angle turning phase-adjusted spinning sidebands (pj-MATPASS) sequence\cite{gan1996improved}) and magic angle spinning (MAS)\cite{polenova2015magic} NMR technique, the spinning sidebands and the anisotropic interactions can effectively be suppressed to attain high-resolution spectra for materials. In the following paragraphs, we discuss the application of NMR in investigating some noteworthy functionalities of pnictogen- and chalcogen-based honeycomb layered oxides.

\textbf{Figure \ref{Figure_35}} shows a 2D pj-MATPASS ${\rm ^{23}}$Na NMR spectrum of alkali-excess $\rm Na_3Ni_{1.5}TeO_6$\cite{grundish2020structural} acquired at 13 kHz and 400 MHz MAS. The NMR spectra of $\rm Na_3Ni_{1.5}TeO_6$ spectra were processed to attain the ${\rm ^{23}}$Na Hahn-Echo NMR spectra shown in \textbf{Figure \ref{Figure_35}b}. Within the crystal structure of $\rm Na_3Ni_{1.5}TeO_6$,\cite{grundish2020structural} Na ions are noted to reside in both the Na layers and their adjacent transition-metal slabs, denoted hereafter respectively as $\rm Na_{\it A} $ and $\rm Na_{\it M} $ sites. ${\rm ^{23}}$Na NMR was thus employed to elucidate the specific locations of the ${\rm Na^{+}}$ ion within the structure of $\rm Na_3Ni_{1.5}TeO_6$. The $\rm Na_3Ni_{1.5}TeO_6$ spectrum attained with the pj-MATPASS pulse sequence (\textbf{Figure \ref{Figure_35}b}) manifests six isotropic Na resonances: four peaks with negative values (negatively-shifted peaks with maximum intensities at --283, --178, --77, 0 ppm) and two positively-shifted peaks with maximum intensities at 477 and 386 ppm. Several of these peaks display asymmetric line shapes due to the quadrupole nature of the ${\rm ^{23}}$Na nucleus (spin $I$ = 3/2). It should be noted that the study did not attempt to perform a detailed fitting of the quadrupole parameters of each environment due to the possible distortions of the line shapes as a result of the use of the pj-MATPASS sequence. 
Despite this, the differences in the quadrupole parameters of the different sites could still be ascertained from the difference in the nutation behaviour of the Hahn-echo spectra. For clarity, pj-MATPASS is effective in suppressing spinning sidebands and also separating both the anisotropic and isotropic resonances,\cite {hung2012isotropic}whereas Hahn-echo pulse sequence is effective in measuring spin-spin relaxation time and baseline correction.\cite {peng2019capacity, liu2019p2} The dominant contribution to the ${\rm ^{23}}$Na NMR shift in \textbf{Figure \ref{Figure_35}b} is related to the Fermi contact interaction in which unpaired spin density is transferred from ${\rm Ni^{2+}}$ to the Na nucleus via an intermediate O ligand. Therefore, the total ${\rm ^{23}}$Na NMR shift observed experimentally can be decomposed into the sum of the individual contributions from each Na--O--${\rm Ni^{2+}}$ bond pathway (connectivity). Based on the Na--O--${\rm Ni^{2+}}$ bond connectivity and the bond angles, the Na resonances with positive values of the chemical shifts (477 and 386 ppm) in \textbf{Figure \ref{Figure_35}b} are assigned to $\rm Na_{\it A} $ sites in the Na layer, with the difference in the chemical shift between the two $\rm Na_{\it A} $ resonances ($i.e.$, 91 ppm) arising from difference in the number of ${\rm Ni^{2+}}$--O--Na pathways. The four sharp peaks at --283, --178, --77 and 0 ppm are assigned to $\rm Na_{\it M} $ sites in the transition-metal layer. These peaks arise from the differences in the local bonding environments of Na ions the transition-metal layer with the surrounding ${\rm Ni^{2+}}$. Therefore, this study adequately evinces the efficacy of NMR data in demonstrating sodium occupancy in the transition-metal layers of $\rm Na_3Ni_{1.5}TeO_6$.\cite{grundish2020structural}

\begin{figure*}[!b]
\centering
  \includegraphics[width=0.9\columnwidth]{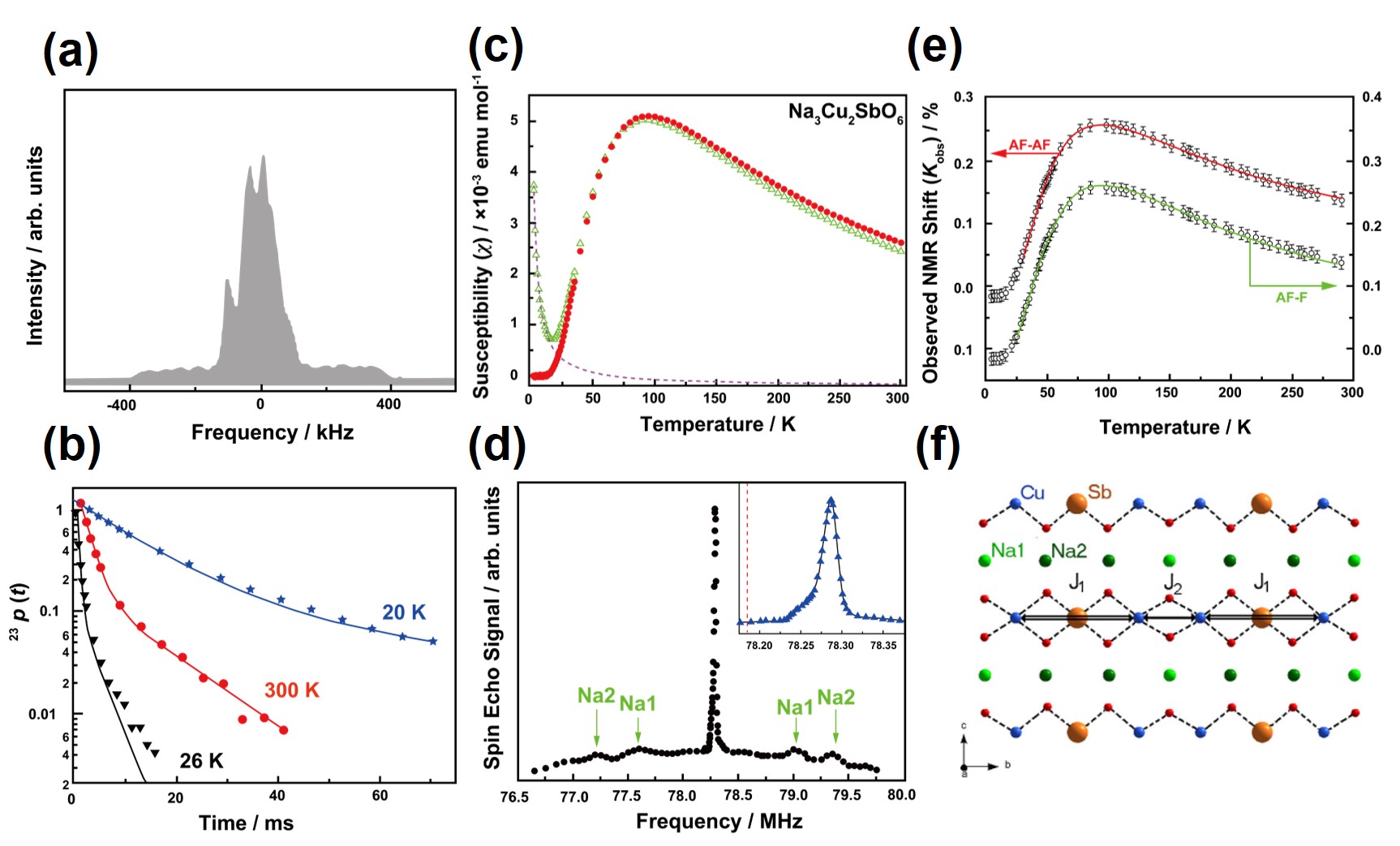}
  \caption{NMR spectra of honeycomb layered $\rm Na_2Ni_2TeO_6$ and $\rm Na_3Cu_2SbO_6$. (a) Fourier-transformed ${\rm ^{23}}$Na NMR spectrum for $\rm Na_2Ni_2TeO_6$ collected at 300 K at a Larmor frequency of 84.670 MHz. The central transition line is affected by the nuclear quadrupole interaction. (b) ${\rm ^{23}}$Na nuclear spin–lattice relaxation curves 
  $^{23}p(t)$ at a central frequency. The solid curves represent the results from least squares fitting using a theoretical multi-exponential function for a central transition line. (c) Magnetic susceptibility of $\rm Na_3Cu_2SbO_6$ in a field of 1 T (open triangles). The dashed curve corresponds to a fit to the Curie-Weiss term with a constant. The resulting spin is shown as the solid circles. After passing this maximum, susceptibility decreases rapidly with further cooling, and an upturn appears below 15 K. (d) Fully resolved ${\rm ^{23}}$Na NMR spectrum for $\rm Na_3Cu_2SbO_6$. ${\rm ^{23}}$Na NMR reference frequency of an aqueous NaBr solution is shown by the vertical line. Owing to electric quadrupole coupling, the ${\rm ^{23}}$Na ($I$ = 3/2) NMR spectrum comprises two satellite lines per site, corresponding to magnetic moment $m$ = ± 1/2 $\rm \leftrightarrow{}$ ± 3/2 transitions. Thus, two non-equivalent Na sites resulting into four distinctive lines. The central transition spectrum of $\rm Na_3Cu_2SbO_6$ measured at room temperature is shown in inset. (e) Temperature dependence of the observed ${\rm ^{23}}$Na NMR shift for $\rm Na_3Cu_2SbO_6$. Fit to the antiferromagnetic(AF)--antiferromagnetic(AF) alternating chain model is shown in solid curve, whereas fit to the antiferromagnetic(AF)--ferromagnetic(F) alternating chain model is represented in dashed curve. (f) An atomistic model of the alternating spin interactions of $\rm Na_3Cu_2SbO_6$. The spin exchange interaction via $J_1$ is illustrated by each double-line. The interaction through a Cu--O--Cu path (defined by $J_2$) is represented in solid lines. Based on this view, the relative strengths of the spin exchange interactions between ${\rm Cu^{2+}}$ (spin =1/2) ions of $\rm Na_3Cu_2SbO_6$  indicate that the strongest super-exchange interaction ($J_1$) occurs along a Cu--O--Sb--O--Cu pathway. The second in magnitude was found to be $J_2$, bridged by one oxygen atom through a Cu--O--Cu path. The other three interactions $J_3$, $J_4$, and $J_5$ are negligibly small. Whether the AF–F or AF–AF alternating chain scenario is appropriate for the realisation of the magnetic property of $\rm Na_3Cu_2SbO_6$ is a topical subject still under debate. Figures (a, b) and Figures (c, d, e, f) reproduced with permission.\cite{Itoh2015, Kuo2012} Copyright 2015 Japan Physical Society and 2012 Elsevier, respectively.}
  \label{Figure_36}
\end{figure*}

In another study, NMR integrating spin-echo signals of various magnetic excitations was used to investigate magnetic properties such as spin frustration effects in honeycomb layered tellurates such as $\rm Na_2Ni_2TeO_6$.\cite{Itoh2015} The Fourier-transformed ${\rm ^{23}}$Na NMR spectrum shown in \textbf{Figure \ref{Figure_36}a},\cite{Itoh2015} indicates the possibility to attain a clear NMR spectrum of $\rm Na_2Ni_2TeO_6$ although its central transition line is affected by a nuclear quadrupole interaction. The nature of the magnetic phase transition was depicted by the ${\rm ^{23}}$Na nuclear spin-lattice relaxation curves (\textbf{Figure \ref{Figure_36}b}). In a similar study, spin-echo NMR spectroscopy was also used to investigate the spin gap behaviour in $\rm Na_3Cu_2SbO_6$ (\textbf{Figure \ref{Figure_36}c}).\cite{Kuo2012} Owing to electric quadrupole coupling, the ${\rm ^{23}}$Na (spin $I$ = 3/2) NMR spectrum of $\rm Na_3Cu_2SbO_6$ (\textbf{Figure \ref{Figure_36}d}) was noted to contain two satellite lines per Na site (denoted as Na1 and Na2), corresponding to magnetic moment $m$ = ± 1/2 $\rm \leftrightarrow{}$ ± 3/2 transitions. As such, the two non-equivalent Na sites are seen to result in four distinctive lines indicated by the arrows in \textbf{Figure \ref{Figure_36}d}. The ${\rm ^{23}}$Na central transition line at room-temperature is shown in the inset of \textbf{Figure \ref{Figure_36}d}. The central transition line shape is mainly featured by the Na2 site, since the site population ratio between Na1 and Na2 in the crystal lattice of $\rm Na_3Cu_2SbO_6$ is in the ration of 1:2 per unit cell.

The ${\rm ^{23}}$Na NMR investigations outlined above provided clear evidence that a spin gap exists in $\rm Na_3Cu_2SbO_6$. The Fermi contact shift (or Knight shift) can be linked to the magnetic susceptibility whose parameters can be obtained by plotting the observed Knight shift (${\rm {\it K}_{obs}}$) against the spin susceptibility ($\rm \chi_{spin}$) or what is referred to as Clogston-Jaccarino\cite{clogston1964interpretation} plot. The antiferromagnetic--ferromagnetic (AF--F) or (antiferromagnetic)--(antiferromagnetic) (AF--AF) chain models based on the atomistic structure (shown in \textbf{Figure \ref{Figure_36}e}) were assessed by invoking spin-lattice relaxation rates and NMR shifts on $\rm Na_3Cu_2SbO_6$ (\textbf{Figure \ref{Figure_36}f}). Altogether, the analyses of NMR spin shifts and spin-lattice relaxation rates revealed the magnetic nature of $\rm Na_3Cu_2SbO_6$ can be well accounted for by alternating the chain scenario atomic model (\textbf{Figure \ref{Figure_36}f}) with two dominant spin exchange interactions (antiferromagnetic ${\rm {\it J}_1}$ and ferromagnetic ${\rm {\it J}_2}$) whose values can be quantitatively determined.\cite{Kuo2012} Moreover, based on spin-lattice relaxation rates, the determined spin gap was found to be identical with the magnetic excitation energy attained from neutron scattering measurements. 

\begin{figure*}[!b]
\centering
  \includegraphics[width=0.7\columnwidth]{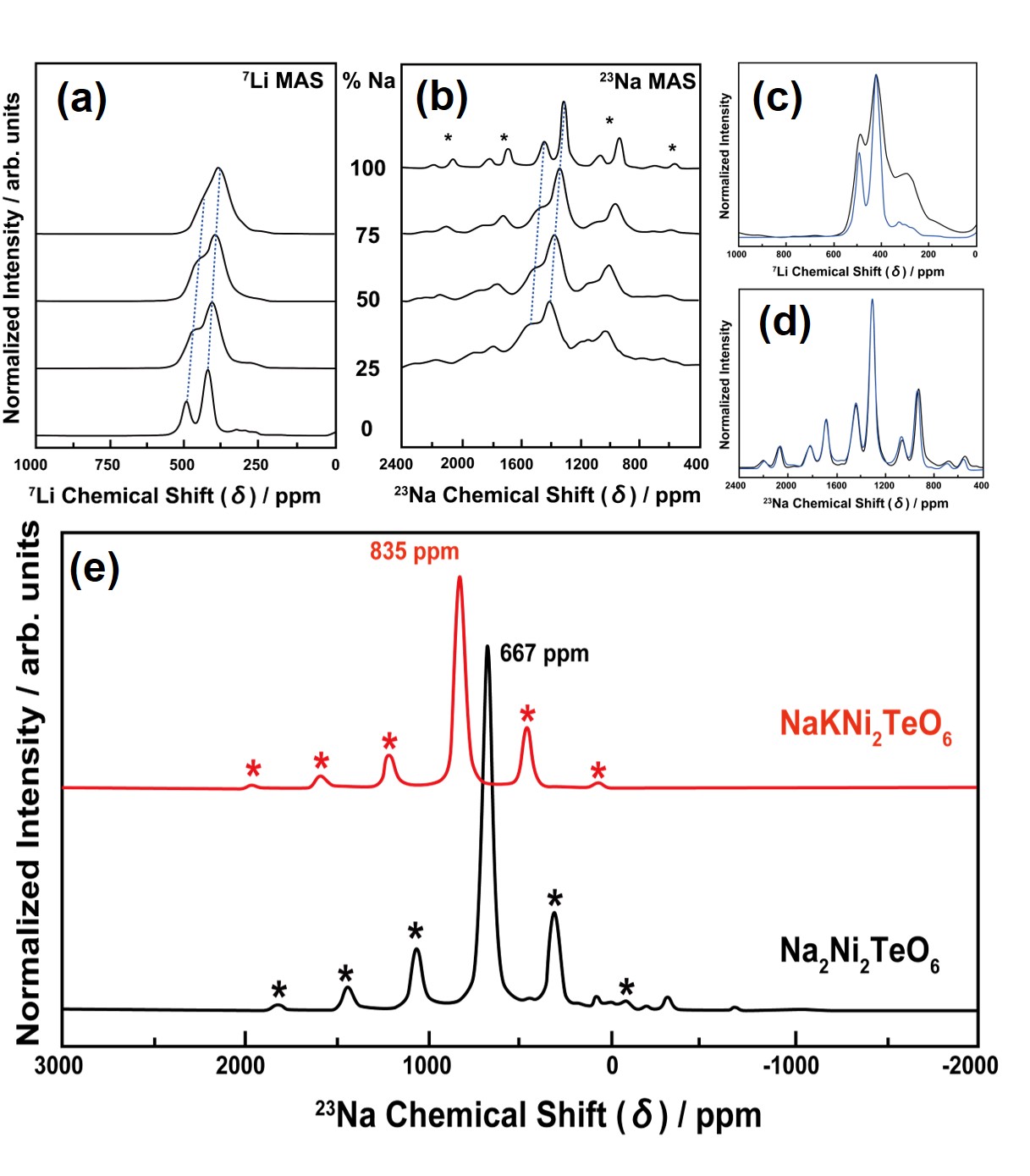}
  \caption{Ascertaining defects and phase purity of honeycomb layered antimonates and tellurates. (a) Solid-state ${\rm ^{7}}$Li NMR spectra of $\rm Li_3Ni_2SbO_6$ and intermediate $\rm Li_{3-{\it x}}Ni_2SbO_6$ compositions, which were synthesised via typical solid-state reaction at high temperatures, after which the samples were rapidly cooled (quenched). The symbol (*) accounts for spinning sidebands. (b) Solid-state ${\rm ^{23}}$Na NMR spectra of quenched $\rm Na_3Ni_2SbO_6$ and intermediate $\rm Na_{3-{\it x}}Li_{\it x}Ni_2SbO_6$ compositions. (b) ${\rm ^{7}}$Li NMR spectra for quenched (black lines) and non-quenched (blue lines) samples of $\rm Li_3Ni_2SbO_6$. There are significant changes in the ${\rm ^{7}}$Li NMR spectra. Non-quenched sample exhibits more defined peaks at 492 and 420 pm and less contribution at 300 ppm (related to honeycomb-ordering alteration or stacking defects). (d) ${\rm ^{23}}$Na NMR spectra for quenched (black lines) and non-quenched (blue lines) samples of $\rm Na_3Ni_2SbO_6$. The lack of disorders was further confirmed by XRD measurements. (e) ${\rm ^{23}}$Na magic angle spinning (MAS)-NMR spectra of $\rm NaKNi_2TeO_6$ (top, in red) and $\rm Na_2Ni_2TeO_6$ (bottom, in black). For $\rm Na_2Ni_2TeO_6$, weak signals are also noticed in some batches (¤) and were assigned to Na atoms residing in the honeycomb transition metal slab, as has been observed in NMR spectra of the alkali-excess $\rm Na_3Ni_{1.5}TeO_6$. The NMR spectra of $\rm Na_2Ni_2TeO_6$ shows the possibility of the existence of Na atoms in the honeycomb transition metal slab. Figures reproduced with permission.\cite{Vallee2019, berthelot2021stacking} Copyright 2019, 2021 American Chemical Society.}
  \label{Figure_37}
\end{figure*}

Solid-state NMR, a spectroscopic variant of NMR, has also been used to obtain information on sodium and lithium local environment of honeycomb layered antimonates $\rm Na_3Ni_2SbO_6$, $\rm Li_3Ni_2SbO_6$ and their intermediate compositions $\rm Li_{3-{\it x}}Ni_2SbO_6$.\cite{Vallee2019} The samples were prepared via solid-state reaction at high temperatures. Some samples were rapidly cooled (quenched) after heat treatment, whilst others were left to cool to room temperature inside the furnace (non-quenched). In principle, the peak shifts for all compounds arise from the Fermi contact interaction between the probed nucleus and the paramagnetic low-spin ${\rm Ni^{2+}}$ ions, implying that the interaction is propagated by chemical bonds and therefore denotes the local electronic structure. Taking the crystal lattice of $\rm Li_3Ni_2SbO_6$ ($C$2/$m$ monoclinic space group) into consideration,\cite{Zvereva2012} Li is located in two distinct crystallographic sites (denoted as 2$d$ and 4$h$, in the Wyckoff notation). Therefore, the two narrow peaks centred at 492 and 420 ppm can be presumed to arise from the Li sites. On the other hand, the broad contribution around 300 ppm is attributed to stacking defects, in congruence with those noted in the layered oxide $\rm Li_2MnO_3$.\cite{breger2005high} \textbf{Figure \ref{Figure_37}b} shows the corresponding ${\rm ^{23}}$Na NMR spectra obtained from $\rm Na_3Ni_2SbO_6$  and its quenched intermediate $\rm Na_{3-{\it x}}Li_{\it x}Ni_2SbO_6$ compositions. The ${\rm ^{23}}$Na NMR spectrum from $\rm Na_3Ni_2SbO_6$ exhibits two contributions assignable to Na ions located in the 2$d$ and 4$h$ Wyckoff sites at 1440 and 1300 ppm, respectively.\cite{Vallee2019}

The ${\rm ^{7}}$Li NMR (\textbf{Figure \ref{Figure_37}a}) and ${\rm ^{23}}$Na NMR (\textbf{Figure \ref{Figure_37}b}) spectra were also used to examine crystallographic information of the intermediate ($\rm Li_{3-{\it x}}Na_{\it x}Ni_2SbO_6$ and $\rm Na_{3-{\it x}}Li_{\it x}Ni_2SbO_6$, respectively) compositions obtained through quenching after the heat treatment. In both cases, the resonance peaks are observed to be significantly broader than the final $\rm Li_3Ni_2SbO_6$ or $\rm Na_3Ni_2SbO_6$ compositions. A closer look into the respective spectra revealed that peak broadness increased as composition stoichiometry moves further away from $\rm Na_3Ni_2SbO_6$ or $\rm Li_3Ni_2SbO_6$. Additionally, weak but significant peak shifts attributed to the variation of the (Na--O and Li--O) interatomic distances are also noted. Shifting towards a lower chemical shift value in ${\rm ^{7}}$Li NMR spectra corresponds to a decrease in the electronic spin transfer from low-spin ${\rm Ni^{2+}}$ to Li nuclei that can be attributed to an increase in the Li--O distances, whereas the displacement to higher chemical shift values in ${\rm ^{23}}$Na NMR spectra is attributed to a decrease in the Na--O distances. Hence, the NMR study performed in these honeycomb layered antimonates demonstrate that the alkali layers in $\rm Li_{3-{\it x}}Na_{\it x}Ni_2SbO_6$ compositions do not correspond to an intergrowth of only Na and Li layers but do contain both alkali ions.

For discernment of the structural configurations engendered by $\rm Li_3Ni_2SbO_6$, $\rm Na_3Ni_2SbO_6$, and their intermediate compositions, it is important to scrutinise their stacking variations. A comparison between the intensity of the resonance peak contributions observed at 300 ppm to those of the two other sharp peaks in the ${\rm ^{7}}$Li NMR (\textbf{Figure \ref{Figure_37}c}) spectra reveals a correlation between the nature of structural defects and the synthesis conditions. Here, the non-quenched $\rm Li_3Ni_2SbO_6$ sample displays sharper resonance peaks arising from the two Li sites with significantly dwindled broad contribution at 300 ppm in vast contrast with the quenched intermediate samples. On the other hand, the differences between the quenched and non-quenched $\rm Na_3Ni_2SbO_6$ samples in the ${\rm ^{23}}$Na NMR spectra (\textbf{Figure \ref{Figure_37}d}) appeared to be very minimal: almost indiscernible. These observations suggest that $\rm Na_3Ni_2SbO_6$ is more easily ordered than $\rm Li_3Ni_2SbO_6$ and is thus less sensitive to quenching.

Similarly, ${\rm ^{23}}$Na NMR has been employed in the analyses of $\rm Na_2Ni_2TeO_6$\cite{berthelot2021stacking} and mixed-alkali compositions such as $\rm NaKNi_2TeO_6$\cite{berthelot2021stacking} to assess their purity and obtain insight into the local structure of the sodium ions. \textbf{Figure \ref{Figure_37}e} shows a comparison between the ${\rm ^{23}}$Na NMR spectra obtained from $\rm Na_2Ni_2TeO_6$ and $\rm NaKNi_2TeO_6$ compositions. The ${\rm ^{23}}$Na NMR spectrum from $\rm Na_2Ni_2TeO_6$ exhibits a major isotropic peak at the chemical shift ${\rm \delta}$ = 677 ppm. Even though the sodium ions in $\rm Na_2Ni_2TeO_6$ reside in crystallographically different sites, their high mobility results in a unique averaged peak that has been similarly observed in previous investigations on other P3- and P2-type sodium-based transition metal layered oxides.\cite{kalapsazova2017, carlier2009, carlier2011}

In the case of $\rm NaKNi_2TeO_6$, a unique isotropic signal located at ${\rm \delta}$ = 825 ppm was observed (\textbf{Figure \ref{Figure_37}e}). The presence of a single resonance for $\rm NaKNi_2TeO_6$ (and accordingly the absence of residual reagent signature at 677 ppm) affirms that the reaction leading to the mixed-alkali composition is complete with no amorphous or crystalline by-products. It is important to mention that attempts to collect ${\rm ^{39}}$K NMR spectra of both $\rm K_2Ni_2TeO_6$ and $\rm NaKNi_2TeO_6$ have not been successful apparently owing to the observation of background ${\rm ^{39}}$K resonances and severe acoustic ringing of the probes used at the very low resonance frequency of ${\rm ^{39}}$K.\cite{Vallee2019}

\begin{figure*}[!t]
\centering
  \includegraphics[width=1.0\columnwidth]{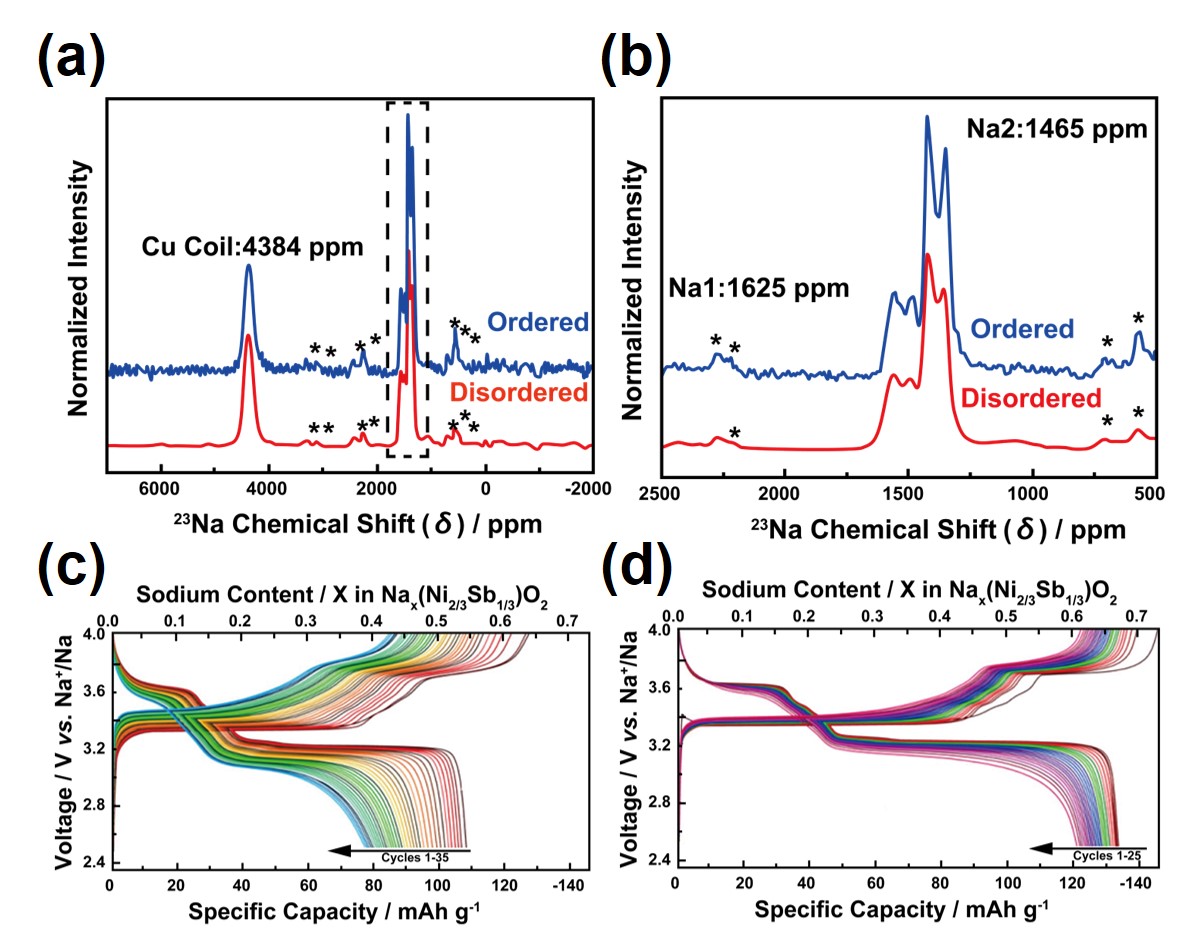}
  \caption{NMR spectra of disordered and ordered polytypes of honeycomb layered antimonate $\rm NaNi_{2/3}Sb_{1/3}O_2$ ($\rm Na_3Ni_2SbO_6$).(a) ${\rm ^{23}}$Na NMR spectra collected for disordered and ordered $\rm NaNi_{2/3}Sb_{1/3}O_2$. Note that an extra resonance peak at 4384 ppm associated with Cu coil emanates from the NMR probe. Asterisks mark the spinning sidebands. (b)  Enlarged image of the ${\rm ^{23}}$Na NMR spectra of disordered and ordered $\rm NaNi_{2/3}Sb_{1/3}O_2$ show in (a). In order to distinguish the various crystallographic sites of Na, site ratios were determined from integrated peak intensities. (c) Voltage (dis)charge profiles of disordered $\rm NaNi_{2/3}Sb_{1/3}O_2$. (d) Voltage (dis)charge profiles of ordered $\rm NaNi_{2/3}Sb_{1/3}O_2$. Electrochemical measurements reveal the ordered polytype to demonstrate higher capacity along with better capacity retention upon subsequent cycling, relative to the disordered polytype. The capacity retention over 30 cycles at a current density equivalent to (dis)charging for 50 hours ($i.e.$, C/50) was found to be 70 \% for the disordered phase and 90 \% for the ordered polytype. Figures reproduced with permission.\cite{Ma2015} Copyright 2015 American Chemical Society.}
  \label{Figure_38}
\end{figure*}

Nonetheless, the resilience of ${\rm ^{23}}$Na against the acoustic ringing of the probes has unlocked the unique solid-state NMR capability of probing the resonances of the quadruple nucleus (spin $I$ = 3/2), as a means to examine the presence of honeycomb ordering at a local scale. \textbf{Figure \ref{Figure_38}a} shows the ${\rm ^{23}}$Na NMR spectra of disordered and ordered polytypes of $\rm NaNi_{2/3}Sb_{1/3}O_2$ ($\rm Na_3Ni_2SbO_6$).\cite{Ma2015} Here, the ${\rm ^{23}}$Na NMR shift is noticeably influenced by the six adjacent octahedral sites occupied by Ni and Sb due to its high sensitivity to the ${\rm ^{23}}$Na local environment. These differences are reflected in the measured values of the ${\rm \eta}$ (asymmetry parameter) and quadrupolar coupling constant. As shown in \textbf{Figures \ref{Figure_38}a} and \textbf{\ref{Figure_38}b}, the observed intensity ratio of the two Na resonances (located at 1625 and 1465 ppm) is close to 2:1 in the ordered structures with the expectations for the two different Na sites (4$h$ and 2$c$ Wyckoff sites) of an ideal honeycomb ordered structure of $\rm Na_3Ni_2SbO_6$ (crystallising in the $C$2/$m$ monoclinic space group).

However, the defects in the honeycomb ordering of Sb and Ni ions in $\rm NaNi_{2/3}Sb_{1/3}O_2$ ($\rm Na_3Ni_2SbO_6$) give rise to varying local environments for Na ions and would thus be expected to generate new resonances in the ${\rm ^{23}}$Na NMR spectrum; the absence of which strongly indicates a firmly structured honeycomb order. The broad resonance at approximately 1100 ppm (highlighted in \textbf{Figure \ref{Figure_38}a}) has been reported to occur due to the presence of a ${\rm Ni^{3+}}$ -containing impurity.\cite{Ma2015} Given that the intensity of this resonance has been observed to remain the same in both disordered and ordered phases, the broadness of their peak cannot be associated with honeycomb structure stacking faults. A comparison between the NMR spectrum of the ordered phase and that of the disordered phase, revealed no substantive differences in the intensity ratio, number, or position of the ${\rm ^{23}}$Na resonances. This clearly indicates that their in-plane honeycomb ordering is fully developed. The importance of preparing ordered honeycomb layered antimonates can be observed by looking at their electrochemical performance. As shown in \textbf{Figures \ref{Figure_38}c} and \textbf{\ref{Figure_38}d}, ordered $\rm NaNi_{2/3}Sb_{1/3}O_2$ electrodes (\textbf{Figure \ref{Figure_38}d}) can facilitate reversible (dis)charged processes at higher capacities more effectively than their disordered polytype (\textbf{Figure \ref{Figure_38}c}).

Altogether, NMR has been used to elucidate various physicochemical properties of pnictogen- and chalcogen-based honeycomb layered oxides, with ${\rm ^{7}}$Li NMR studies (on $\rm Li_3Cu_2SbO_6$ and $\rm Li_3Ni_2SbO_6$, non-stoichiometric $\rm Li_{4/5}Ni_{3/5}Sb_{2/5}O_2$)\cite{vavilova2021, vavilova2021spin, Vallee2019} and ${\rm ^{23}}$Na NMR studies (on $\rm Na_2Ni_2TeO_6$, $\rm Na_2Zn_2TeO_6$, $\rm Na_2Co_2TeO_6$, $\rm NaKNi_2TeO_6$, $\rm Na_3Ni_{1.5}TeO_6$, $\rm Na_3Ni_2SbO_6$ and $\rm Na_3Cu_2SbO_6$)\cite{Vallee2019, berthelot2021stacking, grundish2020structural, Kuo2012, Itoh2015, Aguesse2016, hempel2023dynamics, Yuan2014, kikuchi2022field} dominating. NMR studies on other nuclides such as ${\rm ^{39}}$K NMR on potassium-based honeycomb layered tellurates may yield further insight on the local environment of large radii atoms, extending those of Na. Moreover, a combination of ${\rm ^{125}}$Te and ${\rm ^{23}}$Na MAS NMR has been utilised to attain local structural information of Te and Na in $\rm Na_2Zn_2TeO_6$, in addition to availing information local structural changes induced upon partial substitution of Zn with Ga in $\rm Na_{2-{\it x}}Zn_{2-{\it x}}Ga_{\it x}TeO_6$ ({\it x} = 0.05, 0.10, 0.15, 0.20).\cite {hempel2022, hempel2023dynamics}

\subsubsection{Electron spin resonance (ESR) spectroscopy}
Electron spin resonance spectroscopy (ESR spectroscopy) or electron paramagnetic resonance spectroscopy (EPR spectroscopy) is a non-destructive, non-invasive, accurate and highly sensitive analytical technique for detecting and characterising paramagnetic elements ($i.e.$ elements that have unpaired electrons). Although the basic mechanisms of this technique are akin to those of NMR, EPR utilises electron spins for characterisation instead of the nuclide spins employed by NMR. In ESR, a material is irradiated with microwave beams which continuously sweep the field till resonance ensues. Resonance occurs at a certain field, which is contingent on the metal ligands, the geometry surrounding the metal centre, and the number of unpaired electrons present. ESR measurements are mostly used as complementary techniques alongside other methodologies suchlike magnetic susceptibility measurements.

In many ways, ESR spectroscopy and NMR spectroscopy are similar in that they both probe into the interactions between electromagnetic radiation and the magnetic moments of particles. In fact, the two spectroscopy techniques can even be combined to form a distinctive spectroscopic technique: electron-nuclear double resonance (ENDOR) spectroscopy. Despite these similarities, a number of differences distinguish the two spectroscopies: (i) ESR relies on the interactions between an external magnetic field and unpaired electrons present in any system it is localised to, unlike NMR which revolves around the nuclei of individual atoms; (ii) ESR is usually performed utilising microwaves in the radio frequencies ranging from 3 to 400 GHz, whereas the electromagnetic radiation employed in NMR is usually confined in the 300-1000 MHz range.; (iii) ESR measurements typically employ constant frequencies with varying magnetic field intensities in contrast with NMR spectroscopic protocols which apply constant magnetic field under varying radio frequencies; (iv) ESR measurements often should be performed at extremely low temperatures (usually below 10 K), owing to the short relaxation times of electron spins compared to nuclei. Thus, ESR measurements  necessitate the utilisation of liquid helium as a coolant; and (v) owing to the higher electromagnetic radiation frequency utilised in ESR compared to NMR, ESR spectroscopy is intrinsically about a thousand times more sensitive than NMR spectroscopy.

As mentioned above, ESR spectroscopy provides unfettered access to information pertinent to unpaired electrons {\it exempli gratia }, their quantities, type, nature, surrounding environment, and behaviour. Indeed, measurements performed on $\rm Li_4FeSbO_6$,\cite{Zvereva2013} $\rm Li_4CrSbO_6$,\cite{mandujano2021} $\rm Li_4CrTeO_6$,\cite{mandujano2021} $\rm Li_3Cu_2SbO_6$,\cite{Koo2016} $\rm Li_3Ni_2SbO_6$\cite{Kurbakov2017} and $\rm Ag_3Co_2SbO_6$,\cite{Zvereva2016} have been instrumental in augmenting the current understanding about the magnetic spin dynamics and local coordination of atoms in the pnictogen and chalcogen class of honeycomb layered oxides. To analyse the information acquired through ESR, the following parameters are typically considered: (i) the `g-factor' values which reflect the orbit level occupied by the electron, (ii) the exchange interactions, which reflect the exchanges between electrons and (iii) line width (related to the transverse relaxation time).

\begin{figure*}[!t]
\centering
  \includegraphics[width=0.8\columnwidth]{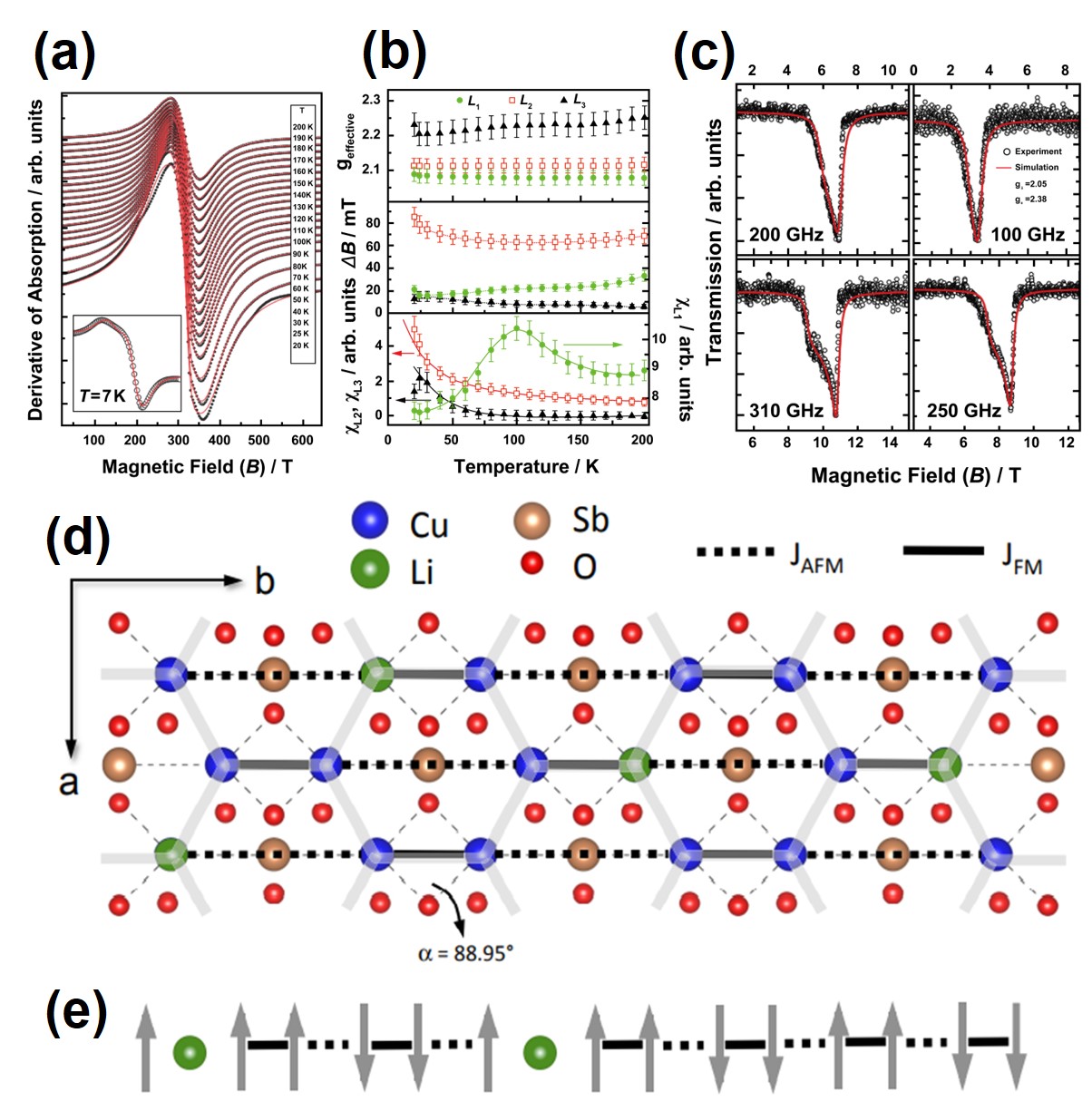}
  \caption{Electron Paramagnetic Resonance Spectroscopy (EPR) analysis of $\rm Li_3Cu_2SbO_6$. (a) Temperature evolution of the EPR spectra. The inset presents a spectrum collected at 7 K. Experimental data are shown in dots whereas the red lines are based on simulations. (b) The temperature dependencies of the effective g-factor (a), the EPR linewidth (b) and the integral EPR intensity (c) for the three resolved components ${\rm {\it L}_1}$, ${\rm {\it L}_2}$ and ${\rm {\it L}_3}$ of the EPR spectra. Both the effective g-factors and the EPR linewidths only weakly depend on temperature down to around 50 K. Below 50 K, the absorption line broadens and gradually changes into an anisotropic powder pattern, which is typical for a material sample with magnetic ${\rm Cu^{2+}}$ ions. (c) High-frequency EPR spectra obtained under various frequencies at 3.5 K. Simulations of the spectra are represented by the red lines with effective g-factors in good agreement with experiment. (d) Rendition of the crystal structure of $\rm Li_3Cu_2SbO_6$ with constituent atoms aligned along the $ab$-plane. Atomic positions were taken from the structural refinement whilst atomic radii are not to scale. The grey hexagon illustrates the underlying honeycomb structure. In order to illustrate inter-site disorder, Li-ions acting as non-magnetic defects are distributed by hand. Solid and dashed lines indicate the dominating magnetic exchange interactions. (e) Minimal magnetic model of the structure shown in (f) showing the AFM--AFM alternating chains. The dotted line is AFM whilst the solid line is FM. Figures reproduced with permission.\cite{Koo2016} Copyright 2016 Physical Society of Japan.}
  \label{Figure_39}
\end{figure*}

To elaborate on such mechanisms, \textbf{Figure \ref{Figure_39}a} shows evolution of the EPR spectra of $\rm Li_3Cu_2SbO_6$ taken at various temperatures.\cite{Koo2016} As shown, the spectra are noted to broaden with increasing temperatures. Note that a sum of three Lorentzian functions associated with three resonance modes (denoted as ${\rm {\it L}_1}$, ${\rm {\it L}_2}$ and ${\rm {\it L}_3}$) was required to accurately describe the line shape in the entire temperature regime. In general, two circular components of the excitant linearly polarised microwave field are considered to fit the experimental ESR spectra owing to the width of the resonance signal. As shown by the red solid lines in \textbf{Figure \ref{Figure_39}a}, the fitted curves appear to be in congruence with the experimental data. \textbf{Figure \ref{Figure_39}b} shows the temperature dependencies of the integral ESR intensity, the ESR linewidth and the effective g-factors of the three resolved components ${\rm {\it L}_1}$, ${\rm {\it L}_2}$ and ${\rm {\it L}_3}$ as derived from the fitting. Only both the effective g-factors and the ESR linewidths weakly depend on temperature down to around 50 K, below which broadening of the absorption line occurs and gradually changes into an asymmetric pattern. This behaviour is typically observed for material samples possessing magnetic ${\rm Cu^{2+}}$ ions (as shown in the inset of \textbf{Figure \ref{Figure_39}a}).

The corresponding g-tensor values were found to be in good agreement with the high-frequency ESR data. To explicitly visualise the difference in the magnetic response for the resolved components, the integral intensity of the ESR spectra (which scales with the number of magnetic spins) was plotted (\textbf{Figure \ref{Figure_39}b}, bottom figure) by double integration of the absorption derivative curves. Both integral ESR intensities $\rm \chi_{L2}$ and $\rm \chi_{L3}$ for the ${\rm {\it L}_2}$ and ${\rm {\it L}_3}$ resonance modes exhibit Curie–Weiss-like behaviour with a vanishing or small Weiss temperature: in other words, $\rm \chi_{L2}$ and $\rm \chi_{L3}$ are associated with quasi-free or weakly interacting spins. Conversely, $\rm \chi_{L1}$ manifests a broad maximum at 100 K resembling a spin-gap behaviour. The ESR parameters for ${\rm {\it L}_2}$ and ${\rm {\it L}_3}$ are, in contrast, observed to increase with decreasing temperature. The similar temperature dependence was concluded to indicate a common origin of both signals: $i.e.$ anisotropic powder signal arising from quasi-free spin contribution. Additionally, representative high-frequency ESR spectra attained at various frequencies at 3.5 K (shown in \textbf{Figure \ref{Figure_39}c}) make it possible to further resolve the magnetic anisotropy innate in $\rm Li_3Cu_2SbO_6$. ESR measurements in conjunction with static magnetic susceptibility measurements revealed a notable number of non-magnetic Li-defects in the Cu-chains of $\rm Li_3Cu_2SbO_6$ (\textbf{Figure \ref{Figure_39}d}), implying a Haldane spin chain model (\textbf{Figure \ref{Figure_39}e}) wherein the edge-states of the finite Haldane chains contribute the quasi-free moments that were experimentally observed.

\begin{figure*}[!t]
\centering
  \includegraphics[width=0.8\columnwidth]{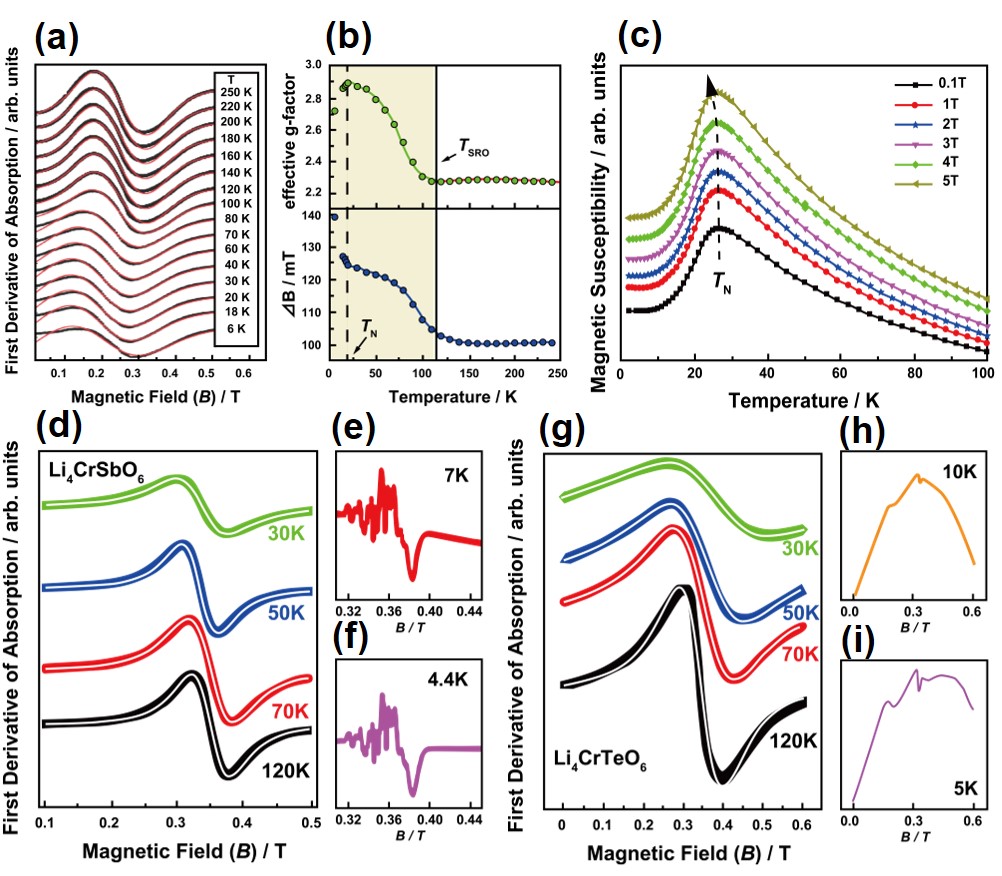}
  \caption{Electronic paramagnetic resonance measurements of $\rm Ag_3Co_2SbO_6$, $\rm Li_4CrSbO_6$ and $\rm Li_4CrTeO_6$. (a) Temperature evolution of the first derivative EPR absorption spectra for $\rm Ag_3Co_2SbO_6$. Experimental data are shown by the black points, whilst the Gaussian fitting made is shown in red lines. EPR spectra in the paramagnetic phase temperature regimes ($i.e.$, temperature T > T${\rm_{\it N}}$ (T: Néel temperature)) display single broad Gaussian shape lines ascribable to ${\rm Co^{2+}}$ ions in octahedral coordination. The shape of the spectra changes noticeably with decreasing temperature. (b) Temperature dependence of the effective g-factor (a dimensionless quantity that characterises the magnetic moment and angular momentum of an atom) and the EPR linewidth for $\rm Ag_3Co_2SbO_6$. The Néel temperature (T${\rm_{\it N}}$) marks the temperature above which antiferromagnetic-to-paramagnetic transition occurs. T${\rm_{\it SRO}}$ denotes the temperature at which short range ordering occurs within the material ($viz$., atomic pairs within the material order over distances comparable to interatomic distances). (c) Temperature dependence of the magnetisation recorded at various external magnetic fields for $\rm Ag_3Co_2SbO_6$. The dashed arrow indicates the shift of T${\rm_{\it N}}$ upon variation of the magnetic field. The magnetisation curves have a slight upward curvature suggesting the presence of a magnetic field induced spin-flop-type transition. (d) EPR signals of $\rm Li_{3.88}Cr_{1.12}SbO_6$ taken at 30, 50, 70 and 120 K. (e) EPR signals of $\rm Li_{3.88}Cr_{1.12}SbO_6$ taken at 7 K. (f) EPR signals of $\rm Li_{3.88}Cr_{1.12}SbO_6$ taken at 4.4 K. (g) EPR signals of $\rm Li_{4.47}Cr_{0.53}TeO_6$ taken at 30, 50, 70 and 120 K. (h) EPR signals of $\rm Li_{4.47}Cr_{0.53}TeO_6$ taken at 10 K. (i) EPR signals of $\rm Li_{4.47}Cr_{0.53}TeO_6$ taken at 5 K. The high temperature signals have a definite profile and are shown fitted by broad Lorentzian curves (white solid lines). Low temperature signals of both the compounds suggest an inhomogeneous magnetic state which qualitatively supports conclusions drawn from neutron diffraction and specific heat measurements. The high temperature EPR signals exhibited in both compounds are indicative of a paramagnetic state. Figures (a,b,c) and Figures (d,e,f,g,h,i) reproduced with permission.\cite{Zvereva2016, mandujano2021} Copyright 2016 Royal Society of Chemistry and 2021 Institute of Physics, respectively.}
  \label{Figure_40}
\end{figure*}

EPR measurements have also been performed to elucidate the magnetic properties of silver Delafossites such as $\rm Ag_3Co_2SbO_6$\cite{Zvereva2016}. \textbf{Figure \ref{Figure_40}a} shows the temperature evolution of the EPR spectra taken for $\rm Ag_3Co_2SbO_6$. EPR spectra in the paramagnetic phase temperature regimes ($i.e.$, temperature T > T${\rm_{\it N}}$ (T: Néel temperature)) display single broad Gaussian shape lines attributable to ${\rm Co^{2+}}$ ions in octahedral coordination geometry. The shape of the spectra changes noticeably with decreasing temperature. Upon cooling the sample, the amplitude of EPR signal increases monotonously. The signal then weakens in amplitude and eventually degrades in the vicinity of T${\rm_{\it N}}$. This is indicative of the opening of a spin gap for the resonance excitation owing to the onset of long-range magnetic order at low temperatures. At high temperatures (beyond 150 K), it was established that the absorption spectra is characterised by essentially temperature-independent values of the linewidth ${\rm \Delta}$B and the effective g-factor (\textbf{Figure \ref{Figure_40}b}). Thereafter, a visible shift of the resonance field along with broadening of the entire spectrum occur upon lowering of the temperature, which is indicative of a large pre-ordering magnetic effect in $\rm Ag_3Co_2SbO_6$ in the broad temperature range of 25–150 K. This phenomenon has been rationalised in terms of a fluctuating internal field which adds to the applied field modifying the resonance conditions, whilst concomitantly slowing up the fluctuation rate of this field can cause continuous broadening of the resonance lines as the temperature decreases. The origin of such fluctuating fields has been traced to the significant short-range correlations, which exist in the system at temperatures typically higher than the long-range order temperature.

On lowering the temperature, EPR spectra usually tends to monotonically broaden as spin correlations develop. Nonetheless, the linewidth (${\rm \Delta}$B) passes through a maximum and thereafter very weakly varies upon approaching T${\rm_{\it N}}$. This behaviour has been ascribed to the depletion of the spin fluctuation density and the saturation of the spin correlation length. The occurrence of strong short-range fluctuations was also noted from the static magnetic susceptibility data exhibiting clear deviation from the Curie–Weiss law around below 150 K (\textbf{Figure \ref{Figure_40}c}).

EPR spectroscopy has also been used to probe the local internal fields, spin relaxations and spin dynamics of honeycomb layered tellurates and antimonates suchlike $\rm Li_4CrTeO_6$\cite{mandujano2021} and $\rm Li_4CrSbO_6$\cite{mandujano2021}. The EPR signals of $\rm Li_{3.88}Cr_{1.12}SbO_6$ (\textbf{Figure \ref{Figure_40}d}) and $\rm Li_{4.47}Cr_{0.53}TeO_6$ (\textbf{Figure \ref{Figure_40}g}) are indicative of a paramagnetism at high temperatures. However, multi-line features are apparent as the temperature is lowered below about 10 K. Particularly, these multi-line features are seen at 4.4 K and 7 K in the case of $\rm Li_{3.88}Cr_{1.12}SbO_6$ (\textbf{Figures \ref{Figure_40}e} and \textbf{\ref{Figure_40}f}), and at 10 K and 5 K for $\rm Li_{4.47}Cr_{0.53}TeO_6$ (\textbf{Figures \ref{Figure_40}h} and \textbf{\ref{Figure_40}i}). In the paramagnetic regime (10 K to 120 K), the EPR signals are narrow and relatively symmetric. Based on EPR measurements complemented with neutron diffraction and specific heat studies, the absence of long-range ordered magnetism in $\rm Li_{3.88}Cr_{1.12}SbO_6$ and $\rm Li_{4.47}Cr_{0.53}TeO_6$ was confirmed and instead, a short-range ordered magnetism entailing Cr clusters was proposed.

In summary, although there are still few reports, EPR/ESR spectroscopy has been used as a complementary technique to explore the local internal fields, spin relaxations, spin dynamics, and defect structures of pnictogen- and chalcogen-based honeycomb layered oxides.\cite{mandujano2021, Zvereva2013, Koo2016, Kurbakov2017, Zvereva2016} 

\subsubsection{M\"{o}ssbauer spectroscopy}
M\"{o}ssbauer effect refers to the recoil-free resonant absorption and emission of gamma radiation by radioactive nuclei bound in a solid. The M\"{o}ssbauer effect results in a spectroscopy with the resolution of the lifetime uncertainty of the excited nuclear state. Typically, this resolution is adequate to analyse the so-called hyperfine energy-splittings of the nucleus caused by electronic fields, rendering M\"{o}ssbauer spectroscopy suitable for providing information about the atomic local environment within a material. M\"{o}ssbauer spectroscopy is seen as a complementary characterisation technique for determining the atomic oxidation state, the chemical environment and symmetry or the magnetic properties of atoms.\cite{dyar2006, gutlich2010}

\begin{figure*}[!t]
\centering
  \includegraphics[width=0.8\columnwidth]{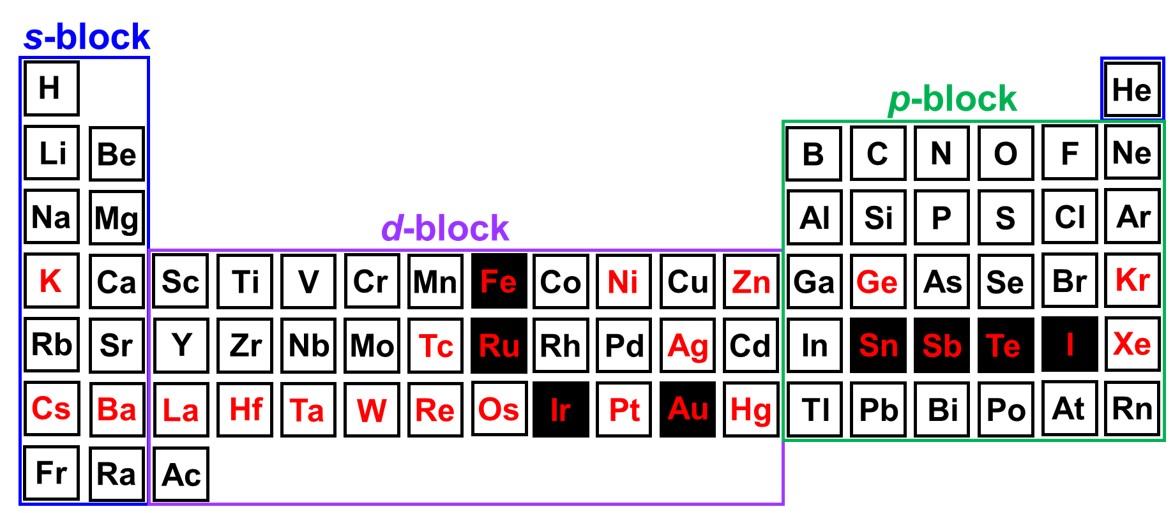}
  \caption{Periodic table of elements with M\"{o}ssbauer isotopes. The M\"{o}ssbauer isotopes are shown in red fonts. Those that are commonly used are shaded in black. Based on several criteria (handling, easy accessibility, transition energy, and suitable lifetime of nuclear excited state), a limited number of elements can be studied by M\"{o}ssbauer spectroscopy.}  \label{Figure_41}
\end{figure*}

As illustrated by the periodic table in \textbf{Figure \ref{Figure_41}}, there exists a number of isotopes with nuclear transitions that can be examined using M\"{o}ssbauer spectroscopy, but most of them only exhibit this phenomenon at low temperatures. In fact, only a few nuclei exhibit the M\"{o}ssbauer effect (for instance, ${\rm ^{197}}$Au, ${\rm ^{153}}$Eu, ${\rm ^{121}}$Sb, ${\rm ^{119}}$Sn, ${\rm ^{57}}$Fe) at room temperatures, thus inhibiting the utility of this technique. At ambient temperatures, only the 23.87 keV of ${\rm ^{119}}$Sn and 14.4 keV transition of ${\rm ^{57}}$Fe have sufficient M\"{o}ssbauer effect probability—the reason why a majority of applications are focussed on ${\rm ^{57}}$Fe.\cite{schunemann2000, gao2007} As such, amongst the compositions encompassed in the pnictogen and chalcogen class of honeycomb layered oxides, the structural characterisation using M\"{o}ssbauer spectroscopy can only be employed on iron-containing honeycomb layered oxides.

\begin{figure*}[!t]
\centering
  \includegraphics[width=0.9\columnwidth]{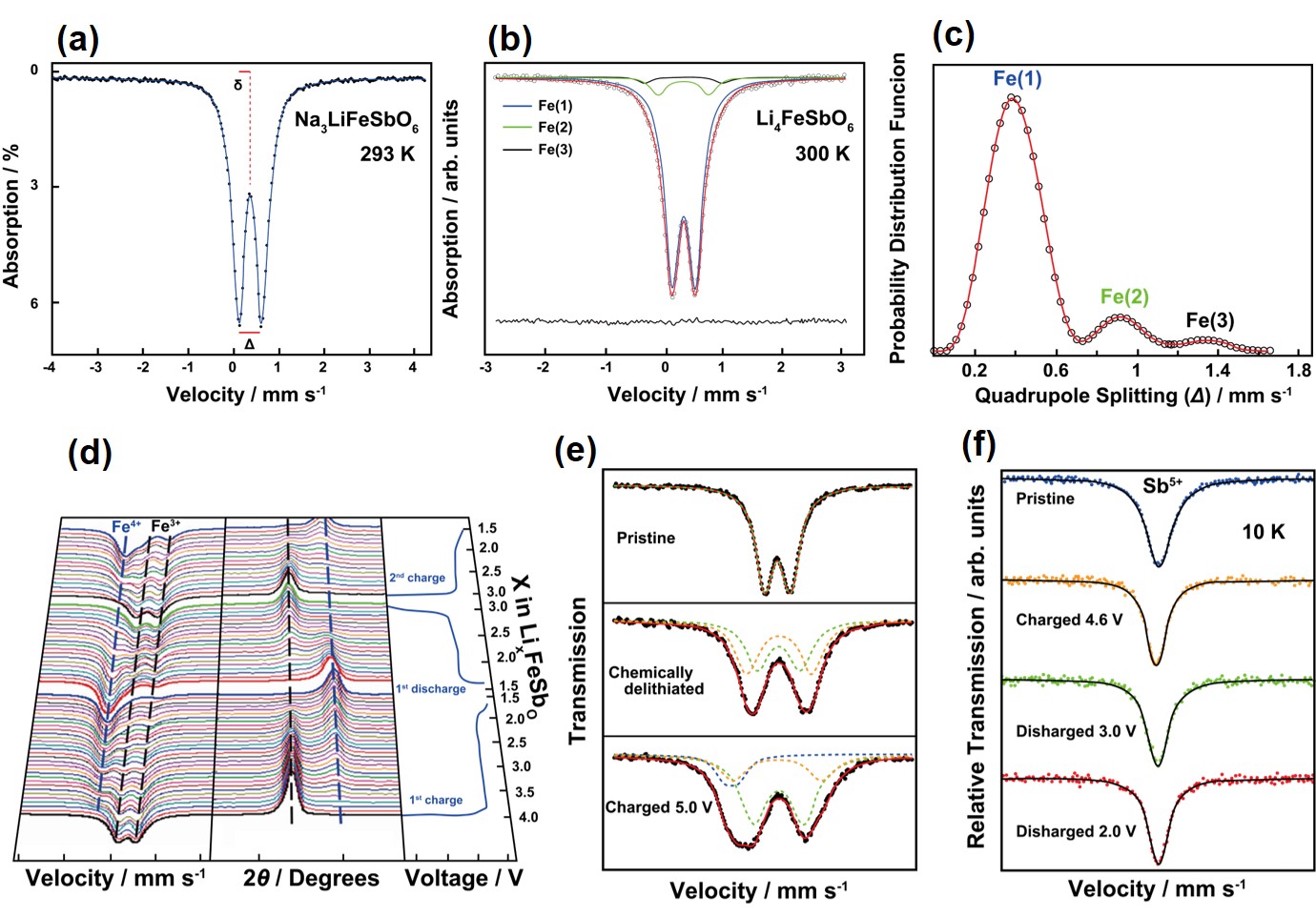}
  \caption{M\"{o}ssbauer spectra of representative honeycomb layered tellurates and antimonates. (a) ${\rm ^{57}}$Fe M\"{o}ssbauer spectra for $\rm Na_3LiFeSbO_6$ (black dotted line) measured at room temperature and the calculated fit (blue solid line) which indicates one paramagnetic doublet. (b) Modelling reconstruction of ${\rm ^{57}}$Fe M\"{o}ssbauer spectra for the $\rm Li_4FeSbO_6$ sample recorded at temperatures 300 K. The experimental spectrum is a superposition of three quadrupole doublets Fe(1), Fe(2) and Fe(3) with the same values of linewidth ($\rm \Gamma_{1}$ = $\rm \Gamma_{2}$ = $\rm \Gamma_{3}$) and isomer shifts ($\rm \delta_{1}$ = $\rm \delta_{2}$ = $\rm \delta_{3}$), but significantly different quadrupole splitting ($\rm \Delta_{i}$). (See \textbf{ Table \ref{Table_11}}) (c) The ${\rm ^{57}}$Fe M\"{o}ssbauer spectrum for the $\rm Li_4FeSbO_6$ sample recorded at 300 K. The resulting function has three peaks, indicating that the iron atoms occupy three non-equivalent crystallographic positions with the average quadrupole splittings $\langle \rm \Delta_1 \rangle$ $\approx$ 0.48 mm s${\rm ^{-1}}$  
  $\langle \rm \Delta_2 \rangle$ $\approx$ 0.95 mm s${\rm ^{-1}}$ and 
  $\langle \rm \Delta_3 \rangle$ $\approx$ 1.37 mm s${\rm ^{-1}}$  respectively. (d) {\it In situ} M\"{o}ssbauer and XRD data obtained for a $\rm Li_4FeSbO_6$ cycled at C/40. It shows combined {\it in situ} ${\rm ^{57}}$Fe M\"{o}ssbauer and XRD data for a sample charged to 4.5 V and discharged to 3.0 V at a current density equivalent to C/40. The ${\rm ^{57}}$Fe M\"{o}ssbauer data shows that the ${\rm Fe^{3+}}$ in the pristine material is converted to an oxidised Fe, labeled ${\rm Fe^{4+}}$. (e) ${\rm ^{57}}$Fe M\"{o}ssbauer data of highly charged samples, along with fits which consists of two ${\rm Fe^{3+}}$ sites in orange and green, and one ${\rm Fe^{4+}}$ site in blue. It shows the Fe spectra for both the samples charged to 5.0 V and the chemically de-lithiated ones. The chemically delithiated sample is entirely made up of Fe in the 3+ state. Both sites were noted to have very large quadrupole splitting indicative of very distorted environments. (f) ${\rm ^{121}}$Sb M\"{o}ssbauer spectra recorded at 10 K for samples at various states of charge, along with the fits using a single ${\rm Sb^{5+}}$ site. Results of the fits are summarised in \textbf{ Table \ref{Table_12}}. Figure (a), Figures (b, c) and Figures (d, e, f) reproduced with permission.\cite{Schmidt2014, Zvereva2013, McCalla2015a} Copyright 2014 Elsevier, 2013 Royal Society of Chemistry and 2015 American Chemical Society, respectively.}
  \label{Figure_42}
\end{figure*}

Similar to NMR spectroscopy, M\"{o}ssbauer spectroscopy probes minute changes in the energy levels of an atomic nucleus in response to its environment. Usually, three types of hyperfine interactions (nuclear interactions) can be observed: (i) hyperfine or magnetic splitting, also referred to as the Zeeman effect; (ii) the electric dipole interaction that causes quadrupole splitting ($\rm \Delta $); and (iii) the electric monopole interaction that gives rise to a chemical shift, also referred to as isomer (isomeric) shift ($\rm \delta $). The isomeric shift ($\rm \delta $) refers to the shift of the centroid of the spectrum from zero velocity, and is given relative to either the source or some standard material such as metallic iron in the case of ${\rm ^{57}}$Fe. In the case of ${\rm ^{57}}$Fe, the quadrupole splitting ($\rm \Delta $) is discerned if there is a separation between the two lines of a ${\rm ^{57}}$Fe doublet. The ${\rm ^{57}}$Fe M\"{o}ssbauer spectra of $\rm Na_3LiFeSbO_6$\cite{Schmidt2014} measured at room temperature (\textbf{Figure \ref{Figure_42}a}) depicts a single paramagnetic doublet whose quadrupole splitting and isomer shift are highlighted by the red lines. Both quadrupole splitting and isomeric shift are customarily represented in terms of the source velocity in millimetres per second (mm s${\rm ^{-1}}$). Quadrupole splitting is used to measure ${\rm Fe^{3+}}$ site distortion, whereas the isomeric shifts are generally associated with the oxidation state of iron and are used to provide information on iron coordination. Additionally, the shapes and widths of the lines can also be used to provide additional structural information of the material under investigation. The ideal M\"{o}ssbauer line shape is the Lorentzian, which is often used to computer-fit experimental data. However, factors such as variations in the local environments or parameter fluctuations, may cause some deviations from ideal Lorentzian shape. In such cases, the data may necessitate the use of other functions, for instance, the Lorentzian, Voigtian (a convolution of Gaussian and Lorentzian functions) or distributions of Lorentzian functions for fitting. As shown in \textbf{Figure \ref{Figure_42}a}, the isomeric shift ($\rm \delta $) of $\rm Na_3LiFeSbO_6$ is attributed to high-spin ${\rm Fe^{3+}}$ in an octahedral coordination. Therefore, this preliminary fit using Lorentzian profile lines allows the characterisation of one doublet assigned to ${\rm ^{57}}$Fe that is in accordance with the expected crystallographic site for $\rm Na_3LiFeSbO_6$.\cite{Schmidt2014} Further analyses of the linewidth revealed the existence of quadrupole splitting distribution which was associated to a local cationic disorder around Fe within the crystal framework of $\rm Na_3LiFeSbO_6$. The mean value of the quadrupolar splitting is rather high, indicative of a slight deformation of the $\rm FeO_6$ octahedron emanating from the ionic radii difference of its neighbouring atoms: ${\rm Sb^{5+}}$ and ${\rm Li^{+}}$.

In an attempt to elucidate the local structural order of Fe atoms in a honeycomb layered oxide, a previous study reported the ${\rm ^{57}}$Fe M\"{o}ssbauer spectroscopic measurements of $\rm Li_4FeSbO_6$\cite{McCalla2015a} shown in \textbf{Figure \ref{Figure_42}b}. Here, the fitted spectrum manifests the existence of an asymmetric paramagnetic doublet with broadened components. The distribution function $p$($\rm \Delta $) of the quadrupole splittings ($\rm \Delta $), shown in \textbf{Figure \ref{Figure_42}c}, has been restored to select the model for interpretation of the spectra. The resulting function $p$($\rm \Delta $) (\textbf{Figure \ref{Figure_42}c}) exhibits three peaks, indicating that the iron atoms occupy three non-equivalent crystallographic positions with average quadrupole splittings that have been listed in \textbf{ Table \ref{Table_11}}. Based on the results of the $p$($\rm \Delta $) profile analysis (\textbf{Figure \ref{Figure_42}c}), the experimental spectrum of $\rm Li_4FeSbO_6$ has been described as a superposition of three quadrupole doublets Fe(1), Fe(2) and Fe(3) with the same values of linewidths ($\rm \Gamma_{1}$ = $\rm \Gamma_{2}$ = $\rm \Gamma_{3}$) and isomeric shifts ($\rm \delta_{1}$ = $\rm \delta_{2}$ = $\rm \delta_{3}$), but significantly different quadrupole splittings ($\rm \Delta_{i}$) as shown in \textbf{ Table \ref{Table_11}}. The isomer shifts of all resolved quadrupole doublets correspond to the high-spin ${\rm Fe^{3+}}$ ($S$ = 5/2, 3${\rm {\it d}^5}$) cations in the $\rm FeO_6$ sites coordinated octahedrally. However, the noticeably different quadrupole splitting parameters $\rm \Delta_{i}$ have been reported to indicate that the ${\rm Fe^{3+}}$ cations in the $\rm Li_4FeSbO_6$ structure are distributed over positions with different local symmetry of anionic polyhedra.

\begin{table*}
\caption{Hyperfine parameters of ${\rm ^{57}}$Fe M\"{o}ssbauer spectra of $\rm Li_4FeSbO_6$. \cite{Zvereva2013}}\label{Table_11}
\begin{center}
\scalebox{1.2}{
\begin{tabular}{cccccc} 
\hline
\textbf{Temperature (K)} & \textbf{Position} & \textbf{$\rm \delta $} & \textbf{$\rm \Delta $} & \textbf{$\rm \Gamma $} & \textbf{$I$}\\ 
 & & \textbf{(mm s${\rm ^{-1}}$)} & \textbf{(mm s${\rm ^{-1}}$)} & \textbf{(mm s${\rm ^{-1}}$)} & \textbf{(mm s${\rm ^{-1}}$)}\\ 
\hline\hline
300 & Fe (1) & 0.33 ± 0.01 & 0.38 ± 0.01 & 0.31 ± 0.01 & 89.1 ± 0.9\\
300 & Fe (2) & 0.33 & 0.86 ± 0.02 & 0.31 & 8.5 ± 0.6\\
300 & Fe (3) & 0.33 & 1.30 ± 0.02 & 0.31 & 2.4 ± 0.4\\
\hline
\end{tabular}}
\end{center}
\end{table*}

${\rm ^{57}}$Fe M\"{o}ssbauer spectroscopic measurements can be complemented with XRD techniques to elucidate the electrochemical reactions occurring during the charging and discharging processes of Fe-based electrode materials. This synergism has been employed to investigate, in real-time, the charge-compensation process during the two-phase electrochemical reaction taking place in a $\rm Li_4FeSbO_6$ electrode (\textbf{Figure \ref{Figure_42}d}). \textbf{Figure \ref{Figure_42}d} shows the combined {\it in situ} ${\rm ^{57}}$Fe M\"{o}ssbauer and XRD data for a $\rm Li_4FeSbO_6$ electrode charged to 4.5 V at C/40 and discharged to 3.0 V. The change in the XRD peaks is indicative of a two-phase transition. Concomitant with the reversible two-phase reaction, the {\it in situ} ${\rm ^{57}}$Fe M\"{o}ssbauer data shows that the ${\rm Fe^{3+}}$ in the pristine material is converted to an oxidised Fe, labelled as ${\rm Fe^{4+}}$ based on the magnetic susceptibility data. The oxidised iron is converted back to ${\rm Fe^{3+}}$, upon discharge, a process which is completed at 3.0 V after the reinsertion of Li. The oxidation state of Fe was further validated not only through electrochemical charging of $\rm Li_4FeSbO_6$, but also through chemical delithiation of $\rm Li_4FeSbO_6$, as shown in \textbf{Figure \ref{Figure_42}e}. 

Additionally, ${\rm ^{121}}$Sb M\"{o}ssbauer spectroscopic measurements have also been performed on $\rm Li_4FeSbO_6$ to assess the changes in the oxidation state upon charging and discharging processes.\cite{McCalla2015a} \textbf{Figure \ref{Figure_42}f} shows the ${\rm ^{121}}$Sb M\"{o}ssbauer data obtained from a pristine $\rm Li_4FeSbO_6$ electrode and electrodes charged to 4.6 V and separately discharged to 3.0 and 2.0 V. \textbf{ Table \ref{Table_12}} shows the corresponding fitted parameters. Although the oxidation state of the constituent antimony does not change, reports have shown the corresponding isomer shifts are influenced by the adjacent oxidised species, resulting in lower values.

\begin{table*}
\caption{Hyperfine parameters of ${\rm ^{121}}$Sb M\"{o}ssbauer spectra of $\rm Li_4FeSbO_6$ samples taken at 10 K. The table shows the ${\rm ^{121}}$Sb M\"{o}ssbauer data for each of the pristine, charged and discharged  electrodes.\cite{McCalla2015a} Sb is in the pentavalent state (${\rm Sb^{5+}}$) during both charging and discharging of $\rm Li_4FeSbO_6$, as affirmed by the isomer shift values. For comparison, the isomer shift values for ${\rm Sb^{3+}}$ are approximately --15 mm s${\rm ^{-1}}$.\cite {lippens2000mossbauer}}\label{Table_12} 
\begin{center}
\scalebox{1}{
\begin{tabular}{ccc} 
\hline
\textbf{Sample} & \textbf{Isomer Shift} & \textbf{Line Width}\\ 

 & \textbf{(mm s${\rm ^{-1}}$)} & \textbf{(mm s${\rm ^{-1}}$)}\\ 
\hline\hline
Pristine & 0.17 ± 0.01 & 4.31 ± 0.03\\
Charge 4.6 V & --0.24 ± 0.01 & 3.28 ± 0.04\\
Discharge 3.0 V & --0.03 ± 0.01 & 3.62 ± 0.03\\
Discharge 2.0 V & 0.08 ± 0.01 & 3.37 ± 0.03\\
\hline
\end{tabular}}
\end{center}
\end{table*}

We emphatically note that reports pertaining M\"{o}ssbauer spectroscopic measurements on pnictogen- and chalcogen-based honeycomb layered oxides remain limited. However, considering that reports on M\"{o}ssbauer spectroscopy on ${\rm ^{125}}$Te,\cite{bresser1993} ${\rm ^{61}}$Ni,\cite{masuda2016} ${\rm ^{67}}$Zn\cite{potzel1993} already exist, we postulate that there is plenty of room to apply M\"{o}ssbauer spectroscopy in the exploration of the diverse local structural properties of honeycomb layered oxides.

\subsubsection{Raman spectroscopy}
Raman spectroscopy is a characterisation technique that observes the rotational, vibrational, and other low-frequency modes in a material. The technique relies on the Raman scattering, or inelasic scattering, of monochromatic light, typically from a laser in the near ultraviolet, near infrared, or visible range. The laser light interacts with phonons, molecular vibrations, or other excitations in the material, resulting in the energy of the laser photons being shifted down or up. The shift in energy avails information relating to the vibrational modes in the material. 

Particularly in the study and characterisation of 2D systems, Raman spectroscopy has emerged as an invaluable technique for its role in unravelling crucial information, suchlike main crystallographic orientations, phase changes, stacking order, number of layers, electronic and phononic properties, defect density and type, functionalisation, mechanical stress, environmental sensing, thermal conductivity, amongst others. Given that Raman spectroscopy utilises light near the visible spectral range, it does not require intricate sample preparation processes and is a non-destructive technique. For concision, we have provided the following references for readers interested in more information on the fundamental physics that govern the Raman effect, its merits, demerits and applications.\cite{zhang2016, graves1989, raman1928, raman1928a}

\begin{figure*}[!b]
\centering
  \includegraphics[width=0.8\columnwidth]{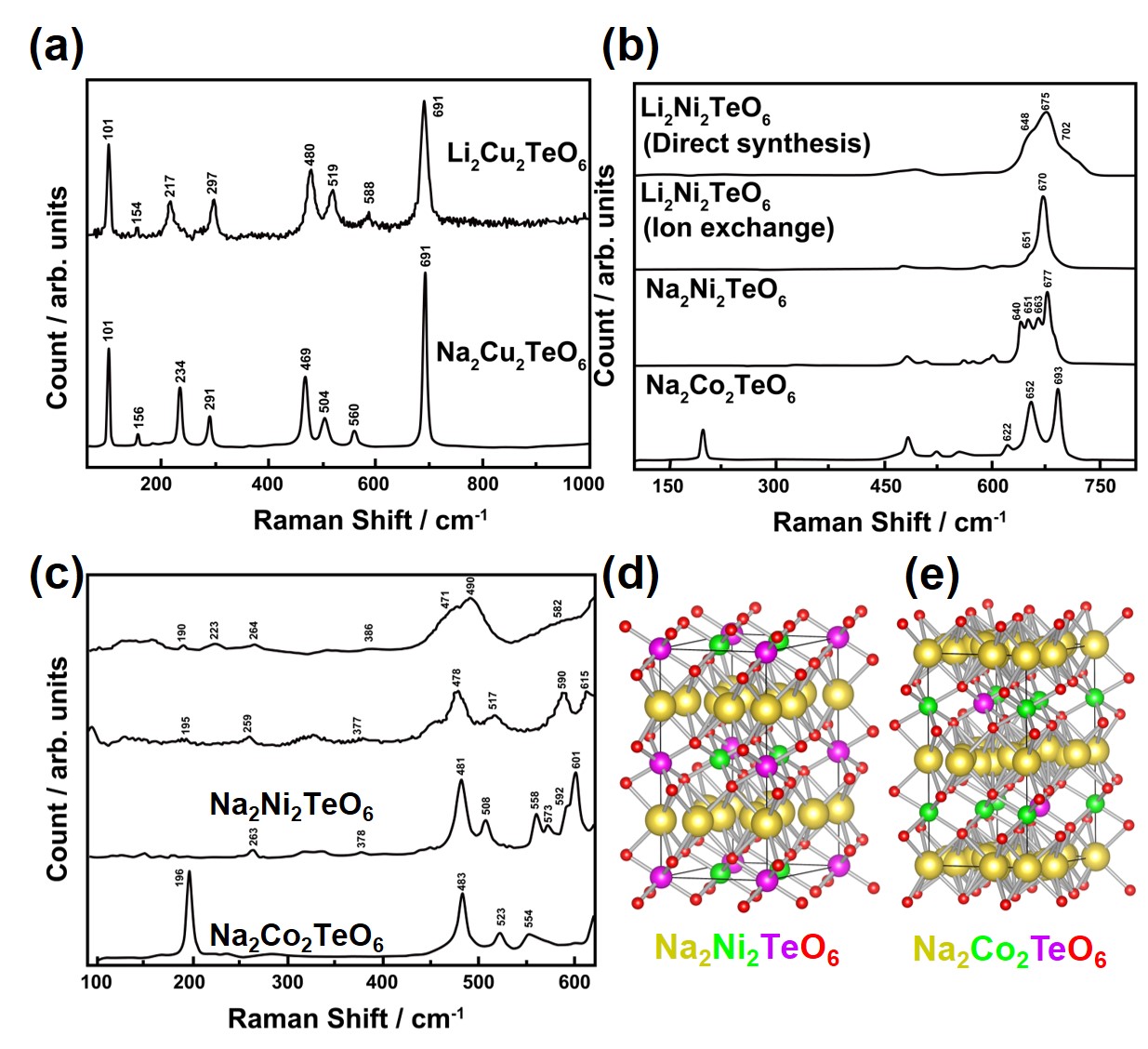}
  \caption{Raman spectra of honeycomb layered tellurates. (a) Comparison of the Raman spectra obtained from $\rm Li_2Cu_2TeO_6$ (prepared via topochemical ion-exchange) and $\rm Na_2Cu_2TeO_6$. Excluding the $\rm TeO_6$ octahedra antisymmetric stretching modes located at 691 cm${\rm ^{-1}}$ and the tilting vibrations of $\rm CuO_6$ centred at 101 cm${\rm ^{-1}}$, all other bonds exhibit Raman peak shift as a result of the larger atomic size of Na. Corresponding peaks analyses are shown in \textbf{ Table \ref{Table_13}}. (b) Raman spectra obtained from $\rm Li_2Ni_2TeO_6$ polytypes, $\rm Na_2Ni_2TeO_6$ and $\rm Na_2Co_2TeO_6$. The Raman spectra for $\rm Li_2Ni_2TeO_6$ (prepared directly via high temperature solid-state reaction) is significantly broadened owing to the presence of structural disorders. $\rm Na_2Ni_2TeO_6$ and $\rm Na_2Co_2TeO_6$ show distinctly varying Raman modes. (c) Enlarged region of the Raman spectra from 100 cm${\rm ^{-1}}$ to 600 cm${\rm ^{-1}}$ wavenumbers for (b). (d) Schematic of the $\rm Na_2Ni_2TeO_6$ crystal structure (crystallising in $P$${\rm 6_3}$/ {\it mmc} centrosymmetric hexagonal space group). Na atoms are in yellow; Ni in green; Te violet and O in red. (e) Schematic of the $\rm Na_2Co_2TeO_6$ crystal structure (crystallising in the non-centrosymmetric space group $P$${\rm 6_3}$22). Figures (a,b,c) reproduced with permission.\cite{Kumar2013} Copyright 2013 Royal Society of Chemistry.}
  \label{Figure_43}
\end{figure*}

Veritably, studies pertaining honeycomb layered tellurates have employed Raman spectroscopy to inspect the local coordination of atoms and bond types in the materials. For example, $\rm Li_2Cu_2TeO_6$\cite{Kumar2013} and $\rm Na_2Cu_2TeO_6$\cite{Kumar2013} prepared via the topochemical ion-exchange route were noted to display sharp and clear peaks attributed to the characteristic Raman bands arising from various stretching, vibration and bending modes of $\rm NaO_6$, $\rm CuO_6$ and $\rm TeO_6$ polyhedra (\textbf{Figure \ref{Figure_43}a}). In another study (\textbf{Figure \ref{Figure_43}b}), the broadening Raman bands visualised during the characterisation of $\rm Li_2Ni_2TeO_6$ synthesised through the high-temperature solid-state reaction route were efficacious in identifying the structural disorders in the heterostructures.\cite{Kumar2013} For clarity, \textbf{Figure \ref{Figure_43}c} shows the spectra displayed at an enlarged scale. As shown, the $\rm TeO_6$ octahedra show symmetric modes at around 560 cm${\rm ^{-1}}$ and 691 cm${\rm ^{-1}}$. Similar $\rm TeO_6$ bending modes are also exhibited at 504 cm${\rm ^{-1}}$ and 469 cm${\rm ^{-1}}$ in the $\rm Na_2Cu_2TeO_6$ spectrum and at 519 cm${\rm ^{-1}}$ and 480 cm${\rm ^{-1}}$ in $\rm Li_2Cu_2TeO_6$ spectrum. The Cu--O stretching and bending modes are centred around 291 cm${\rm ^{-1}}$ and 234 cm${\rm ^{-1}}$ in $\rm Na_2Cu_2TeO_6$, and 297 cm${\rm ^{-1}}$ and 217 cm${\rm ^{-1}}$ in the ordered $\rm Li_2Ni_2TeO_6$ sample prepared directly via topochemical ion-exchange route. At around 100 cm${\rm ^{-1}}$, vibrations ascribable to the tilting of $\rm CuO_6$ octahedra are also observed.

The distinct variations in the Raman modes obtained from $\rm Na_2Ni_2TeO_6$ and $\rm Na_2Co_2TeO_6$ (\textbf{ Table \ref{Table_13}}) are ascribed to the varying local site symmetry of $\rm TeO_6$ octahedra. As shown in \textbf{Figures \ref{Figure_43}d} and \textbf{\ref{Figure_43}e}, the $\rm Na_2Ni_2TeO_6$ and $\rm Na_2Co_2TeO_6$ samples appear to crystallise in the $P\rm 6_3$/{\it mmc} centrosymmetric hexagonal and the $P\rm 6_3 22$ non-centrosymmetric hexagonal space groups, respectively. 

According to Schoenflies notation, the local site symmetry of the $\rm TeO_6$ octahedra in the $\rm Na_2Co_2TeO_6$ and $\rm Na_2Ni_2TeO_6$ crystals are designated as ${\rm {\it D}_3}$ and ${\rm {\it D}_{3d}}$, respectively.\cite{Kumar2013} The symmetries are further confirmed by the corresponding symmetric stretching modes of $\rm TeO_6$. The Raman spectra of $\rm Li_2Ni_2TeO_6$ (prepared via topochemical ion-exchange route) also demonstrated the the Li--O stretching modes at 259 cm${\rm ^{-1}}$ and the O--Ni--O stretching vibration of $\rm NiO_6$ located at 590 cm${\rm ^{-1}}$. The disordered nature of $\rm Li_2Ni_2TeO_6$ (prepared directly via high-temperature solid-state reaction route) is further confirmed by the Raman spectra as visualised by the explicit broadness of the bands, corresponding to $\rm TeO_6$ symmetric stretching vibrations.

\begin{table*}
\caption{Raman modes for $\rm Li_2Cu_2TeO_6$, $\rm Li_2Ni_2TeO_6$ polytypes ((i) and (ii)), $\rm Na_2Cu_2TeO_6$ and $\rm Na_2Ni_2TeO_6$.\cite{Kumar2013} $\rm Li_2Ni_2TeO_6$ prepared via direct solid-state reaction at high temperatures is denoted as $\rm Li_2Ni_2TeO_6$ (i) whilst that prepared via topochemical ion-exchange at low temperatures is denoted as $\rm Li_2Ni_2TeO_6$ (ii).}\label{Table_13}
\begin{center}
\scalebox{0.8}{
\begin{tabular}{cccccl} 
\hline
\textbf{Wavenumber (cm${\rm ^{-1}}$)} & & & & & \\ 

 \textbf{$\rm Li_2Cu_2TeO_6$} & \textbf{$\rm Li_2Ni_2TeO_6$ (i)} & $\rm Li_2Ni_2TeO_6$ (ii) & $\rm Na_2Cu_2TeO_6$ & $\rm Na_2Ni_2TeO_6$ & Modes\\ 
\hline\hline
 & & 702 & & & $\rm TeO_6$ antisymmetric stretching modes\\
691 & 670 & 675 & 691 & 677 & $\rm TeO_6$ symmetric stretching modes\\
 & 651 & 648 & & 663, 651 & $\rm TeO_6$ bending modes\\
 & & & & 640 & Stretching mode of $\rm NiO_6$ octahedra\\
 & 615 & & & 601 & $\rm TeO_6$ symmetric stretching modes\\
 & 590 & 582 & & 592, 558 & O--Ni--O stretching vibration of $\rm NiO_6$ octahedra\\
 588 & & & 560 & & $\rm TeO_6$ stretching modes\\
 480, 519 & 517, 487 & 490, 471 & 469, 504 & 504, 481 & $\rm TeO_6$ bending modes\\
 & 377 & 386 & & 378 & $\rm TeO_6$ stretching modes\\
 297 & & & 291 & & Cu-O stretching vibrations\\
 & 259 & 264 & & 263 & Na/Li--O stretching modes\\
 & & 223 & & & Li--O stretching modes\\
 217 & & & 234 & & Vibrations belonging to O--Cu--O bonds\\
 154 & & & 156 & & O--$A$--O bonds ($A$ = Na or Li)\\
 101 & & & 101 & & Tilting vibrations of $\rm CuO_6$ octahedra\\
\hline
\end{tabular}}
\end{center}
\end{table*}

Raman bands resulting from the bending of O--$M$--O ($M$ = Na, Li) and O--Cu--O are observed between 100 and 800 cm${\rm ^{-1}}$, as depicted by the enlarged spectra in \textbf{Figure \ref{Figure_43}c}. The spectra also exhibit the symmetric stretching modes of $\rm TeO_6$ octahedra at 677 and 601 cm${\rm ^{-1}}$. The bending modes of $\rm TeO_6$ octahedra are centered at 481, 507, 651, and 665 cm${\rm ^{-1}}$. The modes in the far-infrared region, $i.e.$, 263 and 150 cm${\rm ^{-1}}$, occur due to the stretching of the Na--O bond. The Raman shift of 560, 593, and 640 cm${\rm ^{-1}}$ accommodate for the stretching vibrational modes of O--Ni--O bonds in $\rm NiO_6$ octahedra. The designation of the various modes arising from the Raman spectra of the honeycomb layered tellurates is summarised in \textbf{ Table \ref{Table_13}}. Raman spectroscopy has more recently been utilised as a complementary tool for XRD to validate the crystal structure of $\rm Na_2Ni_2TeO_6$.\cite{pati2022unraveling}

Altogether, the few inquests that have employed Raman spectroscopy to probe the pnictogen and chalcogen class of honeycomb layered oxides have illuminated this technique as a potent tool for examining disordered frameworks. Despite the crystallographic prospects represented using Raman spectroscopy, the technique is encumbered by two limitations: its limited spatial resolution and its relatively weak signal response. Nonetheless, advanced Raman spectroscopy techniques, suchlike tip-enhanced Raman spectroscopy (TERS)\cite{verma2017}, stimulated Raman scattering (SRS)\cite{prince2017} and coherent anti-Stokes spectroscopy (CARS)\cite{el2011coherent}, have recently been proffered as measures against the aforementioned limitations. Furthermore, the use of Raman spectroscopy in combination with related characterisation techniques has recently been shown to permit a more comprehensive and analytic insight of structural disorders of materials, for instance, honeycomb layered oxides.

\subsubsection{Ultraviolet-visible (UV-Vis) diffuse reflectance spectroscopy and photoluminescence spectroscopy}
The correlation between the optical properties of inorganic compounds and their electronic configurations is a well-established concept that has been explored in several studies. Principally, the colour of a material combined with knowledge of its composition and structure has been known to aid in elucidating its pertinent electronic interactions such as band gaps. In this light, the colour transitions envisioned in the pnictogen and chalcogen class of honeycomb layered oxides has spurred expeditions into their band gaps through analyses of their optical properties. Compounds such as $\rm Na_2{\it M}_2TeO_6$ ($\rm {\it M} = Co, Cu, Mg, Zn $)\cite{Evstigneeva2011} have been found to possess non-centrosymmetric coordination environments that can lead to acentric structures with fascinating functionalities, such as nonlinear optical activity, pyroelectricity, and piezoelectricity.

\begin{figure*}[!b]
\centering
  \includegraphics[width=0.95\columnwidth]{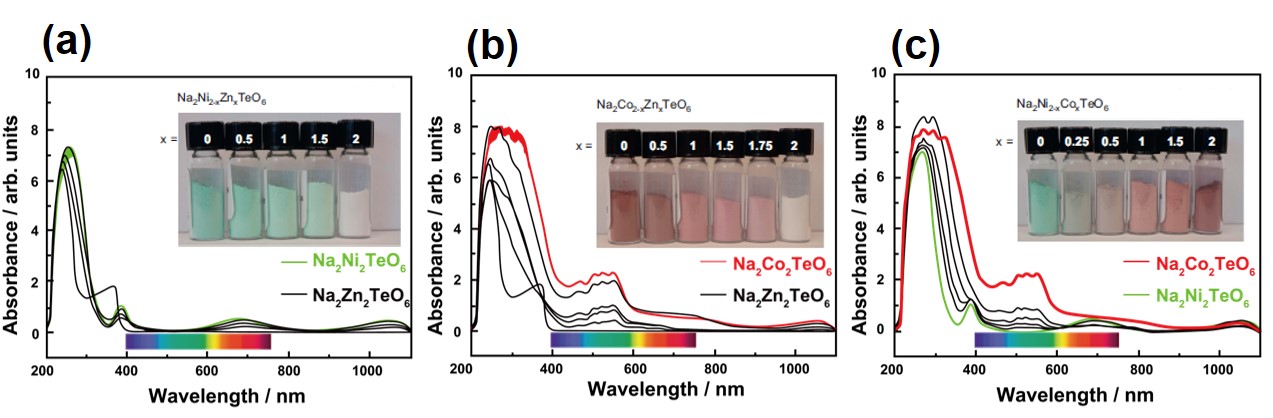}
  \caption{Ultra-violet (UV)-visible diffuse reflectance spectra of honeycomb layered antimonates. Colour variation of the sample powders and corresponding diffuse reflectance spectra for (a) $\rm Na_2Ni_{2-{\it x}}Zn_{\it x}TeO_6$, (b) $\rm Na_2Co_{2-{\it x}}Zn_{\it x}TeO_6$ and (c) $\rm Na_2Ni_{2-{\it x}}Co_{\it x}TeO_6$ solid-solutions. Figures reproduced with permission.\cite{berthelot2012studies} Copyright 2012 Elsevier.}
  \label{Figure_44}
\end{figure*}

In a bid to understand the band gap energy of materials in the present class of honeycomb layered oxides, diffuse reflectance spectroscopy\cite{simmons1975, kortum1963} has been used to analyse compounds such as $\rm Na_3{\it M}_2SbO_6$ ($\rm {\it M} = Cu, Mg, Ni, Zn$),\cite{Schmidt2013} $\rm Na_3Fe{\it M}SbO_6$ ($\rm {\it M} = Mg, Zn, Ni, Co, Fe $),\cite{yadav2022investigation} $\rm Na_3LiFeSbO_6$,\cite{Schmidt2014} $\rm Na_4FeSbO_6$,\cite{Schmidt2014} $\rm Li_3Mn_2SbO_6$,\cite{yadav2022influence}\\
$\rm Li_4CoSbO_6$,\cite{yadav2022influence} $\rm Na_3Mn{\it M}SbO_6$ ($\rm {\it M} = Mg, Zn, Ni, Co, Mn $),\cite{yadav2022investigation} $\rm Cu_3NaFeSbO_6$,\cite{sethi2022cuprous} $\rm Cu_3Cu_{0.7}Na_{0.3}FeSbO_6$,\cite{sethi2022cuprous} $\rm Cu_3LiFeSbO_6$,\cite{sethi2022cuprous} $\rm Li_3Co{\it M}SbO_6$ ($\rm {\it M} = Mg, Zn, Ni$),\cite{pal2022insights} $\rm Li_3Mn{\it M}SbO_6$ ($\rm {\it M} = Mg, Zn, Ni, Co$),\cite{pal2022insights} $\rm Li_4{\it M}SbO_6$ ($\rm {\it M} = Ga, Al, Fe, Mn, Cr$),\cite {Bhardwaj2014} $\rm Ag_3Li{\it M}SbO_6$ ($\rm {\it M} = Ga, Al, Fe, Mn, Cr$)\cite {Bhardwaj2014} and $\rm Na_2{\it M}_2TeO_6$ ($\rm {\it M} = Zn, Ni, Co$)\cite{Berthelot2012a}. \textbf{Figure \ref{Figure_44}} shows the diffuse reflectance spectra of $\rm Na_2Ni_{2-{\it x}}Zn_{\it x}TeO_6$ (\textbf{Figure \ref{Figure_44}a}), $\rm Na_2Co_{2-{\it x}}Zn_{\it x}TeO_6$ (\textbf{Figure \ref{Figure_44}b}) and $\rm Na_2Ni_{2-{\it x}}Co_{\it x}TeO_6$ (\textbf{Figure \ref{Figure_44}c}) solid-solutions.

The reflectance spectra of $\rm Na_2Co_2TeO_6$ (\textbf{Figures \ref{Figure_44}b} and \textbf{\ref{Figure_44}c)} show the presence of two distinct absorption bands; an absorption band located around 500 nm (green region) and a main absorption band centred around 300 nm (ultra-violet (UV) region). A comparison of the reflectance spectra shows the main absorption band is displaced to lower wavelength values without any decline in intensity whereas the second band decreases in intensity without shifting to lower wavelength values, as the composition moves towards the solid-solutions. 
\begin{figure*}[!b]
\centering
  \includegraphics[width=1\columnwidth]{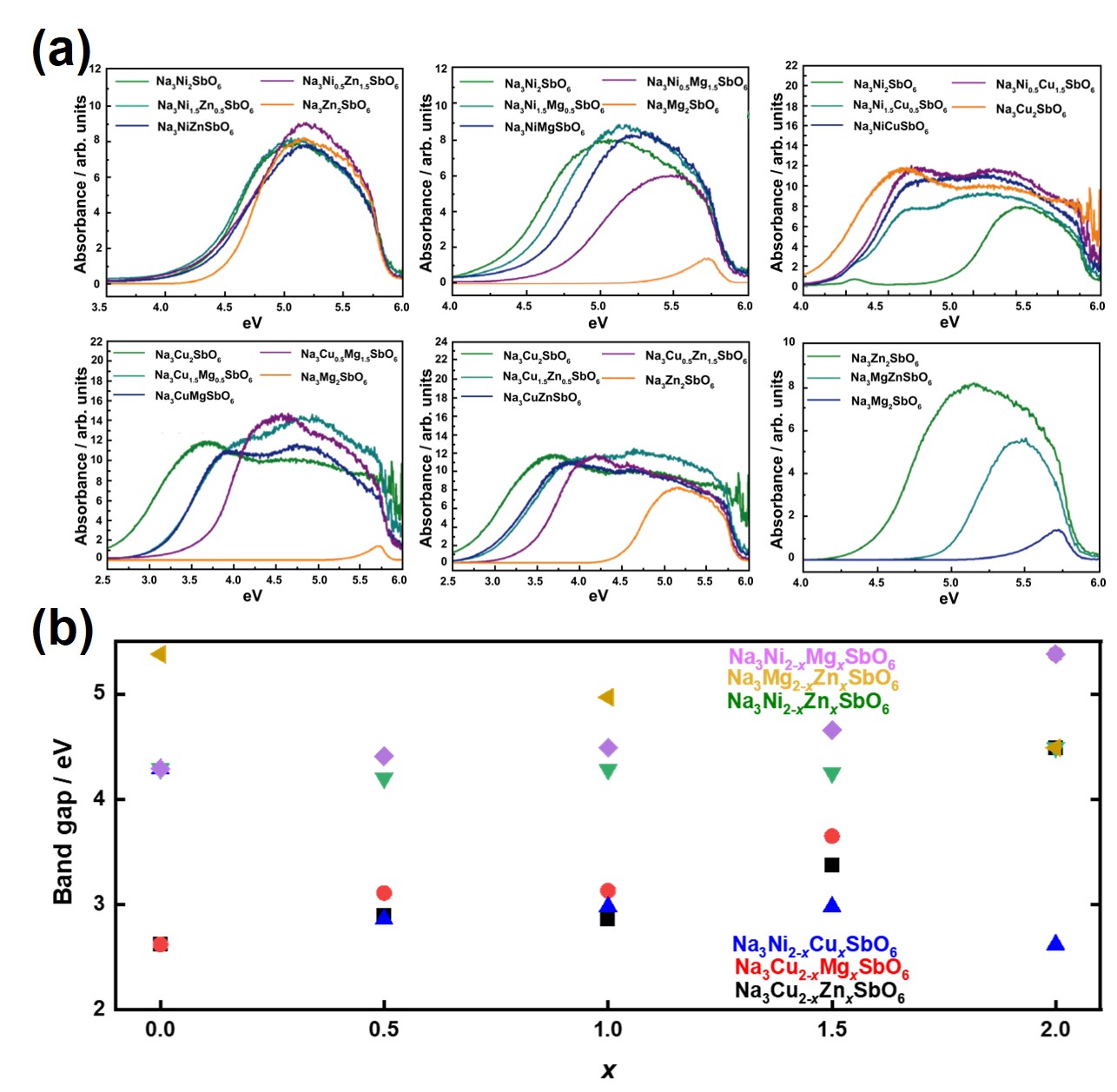}
  \caption{Ultra-violet (UV)–visible diffuse reflectance spectra of honeycomb layered antimonates. (a) Transformed diffuse reflectance spectra (in absorbance versus eV) for each $\rm Na_3{\it M}_{2-{\it x}}{\it M'}_{\it x}SbO_6$ ($\rm {\it M}, {\it M'} =  Zn, Ni, Mg, Cu $) solid solution. The band gaps were extrapolated from the initial absorption onset to get the $x$-intercept. (b) Estimated band gaps ${\rm {\it E}_{g}}$ (eV) extrapolated from transformed diffuse reflectance measurements of $\rm Na_3{\it M}_{2-{\it x}}{\it M'}_{\it x}SbO_6$ ($\rm {\it M}, {\it M'} =  Zn, Ni, Mg, Cu $) solid-solutions. The band gaps appear to be compositionally dependent. Figure (a) reproduced with permission.\cite{Schmidt2013} Copyright 2013 American Chemical Society.}
  \label{Figure_45}
\end{figure*}
Diffuse reflectance spectroscopy has further been employed for honeycomb layered antimonates and their solid-solution compositions entailing $\rm Na_3{\it M}_2SbO_6$ ($\rm {\it M} = Cu, Mg, Ni, Zn $), as shown in \textbf{Figure \ref{Figure_45}a}. Plotting the absorbance against the energy (eV) to create what is known as Tauc plot, allows the linear region to be extrapolated to the $x$-axis to estimate the band gap energy ($i.e.$, the energy separating the bottom of the conduction band and the top of the valence band). In principle, band gap values of materials are estimated using Kubelka-Munk equation\cite{kubelka1931, kubelka1948}: $(\alpha h\nu)^{1/n} = A(h\nu - E_g)$. Here, $A$ is a proportionality constant, $\rm \alpha $ is the absorption coefficient that can be attained using the Beer-Lambert\cite{swinehart1962} law, $h \simeq 6.626\times 10^{-34}$ Js is Planck's constant, $\nu$ is the frequency of the absorbed photon/electromagnetic field, $E_g$ is the band gap energy of the material and $n$ depends on the nature of electronic transitions.

\textbf{Figure \ref{Figure_45}a} shows the Tauc plots of $\rm Na_3{\it M}_2SbO_6$ ($\rm {\it M} = Zn, Ni, Mg, Cu $)\cite{Schmidt2013} and their solid-solution compositions. Plots for the estimated band gaps of the solid-solution compositions ($\rm Na_3{\it M}_{2-{\it x}}{\it M'}_{\it x}SbO_6$ ($\rm {\it M'}, {\it M} = Cu, Mg, Ni, Zn $)) derived from the Tauc plots are provided in \textbf{Figure \ref{Figure_45}b}. Here, the plots display large band gaps that indicate the materials have insulating properties. In spite of that, the band gaps appear to be tuneable depending on the composition of these honeycomb layered oxides.
\begin{figure*}[!t]
\centering
  \includegraphics[width=0.8\columnwidth]{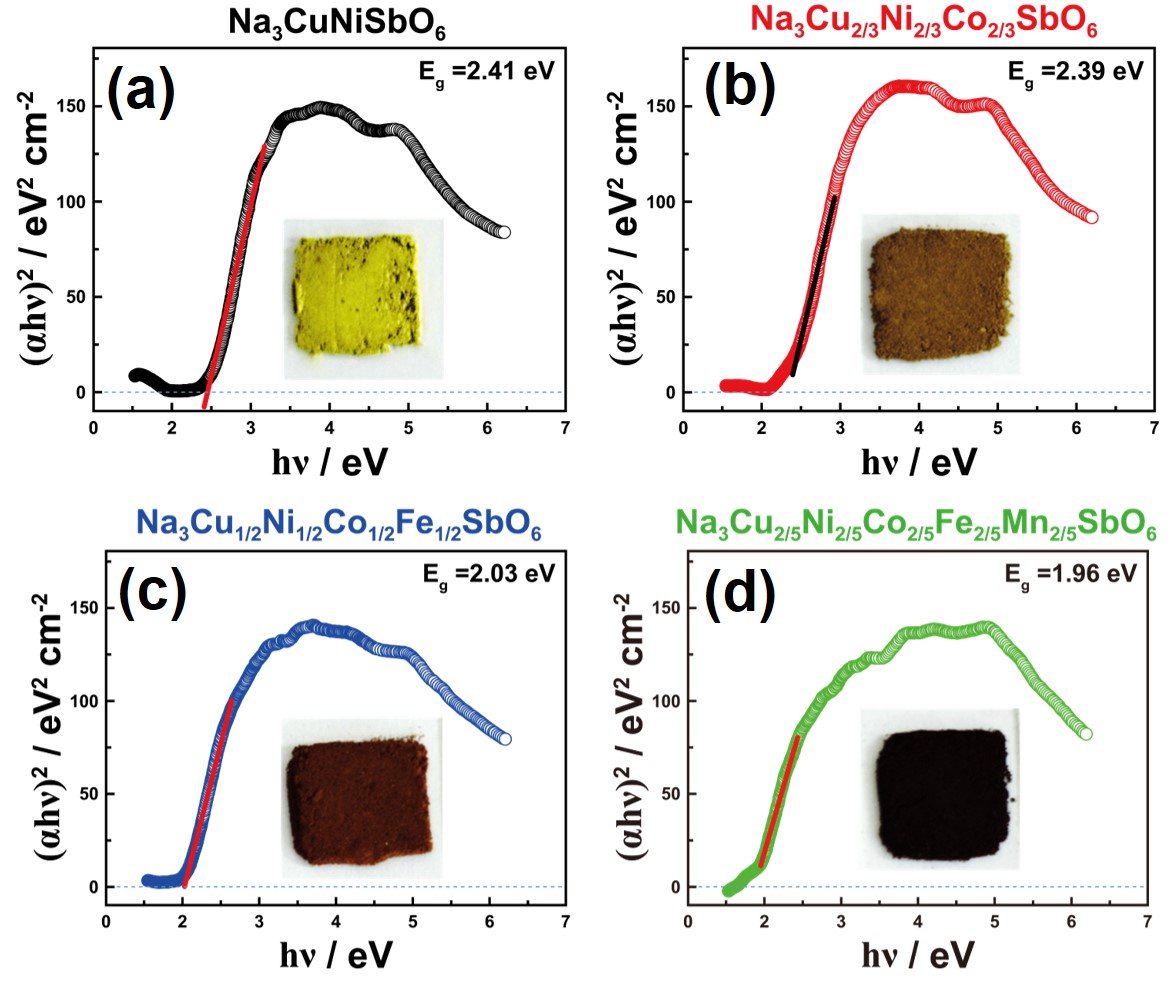}
  \caption{Tauc plots of multicomponent `high entropy' pnictogen- and chalcogen-based honeycomb layered oxides. Tauc plot for (a) $\rm Na_3CuNiSbO_6$, (b) $\rm Na_3Cu_{2/3}Ni_{2/3}Co_{2/3}SbO_6$, (c) $\rm Na_3Cu_{1/2}Ni_{1/2}Co_{1/2}Fe_{1/2}SbO_6$ and (d) $\rm Na_3Cu_{2/5}Ni_{2/5}Co_{2/5}Fe_{2/5}Mn_{2/5}SbO_6$. The linear portion of the curves are interpolated to estimate the band gap values (${\rm {\it E}_{g}}$) shown in the plots. Colours of the compounds are represented in each respective plot and appear to take darker hues with increasing number of elements (increased entropy). Figures reproduced with permission.\cite{karati2021band} Copyright 2021 Elsevier.}
  \label{Figure_46}
\end{figure*}
In fact, band gap tunability is a widely exploited strategy employed in the engineering of visible light absorbing photocatalysts using oxides. Recently multicomponent honeycomb layered oxides such as $\rm Na_3CuNiSbO_6$,\cite{karati2021band} $\rm Na_3Cu_{2/3}Ni_{2/3}Co_{2/3}SbO_6$,\cite{karati2021band} $\rm Na_3Cu_{1/2}Ni_{1/2}Co_{1/2}Fe_{1/2}SbO_6$\cite{karati2021band} and $\rm Na_3Cu_{2/5}Ni_{2/5}Co_{2/5}Fe_{2/5}Mn_{2/5}SbO_6$\cite{karati2021band} have been shown to exhibit band energies suitable for water-splitting reactions 
earmarking these materials for visible light photocatalyst 
applications. \textbf{Figure \ref{Figure_46}} shows the Tauc plots of the multicomponent `high-entropy' honeycomb layered oxides. After extrapolating the straight line towards $x$-axis of the Tauc plots obtained for $\rm Na_3Cu_{2/5}Ni_{2/5}Co_{2/5}Fe_{2/5}Mn_{2/5}SbO_6$, $\rm Na_3Cu_{1/2}Ni_{1/2}Co_{1/2}Fe_{1/2}SbO_6$, $\rm Na_3Cu_{2/3}Ni_{2/3}Co_{2/3}SbO_6$ and $\rm Na_3CuNiSbO_6$, the direct band gap calculated using the Kubelka-Munk relation was found out to be 1.96, 2.03, 2.39 and 2.41 eV, respectively.\cite{karati2021band} The band gaps of these honeycomb layered oxides are observed to decrease continuously from 2.41 eV to 1.96 eV with increments in the transition metal species. This observation has been ascribed to the presence of impurity states which are responsible for inducing non-allowed or allowed transitions in the compounds. These multicomponent honeycomb layered oxides show photocatalytic properties (for instance, $\rm Na_3Cu_{2/5}Ni_{2/5}Co_{2/5}Fe_{2/5}Mn_{2/5}SbO_6$ as shown in \textbf{Figure \ref{Figure_14}}) attributed to the increased entropy in the system which causes a decrease in the optical band-gap thereby triggering changes in the band energetics. Based on the diffuse reflectance measurements of these honeycomb layered antimonates, it can be deduced that the introduction of multiple transition metal elements in the honeycomb slab in equimolar proportions (referred to as `high-entropy' approach) is a feasible approach for tailoring the optical band gaps of these materials and can be extended to related materials—auguring the prospects of new photocatalysts that utilise visible light to split water for $\rm H_2$ production.

Photoluminescence spectroscopy\cite{anpo1999} is another non-destructive technique that has garnered traction in analyses investigating the distribution of energies accompanying photo-absorption and photoemission processes as well as the efficiencies and temporal characteristics of the processes in selected materials. The non-destructive nature of this technique makes a providential approach for gathering rich electronic information {\it inter alia} about defects, band structure, impurity levels, electron-hole relaxation and recombination processes, amongst others. Additionally, the photoluminescence spectra obtained at low sample temperatures have been instrumental in revealing spectral peaks associated with impurities contained within the material. As such, this technique is capable of identifying extremely low concentrations of unintentional and intentional impurities that can profoundly affect material quality and functionalities. It is also worth noting that photoluminescence characterisation can be applied in concert with other spectroscopic analyses such as UV-vis diffuse reflectance to facilitate a more wholesome optical elucidation.

\begin{figure*}[!t]
\centering
  \includegraphics[width=0.85\columnwidth]{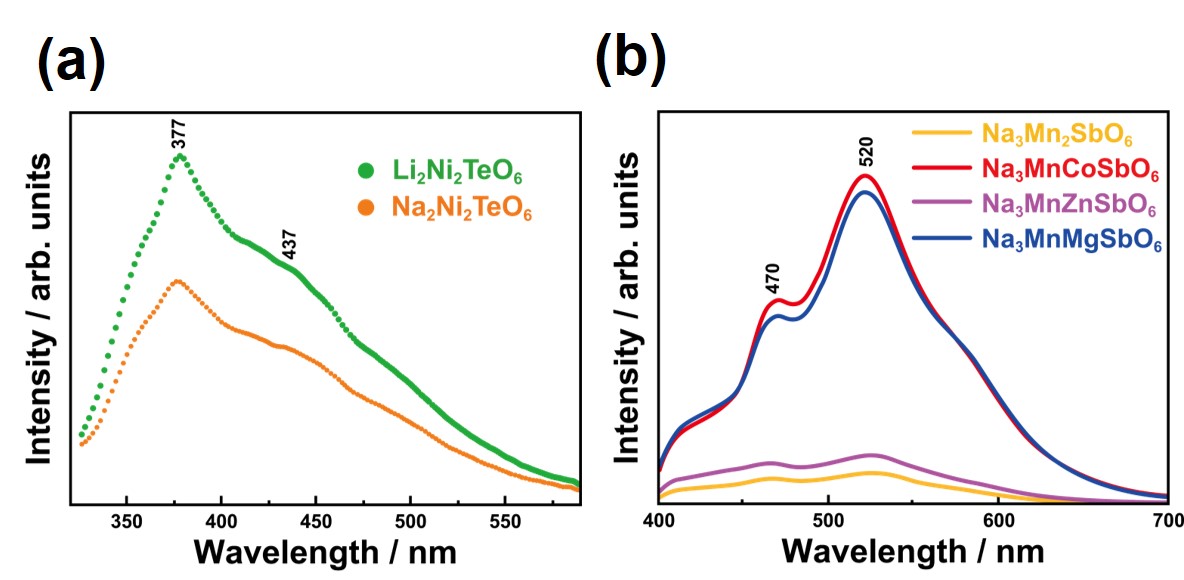}
  \caption{Photoluminescence spectra of honeycomb layered tellurates and antimonates. (a) Photoluminescence spectra of $\rm Na_2Ni_2TeO_6$ (plotted with stars) and $\rm Li_2Ni_2TeO_6$ (prepared via ion-exchange from $\rm Na_2Ni_2TeO_6$ as precursor) (plotted with circles) taken at an excitation wavelength $\rm \lambda $ = 310 nm. (b) Photoluminescence spectra of $\rm Na_3Mn{\it M}SbO_6$ ($\rm {\it M} = Mn, Co, Zn, Mg $). Figure (a) and Figure (b) reproduced with permission.\cite{Kumar2013, yadav2022investigation} Copyright 2013 Royal Society of Chemistry and 2022 Elsevier, respectively.}
  \label{Figure_47}
\end{figure*}

Photoluminescence spectroscopy has been employed in the characterisation of honeycomb layered compositions such as $\rm {\it A}_2Ni_2TeO_6$ ($A = \rm Li, Na $),\cite{Kumar2013} as shown in \textbf{Figure \ref{Figure_47}a}. Here, although both materials displayed a broad photoluminescence spectrum with a maximum around 377 nm, the intensity of the spectrum from $\rm Li_2Ni_2TeO_6$ (prepared via topochemical ion-exchange) was observed to be more intense than that of $\rm Na_2Ni_2TeO_6$. The high intensity was ascribed to the presence of structural disorders (defects) in the metastable $\rm Li_2Ni_2TeO_6$. As such, the introduction of defects can be deduced to be an effective strategy tuning the visible radiation emission in honeycomb layered oxides. In fact, the metastable $\rm Li_2Ni_2TeO_6$ (prepared via topochemical ion-exchange) has been deemed a possible emitter of visible radiation at room-temperature.\cite{Kumar2013}

Similarly, photoluminescence characterisation has been employed on honeycomb layered antimonates such as $\rm Li_4{\it M}SbO_6$ ($\rm {\it M} = Mn, Cr $)\cite {Bhardwaj2014} and $\rm Na_3Mn{\it M}SbO_6$ ($\rm {\it M} = Co, Mg $),\cite{yadav2022investigation} revealing interesting optical characteristics such as the occurrence of green emissions. \textbf{Figure \ref{Figure_47}b} shows the photoluminescence spectra of $\rm Na_3Mn{\it M}SbO_6$ ($\rm {\it M} = Mn, Co, Zn, Mg $). Here, the intensities of the bands emerging around 520 nm are noticeably enhanced by the partial substitution of ${\rm Mn^{2+}}$ by ${\rm Co^{2+}}$, ${\rm Mg^{2+}}$, or ${\rm Zn^{2+}}$ ions. $\rm Na_3Mn{\it M}SbO_6$ ($\rm {\it M} = Mn, Co, Zn, Mg $)\cite{yadav2022investigation} demonstrate typical emissions at 520 and 470 nm, arising (in the Russell–Saunders notation) from the transitions ${\rm ^{4}}$T${\rm _{1g}}$ (${\rm ^{4}}$G) → ${\rm ^{6}}$A${\rm _{1g}}$ (${\rm ^{6}}$S) and ${\rm ^{4}}$T${\rm _{2g}}$ (${\rm ^{4}}$G) → ${\rm ^{6}}$A${\rm _{1g}}$ (${\rm ^{6}}$S), respectively corresponding to the excitation wavelength of 380 nm.

A significant class of luminescent materials for usage in light emitting diodes constitute non-rare earth transition metal activated phosphors, particularly those based on manganese. ${\rm Mn^{2+}}$ ions have been known to emit green emissions in a wide range of host lattices and are routinely utilised as luminescent activators. In the field of white light-emitting diodes (LEDs), on the other hand, ${\rm Mn^{4+}}$-doped oxides are accepted red phosphors. Therefore, explorations into the optical properties in ${\rm Mn^{2+}}$ based honeycomb layered oxides are especially significant for such functionalities and provides a platform for future expeditions into new phosphor materials. 

\newpage

\section{\label{Section: Summary} Summary, perspective and challenges}
In the quest to elaborate the fundamental atomistic mechanisms and explore material functionalities, this review highlights the breakthroughs achieved in the design and development of pnictogen- and chalcogen-based honeycomb layered oxides from synthesis to characterisation perspective. Here, we elucidate the vast crystal chemistry space covered by this class of materials whilst outlining specific characterisation techniques pertinent to the exploration of their assorted functionalities. Looking into the milestones covered by this pnictogen and chalcogen class of honeycomb layered oxides over the last few decades, a few questions linger: `Is this class of materials merely a  {\it du jour } topic hyperbolised by the academic world?' and more importantly, `Have we reached the end of line for advancing these materials?'.

For a long time, the pnictogen and chalcogen class of honeycomb layered oxides has been beheld as structures entailing monolayer arrangement of monovalent atoms (such as Cu, Ag, K, Na, Li, {\it et cetera}) sandwiched between metal slabs.\cite{kanyolo2021honeycomb} However, the recent success in the synthesis of $\rm BaNi_2TeO_6$\cite{song2022influence} has rekindled new prospects of honeycomb layered oxides comprising divalent cations such as $\rm Mg{\it M}_2TeO_6$ ($\rm {\it M} = Ni, Co, Cu, Zn, Mg$), $\rm Ca{\it M}_2TeO_6$ and $\rm Sr{\it M}_2TeO_6$, as discussed in \textbf{Section 3}. Besides, this discovery has proffered the possibility to extend the design of honeycomb layered oxides to materials that encompass actinides such as La sandwiched between the slabs to form exotic layered compounds such as $\rm La{\it M}_{1/3}Sb_{5/3}O_6$ ($\rm {\it M} = Co, Ni $ and $\rm Cu$) with the honeycomb layered structure.\cite{ellert2021magnetic} Outlook into these materials is further stoked by the serendipitous discovery of bilayer arrangement of coinage metal atoms such as Ag in global $\rm Ag_2{\it M}_2TeO_6$ ($\rm {\it M} = Co, Cu, Ni, Zn, Mg$) compositions that is bound to augment their compositional tuneability and structural versatility.\cite{masese2023honeycomb, kanyolo2022advances2} For instance, their saturation processes (\textbf{Figure \ref{Figure_48}c}) are expected to mimic the intercalation processes of monolayered frameworks such as $\rm Li_4NiTeO_6$, $\rm Na_2Ni_2TeO_6$, $\rm Na_3Ni_2SbO_6$, $\rm Na_3Ni_2BiO_6$ and $\rm K_2Ni_2TeO_6$, whereby the redox processes upon de-intercalation of $A =$ Li, Na, K atoms during cycling appear to initially involve only $\rm Ni$ atoms as shown in \textbf{Figures \ref{Figure_30}} and \textbf{\ref{Figure_31}}.\cite{Sathiya2013, Evstigneeva2011, Masese2018, Kim2020, Bhange2017} 
It is likely that their cycling will follow the sequence: de-saturation $\rightarrow$ (monolayer-bilayer) phase transition $\rightarrow$ de-intercalation, which can be tracked by the oxidation states of Ni. Moreover, unlike electro-weak theory\cite{Zee2010}, 
since the photon plays the role of the U($1$) field in the SU($2$)$\times$U(1) symmetry breaking, it leads to a mass term and hence Ag--Ag' pair correlations dual to superconductivity. Nonetheless, since such pairing of the lattice atoms can be interpreted as a Peierls transition in the context of dimerisation\cite{bendazzoli2004huckel, garcia1992dimerization}, instead of superconductivity, the pairing is expected to lead to a metal-insulator phase transition.\cite{stewart2012evidence} Indeed the superconducting and insulating phases are related by duality between the magnetic and electric fields in Maxwell's equations. In the case of the bilayers, the electric Meissner (Coulomb blockade\cite{kanyolo2021renormalization}) effect hinders the extraction of Ag cations unless a critical voltage is applied, which is equivalent to the activation energy needed to break the pairs. In addition, it is important to ascertain the diffusion processes for bilayered materials, which could vastly differ from the conventional monolayered frameworks. For instance, it is 
feasible that the diffusion coefficients of the layers differ from each other, leading to the Kirkendall effect.\cite{paul2014thermodynamics} 
Lastly, the enigmatic mechanisms behind the metallophilicity accredited for the bilayer arrangement of coinage metal atoms, is predicted to awaken scientific interest amongst chemists, physicists, and material scientists in equal measures. 

By the same token, interest in the superior electrochemical performance seen in alkali-excess compositions such as $\rm Na_3Ni_{1.5}TeO_6$\cite{grundish2020structural} as well as tuneable magnetic properties found in $\rm Na_{2.4}Zn_2TeO_6$,\cite {samarakoon2021static} $\rm Li_{4.47}Cr_{0.53}TeO_6$\cite{mandujano2021} and $\rm Li_{3.88}Cr_{1.12}SbO_6$\cite{mandujano2021} is seen as a herald to the further exploration of compositions such as $\rm {\it A}_{2+{\it x}}{\it M}_{(1-{\it x}/2)}TeO_6$ (where $A$ = Na, K, {\it etc.}; $M$= transition metal or $s$-block element; $0 < x < 1$). In fact, preliminary investigations have already shown the potential to design similar mixed-alkali compositions such as $\rm K_3Ni_{1.5}TeO_6$ as well as non-stoichiometric oxide compositions such as $\rm Li_{0.8}Ni_{0.6}Sb_{0.4}O_2$\cite{vavilova2021} (or equivalently as $\rm Li_{2.4}Ni_{1.8}Sb_{1.2}O_6$)—essential landmarks in the expansion of the present compositional chemistry space.

On another front, incipient investigations into honeycomb layered oxides encompassing mixed chalcogen or pnictogen combinations such as $\rm Na_3Ni_2Bi_{0.5}Sb_{0.5}O_6$ have revealed vast improvements in their resultant electrochemical performance compared to their parent compounds such as $\rm Na_3Ni_2BiO_6$ and $\rm Na_3Ni_2SbO_6$.\cite{Bhange2017, Liu2016, Zheng2016, Wang2019a, Yuan2014, Kee2016, Kee2020, Aguesse2016} Equally important, progress in mixed alkali honeycomb layered composition such as $\rm NaKNi_2TeO_6$\cite{masese2021mixed, berthelot2021stacking} has unravelled new expedition directions into the rich crystal chemistry and structural versatility of the broad class of pnictogen- and chalcogen-based honeycomb layered oxides. In effect, such inquests have already borne fruit with our preliminary design of exotic mixed alkali compositions such as $\rm Li_{2/3}Na_{2/3}K_{2/3}{\it M}_2TeO_6$ (where $M$ = Ni, Mg, Co, Cu, Zn). The investigation of the complete phase diagram of related mixed alkali compositions, for instance, $\rm Li_2{\it M}_2TeO_6$--$\rm Na_2{\it M}_2TeO_6$--$\rm K_2{\it M}_2TeO_6$ ternary phase diagram is a subject worthy of pursuit as a guideline for predicting emergent compositional configurations.

The development of high-entropy honeycomb layered oxide compositions has also emerged as another strategy for engineering advanced materials with unique properties unachievable by the conventional honeycomb layered oxide materials. In fact, layered oxides with multi-component transition metals, generally in equiatomic concentrations, are garnering traction in the design of property-oriented novel materials. This class of layered oxides, dubbed as `high-entropy oxides',\cite{berardan2016room, rost2015entropy} has shown superb aptness for accommodating multiple metal cations (in the slabs) into single-phase crystal structures to engender exotic interactions that lead to intriguing novel and unprecedented properties.\cite{sarkar2019high, zhao2020high} From an application perspective, high-entropy layered oxides such as $\rm NaNi_{0.12}Cu_{0.12}Mg_{0.12}Fe_{0.15}Co_{0.15}Mn_{0.1}Ti_{0.1}Sn_{0.1}Sb_{0.04}O_2$,\cite{zhao2020high} $\rm NaNi_{1/4}Co_{1/4} Fe_{1/4} Mn_{1/8}Ti_{1/8} O_2$\cite{yue2015quinary} and $\rm Na_{0.8}Ni_{1/5}Fe_{1/5}Co_{1/5}Mn_{1/5}Ti_{1/5} O_2$,\cite{anang2019} amongst others,\cite{walczak2022, wang2020lithium} have been shown to exhibit high-rate capabilities and excellent capacity retention – key requisites for high-power density cathode materials for rechargeable sodium-ion batteries. Furthermore, the electrochemical behaviour of the high-entropy layered oxides has been noted to hinge on each metal cation present in the slab, thus providing new avenues for tailoring the electrochemical properties through the alteration of elemental composition. Besides their potential in electrochemical applications, high-entropy honeycomb layered oxides such as $\rm Na_3Cu_{2/5}Ni_{2/5}Co_{2/5}Fe_{2/5}Mn_{2/5}SbO_6$\cite{karati2021band} have been demonstrated to have good photocatalytic properties begotten from their lower band gap in comparison to binary and ternary antimonate compositions such as $\rm Na_3CuNiSbO_6$\cite{karati2021band} and $\rm Na_3Cu_{2/3}Ni_{2/3}Co_{2/3}SbO_6$.\cite{karati2021band} This also highlights the `high-entropy' strategy as effective approach to tailoring the optical band gap of honeycomb layered oxides. Furthermore, the `high-entropy' approach can be utilised to design new honeycomb layered oxides that are stable against moisture or tweak the composition of hygroscopic honeycomb layered oxides suchlike $\rm Na_3{\it M}_2BiO_6$ ($M$ = Mg, Zn)\cite {Seibel2013} and $\rm K_2Ni_2TeO_6$.\cite {masese2021topological, Masese2018, Masese2019} Preliminary investigations have shown the feasibility to design multinary compositions (such as $\rm K_2Ni_{0.8}Co_{0.1} Fe_{0.1}Mn_{0.1}Zn_{0.1}Mg_{0.1}Cu_{0.1}TeO_6$ and $\rm Na_3Ni_{1/3}Co_{1/3}Zn_{1/3}Mg_{1/3}Cu_{1/3}Mn_{1/3}SbO_6$), envisaging new avenues to expand the material composition of high-entropy honeycomb layered oxides.

Apropos of synthesis, potential for new development methodologies is immense. For instance, high pressure synthetic routes and the less explored high-temperature metathetic routes can be exploited in the design of new structures. Moreover, exploratory synthesis routes can be augmented with computational tools spanning from density functional theory (DFT) calculations to machine learning to expedite the design of new honeycomb layered oxide compositions. For example, the DFT approach has recently been employed to predict {\it in silico} new structural frameworks of honeycomb layered tellurates such as $\rm Ag_2Ni_2TeO_6$, $\rm Cu_2Ni_2TeO_6$ and $\rm Au_2Ni_2TeO_6$,\cite{tada2022implications} envisaging their applications in various realms spanning from optics, catalysis, and energy storage to analogue condensed matter systems of 2D quantum gravity.\cite{kanyolo2022cationic, masese2023honeycomb, Kanyolo2020, kanyolo2022advances2, kanyolo2021honeycomb, masesemaths, kanyolo2021reproducing, kanyolo2022local, kanyolo2021partition}

Alternative synthesis methodologies such as single crystal growth (using the flux method) of honeycomb layered tellurates such as $\rm Na_2Ni_2TeO_6$\cite{murtaza2021magnetic, xiao2021magnetic} and antimonates such as $\rm Na_3Co_2SbO_6$\cite{Yan2019} and $\rm Na_3Cu_2SbO_6$\cite{Miura2008} have enabled an accurate determination of the intrinsic physicochemical properties via several spectroscopic techniques and magnetic susceptibility measurements. 
Despite this, the sizes of single crystals of honeycomb layered oxides grown using the flux method are not large enough for undertaking bulk transport measurements such as Nernst effect, superconductivity and thermal conductivity measurements. Alternative methods, such as the chemical vapour transport (CVT), can also be used to grow a vast majority of crystals grown using the flux method. Larger sizes of single crystals can be grown using the CVT technique, a feat that is worthy of pursuit. The syntheses of single crystals of $\rm Cu_5SbO_6$\cite {reya2011crystal} and $\rm Na_3Co_2SbO_6$\cite {Yan2019} envision the feasibility of adopting the CVT technique to yield single crystals of related pnictogen- and chalcogen-based honeycomb layered oxides.

Efforts to advance the discovery pace of honeycomb layered oxides have been steered by advances in microscopic imaging and spectroscopic techniques that have been instrumental in unveiling, at the atomic scale, the fundamental physics and chemistry governing these materials. In particular, high-resolution transmission electron microscopy (TEM) has revealed a plenitude of dislocations and topological defects innate in honeycomb layered oxides such as $\rm NaKNi_2TeO_6$, $\rm K_2Ni_2TeO_6$ and $\rm Na_2Ni_2TeO_6$,\cite{masese2021mixed, masese2021topological, masese2021unveiling} thereby expanding the pedagogical scope of these topological materials. Despite the efficacy of these techniques, beam-sensitivity of some honeycomb layered oxide materials may inhibit the sub-\AA ngstr\"{o}m resolution of some topological defects. As a solution, performing TEM with the aid of detectors that possess satisfactory signal-to-noise ratios and higher collection rates at lower beam currents can help reduce sample damage due to high electron dose rates. With the advent of state-of-the-art aberration-corrected (scanning) transmission electron microscopes (TEM/STEM) that are equipped with ultrafast direct detection device (DDD) camera, atomic-resolution low-dose imaging of honeycomb layered oxide nanostructures with unprecedented clarity is possible. Moreover, real-time nanobeam electron diffraction (NBED) mapping in {\it in situ} STEM (also referred to as 4D-STEM\cite{ophus2014recording, ciston20194d, yang20154d}) has emerged, where the speed at which the TEM camera can read-out images taken has been significantly improved via the developments of both versatile software for treatment of big data and fast DDD cameras. In cases where sufficient signal-to-noise ratios cannot be attained without beam damage, observation of the honeycomb layered oxides using cryogenic TEM can be an effective route.\cite {li2017atomic, zhang2018atomic, wang2017new}

Worthy of mention, spectroscopic techniques such as X-ray absorption spectroscopy (XAS) and nuclear magnetic resonance (NMR) have been invaluable in enriching our understanding on the local electronic and magnetic structures of honeycomb layered oxide materials. For instance, the dormant role of pnictogens (such as Bi and Sb) and chalcogens such as Te during charge compensation processes involving $\rm Na_3Ni_2SbO_6$,\cite{Kim2020} $\rm Li_4FeSbO_6$,\cite{McCalla2015a} $\rm Na_3Ni_2BiO_6$\cite{Bhange2017} and $\rm K_2Ni_2TeO_6$\cite{Masese2018} have been ascertained through Sb $K$-edge, Bi ${\rm {\it L}_{3}}$-edge and Te ${\rm {\it L}_{3}}$-edge XAS measurements due to their high bulk sensitivities that surpass other spectroscopic techniques such as the surface-sensitive X-ray photoelectron spectroscopy (XPS). Complimenting XAS measurements with NMR can further yield unequivocal information relating to the overall local structural changes taking place in such materials during operation, for instance, alkali-ion intercalation and de-intercalation processes during battery operation. NMR studies of exotic nuclei such as ${\rm ^{125}}$Te, ${\rm ^{39}}$K, ${\rm ^{209}}$Bi, ${\rm ^{121}}$Sb, {\it et cetera} may be unwieldy for reasons mentioned in \textbf{Section 5}. Nevertheless, hyperpolarisation techniques (most notably dynamic nuclear polarisation (DNP)) and hardware advances, can avail new avenues for assessing exotic nuclides. Whilst these techniques do not entirely circumvent the effects of large quadrupole moments of such nuclides, they offer the promise of substantial sensitivity increases for isotopes found at low concentration in a sample of interest or for isotopes with low natural abundances ($e.g.$, ${\rm ^{17}}$O, ${\rm ^{43}}$Ca).  In addition, {\it Operando} or {\it in situ} measurements will be pivotal to clarifying short-lived phase transitions and stepwise electrochemical reaction kinetics, which would otherwise be missed from {\it ex situ} measurements. 

Materials that operate either solely through oxygen redox processes or via cumulative cation and anionic redox have in recent years attracted tremendous attention in the battery field. M\"{o}ssbauer measurements augmenting magnetic susceptibility measurements have made it possible to examine the oxygen redox process occurring in $\rm Li_4FeSbO_6$ in efforts to illuminate the mechanics behind their high voltage electrochemistry.\cite{McCalla2015a} Complementary techniques such as resonant inelastic X-ray scattering (RIXS)\cite{chen2014recent, bergmann2009x, rovezzi2014hard} can also be used to elucidate the nature of the redox processes in $\rm Li_4FeSbO_6$ that have been posited to entail peroxo-like species. Moreover, probing directly the local environment of O in materials such as $\rm Li_4FeSbO_6$ could be essential in conferring deeper understanding of their O redox behaviour. Whilst we are cognisant that the low natural abundance of ${\rm ^{17}}$O nucleus and its large quadrupole moment ($I$ = 5/2) may render it difficult to attain a high-resolution NMR spectrum, let alone directly connecting to the paramagnetic Fe ions in the material, it is possible to explore the O environment through performing ${\rm ^{17}}$O magic-angle spinning (MAS) solid-state NMR of ${\rm ^{17}}$O enriched samples, as has been demonstrated in layered oxides such as $\rm Li_2MnO_3$.\cite {seymour2016}

Diffraction methods based on X-rays, neutrons or electrons have been used to elucidate the average crystal structures of honeycomb layered oxides. However, the presence of stacking defects has been noted to thwart efforts towards gaining reliable structural characterisation using these diffraction techniques. Nevertheless, the neutron scattering or X-ray diffraction data such as the diffuse features from short-range correlations and Bragg scattering data can be utilised to construct the pair distribution function (PDF) of interatomic distances in such materials, as has been done with $\rm Na_3Ni_2SbO_6$.\cite{Ma2015} PDF fitting can be a potent tool for unveiling short range structural correlations in crystalline materials, including those that are disordered or have imperfections/defects. Additionally, complementary information for analysing the X-ray- or neutron- PDFs can be obtained via other techniques, such as electron energy loss spectroscopy (EELS) combined with STEM, MAS NMR and XAS, which not only provide information on the constituent compositions but also the local atomic structures. Defects such as stacking faults can also be modelled using programs such as DIFFaX or FAULTS,\cite{casas2016faults, treacy1991general, casas2007microstructural, casas2006faults} which can simulate experimental diffraction patterns that can be used to qualitatively determine the amount of stacking faults in a compound, as has been done for crystal structural elucidation of mixed alkali $\rm NaKNi_2TeO_6$.\cite{masese2021mixed, berthelot2021stacking}

The fast ${\rm Na^{+}}$ ionic conductivities displayed by honeycomb layered oxides (such as $\rm Na_2Ni_2TeO_6$\cite{Evstigneeva2011, Bera2020},\\
$\rm Na_2Zn_2TeO_6$,\cite{Li2018Chemistry} $\rm Na_{1.9}Zn_{1.9}Ga_{0.1}TeO_6$\cite{Li2018Chemistry} and $\rm Na_2Mg_2TeO_6$\cite{Li2018ACSApplMater, Evstigneeva2011}) have also driven explorations using spectroscopic techniques that probe the ionic diffusion of such materials. For instance, electrochemical impedance spectroscopy (EIS) and galvanostatic intermittent titration techniques are widely used macroscopic techniques for ascertaining the long-range diffusion kinetics and providing bulk values of conductivity: such as the contributions from grain boundaries and interfacial layers, each of which can have profound effects on the determined diffusion values. As a measure against such sensitivities to surface effects or grain boundaries, microscopic techniques suchlike quasi-elastic neutron scattering (via advanced Fourier difference maps analysis or maximum entropy methods) and NMR can be applied to establish ion transport at the atomic scale. However, it should be mentioned that some of these techniques are yet to be exploited with honeycomb layered oxides but can nonetheless bring new dimensions into current understanding of the electrochemical capabilities of such materials. Unfortunately, the presence of magnetic ions in the honeycomb layered oxides can interfere with the spin-lattice relaxation rate, which may complicate the diffusion analysis using spectroscopic techniques such as NMR. Positive muon spin relaxation ($\mu^{+}$SR) spectroscopy can allow to probe and model ionic diffusion in materials that possess magnetic ions, since both nuclear and electronic contributions to the muon depolarisation can be decoupled (or separated), making $\mu$SR spectroscopy a potent technique for the microscopic study of the ionic diffusion in crystalline materials, as has been done for $\rm K_2Ni_2TeO_6$.\cite{matsubara2020magnetism}

Correspondingly, theoretical simulation methods (such as molecular dynamics (MD) and {\it ab initio} calculations) have played essential roles in expanding knowledge on the mechanics behind cation diffusion within the honeycomb layered oxides such as $\rm NaKNi_2TeO_6$,\cite{masese2021mixed}$\rm Na_2LiFeTeO_6$,\cite{sau2022ring} $\rm Na_2Ni_2TeO_6$,
\cite{sau2016ion1, sau2015ion1, sau2016ion, Sau2015, Sau2016} $\rm Na_2Mg_2TeO_6$,\cite{sau2015ion1} $\rm Li_3Co_2SbO_6$,\cite{chakrabarti2023} $\rm Na_2Zn_2TeO_6$, \cite {Bianchini2019} $\rm K_2Ni_2TeO_6$, {\it etc}.  Despite this, difficulty in finding reliable sets of inter-atomic potential parameters that can faithfully capture the diffusion aspects of such materials poses a critical hurdle for MD techniques. Recently, the Vashishta-Rahman inter-atomic potential has been demonstrated to be accurate in reproducing various transport and structural properties of honeycomb layered tellurates, which are consistent with those observed experimentally (observables).\cite {Evstigneeva2011}

The rich catalytic,\cite{karati2021band, Kadari2016} topological,\cite{kanyolo2022cationic, masese2023honeycomb, Kanyolo2020, kanyolo2022advances2, kanyolo2021honeycomb, masesemaths, kanyolo2021reproducing, kanyolo2022local, kanyolo2021partition, masese2021mixed} magnetic,\cite{Zvereva2013, Derakhshan2007, Viciu2007, Morimoto2007, Derakhshan2008, Koo2008, Kumar2012, Schmidt2014, Sankar2014, Xiao2019, Zvereva2015a, Itoh2015, Zvereva2015b, Koo2016, Zvereva2016, Karna2017, Zvereva2017, Stavropoulos2019, Korshunov2020, Yao2020, Li2019a, Miura2006, Schmitt2006, Miura2007, Miura2008, Li2010, Kuo2012, Roudebush2013a, Zhang2014, Jeevanesan2014, Lefrancois2016, Wong2016, Werner2017, Scheie2019, Schmitt2014, Kurbakov2020, motome2020, Nalbandyan2013, Zvereva2012, Bera2017, Kurbakov2017, Werner2019, Ramlau2014, Stratan2019, Bieringer2000, kumar2022, vasilchikova2022magnetic, bera2022magnetism, fu2023signatures, vavilova2023, xiang2023, guang2023, fu2023a, liu2023, yao2023, yadav2023, khanom2023} 
electronic and electrochemical\cite{Masese2018, Kumar2012, Evstigneeva2011, Nalbandyan2013a, Sau2015, Sau2016, Li2018Chemistry, Li2018ACSApplMater, Wu2018, Deng2019, Bianchini2019, Wu2020, Dubey2020, Sathiya2013, Grundish2019, Yang2017, Yuan2014, Masese2018, Bhange2017, Masese2019, Yoshii2019, Wang2019b, Zheng2016, Ma2015, Kee2016, Aguesse2016, Kee2020, Wang2018, Bao2014, Kim2017, Huang2020, Wang2019a, Paharik2017, Dai2017, Han2016, Zhao2019, McCalla2015a, feng2022suppression, Xiao2020a, teshima2022, mukherjee2023, luke2023, wang2023} properties of pnictogen- and chalcogen-based honeycomb layered oxides sufficiently attest to the need for continued discovery, optimisation and characterisation of new honeycomb layered oxide materials. Besides their unique crystalline structure, their inherent structural symmetries which facilitate 2D atomistic interactions to dominate the honeycomb layered heterostructures serve as an impetus to the exploration of unconventional magnetic phenomena such as Heisenberg-Kitaev interactions\cite{Kitaev2006, kanyolo2021honeycomb} as well as new-fangled emergent properties such as quantum geometries and topologies\cite{kanyolo2022cationic, masese2023honeycomb, Kanyolo2020, kanyolo2022advances2, kanyolo2021honeycomb, masesemaths, kanyolo2021reproducing, kanyolo2022local, kanyolo2021partition} most of which squarely cut across the various fields displayed in \textbf{Figure \ref{Figure_5}}, whilst potential applications are detailed in \textbf{Table \ref{Table_14}}.

\begin{table*}
\caption{Non-exhaustive list of potential applications for exemplar pnictogen- and chalcogen-based honeycomb layered oxides. Asterisks (*) before a chemical formula indicates that the material slightly strays from the strict definition of pnictogen- and chalcogen-based honeycomb layered oxides provided in Section \ref{Section: Introduction}, whilst dagger ($^\dagger$) indicates the applications suggested in the review.}\label{Table_14}
\begin{center}
\scalebox{0.7}{
{\color{black}\begin{tabular}{ll} 
\hline
\textbf{Potential application} & \textbf{Honeycomb layered oxide composition(s)}\\ 
\hline\hline
& \\
\textbf{Photocatalysts} & $\rm Na_3Cu_{2/3}Ni_{2/3}Co_{2/3}SbO_6$,\cite{karati2021band} $\rm Na_3Cu_{1/2}Ni_{1/2}Co_{1/2}Fe_{1/2}SbO_6$\cite{karati2021band}, $\rm Na_3Cu_{2/5}Ni_{2/5}Co_{2/5}Fe_{2/5}Mn_{2/5}SbO_6$,\cite{karati2021band}\\
 & $\rm Na_3CuNiSbO_6$\cite{karati2021band}, $\rm Na_2Ni_2TeO_6$\cite{Kadari2016}, $^\dagger\rm Na_3Ni_{1/3}Co_{1/3}Zn_{1/3}Mg_{1/3}Cu_{1/3}Mn_{1/3}SbO_6$\\
 & \\
\textbf{Photoluminescence} & $\rm Li_4{\it M}SbO_6$ ($\rm {\it M} = Mn, Cr $)\cite {Bhardwaj2014}, $\rm Na_3Mn{\it M}SbO_6$ ($\rm {\it M} = Co, Mg $)\cite{yadav2022investigation}\\
 & \\
\textbf{Spintronics:} Kitaev materials & $\rm Na_2Co_2TeO_6$\cite{sanders2021dominant}, $\rm Na_3Co_2SbO_6$\cite{sanders2021dominant, fu2023signatures, vavilova2023}, $^\dagger\rm Ag_3Co_2SbO_6$, $^\dagger\rm Ag_3Co_2BiO_6$, $^\dagger\rm K_2Co_2TeO_6$, $^\dagger\rm BaCo_2TeO_6$,\\
 & $^\dagger\rm Ca_{\it x}Co_2TeO_6$ ($0 < x \leq 1$), $^\dagger\rm Ag_2Co_2TeO_6$\\
 & \\
\textbf{Spintronics:} magneto-resistance/ & ${\rm Ag}_2M\rm O_2$ ($M = \rm Cr, Ni, Co, Mn, Fe$)\cite{taniguchi2018, taniguchi2020}, ${\rm Ag_6}M_2\rm TeO_6$ ($M = \rm Co, Cu, Ni, Zn, Mg$)\cite{masese2023honeycomb}\\
 partial magnetic order & \\
& \\
\textbf{2D quantum gravity analogues} & $A_2M_2\rm TeO_6$ ($A = \rm Na, K$ \textit{, etc}, $M = \rm Ni, Co, Cu$ \textit{, etc}), Ordered ${\rm Li_2}M_2\rm TeO_6$ ($M = \rm Ni, Co, Cu$ \textit{, etc}),\\
(gauge/gravity duality\cite{kanyolo2022advances2}) &  Ordered ${\rm Li_3}M_2\rm SbO_6$, Ordered ${\rm Li_3}M_2\rm BiO_6$, ${\rm Ag_3}M_2\rm SbO_6$ ($M = \rm Co, Cu, Ni$\textit{,etc}), ${\rm Ag_3}M_2\rm BiO_6$, ${\rm Ag_2}M_2\rm TeO_6$, \\
 & $^\dagger{\rm Ca_{\it x}}M_2\rm TeO_6$ ($M = \rm Ni, Co, Cu$; $0 < x \leq 1$)\cite{kanyolo2022cationic, Kanyolo2020, kanyolo2021partition}\\
 & \\
\textbf{Peierls (semi)conductors/insulators\cite{kanyolo2022advances2}}
& *$\rm Ag_2F$\cite{andres1966superconductivity}, *${\rm Ag}_2M\rm O_2$ ($M = \rm Co, Ni, Cr, Mn, Fe$)\cite{yoshida2020static, schreyer2002synthesis, matsuda2012partially, ji2010orbital}, *$\rm Ag_{16}B_4O_{10}$\cite{kovalevskiy2020uncommon}, ${\rm Ag_6}M_2\rm TeO_6$ ($M = \rm Co, Ni, Zn, Mg$)\cite{masese2023honeycomb}\\
(pseudo-spin models\cite{kanyolo2022cationic, tada2022implications, kanyolo2023honeycomb, kanyolo2023pseudo}) & \\
 & \\ 
\textbf{Energy Storage:} cathode materials & $A_xM_yD\rm O_6$ ($A = \rm Li, Na, K, NaK, Ag, Cu, Ba$ \textit{etc} ($1 \leq x \leq 4$), $M = \rm Ni, Cu, Cr, Co, Mg, Fe, Mn$ \textit{etc} ($1 \leq y \leq 2$),\\
& and $D = \rm Te, Bi, Sb$ \textit{etc})\cite{Seibel2013, Liu2016, Bhange2017, Berthelot2012, Grundish2019, Xu2005, Stratan2019, Smirnova2005, Schmidt2013, Yan2019, Yadav2019, Stratan2019, Brown2019, Greaves1990, Zvereva2013, Bhardwaj2014, Masese2018, masese2021topological, song2022influence, Kurbakov2020, Bera2017, masese2021unveiling, masese2021topological, Kimber2010, Berthelot2012, Ramlau2014, Nagarajan2002, zvereva2016d, reya2011crystal, masese2021mixed}\\
 & \\
\textbf{Energy Storage:} solid-state electrolytes & $\rm Na_2Mg_2TeO_6$\cite{Li2018ACSApplMater}, $\rm Na_2Zn_2TeO_6$\cite{Deng2019, Wu2020, teshima2022, teshima2022electrical, itaya2021sintering, Li2018Chemistry, Wu2018}, $\rm K_2Mg_2TeO_6$\cite{Masese2018}, $\rm NaKMg_2TeO_6$\cite{masese2021mixed}, $\rm NaKZn_2TeO_6$\cite{masese2021mixed},\\
 & ${\rm Ag_2}M_2\rm TeO_6$ ($M = \rm Zn, Mg$)\cite{masese2023honeycomb}, $^\dagger{\rm Ca_{\it x}}M_2\rm TeO_6$ ($M = \rm Mg, Zn$; $0 < x \leq 1$)\\
& \\ 
\hline
\end{tabular}}}
\end{center}
\end{table*}

For instance, when a specific honeycomb layered oxide is used as a feasible cathode material for next-generation battery applications\cite{Masese2018, masese2022road, kanyolo2022cationic, kanyolo2022advances2, kanyolo2021partition, kanyolo2021honeycomb, Kanyolo2020}, cationic charge conservation in 2D implies de-intercalation/
intercalation processes to a large extent are governed by the so-called Chern-Simons gauge theory\cite{Zee2010}, predicting that conductance under single-layer high-resolution measurements should exhibit clear-cut discrete-like behaviour proportional to the number of cations/vacancies being extracted/created.\cite{kanyolo2022cationic} Whilst quantitatively different, such cation/vacancy correlations imply that the gauge theory describing the dynamics of cations must be qualitatively the same as/dual to another theory describing the creation of vacancies in the cathode. Since vacancies are geometric/topological features in the lattice, this requires a duality between the gauge theory of cations and the geometric/topological theory of vacancies, a gauge-gravity (geometry) duality, where the gravity theory is the well-known 2D Louville conformal field theory (CFT).\cite{kanyolo2022advances2, zamolodchikov1996conformal} The theoretical descriptions for such phenomena have been found to be useful in information theory of black holes, suggesting the ultimate use of honeycomb layered materials as analogue gravitational systems within the emerging field of quantum materials.\cite{Kanyolo2020, kanyolo2022cationic, kanyolo2022local, cava2021introduction} 

Moreover, since the characteristics of any geometric theory can be dependent on the number of dimensions, this duality may offer an avenue to engineer specific behaviour by altering the dimensions of the prevalent lattices in honeycomb layered materials. Moreover, in 3\textit{d} transition metal lattices within the slabs of honeycomb layered $\rm Na_3Co_2SbO_6$, 2D exotic magnetic behaviour such as Kitaev quantum spin liquid is thought to be governed by a trigonal crystal field along the $c$-axis that couples to pseudo-spin of $\rm Co^{2+}$ honeycomb lattice in the slabs, and is therefore tunable by strain.\cite{liu2020, csiszar2005controlling} Due to the fusion rules associated with the non-abelian statistics of anyons in the exactly soluble model, Kitaev materials are poised to have applications in quantum computing technologies.\cite{kitaev2006anyons, fu2023signatures}  

Another precursor to such engineering can be identified within the context of Ag bilayered materials such as $\rm Ag_2^{1/2+}F^{1-}$ $= \rm Ag^{2+}Ag^{1-}F^{1-}$, $\rm Ag_2^{1/2+}Ni^{3+}O_2^{2-}$ $ = \rm Ag^{2+}Ag^{1-}Ni^{3+}O_2^{2-}$ and more recently $2\rm Ag_6^{1/2+}Ni_2^{2+}Te^{5+}O_6^{2-}$ $= \rm Ag_6^{2+}Ag_6^{1-}Ni_4^{2+}Te^{4+}Te^{6+}O_{12}^{2-}$,\cite{johannes2007formation, yoshida2006spin, masese2023honeycomb} where the Liouville CFT describes the critical point of a monolayer-bilayer phase transition as the bipartite Ag honeycomb lattice is understood to bifurcate into 
a pair of adjacent Ag hexagonal lattices.\cite{masese2023honeycomb, kanyolo2022advances2, kanyolo2022cationic, tada2022implications} The monolayer-bilayer phase transition relies on the affinity of the Ag atom, treated as a fermion, to form anomalous oxidation states of Ag cations (namely $\rm Ag^{2+}$, $\rm Ag^{1-}$ alongside the conventional $\rm Ag^{1+}$), whereby $\rm Ag^{2+}$ and $\rm Ag^{1-}$ oxidation states are degenerate thus rendering the bipartite honeycomb lattice unstable unless this degeneracy is lifted by \textit{e.g.} inducing a mass term in the Lagrangian proportional to the conformal dimension, $\Delta(n) = (n - 2)/2$ responsible for the bifurcation, whereby $n = 2, 3$ is the number of space dimensions (degeneracy in the 2D lattice must be lifted by distortion/bifurcation to achieve enhanced stability -- a theorem analogous to Peierls' in one-dimension (1D).\cite{peierls1955quantum, garcia1992dimerization, peierls1979surprises}

Whilst such anomalous Ag oxidation states have already been observed separately in \textit{e.g.} the anti-ferromagnetic material $\rm Ag^{2+}F_2^{1-}$ and silver clusters ${\rm Ag}^{1-}_N$ ($N = 1$)\cite{kurzydlowski2021fluorides, grzelak2017metal, schneider2005unusual, dixon1996photoelectron, ho1990photoelectron, minamikawa2022electron, minamikawa2022correction}, presently their existence in bilayered materials can only be inferred from the exhibited subvalent states ($\rm Ag_2^{1/2+} = Ag^{2+}Ag^{1-}$, $\rm Ag_3^{2/3+} = Ag^{2+}Ag^{1-}Ag^{1+}$ \textit{etc}.) observed in all known bilayered frameworks. Indeed, since all experimentally reported Ag cations within bilayered structures are so far subvalent, no mass term is expected for $\rm Ag^{1+}$ in the formalism in the 2D honeycomb lattice (for instance, such a mass term would correspond to a Majorana mass term). This observation has inspired the introduction of emergent SU($2$)$\times$U($1$) gauge interactions between the $\rm Ag^{2+}$, $\rm Ag^{1-}$ and $\rm Ag^{1+}$ fermions, analogous to the lepton interactions in the standard model of particle physics\cite{weinberg1967model, masese2023honeycomb}, where $\rm Ag^{2+}$ oxidation state is a right-handed fermion whereas $\rm Ag^{1-}$ and $\rm Ag^{1+}$ are both left-handed with opposite isospin. The broken SU($2$)$\times$U($1$) gauge interactions generate a Dirac mass term between $\rm Ag^{2+}$ and $\rm Ag^{1-}$ (observed as the \textit{effective} cation with the subvalent state, $\rm Ag^{1/2+}$) interpreted as a stronger argentophilic bond neither present between two $\rm Ag^{1+}$ cations nor metallic bonds in elemental $\rm Ag$, thus further stabilising the Ag bilayered structure.\cite{masese2023honeycomb} Incidentally, due to their varied nature of origin, this explains the shorter argentophilic bonds ($\leq 2.83$ \AA) reported in Ag bilayers compared to the metallic bond in elemental silver ($\leq 2.89$ \AA).\cite{masese2023honeycomb}

Finally, this review abundantly affirms that the current knowledge on honeycomb layered oxides is a drop in an ocean full of unparalleled prospects that augur a new age of technology. However, these new heights demand unremitting research endeavours from wide-reaching scientific communities featuring physicists, material chemists, theorists, mathematicians and electrochemists. As to questions posed earlier in this section, the following quote will suffice: `{\it One thing we do not seem to learn from experience, is that \textbf{we do not often learn from experience alone}}'.\cite {zaslavsky2011constructing, watson2006mathematics}

\newpage
\section*{Acknowledgements}
The authors would like to acknowledge the financial support of AIST Edge Runners Funding, TEPCO Memorial Foundation, Japan Society for the Promotion of Science (JSPS KAKENHI Grant Number 23K04922) and Iketani Science and Technology Foundation. The authors also acknowledge fruitful discussions with D. Ntara during the cradle of the ideas herein. Both authors are grateful for the unwavering support from their family members (T. M.: Ishii Family, Sakaguchi Family and Masese Family; G. M. K.: Ngumbi Family. 

\newpage

\section*{Conflicts of interest}
There are no conflicts to declare.

\newpage

\bibliography{rsc} 
\bibliographystyle{rsc} 

\end{document}